\documentclass[runningheads,epsf]{svmult}
\usepackage{amssymb}
\usepackage{epsfig}
\usepackage{cite}
\usepackage{rotating}
\usepackage[nocolon]{makesdx}

\usepackage{graphicx}
\usepackage{makeidx}
\usepackage{multicol}
\usepackage{subeqnar}
\usepackage{cropmark}
\usepackage{physmult}

\makesymbolindex

\def\be{\begin{equation}}
\def\ee{\end{equation}}
\def\ba{\begin{eqnarray}}
\def\ea{\end{eqnarray}}

\def\LSCO{La$_{2-x}$Sr$_x$CuO$_4$}

\def\YBCO{YBa$_2$Cu$_3$O$_{7-\delta}$}
\def\BKBO{BaKBiO}
\def\BSCCO{Bi$_2$Sr$_2$CaCu$_2$O$_{8+\delta}$}
\def\C60{A$_x$C$_{60}$}
\def\LNSCO{La$_{1.6-x}$Nd$_{0.4}$Sr$_x$CuO$_{4}$}

\def\BKBO{BaKBiO}
\def\Y248{YBa$_2$Cu$_4$O$_8$}

\begin{document}

\title*{Concepts in High Temperature Superconductivity}
\author{E.~W.~Carlson, V.~J.~Emery, S.~A.~Kivelson, and D.~Orgad}
\date{\today}
\maketitle

\titlerunning{Concepts in High Temperature Superconductivity}

\section*{Preface}

It is the purpose of this paper to explore the theory of high temperature
superconductivity.  Much of the motivation for this comes from the
study of cuprate high temperature superconductors.  However, we do not
focus in great detail on the  remarkable and exciting
physics that has been discovered in these materials. Rather, we focus on the
core theoretical issues associated with the mechanism of high temperature
superconductivity.
Although our discussions
of theoretical issues in a strongly correlated superconductor
are intended to be self contained and
pedagogically complete, our discussions of experiments in the cuprates
are, unfortunately, considerably more truncated and impressionistic.

Our primary focus is on physics at intermediate temperature scales of
order $T_c$ (as well as the somewhat larger
``pseudogap'' temperature) and energies of order the gap maximum,
$\Delta_0$.   Consequently (and reluctantly) we have omitted any detailed
\sindex{tc}{$T_c$}{Superconducting transition temperature}
\sindex{zzdeltao}{$\Delta_0$}{Superconducting gap maximum}
discussion of a number of fascinating topics in cuprate
superconductivity, including the low energy physics associated with
nodal quasiparticles,
the properties of the vortex matter which results from the application of a
magnetic field,  the effects
of disorder, and a host of material specific issues. 
This paper is long enough as it is!
\newpage

\subsection*{\protect\bigskip {\protect\Large Contents}}

\textbf{1. Introduction {\vspace{-0.71cm}\begin{flushright} 7 \end{flushright}}}
\vspace{-0.1cm}
\noindent{\em We highlight our main themes:  mesoscale structure
and the need for a kinetic energy driven mechanism.}

\vspace{0.4cm}
\noindent\textbf{2. High Temperature Superconductivity is Hard to Attain} 
\textbf{\vspace{-0.71cm}\begin{flushright} 9 \end{flushright}}
\noindent
\vspace{-0.4cm}

\noindent{\em We explore the reasons why high temperature superconductivity 
is so  difficult to achieve from the perspective of the BCS-Eliashberg 
approach. Because of retardation, increasing the frequency of the 
intermediate boson cannot significantly raise $T_{c}$. Strong coupling 
tends to reduce the phase ordering temperature while promoting competing 
instabilities.}
\vspace{0.1cm}

2.1 Effects of the Coulomb repulsion and retardation on pairing
\vspace{-0.72cm}\begin{flushright} 10 \end{flushright}\vspace{-0.23cm}

2.2 Pairing {\it vs.} phase ordering
\vspace{-0.7cm}\begin{flushright} 12 \end{flushright}\vspace{-0.23cm}

2.3 Competing orders
\vspace{-0.7cm}\begin{flushright} 13 \end{flushright}\vspace{-0.23cm}

\vspace{0.4cm}
\noindent\textbf{3 Superconductivity in the Cuprates: General Considerations}
\textbf{\vspace{-0.71cm}\begin{flushright} 15 \end{flushright}}
\noindent
\vspace{-0.4cm}

\noindent{\em Some of the most important experimental facts concerning the
cuprate high temperature superconductors are described with particular
emphasis on those which indicate the need for a new approach to the mechanism
of high temperature superconductivity.  A perspective on the 
pseudogap phenomena and the origin of $d$-wave-like pairing is
presented.}
\vspace{0.1cm}

3.1 A Fermi surface instability requires a Fermi surface
\vspace{-0.71cm}\begin{flushright} 17 \end{flushright}\vspace{-0.23cm}

3.2 There is no room for retardation
\vspace{-0.71cm}\begin{flushright} 17 \end{flushright}\vspace{-0.23cm}

3.3 Pairing is collective!
\vspace{-0.71cm}\begin{flushright} 18 \end{flushright}\vspace{-0.23cm}

3.4 What determines the symmetry of the pair wavefunction?
\vspace{-0.71cm}\begin{flushright} 19 \end{flushright}\vspace{-0.23cm}

3.5 What does the pseudogap mean?
\vspace{-0.71cm}\begin{flushright} 21 \end{flushright}\vspace{-0.23cm}

\hspace{0.6cm}{3.5.1 What experiments define the pseudogap?}
\vspace{-0.71cm}\begin{flushright} 21 \end{flushright}\vspace{-0.23cm}

\hspace{0.6cm}{3.5.2 What does the pseudogap imply for theory?}
\vspace{-0.71cm}\begin{flushright} 27 \end{flushright}\vspace{-0.23cm}

\vspace{0.4cm}
\noindent\textbf{4.  Preview: Our View of the Phase Diagram}
\textbf{\vspace{-0.68cm}\begin{flushright} 30 \end{flushright}}
\vspace{0.1cm}
\noindent{\em We briefly sketch our view of how the interplay between stripe 
and superconducting order leads to high temperature superconductivity, various
pseudogap phenomena, and non-Fermi liquid behaviors that resemble the physics 
of the 1D electron gas. This serves as a trailer for Section 13.}

\vspace{0.4cm}
\noindent\textbf{5. Quasi-1D Superconductors}
\textbf{\vspace{-0.71cm}\begin{flushright} 32 \end{flushright}}
\vspace{0cm}

\noindent{\em The well developed theory of quasi-one dimensional
superconductors is int- \newline roduced as the best theoretical laboratory 
for the study of strongly correlated}
\newpage
\noindent{\em electron fluids. The normal state is a non-Fermi 
liquid, in which the electron is fractionalized. It can exhibit a broad 
pseudogap regime for temperatures above $T_{c}$ but below the high temperature
Tomonaga-Luttinger liquid regime. $T_c$ marks a point of dimensional 
crossover, where familiar electron quasiparticles appear with the 
onset of long range superconducting phase coherence.}
\vspace{0.2cm}

5.1 Elementary excitations of the 1DEG
\vspace{-0.7cm}\begin{flushright} 33 \end{flushright}\vspace{-0.2cm}

5.2 Spectral functions of the 1DEG---signatures of fractionalization
\vspace{-0.71cm}\begin{flushright} 39 \end{flushright}\vspace{-0.2cm}

5.3 Dimensional crossover in a quasi-1D superconductor
\vspace{-0.71cm}\begin{flushright} 45 \end{flushright}\vspace{-0.2cm}

\hspace{0.6cm}{5.3.1 Interchain coupling and the onset of order}
\vspace{-0.71cm}\begin{flushright} 45 \end{flushright}\vspace{-0.2cm}

\hspace{0.6cm}{5.3.2 Emergence of the quasiparticle in the ordered state}
\vspace{-0.71cm}\begin{flushright} 47 \end{flushright}\vspace{-0.2cm}

5.4 Alternative routes to dimensional crossover
\vspace{-0.71cm}\begin{flushright} 50 \end{flushright}\vspace{-0.2cm}

\vspace{0.4cm}
\noindent\textbf{6. Quasi-1D Physics in a Dynamical Stripe Array}
\textbf{\vspace{-0.7cm}\begin{flushright} 50 \end{flushright}}
\vspace{-0.1cm}

\noindent{\em An interesting generalization of the quasi-1D system occurs
when the background geometry on which the constituent 1DEG's reside is
itself dynamically fluctuating.  This situation arises in conducting
stripe phases.}
\vspace{0.15cm}

6.1 Ordering in the presence of quasi-static stripe fluctuations
\vspace{-0.71cm}\begin{flushright} 51 \end{flushright}\vspace{-0.2cm}

6.2 The general smectic fixed point
\vspace{-0.71cm}\begin{flushright} 53 \end{flushright}\vspace{-0.2cm}

\vspace{0.4cm}
\noindent\textbf{7. Electron Fractionalization in $D>1$ as a Mechanism of High
Temperature Superconductivity}
\textbf{\vspace{-0.7cm}\begin{flushright} 55 \end{flushright}}
\vspace{-0.1cm}

\noindent{\em Spin-charge separation offers an attractive route to 
high temperature superconductivity. It occurs robustly in 1D, but is now 
known to occur in higher dimensions as well, although seemingly only under 
very special circumstances.}
\vspace{0.15cm}

7.1 RVB and spin-charge separation in two dimensions
\vspace{-0.71cm}\begin{flushright} 56 \end{flushright}\vspace{-0.2cm}

7.2 Is an insulating spin liquid ground state possible in $D>1$?
\vspace{-0.71cm}\begin{flushright} 57 \end{flushright}\vspace{-0.2cm}

7.3 Topological order and electron fractionalization
\vspace{-0.71cm}\begin{flushright} 59 \end{flushright}\vspace{-0.2cm}

\vspace{0.4cm}
\noindent\textbf{8. Superconductors with Small Superfluid Density}
\textbf{\vspace{-0.7cm}\begin{flushright} 59 \end{flushright}}
\vspace{-0.1cm}

\noindent{\em In contrast to conventional superconductors, 
in superconductors with small
superfluid density, fluctuations of the phase of the superconducting
order parameter affect the properties of the system in profound ways.}
\vspace{0.15cm} 

8.1 What ground state properties predict $T_c$?
\vspace{-0.71cm}\begin{flushright} 59 \end{flushright}\vspace{-0.2cm}

8.2 An illustrative example: granular superconductors
\vspace{-0.71cm}\begin{flushright} 62 \end{flushright}\vspace{-0.2cm}

8.3 Classical phase fluctuations
\vspace{-0.71cm}\begin{flushright} 66 \end{flushright}\vspace{-0.2cm}

\hspace{0.6cm}{8.3.1 Superconductors and classical XY models}
\vspace{-0.71cm}\begin{flushright} 66 \end{flushright}\vspace{-0.2cm}

\hspace{0.6cm}{8.3.2 Properties of classical XY models}
\vspace{-0.71cm}\begin{flushright} 67 \end{flushright}\vspace{-0.2cm}

8.4 Quantum considerations
\vspace{-0.71cm}\begin{flushright} 70 \end{flushright}\vspace{-0.2cm}
\newpage
8.5 Applicability to the cuprates
\vspace{-0.71cm}\begin{flushright} 71 \end{flushright}\vspace{-0.2cm}

\hspace{0.6cm}{8.5.1 $T_c$ is unrelated to the gap in underdoped cuprates}
\vspace{-0.71cm}\begin{flushright} 72 \end{flushright}\vspace{-0.2cm}

\hspace{0.6cm}{8.5.2 $T_c$ is set by the superfluid density in underdoped cuprates}
\vspace{-0.71cm}\begin{flushright} 72 \end{flushright}\vspace{-0.2cm}

\hspace{0.6cm}{8.5.3 Experimental signatures of phase fluctuations}
\vspace{-0.71cm}\begin{flushright} 72 \end{flushright}\vspace{-0.2cm}

\vspace{0.4cm}
\noindent\textbf{9. Lessons from Weak Coupling}
\textbf{\vspace{-0.7cm}\begin{flushright} 73 \end{flushright}}
\vspace{-0.1cm}

\noindent{\em The weak coupling renormalization group approach to the
Fermi liquid and the 1DEG is presented. The role of retardation, the
physics of the Coulomb pseudopotential, and the nonrenormalization of
the electron-phonon coupling in a BCS superconductor are systematically 
derived. The strong renormalization of the electron-phonon
interaction in the 1DEG is contrasted with this---it is suggested that
this may be a more general feature of non-Fermi liquids.}
\vspace{0.15cm}

9.1 Perturbative RG approach in $D > 1$
\vspace{-0.71cm}\begin{flushright} 73 \end{flushright}\vspace{-0.2cm}

9.2 Perturbative RG approach in $D = 1$
\vspace{-0.71cm}\begin{flushright} 77 \end{flushright}\vspace{-0.2cm}

\hspace{0.6cm}{9.2.1 The one loop beta function}
\vspace{-0.71cm}\begin{flushright} 77 \end{flushright}\vspace{-0.2cm}

\hspace{0.6cm}{9.2.2 Away from half filling}
\vspace{-0.71cm}\begin{flushright} 78 \end{flushright}\vspace{-0.2cm}

\hspace{0.6cm}{9.2.3 Half filling}
\vspace{-0.71cm}\begin{flushright} 80 \end{flushright}\vspace{-0.2cm}

\vspace{0.4cm}
\noindent\textbf{10. Lessons from Strong Coupling}
\textbf{\vspace{-0.7cm}\begin{flushright} 80 \end{flushright}}
\vspace{-0.1cm}

\noindent{\em In certain special cases, well controlled analytic results
can be obtained in the limit in which the bare electron-electron and/or
electron-phonon interactions are strong. We discuss several such
cases, and in particular we demonstrate a theoretically well established 
mechanism in one dimension, the ``spin gap proximity effect,'' by which 
strong repulsive interactions between electrons can result in a large and 
robust spin gap and strongly enhanced local superconducting correlations. 
We propose this as the paradigmatic mechanism of high temperature 
superconductivity.}
\vspace{0.15cm}

10.1 The Holstein model of interacting electrons and phonons
\vspace{-0.71cm}\begin{flushright} 80 \end{flushright}\vspace{-0.2cm}

\hspace{0.6cm}{10.1.1 Adiabatic limit: $E_F \gg \omega_D$}
\vspace{-0.71cm}\begin{flushright} 81 \end{flushright}\vspace{-0.2cm}

\hspace{0.6cm}{10.1.2 Inverse adiabatic limit; negative $U$ Hubbard model}
\vspace{-0.71cm}\begin{flushright} 81 \end{flushright}\vspace{-0.2cm}

\hspace{0.6cm}{10.1.3 Large $U_{eff}$:  bipolarons}
\vspace{-0.71cm}\begin{flushright} 82 \end{flushright}\vspace{-0.2cm}

10.2 Insulating quantum antiferromagnets
\vspace{-0.71cm}\begin{flushright} 83 \end{flushright}\vspace{-0.2cm}

\hspace{0.6cm}{10.2.1 Quantum antiferromagnets in more than one dimension}
\vspace{-0.71cm}\begin{flushright} 83 \end{flushright}\vspace{-0.2cm}

\hspace{0.6cm}{10.2.2 Spin gap in even leg Heisenberg ladders}
\vspace{-0.71cm}\begin{flushright} 85 \end{flushright}\vspace{-0.2cm}

10.3 The isolated square
\vspace{-0.71cm}\begin{flushright} 87 \end{flushright}\vspace{-0.2cm}

10.4 The spin gap proximity effect mechanism
\vspace{-0.71cm}\begin{flushright} 90 \end{flushright}\vspace{-0.2cm}

\vspace{0.4cm}
\noindent\textbf{11. Lessons from Numerical Studies of Hubbard and Related {Models}}
\textbf{\vspace{-0.7cm}\begin{flushright} 92 \end{flushright}}
\vspace{-0.1cm}

\newpage

\noindent{\em The careful use of numerical studies to understand the
physics on scales relevant to the mechanism of high temperature
superconductivity is advocated.}
\vspace{0.15cm}

11.1 Properties of doped ladders
\vspace{-0.7cm}\begin{flushright} 94 \end{flushright}\vspace{-0.2cm}

\hspace{0.75cm}{11.1.1 Spin gap and pairing correlations}
\vspace{-0.7cm}\begin{flushright} 94 \end{flushright}\vspace{-0.2cm}

\hspace{0.75cm}{11.1.2 Phase separation and stripe formation in ladders}
\vspace{-0.7cm}\begin{flushright} 103 \end{flushright}\vspace{-0.2cm}

11.2 Properties of the two dimensional $t-J$ model
\vspace{-0.7cm}\begin{flushright} 106 \end{flushright}\vspace{-0.2cm}

\hspace{0.75cm}{11.2.1 Phase separation and stripe formation}
\vspace{-0.7cm}\begin{flushright} 106 \end{flushright}\vspace{-0.2cm}

\hspace{0.75cm}{11.2.2 Superconductivity and stripes}
\vspace{-0.7cm}\begin{flushright} 111 \end{flushright}\vspace{-0.2cm}

\vspace{0.4cm}
\noindent\textbf{12. Doped Antiferromagnets}
\textbf{\vspace{-0.7cm}\begin{flushright} 113 \end{flushright}}
\vspace{-0.1cm}

\noindent{\em There are many indications that the cuprate superconductors
should be viewed as doped antiferromagnetic insulators. The motion of dilute
holes in an aniterromagnet is highly frustrated, and attempts to understand
the implications of this problem correspondingly frustrating. However, one
generic solution is macroscopic or microscopic phase separation into 
hole poor antiferromagnetic regions and hole rich metallic regions.}
\vspace{0.15cm}

12.1 Frustration of the motion of dilute holes in an antiferromagnet
\vspace{-0.7cm}\begin{flushright} 114 \end{flushright}\vspace{-0.2cm}

\hspace{0.75cm}{12.1.1 One hole in an antiferromagnet}
\vspace{-0.7cm}\begin{flushright} 116 \end{flushright}\vspace{-0.2cm}

\hspace{0.75cm}{12.1.2 Two holes in an antiferromagnet}
\vspace{-0.7cm}\begin{flushright} 116 \end{flushright}\vspace{-0.2cm}

\hspace{0.75cm}{12.1.1 Many holes: phase separation}
\vspace{-0.7cm}\begin{flushright} 118 \end{flushright}\vspace{-0.2cm}

12.2 Coulomb frustrated phase separation and stripes
\vspace{-0.7cm}\begin{flushright} 121 \end{flushright}\vspace{-0.2cm}

12.3 Avoided critical phenomena
\vspace{-0.7cm}\begin{flushright} 123 \end{flushright}\vspace{-0.2cm}

12.4 The cuprates as doped antiferromagnets
\vspace{-0.7cm}\begin{flushright} 125 \end{flushright}\vspace{-0.2cm}

\hspace{0.75cm}{12.4.1 General considerations}
\vspace{-0.7cm}\begin{flushright} 125 \end{flushright}\vspace{-0.2cm}

\hspace{0.75cm}{12.4.2 Stripes}
\vspace{-0.7cm}\begin{flushright} 126 \end{flushright}\vspace{-0.2cm}

12.5 Additional considerations and alternative perspectives
\vspace{-0.7cm}\begin{flushright} 127 \end{flushright}\vspace{-0.2cm}

\hspace{0.75cm}{12.5.1 Phonons}
\vspace{-0.7cm}\begin{flushright} 127 \end{flushright}\vspace{-0.2cm}

\hspace{0.75cm}{12.5.2 Spin-Peierls order}
\vspace{-0.7cm}\begin{flushright} 127 \end{flushright}\vspace{-0.2cm}

\hspace{0.75cm}{12.5.3 Stripes in other systems}
\vspace{-0.7cm}\begin{flushright} 128 \end{flushright}\vspace{-0.2cm}

\vspace{0.4cm}
\noindent\textbf{13. Stripes and High Temperature Superconductivity}
\textbf{\vspace{-0.7cm}\begin{flushright} 128 \end{flushright}}
\vspace{-0.1cm}

\noindent{\em We present a coherent view---our view---of high
temperature superconductivity in the cuprate superconductors.
This section is more broadly phenomenological than is the rest of this paper.}
\vspace{0.15cm}

13.1 Experimental signatures of stripes
\vspace{-0.7cm}\begin{flushright} 130 \end{flushright}\vspace{-0.2cm}

\hspace{0.75cm}{13.1.1 Where do stripes occur in the phase diagram?}
\vspace{-0.7cm}\begin{flushright} 130 \end{flushright}\vspace{-0.2cm}

13.2 Stripe crystals, fluids, and electronic liquid crystals
\vspace{-0.7cm}\begin{flushright} 134 \end{flushright}\vspace{-0.2cm}

\newpage

13.3 Our view of the phase diagram---Reprise
\vspace{-0.7cm}\begin{flushright} 137 \end{flushright}\vspace{-0.2cm}

\hspace{0.75cm}{13.3.1 Pseudogap scales}
\vspace{-0.7cm}\begin{flushright} 138 \end{flushright}\vspace{-0.2cm}

\hspace{0.75cm}{13.3.1 Dimensional crossovers}
\vspace{-0.7cm}\begin{flushright} 138 \end{flushright}\vspace{-0.2cm}

\hspace{0.75cm}{13.3.2 The cuprates as quasi-1D superconductors}
\vspace{-0.7cm}\begin{flushright} 139 \end{flushright}\vspace{-0.2cm}

\hspace{0.75cm}{13.3.3 Inherent competition}
\vspace{-0.7cm}\begin{flushright} 140 \end{flushright}\vspace{-0.2cm}

13.4 Some open questions
\vspace{-0.66cm}\begin{flushright} 141 \end{flushright}\vspace{-0.2cm}

\hspace{0.75cm}{13.4.1 Are stripes universal in the cuprate superconductors?}
\vspace{-0.7cm}\begin{flushright} 141 \end{flushright}\vspace{-0.2cm}

\hspace{0.75cm}{13.4.2 Are stripes an unimportant low temperature complication?}\\
\vspace{-0.7cm}\begin{flushright} 142 \end{flushright}\vspace{-0.2cm}

\hspace{0.75cm}{13.4.3 Are the length and time scales reasonable?}
\vspace{-0.7cm}\begin{flushright} 143 \end{flushright}\vspace{-0.2cm}

\hspace{0.75cm}{13.4.4 Are stripes conducting or insulating?}
\vspace{-0.7cm}\begin{flushright} 143 \end{flushright}\vspace{-0.2cm}

\hspace{0.75cm}{13.4.5 Are stripes good or bad for superconductivity?}
\vspace{-0.7cm}\begin{flushright} 144 \end{flushright}\vspace{-0.2cm}

\hspace{0.75cm}{13.4.6 Do stripes produce pairing?}
\vspace{-0.7cm}\begin{flushright} 145 \end{flushright}\vspace{-0.2cm}

\hspace{0.75cm}{13.4.7 Do stripes really make the electronic structure quasi-1D?}
\vspace{-0.7cm}\begin{flushright} 146 \end{flushright}\vspace{-0.2cm}

\hspace{0.75cm}{13.4.8 What about overdoping?}
\vspace{-0.7cm}\begin{flushright} 147 \end{flushright}\vspace{-0.2cm}

\hspace{0.75cm}{13.4.9 How large is the regime of substantial fluctuation superconductivity?}
\vspace{-0.7cm}\begin{flushright} 148 \end{flushright}\vspace{-0.2cm}

\hspace{0.75cm}{13.4.10 What about phonons?}
\vspace{-0.7cm}\begin{flushright} 149 \end{flushright}\vspace{-0.2cm}

\hspace{0.75cm}{13.4.11 What are the effects of quenched disorder?}
\vspace{-0.7cm}\begin{flushright} 149 \end{flushright}\vspace{-0.2cm}

\vspace{0.4cm}
\noindent\textbf{List of Symbols}
\textbf{\vspace{-0.7cm}\begin{flushright} 151 \end{flushright}}
\vspace{-0.1cm}

\newpage

\section{Introduction}
\label{introduction}

Conventional superconductors are good metals in their normal states,
\marginpar{\em
The virtues of BCS theory are extolled.}
and are well described by Fermi liquid theory.  
They also exhibit a hierarchy of energy scales, 
$E_F \gg \hbar \omega_D \gg k_B T_c$,  
where $E_F$ and $\hbar \omega_D$ are the
Fermi   and Debye energies, respectively, 
and $T_c$ is the superconducting transition temperature.  
\sindex{ef}{$E_F$}{Fermi energy}
\sindex{zzomega}{$\omega_D$}{Debye frequency}
\sindex{kb}{$k_B$}{Boltzmann's constant}
Moreover, 
one typically does not have to think about the interplay between 
superconductivity and any other sort of collective ordering,
since in most cases 
the only weak coupling instability of a Fermi liquid
is to superconductivity.
These reasons underlie the success of the BCS-Eliashberg-Migdal theory
in describing metallic superconductors\cite{schriefferbook}.  

By contrast, the cuprate high temperature superconductors \cite{mueller}
(and various
\marginpar{\em The assumptions of BCS theory are violated by the 
high temperature superconductors.}
other newly discovered materials with high superconducting transition
temperatures) are highly correlated ``bad
metals,''\cite{badmetals,andersonconstraints}  with normal state
properties that are not at all those of a Fermi liquid.  
There is compelling evidence that they are better thought of as
doped Mott insulators, rather than as strongly interacting versions of
conventional metals\cite{pwa87,pnas,palreview}.  
The cuprates also exhibit numerous types of 
low temperature order
which interact strongly with the superconductivity,  
the most prominent being antiferromagnetism 
and the unidirectional charge and spin density wave ``stripe'' order.
These orders can compete or coexist with superconductivity.  
Furthermore, whereas phase fluctuations of the superconducting order parameter
are negligibly small in conventional superconductors, 
fluctuation effects are of order one in the high temperature 
superconductors because of their  much smaller superfluid stiffness.  

Apparently, none of this complicates the fundamental character of the
superconducting order parameter:  it is still a charge 2e scalar field, 
although it transforms according to a nontrivial 
representation of the point group symmetry of the
crystal---it is a ``$d$-wave superconductor.''
At asymptotically low temperatures and energies, there is every
reason to expect
that the physics is dominated by nodal
quasiparticles that are similar to those that 
one might find in a BCS superconductor of the same
symmetry.
Indeed, there is considerable direct experimental evidence that this
expectation is
realized\cite{taillefernodal,bonnnodal,kamnodal,junodnodal}.    However,
the failure of Fermi liquid theory to describe the normal state and the
presence of competing orders necessitates an entirely different approach
to understanding  much of the physics, especially at intermediate scales
of order $k_BT_c$,  
which is the relevant scale for 
the mechanism of high temperature superconductivity.  

It is the purpose of this paper to address the physics of
\marginpar{\em The purpose of this paper.}
high  temperature superconductivity
at these intermediate scales.  
We pay particular attention to the problem of charge dynamics in doped
Mott insulators.  We also stress 
the physics of quasi-one dimensional superconductors, in part
because that is 
the one {\em theoretically} well understood limit in which
superconductivity emerges from a non-Fermi liquid
normal state.  To the extent that the 
physics evolves adiabatically from the quasi-one 
to the quasi-two dimensional limit, 
this case provides considerable insight into the
actual problem of interest.  
The soundness of this approach can be argued from the
observation that \YBCO \,(YBCO) (which is strongly orthorhombic)
exhibits very similar physics to that of the more tetragonal
cuprates.
Since the conductivity and the superfluid density in YBCO 
exhibit a factor of
$2$ or greater anisotropy within the plane, \cite{ybcoanis,anis248} 
this material is already part way toward the
quasi-one dimensional limit without substantial changes in the physics! 
In the second place, because of the delicate interplay
between stripe and superconducting orders observed in the
cuprates, it is reasonable to speculate that 
the electronic structure may be literally quasi-one
dimensional at the local level, even when little of this
anisotropy is apparent at the macroscopic scale.  

A prominent theme of this article is the role of
{\em mesoscale structure} \cite{C60}.
\marginpar{\em Mesoscale electronic structure is emphasized.}
Because the kinetic energy is strongly dominant
in good metals, their wavefunctions are very rigid and  
hence the electron density is highly homogeneous in real space, 
even in the presence of a spatially varying external
potential ({\it e.g.} disorder).  
In a highly correlated system, the electronic structure  
is much more prone to inhomogeneity\cite{frustrated,castro1,dagotto2},
and intermediate scale structures (stripes are an example)
are likely an integral piece of the physics. Indeed, based on 
the systematics of local superconducting correlations  
in exact solutions of various limiting models and
in numerical ``experiments'' on $t-J$ and Hubbard models, we have come to the
conclusion that mesoscale structure may be {\em essential} 
to a mechanism of high temperature superconducting pairing.  
(See Sections \ref{strong} and \ref{numerical}.) This is a
potentially important guiding principle in the search for new high
temperature superconductors.

This is related to a 
\marginpar{\em A kinetic energy driven mechanism is called for.}
concept that we believe is central to 
the mechanism of high temperature superconductivity:  
the condensation is driven by a lowering of kinetic energy.
A Fermi liquid normal state is essentially 
the ground state of the electron kinetic energy, so
any superconducting state which emerges from it 
must have {\em higher} kinetic energy.  
The energy gain which powers the
superconducting transition from a Fermi liquid  must therefore be energy
of  interaction---this
underlies any BCS-like approach to the problem. 
In the opposite limit of strong repulsive interactions between electrons, 
the normal state has high kinetic energy.
It is thus possible to conceive of a 
{\em kinetic energy driven mechanism of superconductivity}, 
in which the strong frustration of the kinetic energy 
is partially relieved upon entering the superconducting
state\cite{andersonadvances,hirsch,spin gap,interlayer,Zhang98,guinea,hirschperspective}. 
Such a mechanism does not require subtle induced attractions, but derives
directly from the strong repulsion between electrons.   
As will be discussed in Section~\ref{strong}, 
the proximity effect in the conventional
theory of superconductivity is a prototypical example of such a kinetic energy driven
mechanism: when a superconductor and a normal metal are placed in contact
with each other, the electrons in the metal pair (even if the
interactions between them are repulsive) in order to lower their 
zero point kinetic energy by delocalizing across the interface.
A related phenomenon, which we have called the
``spin gap proximity effect''\cite{spin gap,spingap2} (see Section~\ref{sgpfx}),
produces strong superconducting correlations in 
$t-J$ and Hubbard ladders \cite{scalapinofisk}, where 
the reduction of kinetic energy transverse to the ladder direction 
drives pairing.  
It is unclear to us whether experiments can unambiguously distinguish 
between a potential energy and a kinetic energy driven 
mechanism.\footnote{Recent papers by Molegraaf {\it et al}\cite{marel} and Santadner-Syro
{\it et al}\cite{nicolkinetic} present very plausible experimental evidence of a kintetic
energy driven mechanism of superconductivity in at least certain high temperature 
superconductors.}   But since 
the interaction between electrons is strongly 
repulsive for the systems in question, 
we feel that the {\em a priori} case for a kinetic
energy driven mechanism is very strong.

Our approach in this article is first to analyze various aspects 
\marginpar{\em The plan of the article is discussed.}
of high temperature
superconductivity as abstract problems in theoretical physics, and 
then to discuss 
their specific application to 
the cuprate high temperature superconductors.\footnote{While
examples of similar behavior can be found in other materials, for
ease of exposition we have focused on this single example.}  We have
also attempted to make each section self contained.
Although many readers no doubt will be drawn to read
this compelling article in its entirety,
we have also tried to make it useful for those readers who are interested in
learning about one or another more specific issue.  
The first eleven sections focus on theoretical issues, 
except for Section~\ref{ourview},
where we briefly sketch the mechanism in light of {\em our} view
of the phase diagram of the cuprate superconductors.
In the final section, we focus more directly on the physics of high temperature
superconductivity in the cuprates, and summarize some of the experimental issues that remain, in our opinion,
unsettled.  Except where dimensional arguments are important, we will henceforth work
with units in which $\hbar=k_B=1$.
\marginpar{\em $\hbar=1$}
\marginpar{\em $k_{B}=1$.}

\section{High Temperature Superconductivity is Hard to Attain}

\addtocontents{toc}{
\noindent{\em We explore the reasons
why high temperature superconductivity is so  difficult to achieve from the
perspective of the  BCS-Eliashberg approach.
Because of 
retardation, 
increasing the frequency of the intermediate boson
cannot significantly raise $T_{c}$.  Strong
coupling tends to reduce 
the phase ordering
temperature, and promote competing instabilities.}
}
\label{hard}

Superconductivity in metals is the result of two distinct quantum
\marginpar{\em Catch 22}
phenomena: pairing and long range phase coherence. In conventional 
homogeneous superconductors, the phase stiffness is so great that
these two phenomena occur simultaneously. On the other hand, in granular 
superconductors and Josephson junction arrays, pairing occurs at the bulk 
transition temperature of the constituent metal, while long range phase 
coherence, if it occurs at all, 
obtains at a much lower temperature characteristic of 
the Josephson coupling between superconducting grains. 
To achieve high temperature superconductivity
requires that both scales be elevated simultaneously.  However, 
given that the bare interactions between electrons are strongly repulsive,
it is somewhat miraculous that electron pairing occurs at all.  
Strong interactions, which might enable pairing at high 
scales, typically also have the effect of strongly suppressing the 
phase stiffness, and moreover typically induce other 
orders\footnote{{\em I.e.} magnetic, structural, {\em etc.}} 
in the system which compete with 
superconductivity.

It is important in any discussion of the theory of high temperature 
superconductivity
\marginpar{\em BCS is not for high $T_c$ superconductivity.}
to have clearly in mind why conventional metallic
superconductors, which are so completely understood in the context of
the Fermi liquid based BCS-Eliashberg theory, rarely 
have $T_{c}$'s above 15K, and never above 30K.  
In this section, we briefly discuss the principal reasons why 
a straightforward extension of the BCS-Eliashberg theory {\em does
not} provide a framework for understanding high temperature 
superconductivity, 
whether in the cuprate superconductors, or in C$_{60}$,
or {\em possibly} even {\BKBO} or MgB$_{2}$.

\subsection{ Effects of the Coulomb repulsion and retardation on pairing}
\label{sec:retardation}

In conventional BCS superconductors, the instantaneous 
interactions between electrons are typically repulsive (or at best 
very weakly attractive)---it is only 
because the phonon induced attraction is retarded that it (barely) dominates 
at low frequencies.
Even if new types of intermediate bosons are invoked to replace 
phonons in a straightforward variant of the BCS mechanism, the 
instantaneous interactions will still be repulsive, so any induced
attraction is typically weak, and only operative at low frequencies.

Strangely enough, the deleterious effects of the Coulomb
\marginpar{\em
Never forget the Coulomb interaction.}
interaction on high temperature superconductivity has been largely
ignored in the theoretical literature. The 
suggestion has been made that high pairing scales can be achieved
by replacing the relatively low frequency phonons which mediate the
pairing in conventional metals by higher frequency bosonic modes,
such as the spin waves in the high temperature
superconductors\cite{pinessfe,bulutscal,scalspin,levinsfe} or the shape
modes\cite{zaanenc60,shc60} of C$_{60}$ molecules.  
However, in most theoretical
treatments of this idea, the Coulomb pseudopotential is either neglected
or treated in a cavalier manner.
That is, models are considered in  which the
instantaneous interactions between electrons are strongly  attractive.  This
is almost
certainly\cite{andersonmorel,scalapinomu,khlebnikov,spin gap,C60} an
unphysical assumption!

In Section \ref{weak}, we use modern renormalization group (RG) 
methods\cite{shankar,polchinski} to derive the conventional expression for
the Coulomb  pseudopotential, and how it enters the effective pairing
interaction at frequencies lower than the Debye frequency, $\omega_{D}$.  
This theory is well controlled so long as $\omega_{D}\ll E_{F}$ and the 
interaction strengths are not too large.  It is worth reflecting 
on a well known, but remarkably 
profound result that emerges from this analysis:
As electronic states are integrated out between the microscopic 
scale $E_{F}$ and the intermediate scale, $\omega_{D}$, the 
electron-phonon interaction is unrenormalized (and so can be well 
estimated from microscopic considerations), but the Coulomb 
repulsion is reduced from a bare value, $\mu$, to a renormalized 
value,
\be
\mu^{*}=\mu/[1+\mu\log(E_{F}/\omega_{D})].
\label{mu*}
\ee 
Here, as is traditional, $\mu$ and $\mu^{*}$ are the dimensionless 
measures of the interaction strength obtained by multiplying the
\sindex{zzmu}{$\mu$}{Bare Coulomb repulsion}
\sindex{zzmua}{$\mu^{*}$}{Renormalized Coulomb repulsion}
interaction strength by the density of states. We define  $\lambda$  
in an analogous manner for the electron-phonon interaction.
\sindex{zzlambda}{$\lambda$}{Bare electron-phonon coupling}
Thus, even if the instantaneous interaction 
is repulsive ({\it i.e.} $\lambda-\mu<0$), the effective
interaction at the scale $\omega_{D}$ 
will nonetheless be attractive ($\lambda-\mu^{*}>0$) 
for $\omega_{D}\ll E_{F}$. Below this scale, 
the standard RG analysis yields the familiar weak coupling estimate of 
the pairing scale $T_{p}$:
\be
T_{p}\sim \omega_{D}\exp[ -1/(\lambda-\mu^{*})].
\ee
\sindex{tp}{$T_p$}{Pairing scale}

\marginpar{\em
Retardation is an essential feature of the BCS mechanism.}
The essential role of retardation is made clear if one considers the
dependence of $T_{p}$ on $\omega_{D}$:  
\be 
\frac{d\log[T_{p}]}{d\log[\omega_{D}]}=1-\left[
\mu^{*}\log\left(\frac{T_{p}}{\omega_{D}}\right)\right]^{2}.
\label{logderiv}
\ee
So long as $\omega_D \ll E_{F}\exp[-(1-\lambda)/\lambda\mu]$, 
we have $\frac{d\log[T_{p}]}{d\log[\omega_{D}]} \approx 1$, and 
$T_{p}$ is a linearly rising function of $\omega_{D}$,
giving rise to the 
conventional isotope effect.\footnote{Recall, for phonons, 
$d\log[\omega_D]/d\log[M]=-1/2$.}  
However, 
when
$\omega_{D}> T_{p}\exp[1/\mu^{*}],$
we have $\frac{d\log[T_{p}]}{d\log[\omega_{D}]} < 0$, and 
$T_p$  becomes a decreasing function of $\omega_{D}$!  
Clearly, unless $\omega_{D}$ is 
exponentially smaller than $E_{F}$, superconducting pairing is 
impossible by the conventional 
mechanism\footnote{In the present discussion we have imagined varying 
$\omega_D$ while keeping fixed the electron-phonon coupling constant,
$\lambda=\frac{C}{M\omega_D^2}=\frac{C}{K}$,
where $C$ is proportional to the (squared) gradient of the electron-ion 
potential and $K$ is the ``spring constant'' between the ions. 
If we consider instead the effect of increasing $\omega_D$ at
fixed $C/M$, it leads 
to a decrease in
$\lambda$ and hence a very rapid suppression  of the pairing scale.}.

This problem is particularly vexing in the cuprate high temperature 
superconductors and similar materials, which have low electron densities, 
and incipient or apparent Mott insulating behavior. This means that
screening of the Coulomb interaction is typically poor, and $\mu$ is 
thus expected to be large. Specifically, from the inverse Fourier 
transform of the $\vec k$ dependent gap function measured\cite{shendessau}
in angle resolved photoemission spectroscopy (ARPES) on {\BSCCO}, 
it is possible to conclude  (at least at the level of the BCS gap equation) 
that the dominant  pairing interactions have a range equal to the
nearest neighbor  copper distance.  
\marginpar{\em Pairing's Bane}
Since this distance is less than the distance 
between doped holes, it is difficult to believe that metallic 
screening is very effective at these distances. From cluster 
calculations and an analysis of various local spectroscopies, 
a crude estimate\cite{spin gap} of the Coulomb repulsion at this 
distance is of order 0.5eV or more.  To obtain pairing 
from a conventional mechanism with relatively little retardation, 
it is necessary that the effective attraction be considerably larger 
than this!

We are therefore led to the conclusion that the only way a BCS mechanism 
can produce a high pairing scale is if the effective attraction, $\lambda$, 
is very large indeed.  This, however, brings other problems with it.

\subsection{  Pairing {\it vs}. phase ordering}

In most cases, it is unphysical to assume the existence of strong 
attractive interactions between electrons.  However, even supposing 
we ignore this, strong attractive interactions bring about other problems
for high temperature superconductivity:  1)  There is a concomitant 
strong reduction of the phase ordering temperature and thus of $T_{c}$.
 2)  There is the possibility of competing orders.  
We discuss the first problem here, and the second in Section~\ref{comporder}.

Strong attractive interactions typically result in a large increase 
in the effective mass, and a 
corresponding reduction of the phase ordering temperature.
Consider, for example, the strong coupling limit of the negative $U$ Hubbard 
\sindex{u}{$U$}{On-site Hubbard interaction}
model \cite{negU} or the Holstein model \cite{holst}, discussed in 
Section \ref{strong}.  
In both cases,
pairs have a large binding energy,  but they typically Bose condense at a 
very low temperature because of the large effective mass  of a tightly 
bound pair---the effective mass is proportional to $|U|$ 
in the Hubbard model and is exponentially large in the Holstein model.
(See Section~\ref{strong}.)

Whereas in conventional superconductors, the bare superfluid stiffness is 
\marginpar{\em Phase ordering is a serious business in the cuprates.}
so great that even a substantial renormalization of the effective mass  would
hardly matter, in the cuprate high temperature superconductors, 
the superfluid stiffness is small,
and a substantial mass  renormalization would be
catastrophic.
The point  can be made most simply by
considering the result of  simple  dimensional analysis. The density of doped
holes per plane in an optimally  doped high temperature superconductor is
approximately $n_{2d}=10^{14}$cm$^{-2}$. Assuming a density of hole pairs 
that is 
\sindex{n2d}{$n_{2d}$}{Density of doped holes per plane}
half this, and taking the rough estimate for the pair effective mass, 
$m^{*}=2m_{e}$, we find a phase ordering scale,
\sindex{ma}{$m^{*}$}{Effective pair mass}
\be
T_{\theta}= \hbar^{2}n_{2d}/2m^{*}\approx 10^{-2}eV\approx 100K \; .
\ee
Since this is in the neighborhood of the actual $T_{c}$,
\sindex{tzztheta}{$T_{\theta}$}{Phase ordering scale}
it clearly implies that any large mass 
renormalization would be incompatible with a high transition 
temperature.  What about conventional superconductors?  A similar 
estimate in a 
$W=10\AA$ thick Pb film gives
$T_{\theta}= \hbar^{2}n_{3d}W/2m^{*}\approx 1eV\approx 10,000K$!
Clearly, phase fluctuations are unimportant in Pb.
This issue is addressed in detail in Section~\ref{phase}.

We have seen how 
 $T_p$ and
\marginpar{\em A general principle is proposed: ``optimal'' $T_c$ occurs as
a crossover.} 
$T_{\theta}$ 
have  opposite dependence on coupling strength.
If this is a general trend, then it
is likely that 
any material in which $T_{c}$ has been optimized 
has effectively been tuned to a crossover point between pairing and 
condensation.  A modification of the material which produces stronger  
effective interactions will increase phase fluctuations and thereby
reduce $T_{c}$,
while weaker interactions will lower 
$T_{c}$ because of pair breaking.  
In Section \ref{phase} 
it will be shown that optimal doping in the cuprate 
superconductors corresponds to precisely this sort of crossover from 
a regime in which $T_{c}$  is determined by phase ordering to a  pairing 
dominated regime.

\subsection{Competing orders}
\label{comporder}

A Fermi liquid is a remarkably robust state of matter.  In the absence of
nesting, it is stable for a range\footnote{As long as the interactions are
not too strong.}  
of repulsive interactions;
the Cooper instability is its
only weak coupling instability.  
The phase diagram of simple metals consists of a high temperature metallic
phase and a low temperature superconducting state.  When the
superconductivity is suppressed by either a magnetic field or appropriate
disorder ({\it e.g.} paramagnetic impurities), the system remains metallic
down to the lowest temperatures.  

The situation becomes considerably more complex for sufficiently strong
interactions between electrons.
In this case, the Fermi liquid
description of the normal or high temperature phase 
breaks down\footnote{Whether
it breaks down for fundamental or practical reasons is unimportant.} and many
possible phases compete.  In addition to metallic and superconducting phases,
one would generally expect various sorts of electronic ``crystalline''
phases, including charge ordered phases ({\it i.e.} a charge 
density wave---CDW---of which the
Wigner crystal is the simplest example) and spin ordered phases 
({\em i.e.} a spin density wave---SDW---of which the N\'{e}el state is
the  simplest example).  

Typically, one thinks of such phases as
insulating, but it is certainly possible for charge and spin order to
coexist with metallic or even superconducting electron transport.  
For example, 
this can occur in a
conventional weak coupling theory 
if the density wave order
opens a gap on only part of the Fermi surface, leaving other parts
gapless\cite{gruner}.  It can also occur in a multicomponent system, in which
the density wave order involves one set of electronic orbitals, and the
conduction occurs through others---this is the traditional understanding of
the coexisting superconducting and magnetic order in the Chevrel
compounds\cite{chevrel}.

Such coexistence is also possible for less conventional orders.
\marginpar{\em ``Stripe'' order}
One particular
class of competing orders is known loosely as ``stripe'' order.
Stripe order refers
to  unidirectional density wave order, {\it i.e.} order which
spontaneously breaks translational  symmetry in one direction but not in
others.  We will refer to charge stripe order, if the broken symmetry
leads to charge density modulations and spin stripe order if the broken
symmetry leads to spin density modulations, as well.  
Charge stripe order
can occur  without spin order, but spin order (in a sense that will be
made precise,  below) implies charge order\cite{zkelandau}. 
Both are known on theoretical and experimental grounds to be 
a prominent feature of
doped Mott insulators in general, and the high temperature
superconductors in 
particular\cite{pnas,tranquadastripes,tranquadareview,zaanennature,sachdevscience,orensteinmillis,baskaran}.  
Each of these
orders can occur in an insulating,  metallic, or superconducting state.

In recent
years there has been considerable theoretical interest in other types of
order  that could be induced by strong interactions.  
{F}rom the perspective of
stripe phases, it is natural to consider various partially melted ``stripe
liquid'' phases, and to classify such phases, in analogy with the
classification of phases of classical liquid crystals, according to their
broken symmetries\cite{kfe}.  For instance, one can imagine a phase that
breaks rotational symmetry (or, in a crystal, the point group symmetry) but
not translational symmetry, {\it i.e. } quantum (ground state) analogues of
nematic or hexatic liquid crystalline phases.  Still more exotic phases, 
such as those with
ground state orbital
currents\cite{varma,cnl,marstonafflecksf,kotliarsf,leeloops,dhleeloops}
or topological order\cite{senthil},
have
also been suggested as the explanation for various observed features of the
phenomenology of the high temperature superconductors.  

Given the complex character of the phase diagram of highly correlated
\marginpar{\em Competition matters...}
electrons, it is clear that the conventional approach to superconductivity,
which focuses solely on the properties of the normal metal and the pure
superconducting phase, is suspect.  A more global approach, which takes into
account some (or all) of the competing phases is called for.  Moreover, even
the term ``competing'' carries with it a prejudice that must not be
accepted without thought.  In a weakly correlated system, in which any low
temperature ordered state occurs as a Fermi surface instability, different
orders generally {\em do}  compete:  if one order produces a gap on part of
the Fermi surface, there are fewer remaining low energy degrees of freedom to
participate in the formation of another type of order.  For highly correlated
\marginpar{\em ...and so does symbiosis.}
electrons, however, the sign of the interaction between different types of
order is less clear. 
It can happen\cite{aeppliandsk} that under one set of circumstances, a given
order tends to enhance superconductivity and under others, to suppress it.

The issue of competing orders, of course, is not new. 
In a Fermi
liquid, strong effective attractions typically lead to
lattice instabilities, charge or spin density wave order, etc.
Here the problem is that the system either becomes an insulator or, if it 
remains metallic, the residual attraction is typically weak.
For instance, lattice instability has been seen to limit the 
superconducting transition temperature of the A15 compounds, the high 
temperature superconductors of a previous generation.
Indeed, the previous generation of BCS based theories which addressed the 
issue always concluded that competing orders suppress
superconductivity\cite{chevrel}.

More recently it has been argued that  
near an instability 
to an ordered state there is a low lying collective 
mode (the incipient Goldstone mode of the ordered phase)
which can play the role of the phonon in a BCS-like mechanism of 
superconductivity\cite{pinessfe,castro2,castro5}. In an interesting variant of 
this idea, it has been argued that 
in the neighborhood of a zero temperature transition to an ordered phase, 
quantum critical fluctuations
can 
mediate superconducting pairing in a more or less traditional
way\cite{sachdevchub,lanzarichqcbcs,lanzarichqcbcs2}. 
There are reasons to expect this type of fluctuation mediated pair binding
to lead to a depression of $T_c$.
If the collective modes are nearly Goldstone modes (as opposed to 
relaxational ``critical modes''), general considerations governing the
couplings  of such modes in the ordered phase imply that the
superconducting transition temperature  is depressed substantially from
any naive estimate by large vertex  corrections\cite{schrieffer}.
Moreover, in a  regime of large fluctuations to a nearby ordered phase,
one generally expects a density of states reduction due to the
development of a  pseudogap;  feeding this psuedogapped  density of
states  back into the BCS-Eliashberg  theory will again result  in a
significant reduction of
$T_{c}$.

\section{ Superconductivity in the 
Cuprates:  General Considerations}
\label{general}

\addtocontents{toc}{
\noindent{\em Some of the most important experimental facts concerning the
cuprate high temperature superconductors are described with particular
emphasis on those which indicate the need for a new approach to the mechanism
of high temperature superconductivity.  A perspective on the 
pseudogap phenomena and the origin of $d$-wave-like pairing is
presented.}
}

While the principal focus of the present article is theoretical, 
the choice of topics and models and the approaches are very much
motivated by our interest in the experimentally observed properties
of the cuprate high temperature superconductors. In this section,
we discuss briefly some of the most dramatic (and least controversial)
aspects of the phenomenology of these materials, and what sorts of
constraints those observations imply for theory.  As we are 
primarily interested in
the origin of high temperature superconductivity, 
we will deal here almost exclusively with experiments in the temperature 
and energy ranges between about $T_{c}/2$ and a few times $T_c$.

Before starting, there are a number of descriptive terms that warrant
definition.  The parent state of each family of the high temperature
superconductors is an antiferromagnetic ``Mott'' insulator with one 
hole (and spin $1/2$) per planar
copper.\footnote{The term ``Mott insulator'' means many things to many
people.   One definition is that a Mott insulator is insulating because
of interactions  between electrons, rather than because a noninteracting
band is filled.   This is not a precise definition. For example, a Mott
insulating state  can arise due to a spontaneously broken symmetry which
increases the  size of the unit cell. However, this is adiabatically
connected to the  weak coupling limit, and can be qualitatively
understood via generalized Hartree-Fock theory. There is still a {\em
quantitative} distinction  between a weak coupling ``simple'' insulator
on the one hand, which has  an insulating gap that is directly related 
to the order parameter which characterizes the broken symmetry, and the
``Mott'' insulator on the other hand, which has an insulating gap which is 
large due to the strong repulsion between electrons.  In the latter case, 
the resistivity begins to grow very large compared to the quantum of 
resistance well above the temperature at which the broken symmetry occurs. 
The undoped cuprate superconductors are clearly Mott insulators in the 
quantitative sense that the insulating gap is of order 2eV, 
while the antiferromagnetic ordering temperatures are around 30 meV.

However, for those who prefer\cite{laughlincritique} a sharp, 
{\em qualitative} distinction, the term ``Mott insulator'' is reserved 
for ``spin liquid'' states which are distinct zero temperature phases of 
matter, do not break symmetries, and cannot be understood in terms of any 
straightforward Hartree-Fock description.  Many such exotic states have 
been theoretically envisaged, including the long~\cite{pwa87,bza} and 
short ranged\cite{krs,rk,rs} RVB liquids, the chiral
spin liquid\cite{kalmeyerlaughlin,wenzee,weigmann}, 
the nodal spin liquid~\cite{nayakfishernodal,nayakfisherdual} and
various other fractionalized states with topological
order\cite{senthilfisher,wentopo}. Very recently, in the first 
``proof of principle,'' a concrete model with a well defined short ranged RVB
{\em phase} has been discovered\cite{shivaji,sondhifradkin}.  }
\begin{figure}[ht!!!]
\begin{center}
\epsfig{figure=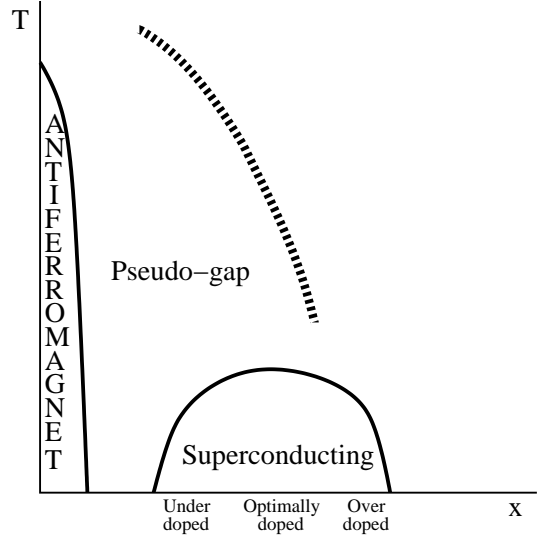,width=0.6\linewidth}
\end{center}
\caption[Schematic phase diagram of high temperature superconductors]
{Schematic phase diagram of a cuprate high temperature superconductor 
as a function of temperature and $x$---the density of doped holes per 
planar Cu. The solid  lines represent phase transitions into the 
antiferromagnetic (AF) and superconducting (SC) states. The dashed line 
marks the openning of a pseudogap (PG). The latter crossover is not 
sharply defined and there is still debate on its position; see 
Refs.~\citen{timuskreview,loramreview}.}
\label{phasediag}
\end{figure}
These insulators are transformed into superconductors by introducing a 
concentration, $x$, of ``doped holes'' into the copper oxide planes.  
As a function of increasing $x$, the antiferromagnetic transition temperature 
is rapidly suppressed to zero, then the superconducting transition 
temperature rises from zero to a maximum and then drops down again. (See Fig.
\ref{phasediag}.)  Where $T_c$ is an increasing function of $x$, the 
materials are said to be ``underdoped.'' They are ``optimally doped'' 
where $T_c$ reaches its maximum at $x\approx 0.15$, and
they are ``overdoped'' for larger $x$.  In the underdoped regime there are a
variety of crossover phenomena observed\cite{timuskreview,loramreview} at 
temperatures above $T_c$ in which various forms of spectral weight at low 
energies are apparently suppressed---these phenomena are associated with 
the opening of a ``psuedogap.'' There are various families of high 
temperature superconductors, all of which have the
same nearly square copper oxide planes, but different structures in the
regions between the planes.  One characteristic that seems to have a fairly
direct connection with $T_c$ is the number of copper-oxide planes that are
close enough to each other that interplane coupling may be significant; $T_c$
seems generally to increase with number of planes
within a homologous series, at least as one
progresses from ``single layer'' to ``bilayer,'' to ``trilayer''
materials\cite{andersonconstraints,leggettlayers}.

\sindex{x}{$x$}{Hole doping}

\subsection{A Fermi surface instability requires a Fermi surface}

As has been stressed, for instance, by Schrieffer\cite{schriefferbook}, 
BCS theory relies heavily on the accuracy with which the normal state is 
described by Fermi liquid theory. BCS superconductivity is a Fermi surface 
instability, which is only a reasonable concept
if there is a well defined Fermi surface. BCS-Eliashberg theory relies on the
dominance of a certain class of diagrams, summed to all orders in 
perturbation theory. This can be justified from phase space considerations
for a Fermi liquid, but need not be valid more generally. To put it most
physically, BCS theory pairs well defined quasiparticles, 
and therefore requires well defined quasiparticles in the normal state.

There is ample evidence that in optimally and underdoped cuprates, at least,
\marginpar{\em
We belabor the need for a non-Fermi liquid based approach.}
there are no well defined quasiparticles in the normal state. This can be
deduced directly from ARPES studies of the single particle spectral
function\cite{andersonnoqp,laughlin,frac,lnsco,valla,fedorov,
bsccoflat,shenprb}, or indirectly from 
an analysis of various spin, current, and density response functions of the
system\cite{badmetals,andersonconstraints}. (Many, though not all, of these 
response functions have been successfully
described\cite{mflph,varmanussinov,varmapnas}  by the ``marginal Fermi liquid''
phenomenology.)  Because we understand the  nature of a Fermi liquid so well, it is
relatively straightforward to  establish that a system is a non-Fermi liquid, at
least in extreme cases. It is much  harder to establish the cause of 
this behavior---it could be due to the proximity of a fundamentally 
new non-Fermi liquid ground state phase of matter,
or it could be because the characteristic coherence temperature, 
below which well defined quasiparticles dominate the physics, is lower 
than the temperatures of interest. Regardless of the reason for the breakdown 
of Fermi liquid theory, a description of the
physics at scales of temperatures comparable to $T_c$ can clearly {\it not} be
based on a quasiparticle description, and thus cannot rely on BCS theory.

\subsection{There is no room for retardation}

As stressed in Section~\ref{sec:retardation}, retardation
plays a pivotal role in the BCS mechanism.  In the typical metallic 
superconductor, the Fermi energy is of order $10$eV, while phonon 
frequencies are of order $10^{-2}$eV,
so $E_F/\omega_D\sim 10^3$!  Since the renormalization of the
Coulomb pseudopotential is logarithmic, this large value of the retardation is
needed.  In the cuprate superconductors, the bandwidth measured in ARPES is
roughly $E_F\approx 0.3$eV---this is a renormalized bandwidth of sorts, 
but this is presumably what determines the quasiparticle dynamics.
Independent of anything else, the induced interaction must clearly be fast 
compared to the gap scale, $\omega_D > 2\Delta_0$, where $\Delta_0$ is the 
magnitude of the superconducting gap. From either ARPES\cite{harrispg,ding} 
or tunnelling\cite{renner}  experiments, we can estimate
$2\Delta_0\approx 0.06$eV. Thus, a rough upper bound
$E_F/\omega_D < E_F/2\Delta_0 \sim 5$ can be established on how retarded an
interaction in the cuprates can possibly be.  That is almost not retarded at
all!

\subsection{Pairing is collective!}

For the most part, the superconducting coherence length, $\xi_0$, cannot be
\sindex{zzxi}{$\xi_0$}{Superconducting coherence length}
directly measured in the high temperature superconductors because, for $T\ll T_c$,
the upper critical field, $H_{c2}$, is too high to access readily.
However, it can be inferred indirectly\cite{ando1d,maggio1995,kap,pan,sonier} 
in various ways, and 
for the most part people have concluded that $\xi_0$ is approximately 2 or 
3 lattice constants in typical optimally doped materials.  
This has lead many people to conclude that these
materials are nearly in a ``real space pairing''
limit\cite{mohitrealspace,levinrealspace,mottbose,mottbipolaron,uemura}, 
in which pairs of holes form actual two particle bound states, 
and then Bose condense at $T_c$. This notion is based on the observation 
that if $x$ is the density of ``doped holes'' per site, then
the number of pairs per coherence area,
$N_p=(1/2)x\pi\xi_0^2/a^2$, is a number which is approximately equal to 1 for
``optimal doping,'' $x\approx 0.15 - 0.20$.  
\sindex{zzpi}{$\pi$}{3.141592653589793238462643\ldots}

However, there are strong {\em a priori} 
and empirical reasons to discard this viewpoint.
\marginpar{\em Real space pairs are dismissed.}  

On theoretical grounds: In a system dominated by strong repulsive interactions
between electrons, it is clear that pairing must be a collective phenomenon. 
The Coulomb interaction between an isolated pair of doped holes would 
seem to be prohibitively large, and it seems unlikely that a strong enough 
effective attraction can emerge to make such a strong binding possible. 
(Some numerical studies of this have been carried out, in the context of 
ladder systems, by Dagotto and collaborators \cite{dagotto3}.)  
Moreover, it is far from clear that the dimensional argument used above 
makes any sense: Why should we only count doped holes in
making this estimate?  What are the rest of the holes
doing all this time?  If we use the density of holes per site 
($1+x$), which is consistent with the area enclosed by the Fermi surface 
seen in ARPES\cite{luttingerarea},
the resulting $N_p$ is an order of magnitude larger than the above 
estimate.\footnote{A theory of real space pairs which includes
all the electrons and the repulsive interactions between them can be
caricatured as a hard core quantum dimer model\cite{rk}. Here the pairing is
collective, due to the high density of pairs. Indeed, $N_p$ involves
all of the electrons (the doped holes are not paired at all), 
but the superfluid density is small, involving only the density of 
doped holes. This contrasts markedly with
the case of clean metallic superconductors where the density of pairs (that is,
the density of electrons whose state is significantly altered by pairing) is
\sindex{nef}{$N(E_F)$}{Density of states at $E_F$}
small, $\sim N(E_F)\Delta_0$, while the superfluid density is large and
involves all the electrons.  There is some evidence that the former situation
in fact pertains to the high temperature superconductors\cite{crossovers}.}

On experimental grounds:  The essential defining feature of real space pairing
is that the chemical potential moves below the bottom of the band. Incipient
real space pairing must thus be associated with significant motion of the
chemical potential toward the band bottom with pairing 
\cite{mohitrealspace,levinrealspace,armen1,armen2}. However, experimentally, 
the chemical potential is found to lie in the middle of the
band, where the enclosed area of the Brilloin zone satisfies Luttinger's
theorem, and no significant motion at $T_c$ (or at any pseudogap temperature
in underdoped materials) has been observed
\cite{allenpinmu1,allenpinmu2,watanabepinmu,inopinmu,marel2}. 
This fact, alone, establishes that the physics is nowhere near 
the real space pairing limit.

\subsection{What determines the symmetry of the pair wavefunction?}

\marginpar{\em Theory has had its triumphs.}
Independent of but contemporary with
the discovery of 
high temperature superconductivity in the cuprates, 
Scalapino, Loh, and Hirsch\cite{hirschdwave}, 
in a prescient work suggested the possibility of
superconductivity in the two dimensional Hubbard model in the 
neighborhood of the antiferromagnetic state at half filling. This 
work, which was in spirit a realization of the 
ideas of Kohn and  Luttinger\cite{kohnluttinger}, concluded that the 
dominant superconducting instability 
should have $d_{(x^{2}-y^{2})}$ symmetry, as opposed to $s$ symmetry.
Immediately after the discovery of high temperature superconductivity, a 
large number of other 
theorists\cite{cuo,kotliarliu,rice,scalapinodwv,pinessfe,castro3,castro4}
came to the same conclusion, based on a 
variety of purely theoretical analyses, although at the
time the experimental evidence of such pairing was ambiguous, at best.
By now it seems very clear that this idea was correct, {\it at least} for a 
majority of the cuprate superconductors,  based on a variety of 
phase sensitive measurements\cite{vanharlingendwv,kirtleydwv,dynesdwv}.  
This represents one of the great triumphs of theory in this field.  
(There are still some experiments which appear to contradict this symmetry 
assignment\cite{klemmswv}, so the subject cannot be said to be 
completely closed, but it seems very unlikely that the basic 
conclusion will be overturned.)

\marginpar{\em
$d$-wave pairing is defined.}
While the names ``$s$'' and ``$d$'' relate to the rotational 
symmetries of free space, it is important to understand 
what is meant by $s$-wave and $d$-wave in a lattice system 
which, in place of continuous rotational symmetry, has the discrete point group 
symmetry of the crystal.  Consequently, the possible pairing 
symmetries correspond to the irreducible representations of the point 
group: singlet orders are even under 
inversion and triplet orders are odd.
In the case of a square crystal
\footnote{The pairing symmetries should really be
classified according to the point group of a tetragonal crystal,
but since the cuprates are quasi-two dimensional, it is conventional,
and probably reasonable, to classify them according to the symmetries
of a square lattice.}, the possible singlet orders (all 
corresponding to one dimensional representations) are 
colloquially called $s$, $d_{(x^{2}-y^{2})}$, $d_{(xy)}$, and 
$g$, and transform
like $1$, $(x^{2}-y^{2})$, 
$(xy)$, and $(x^{2}-y^{2})(xy)$, respectively.  
As a function of angle, the gap 
parameter in an $s$-wave 
order always has a unique sign, the $d$-wave gap changes sign four 
times, and the $g$-wave changes sign 8 times.
A fifth type of order is sometimes 
discussed, called extended-$s$, in which the gap function changes sign 
as a function of the magnitude of $\vec k$, rather than as a function 
of its direction---this is not a true symmetry 
classification, and in any generic model there is always finite mixing 
between $s$ and extended $s$.  

\marginpar{\em ``$d$-wave-like'' pairing is defined.}
In crystals with lower symmetry, there are 
fewer truly distinct irreducible representations. For instance, if 
the square lattice is replaced with a rectangular one, the distinction 
between $s$ and $d_{(x^{2}-y^{2})}$ is lost (they mix), as is that 
between  $d_{(xy)}$ and $g$.  On the other hand, if the elementary 
squares are sheared to form rhombuses, then the $s$ and $d_{(xy)}$ 
symmetries are mixed, as are  $d_{(x^{2}-y^{2})}$, and $g$.  Both of 
these lower symmetries correspond to a form of orthorhombic distortion
observed in the cuprates---the former is the correct symmetry
group for {\YBCO} and the latter for {\LSCO}. However, so long as
the physics does not change fundamentally as the lattice symmetry is
reduced, it is reasonable to classify order parameters as ``$d$-wave-like'' or
``$s$-wave-like.''  We define an order parameter as being $d$-wave-like if it
changes sign under 90$^o$ rotation, although it is only a true $d$-wave if its
magnitude is invariant under this transformation. Conversely, 
it is $s$-wave-like if its sign does not change under this rotation, 
or when reflected through any approximate symmetry plane. 
In almost all cases what is really being seen in phase sensitive 
measurements on the cuprates is that the order parameter is $d$-wave-like. 
(It is worth noting that in $t-J$ and Hubbard ladders,
$d$-wave-like pairing is the dominant form of pairing observed in both 
analytic and numerical studies, as discussed below.)

There is a widespread belief that $d$-wave symmetry follows directly 
\marginpar{\em Strong repulsion does not necessarily lead to d-wave pairing.}
from the presence of strong short range interactions between 
electrons, irrespective of details such as band structure.  The essential 
idea here follows from the observation that the pair wavefunction, 
at the level of BCS mean field theory, is expressed in terms of the gap 
parameter, $\Delta_{\vec k}$, and the quasiparticle spectrum, 
$E_{\vec k}$, as
\be
\phi_{pair}(\vec r)=\sum_{\vec k}\frac{1}{L^{d/2}}  
e^{i\vec k\cdot \vec r} \frac {\Delta_{\vec k}} {2E_{\vec k}} \; .
\label{phipair}
\ee
\sindex{zzdeltak}{$\Delta_{\vec k}$}{BCS gap parameter}
In the presence of strong short range repulsion (and weaker longer range
attraction) between electrons, 
it is favorable for $\phi_{pair}$ to vanish at $\vec r=\vec 0$, which 
\sindex{d}{$d$}{Dimension}
it does automatically if the pairing is not $s$-wave. 
While this argument makes some physical sense, it is 
ultimately {\em wrong}. In the limit of dilute electrons, where the coherence 
length is much smaller than the inter-electron distance,
the pairing problem reduces to a two particle problem. 
It is well known that in the continuum the lowest energy two particle 
spin singlet bound state is nodeless. Given certain mild conditions on 
the band structure one can also prove it on the lattice 
\footnote{This is true under conditions that the hopping matrix, {\it i.e.} 
the band structure, satisfies a Peron-Frobenius condition.}. 
Therefore, in this limit, the order parameter is necessarily $s$-wave-like! 

The above discrepancy teaches us that it is the presence of the kinematical 
constraints imposed by the Fermi sea that allows for 
non $s$-wave pairing.  The ultimate pairing symmetry is a reflection of 
the distribution in momentum space of the low energy single particle 
spectral weight. The reason for this is clear within BCS theory where 
the energy gain, which drives the transition, comes from the 
interaction term 
\be
{\rm Potential} \ \ {\rm Energy} =
\sum_{\vec k,\vec k'} 
V_{\vec k,\vec k'}\frac{\Delta_{\vec k}} {2E_{\vec k}}
\frac{\Delta_{\vec k'}} {2E_{\vec k'}} \; ,
\ee
\sindex{vkk}{$V_{\vec k,\vec k'}$}{BCS pair potential}
which is maximized by a gap function that peaks in regions of high 
density of states unless the pairing potential that connects these 
regions is particularly small. (Although we do not know of an explicit 
justification of this argument for a non-BCS theory, for example one 
which is driven by gain in kinetic energy, we feel that the physical 
consideration behind it is robust.)

Finally, there is another issue which is related to order parameter 
\marginpar{\em Nodal quasiparticles do not a $d$-wave mean.}
symmetry in a manner that is more complex than is usually thought---this 
is the issue of the existence of nodal quasiparticles. While nodal 
quasiparticles are natural in a
$d$-wave superconductor, $d$-wave superconductors can be nodeless, and $s$-wave
superconductors can be nodal.  To see this, it is possible to work 
entirely in the weak coupling limit where BCS theory is reliable.  
The quasiparticle excitation
spectrum can thus be expressed as
\be
E_{\vec k}=\sqrt{\varepsilon_{\vec k}^2+\Delta_{\vec k}^2} \; ,
\ee
where $\varepsilon_{\vec k}$ is the quasiparticle dispersion in the normal
state (measured from the Fermi energy). 
Nodal quasiparticles occur wherever the Fermi surface, that is the 
locus of points where $\varepsilon_{\vec k}=0$, crosses a line of gap nodes, 
the locus
of points where $\Delta_{\vec k}=0$.  If the Fermi surface is closed around 
the origin, $\vec k=0$, or about the Brilloin zone center, 
$\vec k=(\pi,\pi)$ (as it is most likely in optimally doped {\BSCCO} 
\cite{arpesdamashen}), then the $d$-wave 
symmetry of $\Delta_{\vec k}=0$ implies the existence of nodes.  
However, if the Fermi surface were closed about $\vec k=(0,\pi)$ 
(and symmetry related points), there would be no
nodal quasiparticles\cite{spbag}. Indeed, it is relatively easy to
characterize \cite{matsprl,sachdevnodal} the quantum phase transition between 
a nodal and nodeless $d$-wave superconductor which occurs as a parameter that 
alters the underlying band structure is varied.  
Conversely, it is possible to have lines of gap nodes for an 
extended $s$-wave superconductor,
and if these cross the Fermi surface, the
superconductor will posses nodal quasiparticles.

\subsection{What does the pseudogap mean?}
\label{sec:pseudogap}

\subsubsection{What experiments define the pseudogap?}
One of the most prominent, and most discussed features of the
cuprate
\marginpar{\em What's so pseudo about the pseudogap?}  
superconductors is a set of crossover
phenomena\cite{timuskreview,loramreview,cnl} 
which are widely observed in underdoped
cuprates and, to various extents, in optimally and even slightly 
overdoped materials. 
Among the experimental probes which are used to locate the 
pseudogap temperature in different materials are: 

\begin{figure}[ht!!!]
\begin{center}
\epsfig{figure=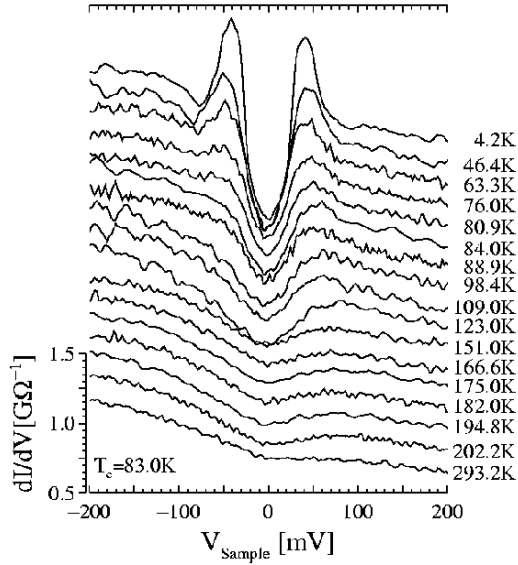,width=0.6\linewidth}
\end{center}
\caption[Tunnelling density of states in underdoped BSCCO]{
Tunnelling density of states in a sample of underdoped {\BSCCO } ($T_c$=83K) 
as a function of temperature. Note that there is no tendency for the gap
to close as $T_c$ is approached from below, but that the sharp 
``coherence peaks'' in the spectrum do vanish at $T_c$. From Ref.~\citen{renner}}
\label{stmpseudo}
\end{figure}

\begin{figure}[ht!!!]
\begin{center}
\epsfig{figure=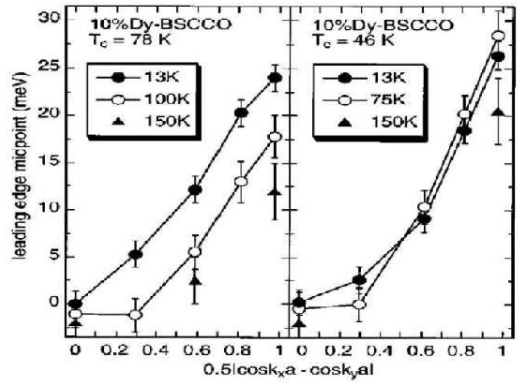,width=0.6\linewidth}
\end{center}
\caption[Angular dependence of the gap in BSCCO]{\label{arpespseudo}
The angular dependence of the gap in the normal and superconducting states 
of underdoped Bi$_2$Sr$_2$Ca$_{1-x}$Dy$_x$Cu$_2$O$_{8+\delta}$ as deduced 
from the leading edge energy of the single hole spectral function 
$A^<(\vec{k},\omega)$ measured by ARPES. 
A straight line in this plot would correspond to the simplest
$d_{x^2-y^2}$ gap, $|\Delta_{\vec k}|=\Delta_0|\cos(k_x)-\cos(k_y)|$. 
{F}rom Ref.~\citen{harrispg}.}
\end{figure}

1)  {\bf ARPES and c-axis tunnelling}: There is a suppression of the 
low energy single particle spectral weight, 
shown in Figs. \ref{stmpseudo} and \ref{arpespseudo} at temperatures 
above $T_c$ as detected, primarily, in c-axis tunnelling\cite{fischerpseudo} 
and ARPES\cite{ding,harrispg} experiments. The scale of energies 
and the momentum dependence of this suppression are very reminiscent of
the $d$-wave superconducting gap observed in the same materials at temperatures 
well below $T_c$. This is highly suggestive of an identification between the 
pseudogap and some form of local superconducting pairing. Although a 
pseudogap energy scale is easily deduced from these experiments, 
it is not so clear to us that an unambiguous temperature scale can be 
cleanly obtained from them.  (The c-axis here, and henceforth, refers to 
the direction perpendicular to the copper-oxide planes, which are also 
referred to, crystallographically, as the ab plane.)

\begin{figure}[ht!!!]
\begin{center}
\epsfig{figure=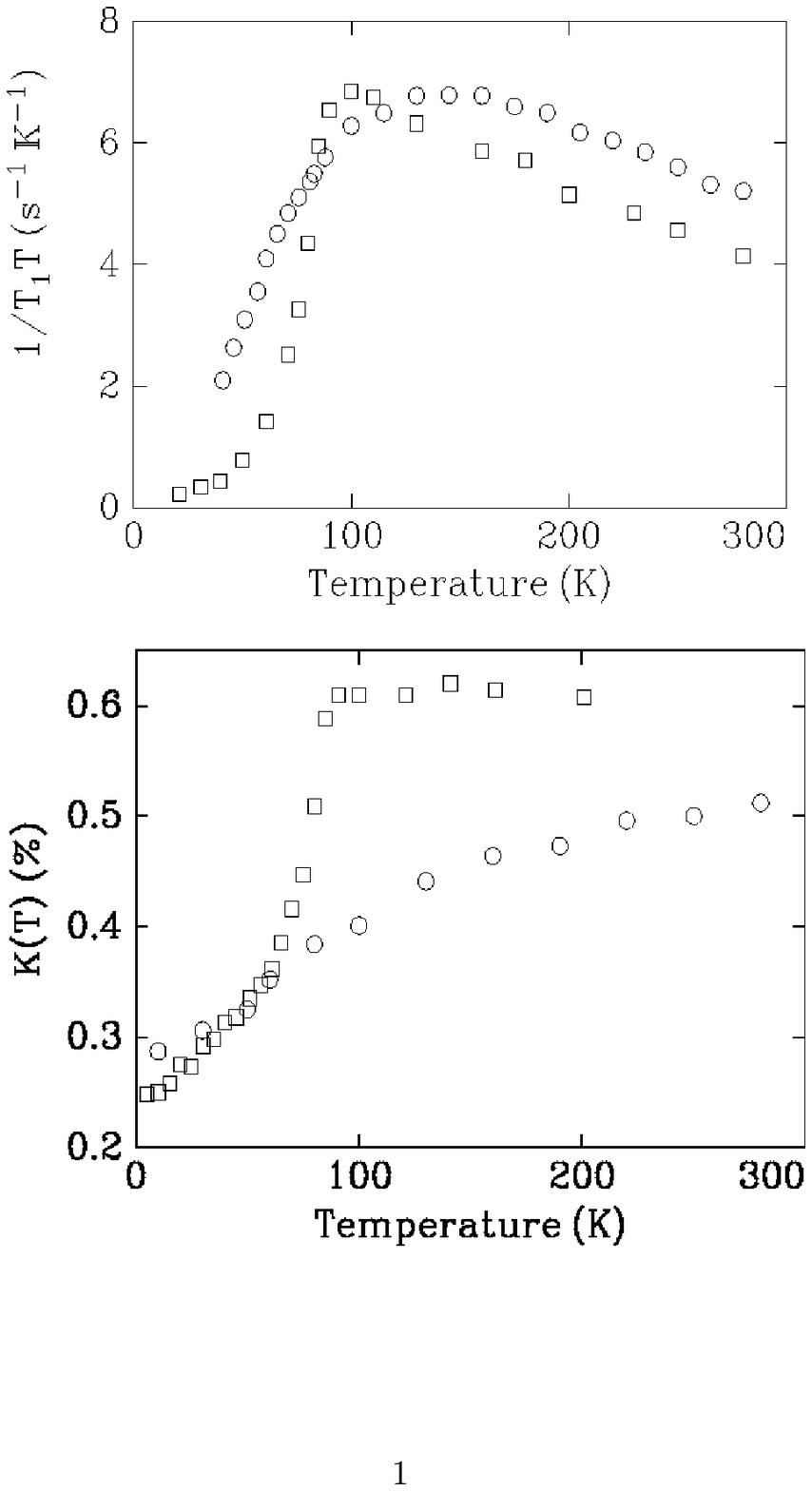,width=0.6\linewidth}
\end{center}
\caption[Relaxation rate and Knight shift in YBCO]{\label{nmrpseudo}
Temperature dependence of the planar $^{63}$Cu relaxation rate $1/T_1T$ and 
Knight shift $K$ in optimally doped YBa$_2$Cu$_3$O$_{6.95}$ (squares) 
and underdoped YBa$_2$Cu$_3$O$_{6.64}$ (circles). 
{From} Ref.~\citen{timuskreview}.}
\end{figure}

2)  {\bf Cu NMR}:  There is a suppression of low energy spin fluctuations 
as detected \cite{hammelwipeout} primarily in Cu NMR. 
In some cases, two rather 
different temperature scales are deduced from these experiments: an upper 
crossover temperature, at which a peak occurs in $\chi^{\prime}$, the 
real part of the uniform spin susceptibility ({\it i.e.} the Knight shift), 
and a lower crossover temperature, below which $1/T_1T$ drops precipitously.  
(See Fig. \ref{nmrpseudo}.) Note that  
$1/T_1T\propto \lim_{\omega\rightarrow 0}\int d\vec k f(\vec k)
\chi^{\prime\prime}(\vec k, \omega)/\omega$, the $\vec k$ averaged density of
states for magnetic excitations, where $f(\vec k)$ is an appropriate form
factor which reflects the local hyperfine coupling.  
\sindex{t1t}{$1/T_1T$}{NMR relaxation rate}
Although the temperature scale deduced from $\chi^{\prime}$ is more or less 
in accordance with the pseudogap scale deduced from a number of other 
spectroscopies, it is actually a measure of the reactive response of the
spin system.  The notion of a gap can be more directly identified with a 
feature in $\chi^{\prime\prime}$.  ({\em A note of warning}: 
while the structure in $1/T_1T$ can be fairly sharp at times, 
the observed maxima in $\chi^{\prime}$ are always very broad and do 
not yield a sharply defined temperature scale without further analysis.)

\begin{figure}[ht!!!]
\begin{center}
\epsfig{figure=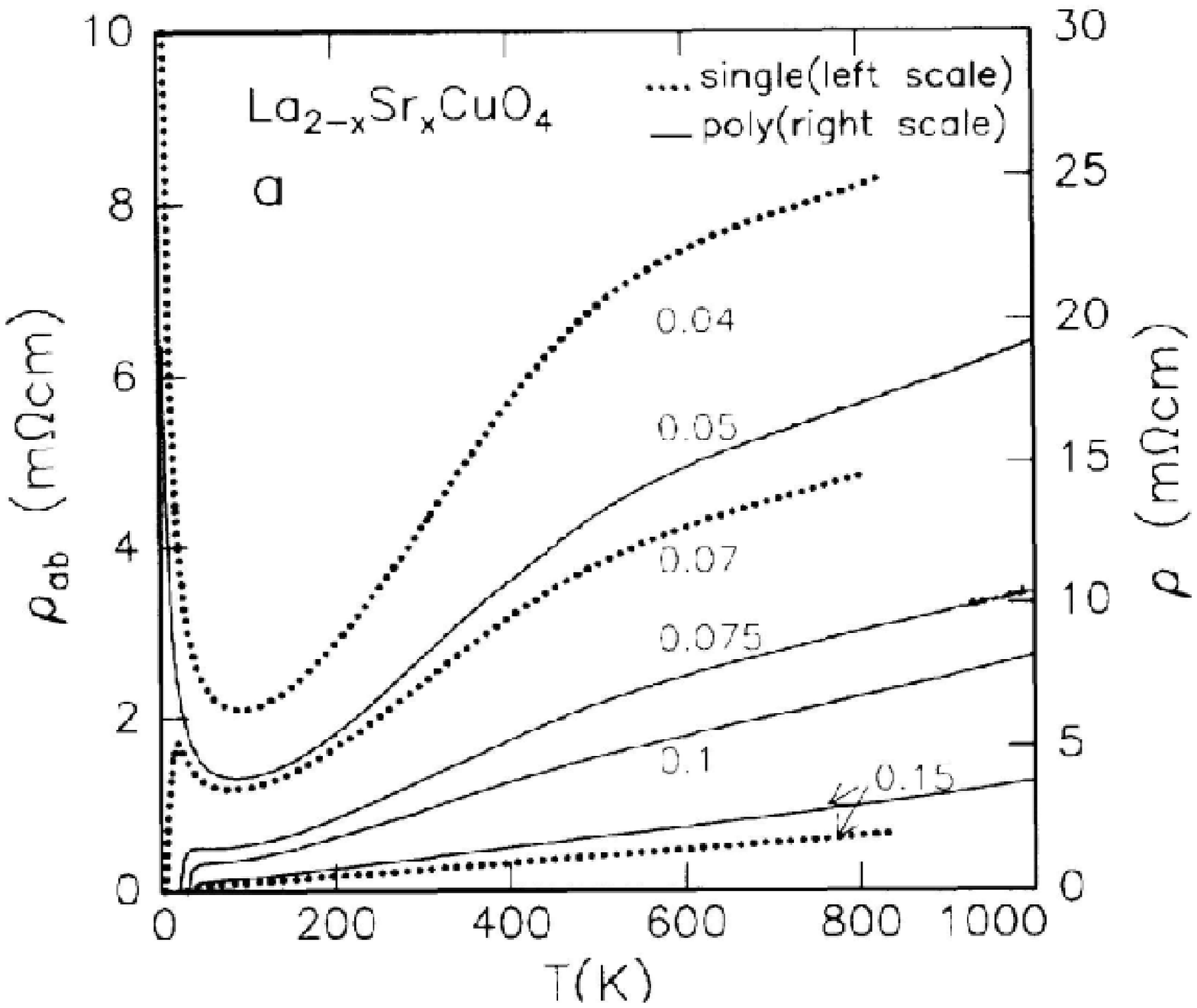,width=0.6\linewidth}
\end{center}
\caption[Longitudinal resistivity in LSCO]{\label{rhopseudo}
The temperature dependence of the longitudinal resistivity in underdoped and 
optimally doped \LSCO. The dotted lines correspond to the in-plane 
resistivity ($\rho_{ab}$) of single crystal films while the solid lines 
depict the resistivity ($\rho$) of polycrystalline samples. The doping levels 
are indicated next to the curves. From Ref.~\citen{takagi}.}
\end{figure}

3)  {\bf Resistivity}: There is a significant deviation\cite{takagi,248rho} 
of the resistivity in the ab plane from the
$T$ linear temperature dependence which is universally
observed at high temperatures.  A pseudogap temperature is then identified as
the point below which $d\rho_{xx}/dT$ deviates (increases) significantly from
its high temperature value. (See Fig. \ref{rhopseudo}.)  In some cases, a 
similar temperature scale can be inferred from a scaling analysis of the Hall
resistance, as well.

\begin{figure}[ht!!!]
\begin{center}
\epsfig{figure=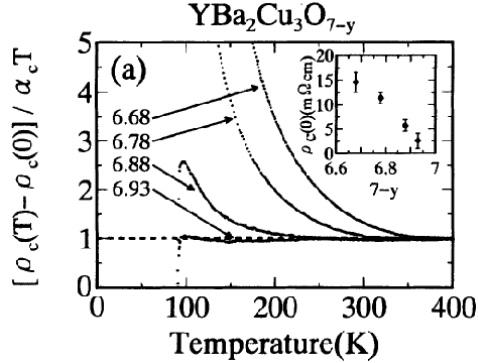,width=0.55\linewidth}
\end{center}
\caption[c-axis resistivity in underdoped YBCO]{\label{caxispg}
The temperature dependence of the c-axis resistivity in underdoped and 
optimally doped \YBCO. Here $\alpha_c$ and $\rho_c(0)$ are the slope and the 
intercept, respectively, when the metalic part of $\rho_c$ is approximated 
by a linear-$T$ behavior. The inset shows how $\rho_c(0)$ varies with oxygen 
content. From Ref.~\cite{takeno3}.}
\end{figure}

The pseudogap also appears in the c-axis resitivity, although in a 
somewhat different manner\cite{takeno3,ando8}. In this direction, the 
pseudogap results in a strong increase in the resistivity, reminiscent 
of the behavior of a narrow gap semiconductor, as shown in Fig.~\ref{caxispg}.
If we imagine that the c-axis transport is dominated by tunnelling events 
between neighboring planes, it is reasonable that a bulk measurement of
$\rho_c$ will reflect the pseudogap  in much the same way as 
the c-axis tunelling does.

\begin{figure}[ht!!!]
\begin{center}
\epsfig{figure=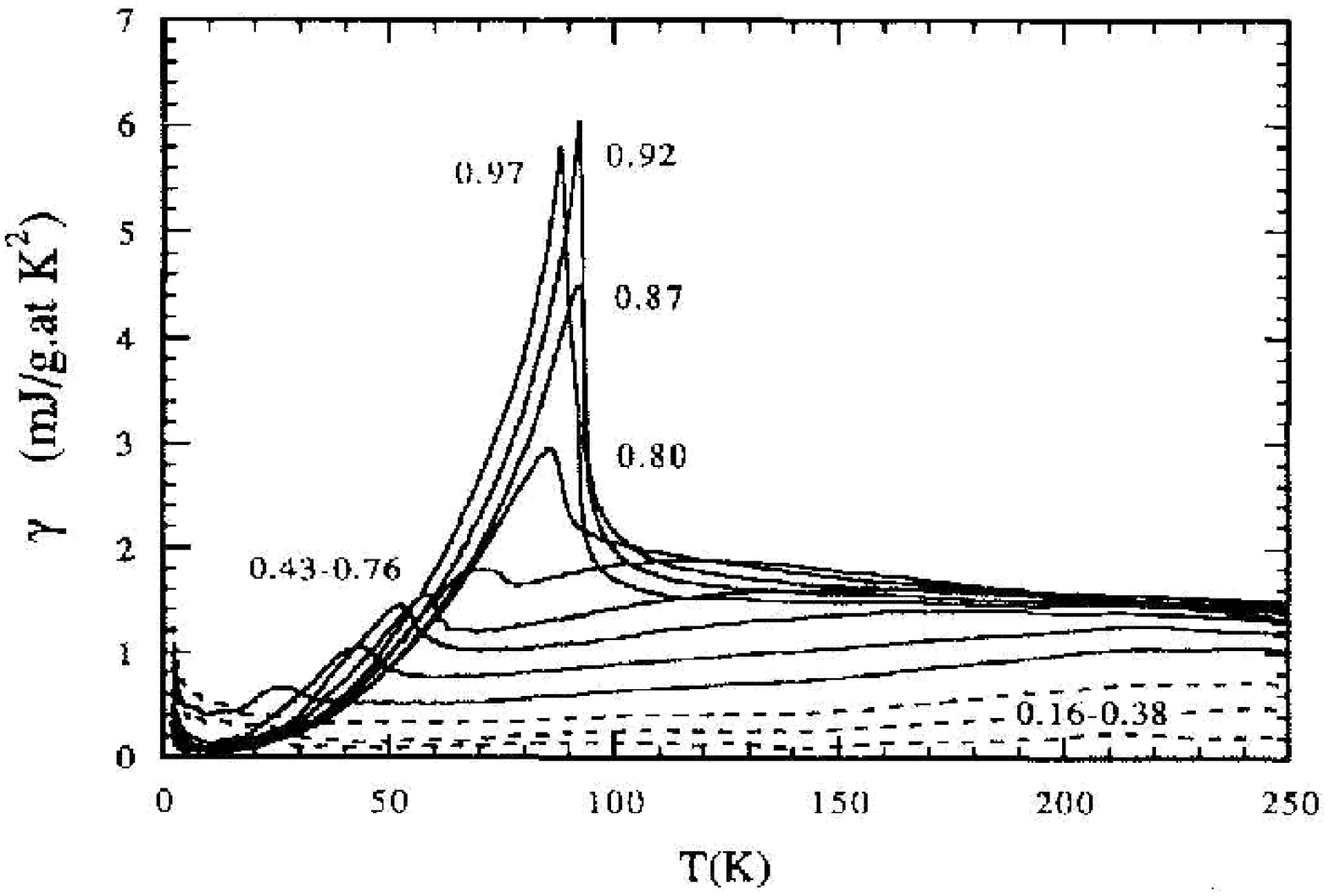,width=0.55\linewidth}
\end{center}
\caption[$\gamma\equiv C_V/T$ in underdoped YBCO]
{\label{cpseudo}Thermal density of ``electronic'' states, 
$\gamma\equiv C_V/T$ as a function of temperature for various oxygen 
concentrations in underdoped YBa$_2$Cu$_3$O$_{6+x}$. 
{F}rom Ref.~\citen{loramprl}. As discussed in \cite{loramprl}, 
a complicated proceedure has been used to subtract the large nonelectronic
component of the measured specific heat.}
\end{figure}
\sindex{t}{$T$}{Temperature}

4)  {\bf Specific heat}:  
There is a suppression of the expected electronic specific heat 
\cite{loramreview}.  
Above the pseudogap scale, the specific heat is generally found to be linear 
in temperature, $C_V \approx\gamma T$,  but below the pseudogap temperature, 
$C_V/T$ begins to decrease with decreasing temperature. 
(See Fig. \ref{cpseudo}.) Interestingly, since the value of
$\gamma$ above the pseudogap temperature appears to be roughly 
doping independent, the drop in the specific at lower temperatures can be
interpreted as a doping dependent loss of entropy, 
$\Delta S(x)\equiv S(x,T)-S(x_{optimal},T)$, with a magnitude which is 
independent of temperature for any $T > T^*$. This is the origin
of the famous (and still not understood) observation of Loram and collaborators 
\cite{loramprl} that there is a large entropy, $k_B/2$, which 
is somehow associated with each doped hole. {\em A word of
warning}: except at the lowest temperatures, the electronic specific 
heat is always a small fraction of the total specific heat, and complicated
empirical subtraction procedures, for which the theoretical justification is
not always clear to us, are necessary to extract the electronic contribution.

\begin{figure}[ht!!!]
\begin{center}
\epsfig{figure=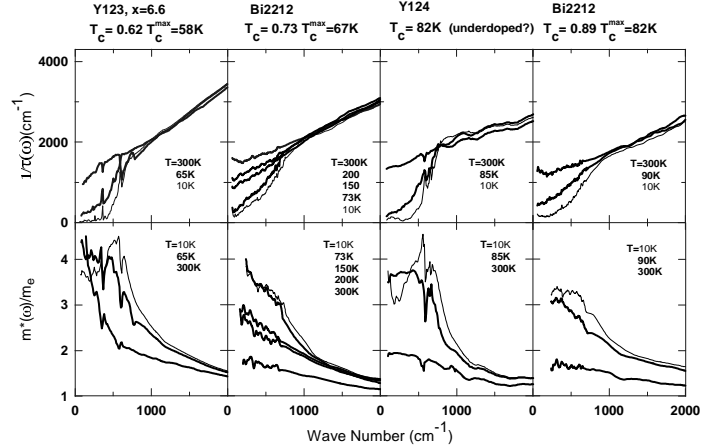,width=0.8\linewidth}
\end{center}
\caption[Lifetime and mass enhancement from IR conductivity]{\label{taupseudo}
Upper panels: Frequency dependent scattering rate for a series of 
underdoped cuprate superconductors above, near and below the superconducting 
transition temperature. Lower pannels: The effective mass enhancement 
$m^*/m_e=1+\lambda(\omega)$. Both are deduced from fitting infrared 
conductivity data to an extended Drude model 
$\sigma=(\omega_P^2/4\pi)/[1/\tau(\omega)-i\omega(1+\lambda(\omega))]$.   
{From} Ref.~\citen{basovpseudo}}
\end{figure}

5)  {\bf Infrared conductivity}:  
There is an anomalous motion of infrared spectral weight to low
energies\cite{orenstein,basovpseudo}.  
The pseudogap is most clearly identified by plotting \cite{basovpseudo} 
the frequency dependent scattering rate, defined either as
$1/\tau^*(\omega) \equiv
\omega\sigma_{ab}^{\prime}(\omega)/\sigma_{ab}^{\prime\prime}(\omega)$, or as
$1/\tau(\omega)=[\omega_P^2/4\pi] {\rm Re}[1/\sigma(\omega)]$ where 
$\omega_P$ is the plasma frequency; the pseudogap is rather harder to pick 
out from the in-plane conductivity, $\sigma_{ab}^{\prime}$, itself. 
At large $\omega$, one generally sees $1/\tau(\omega)\approx A\omega$, 
and it then drops to much smaller values, $1/\tau\ll \omega$, below a 
characteristic pseudogap frequency, see Fig.~\ref{taupseudo}.  
($A$ is generally a bit larger than 1 in underdoped materials and roughly
equal to 1 in optimally doped ones.)
\sindex{zzomega}{$\omega$}{Frequency}
\sindex{zzomegap}{$\omega_{P}$}{Plasma frequency}

\begin{figure}[ht!!!]
\begin{center}
\epsfig{figure=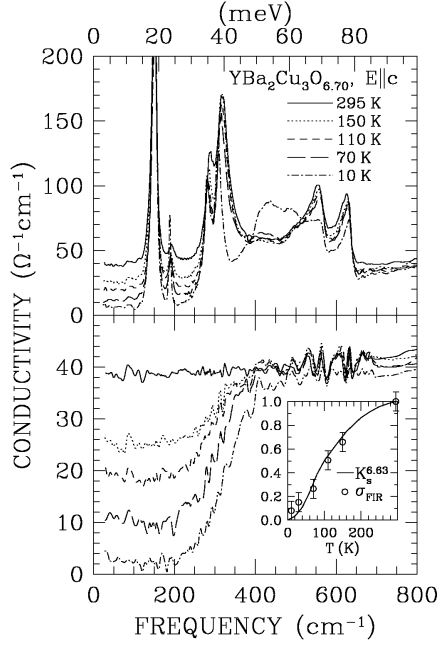,width=0.6\linewidth}
\end{center}
\caption[c-axis optical conductivity in underdoped YBCO]{\label{caxispseudo}
The $c$-axis optical conductivity of underdoped {\YBCO} ($T_c=63$K) as a 
function of temperature (top panel). The optical conductivity after the 
substraction of the phonon features is presented in the lower panel. 
The inset compares the low frequency conductivity with the Knight shift. 
{From} Ref.~\citen{homes}.}
\end{figure}

While in optimally doped materials, this
manifestation of a pseudogap is only observed at temperatures less than $T_c$,
in underdoped materials, it is seen to persist well above $T_c$, and
indeed to be not strongly temperature dependent near $T_c$.  
A characteristic
pseudogap energy is easily identified from this data, but, again, it is not
clear to us to what extent it is possible to identify a clear pseudogap temperature
from this data. A pseudogap can also be deduced directly 
\cite{homes,basovcaxis} from an analysis of $\sigma_c^{\prime}(\omega)$, 
where it manifests itself as a suppressed response at low frequencies, 
as shown in Fig.~\ref{caxispseudo}. 

\begin{figure}[ht!!!]
\begin{center}
\epsfig{figure=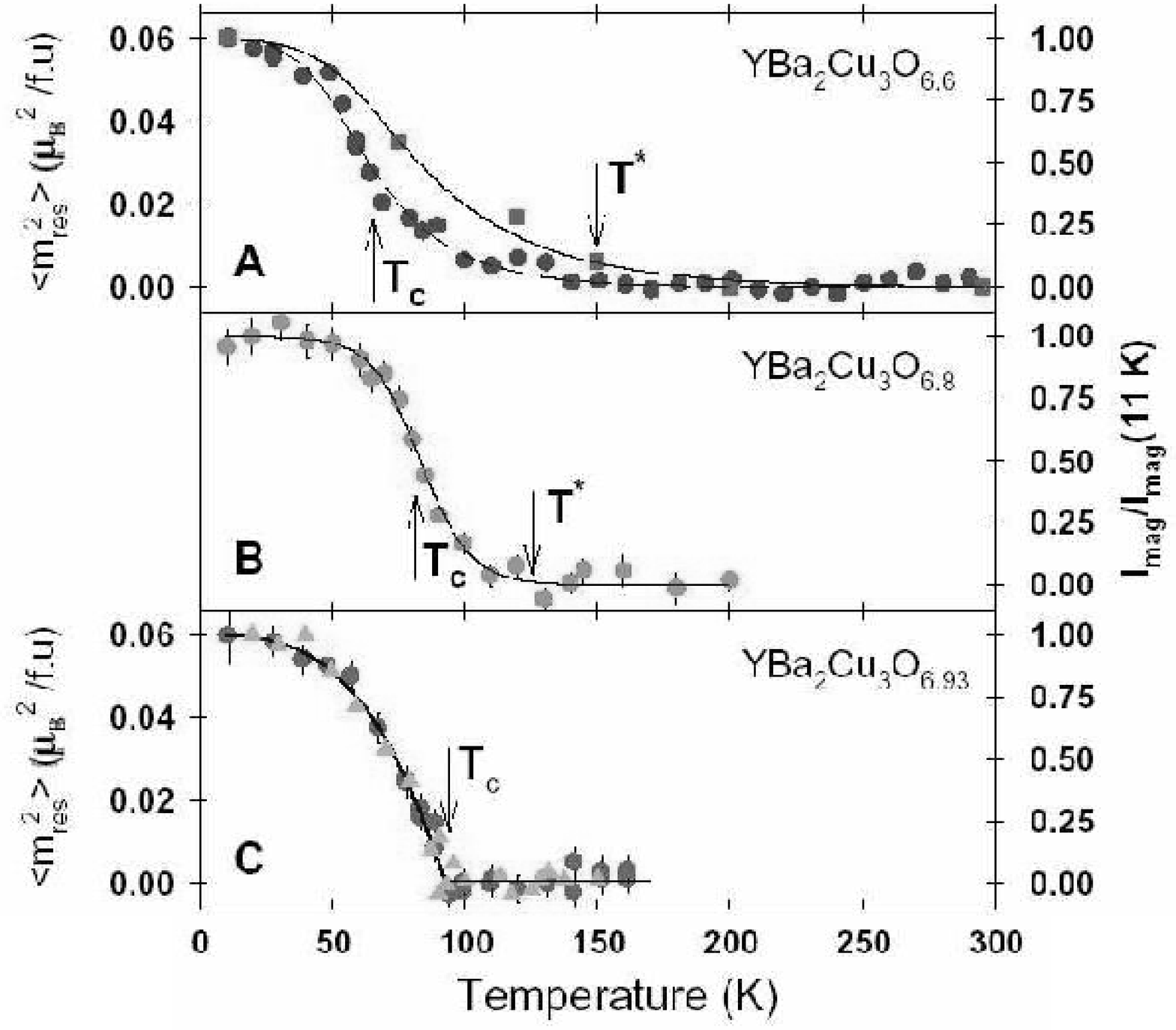,width=0.7\linewidth}
\end{center}
\caption[Resonant peak in underdoped YBCO]{\label{neutronpseudo}
The temperature dependence of the intensity of the so called resonant 
peak observed in neutron scattering in underdoped {\YBCO}. 
{From} Ref.~\citen{daimook}}
\end{figure}

6)  {\bf Inelastic neutron scattering}:  
There are temperature dependent changes in the dynamic spin structure factor
as measured by inelastic neutron scattering. Here, both features associated
with low energy incommensurate magnetic correlations (possibly associated with
stripes) \cite{ybco6.7} and the so-called ``resonant peak'' are found to 
emerge below a temperature which is very close to $T_c$ in optimally doped 
materials, but which rises considerably above $T_c$ in underdoped materials 
\cite {daimook}. (See Fig. \ref{neutronpseudo}.)

\subsubsection{What does the pseudogap imply for theory?}

It is generally accepted that the pseudogap, in one way or another, 
reflects the collective physics associated with the growth of electronic 
correlations. This, more than any other aspect
of the data, has focused attention on theories of the collective variables
representing the order parameters of various possible broken symmetry
states\cite{batloggemery,cnl,wenlee,dimxover,kfe,levi,senthilfisher,varma,spin gap,castro5,dhlphase,subirold,UL,SO5,assa1,assa2,Ye,abanov,baskaran}.
Among these theories, there are two rather different classes of ways 
to interpret the pseudogap phenomena.  

{\bf 1)}  It is well known that fluctuation effects can produce local order
which, under appropriate circumstances, can extend well into the disordered
phase. Such fluctuations produce in the disordered phase some of the local
characteristics of the ordered phase, and if there is a gap in the ordered
phase, a pseduogap as a fluctuation effect is eminently reasonable---see Fig.
\ref{phasediag}. As is discussed in Section \ref{phase}, 
the small superfluid density of the 
cuprates leads to the unavoidable conclusion that superconducting 
fluctuations are an order 1 effect in these materials, so it is quite 
reasonable to associate some pseudogap phenomena with these fluctuations. 
However, as the system is progressively underdoped, it gets closer 
and closer to the antiferromagnetic insulating state, and indeed there is 
fairly direct NMR evidence of increasingly strong local antiferromagnetic 
correlations\cite{takigawa}. It is thus plausible that there are significant 
effects of antiferromagnetic fluctuations, and since the antiferromagnetic 
state also has a gap, one might expect these fluctuations to contribute to 
the pseudogap phenomena as well. There are significant incommensurate charge 
and spin density (stripe) fluctuations observed directly in scattering 
experiments on a variety of underdoped materials
\cite{mook,ybco,ybco6.7,tranquada,tranquadareview}, 
as well as the occasional stripe ordered
phase\cite{mookchg,lsco1d,yshift2,wells1,birgeneausmallx}.   These fluctuations, too,
certainly contribute to the observed pseudogap phenomena. Finally, fluctuations
associated with  more exotic phases, especially the ``staggered flux phase'' 
(which we will discuss momentarily) have been proposed\cite{nagaosalee,wenlee} 
as contributing to the pseudogap as well.

There has been a tremendous amount of controversy in the literature concerning
\marginpar{\em Crossovers can be murky.}
which of these various fluctuation effects best account for the
observed pseudogap phenomena. Critical phenomena, which are clearly 
associated with the phase fluctuations of the superconducting order 
parameter, have  been observed
\cite{meingast,corson,bonnandhardy,xypasler} 
in regions that extend between 10\% to 40\% above and below the superconducting $T_c$ 
in optimally and underdoped samples of {\YBCO} and {\BSCCO}; in our opinion, 
the dominance of superconducting fluctuations in this substantial range of 
temperatures is now beyond question. However, pseudogap phenomena are clearly 
observed in a much larger range of temperatures. 
Even if fluctuation effects are ultimately the
correct explanation  for all the pseudogap phenomena, there may not truly be one type
of  fluctuation which  dominates the physics over the entire range of temperatures.

\begin{figure}[ht!!!]
\begin{center}
\epsfig{figure=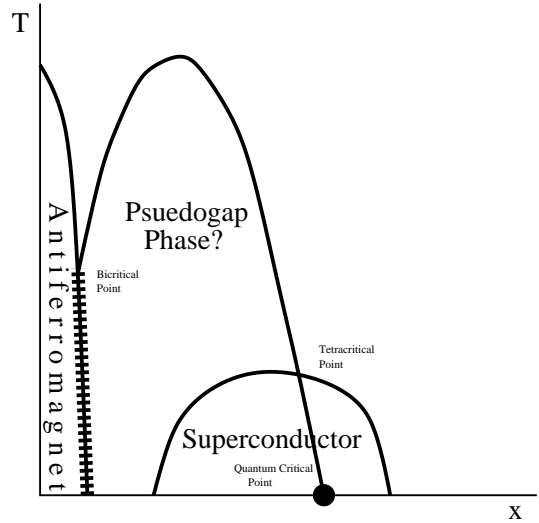,width=0.6\linewidth}
\end{center}
\caption[Phase diagram:  the pseudogap as a distinct phase]{\label{pgideas}
There are many ideas concerning the meaning of the pseudogap.
Defined purely phenomenologically, as shown in Fig. \ref{phasediag}, it is 
a region in which there is a general reduction in the density of low energy 
excitations, and hence is bounded by an ill-defined crossover line. It is 
also possible that, to some extent, the pseudogap reflects the 
presence  of a broken symmetry, in which case it must be bounded by a precise 
phase boundary, as shown in the present figure. There are many ways such a
pseudogap phase could interact with the other well established phases.
For purposes of illustration, we have shown a tetracritical and a bicritical 
point where the pseudogap meets, respectively, the superconducting and 
antiferromagnetic phases. One consequence of the assumption that the 
transition into the pseudogap phase is continuous is the exisence of a 
quantum critical point (indicated by the heavy circle) somewhere under the
superconducting dome. See, for example, Refs. 
\cite{spin gap,kfe,varmaqcp,cnl,castro5}.}
\end{figure}  

To illustrate this point explicitly, consider a
\marginpar{\em One cannot always tell a fluctuating superconductor from a
fluctuating insulator!} one dimensional electron gas (at an incommensurate
density) with weak attractive backscattering interactions.  
(See Section \ref{1D}.) If the backscattering interactions are attractive
($g_1<0$), they produce a spin gap $\Delta_s$. This gap persists 
\sindex{zzdeltas}{$\Delta_s$}{Spin gap}
as a pseudogap in the spectrum up to temperatures of order
$\Delta_s/2$. Now, because of the nature of fluctuations in one dimension, 
the system can never actually order at any finite temperature.
However, there is a very real sense in which one can view the pseudogap as an
effect of superconducting fluctuations, since at low temperatures, the
superconducting susceptibility is 
proportional to $\Delta_s$. The problem is that one can equally well view the 
pseudogap as an effect of CDW fluctuations. One could arbitrarily declare 
that where the CDW susceptibility is the most divergent, the pseudogap should 
be viewed as an effect of local CDW order, while when the superconducting 
susceptibility is more divergent, it is an effect of local pairing. 
However, this position is untenable; by varying the strength of the 
forward scattering ($g_2$), it is possible to pass smoothly from one 
regime to the other without changing $\Delta_s$ in any way !

{\bf 2)}  There are several theoretical proposals\cite{kfe,varma,cnl} on the
table which suggest that there is a heretofore undetected electronic phase
transition in underdoped materials with a transition temperature
well above the superconducting $T_c$. As a function of doping, this
transition temperature is pictured as decreasing, and tending to
zero at a quantum critical point somewhere in the neighborhood of
optimal doping, as shown schematically in Fig. \ref{pgideas}. 
If such a transition occurs, it would be natural to associate
\marginpar{\em Covert phase transitions are considered.}
at least some of the observed pseudogap phenomena with it.  Since these
scenarios involve a new broken symmetry, they make predictions which are, in
principle, sharply defined and falsifiable by experiment. However, there is an
important piece of phenomenology which these theories must address:  if there
is a phase transition underlying  pseudogap formation, why hasn't direct
thermodynamic evidence ({\it i.e.} nonanalytic behavior of the specific heat,
the susceptibility, or some other correlation function of the system) been seen
in existing experiments? Possible answers to this question typically invoke
disorder broadening of the proposed phase transition\cite{cnl}, rounding of 
the transition by a symmetry breaking field\cite{kfe}, or possibly the 
intrinsic weakness of the thermodynamic signatures of the transition under 
discussion\cite{varma,sudip8vertex}.

In any case, although these proposals are interesting
in their own right, and potentially important for the interpretation of
experiment, they are only indirectly related to the theory of high temperature
superconductivity, which is our principal focus in this article.  For this
reason, we will not further pursue this discussion here.

\section{Preview:  Our View of the Phase Diagram}
\label{ourview}

Clearly, the pseudogap phenomena described above are just the tip of
the iceberg, and any understanding of the physics of the cuprate high
temperature superconductors will necessarily be complicated. 
For this reason, we have arranged this article to focus
primarily on high temperature superconducitivity as an abstract
theoretical issue, and only really discuss how these ideas apply to the
cuprates in Section~\ref{finalchapter}. However, to orient the
reader, we will take a moment here to briefly sketch {\em our}
understanding of how these abstract issues determine the behavior, 
especially the high temperature superconductivity of the cuprates.  

\begin{figure}[ht!!!]
\begin{center}
\epsfig{figure=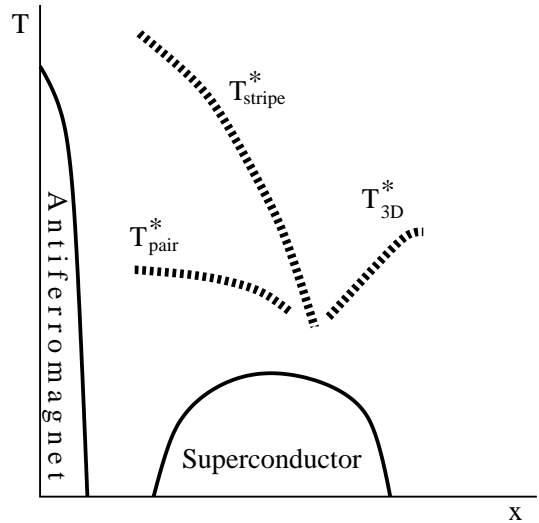, width=.6 \linewidth}
\end{center}
\caption[Phase diagram within the stripes scenario]{\label{fig:ourview}
Phase diagram as a function of temperature and doping within the stripes 
scenario discussed here.}
\end{figure}

Fig.~\ref{fig:ourview} is a schematic representation of the 
temperature {\it vs.} doping phase diagram of a representative cuprate.  
There are four energy scales relevant to the mechanism of 
superconductivity, marked as 
$T^{*}_{stripe}$, $T^{*}_{pair}$, $T_{3D}^*$ 
and $T_{c}$.  Away from  the peak of the superconducting 
dome, these energy scales are often well separated. At least some of the
pseudogap phenomena are, presumably, associated with the two
crossover scales, $T^*_{pair}$ and $T^*_{stripe}$.
\sindex{tstarstripe}{$T^{*}_{stripe}$}{Crossover temperature at which charge
  stripes form}
\sindex{tstarpair}{$T^{*}_{pair}$}{Crossover temperature at which pairs form}

{\bf Stripe Formation} $T^{*}_{stripe}$:
The kinetic energy of doped holes is frustrated in an antiferromagnet. 
As the temperature is lowered through $T^{*}_{stripe}$, the doped holes
are effectively ejected from the antiferromagnet to form metallic regions, 
thus relieving some of this frustration. Being charged objects, the holes
can only phase separate on short length scales, since the Coulomb
repulsion enforces charge homogeneity at long length scales. As a
result, at $T^{*}_{stripe}$, the material develops significant one dimensional
charge modulations, which we refer to as charge stripes. This can be an
actual phase transition ({\it e.g.} to a ``nematic phase''), or a crossover scale at
which significant local charge stripe correlations develop.  

{\bf Pair Formation} $T^{*}_{pair}$:
While stripe formation permits hole delocalization in one direction,
hole motion transverse to the stripe is still restricted. It is thus
favorable, under appropriate circumstances, for the holes to pair
so that the
pairs can spread out somewhat into the antiferromagnetic neighborhood of
the stripe. This ``spin gap proximity effect''\cite{spin gap} (see Section~\ref{sgpfx}), 
which is much like the proximity effect at the interface between a 
normal metal and a conventional superconductor, results in the opening 
of a spin gap and an enhancement of the superconducting susceptibility 
on a single stripe. 
In other words, $T^*_{pair}$ marks a crossover below which the
superconducting order parameter amplitude (and therefore a 
superconducting pseudo gap) has developed, but without global phase coherence.  

{\bf Superconductivity} $T_{c}$:
Superconducting long ranged order onsets as the phase of 
the superconducting order parameter on each charge stripe 
becomes correlated across the sample. Since it is triggered 
by Josephson tunnelling between stripes, this is a kinetic energy 
driven phase ordering transition. 

{\bf Dimensional Crossover} $T^*_{3D}$:  
Superconducting long range order 
implies coherence in all three dimensions, and hence the existence of well 
defined electron-like quasiparticles\cite{newjohnson,interlayer,dimxover}. 
Where the stripe order is sufficiently strong (in the underdoped regime),
the dimensional crossover to 3D physics is directly associated with the 
onset of superconducting order. However, in overdoped materials, where the 
electron dynamics is less strongly influenced by stripe formation, we expect 
the dimensional crossover
to occur well above $T_c$.  (See Section~\ref{1D}.)

\section{Quasi-1D Superconductors}
\label{1D}

\addtocontents{toc}{

\noindent{\em The well developed theory of quasi-one dimensional
superconductors is introduced as the best theoretical
laboratory  for the study
of strongly correlated electron fluids. 
The normal state is a non-Fermi liquid, in which the electron is 
fractionalized. It can exhibit a broad pseudogap regime 
for temperatures above $T_{c}$ but below the high temperature
Tomonaga-Luttinger liquid regime.
$T_c$ marks a point of dimensional 
crossover, where familiar electron quasiparticles appear with the 
onset of long range superconducting phase coherence.  } }

In this section we address the physics of the one dimensional 
electron gas and quasi-one dimensional systems consisting of 
higher dimensional arrays of weakly coupled chains. Our motivation is 
twofold. Firstly, these systems offer a concrete realization
of various non-Fermi liquid phenomena 
and are amenable to {\em controlled} theoretical treatments. 
As such they constitute a unique theoretical 
laboratory for studying strong correlations. In particular, for whatever 
reason,much of the experimentally observed behavior of the cuprate
superconductors is strongly reminiscent\cite{andersonnoqp,dimxover,frac}  of a
quasi-1D superconductor. Secondly, we are  motivated by a growing body of
experimental evidence for the existence  of electron smectic and nematic 
phases in the high temperature  superconductors, manganites and quantum Hall
systems\cite{pnas,zaanenscience,ando7,eisenstein,foglerreview,foglerliqcryst}.
It is possible that these materials actually are quasi 1D on a local scale.   

Our emphasis will be on quasi-one dimensional superconductors, 
the different unconventional signatures they exhibit as a 
function of temperature, and the conditions for their expression and 
stability. We will,
however, include some discussion of other  quasi-one dimensional phases
which typically tend to suppress  superconductivity.  
It is also worth noting that, for the most part, the discussion is
simply generalized to quasi-1D systems with different types of order,
including quasi-1D CDW insulators.  

\subsection{Elementary excitations of the 1DEG}
\label{1dcompete}

We begin by considering the continuum model of an interacting 
one dimensional electron gas (1DEG). It consists of approximating the 
1DEG by a pair of linearly dispersing branches of left ($\eta=-1$) and 
right ($\eta=1$) moving spin $1/2$ ($\sigma=\pm 1$ denotes the $z$ spin 
component) fermions constructed around the left and right Fermi points of 
the 1DEG. This approximation correctly describes the physics in the limit 
of low energy and long wavelength where the only important processes are 
those involving electrons in the vicinity of the Fermi points. The 
Hamiltonian density of the model is 
\begin{eqnarray}
\label{hamiltonian1}
\nonumber
{\cal H}=&-&i v_F\sum_{\eta,\sigma=\pm 1}\eta \psi_{\eta,\sigma}^{\dagger}
\partial_x\psi_{\eta,\sigma} \\
\nonumber
&+& \frac{g_4}{2}\sum_{\eta,\sigma=\pm 1}
\psi_{\eta,\sigma}^\dagger\psi_{\eta,-\sigma}^\dagger\psi_{\eta,-\sigma}
\psi_{\eta,\sigma}\\
\nonumber
&+&g_2\sum_{\sigma,\sigma'=\pm 1}\psi_{1,\sigma}^\dagger\psi_{-1,\sigma'}
^\dagger\psi_{-1,\sigma'}\psi_{1,\sigma} \\
\nonumber
&+&g_{1\parallel}\sum_{\sigma=\pm 1}\psi_{1,\sigma}^\dagger\psi_{-1,\sigma}
^\dagger\psi_{1,\sigma}\psi_{-1,\sigma} \\ 
&+&g_{1\perp}\sum_{\sigma=\pm 1}\psi_{1,\sigma}^\dagger
\psi_{-1,-\sigma}^\dagger\psi_{1,-\sigma}\psi_{-1,\sigma} \; , \;\;
\end{eqnarray} 
where, {\em e.g.}, $\psi_{1,1}$ destroys a right moving electron of spin $1/2$. 
The $g_4$ term describes forward scattering events of electrons 
in a single branch. The $g_2$ term corresponds to similar events but 
involving electrons on both branches. Finally, the $g_{1\parallel}$ and 
$g_{1\perp}$ terms allow for backscattering from one branch to the other. 
The system is invariant under $SU(2)$ spin rotations provided $g_{1\parallel}
=g_{1\perp}=g_1$. In the following we consider mostly this case. 
\sindex{vf}{$v_F$}{Fermi velocity}
\sindex{zzpsi}{$\psi_{\eta,\sigma}^{\dagger}$}{Fermion creation operator in the 1DEG}
\sindex{g2}{$g_2$}{Forward scattering on both branches}
\sindex{g1perp}{$g_{1\perp}$}{Backscattering of opposite spin particles}
\sindex{g1para}{$g_{1\parallel}$}{Backscattering of same spin particles}
\sindex{g1}{$g_1$}{Backscattering}
\sindex{g4}{$g_4$}{Forward scattering within same branch}
\sindex{g3}{$g_3$}{Umklapp scattering}
\sindex{g}{$G$}{Reciprocal lattice vector}
\sindex{kf}{$k_F$}{Fermi wavevector}

Umklapp processes of the form 
\begin{eqnarray}
\nonumber
g_3\psi_{-1,\uparrow}^\dagger
\psi_{-1,\downarrow}^\dagger\psi_{1,\downarrow}\psi_{1,\uparrow}
e^{i(4k_F-G)x} + {\rm H.c.} \; ,
\end{eqnarray}
are important only when $4k_F$ equals a reciprocal 
lattice vector $G$. When the 1DEG is incommensurate ($4k_F\neq G$), 
the rapid phase oscillations in this term render it irrelevant 
in the renormalization group sense. We will assume such incommensurability 
and correspondingly ignore this term. We will also neglect single particle
scattering  between branches (for example due to disorder) and terms that
do not  conserve the $z$ component of the spin.

It is important to stress\cite{emeryreview} that in considering this
model we are focusing on the long distance physics that can be
precisely derived from an effective field theory.  However, all the
coupling constants that appear in Eq.
\ref{hamiltonian1} are effective parameters which implicitly include much
of the high energy physics.  For instance, the bare velocity which
enters the model, $v_F$, is not necessarily simply related to the dispersion of
the band electrons in a zeroth order, noninteracting model, but instead
includes all sorts of finite renormalizations due to the interactions. 
The weak coupling perturbative renormalization group treatment of this
model is discussed in Section \ref{weak}, below;  the most important
result from this analysis is that the Fermi liquid fixed point is always
unstable, so that an entirely new, nonperturbative method must be
employed to reveal the low energy  physics.

Fortunately, such a solution is possible;  the Hamiltonian in 
\marginpar{\em Bosonization}
Eq.~(\ref{hamiltonian1}) is equivalent to a model of two independent  bosonic
fields, one representing the charge and the other the spin degrees  of
freedom in the system. (For reviews and recent perspectives see Refs. 
\citen{emeryreview,solyom,fradkinbook,shankar,vondelft98,tsvelikbook,
schulzreview,voitreview}.) 
The two representations are related via the bosonization identity
\begin{equation}
\label{bidentity}
\psi_{\eta,\sigma}=\frac{1}{\sqrt{2\pi a}}F_{\eta,\sigma}\exp[
-i\Phi_{\eta,\sigma}(x)] \; ,
\end{equation}
which expresses the fermionic fields in terms of self dual fields 
$\Phi_{\eta,\sigma}(x)$ obeying $[\Phi_{\eta,\sigma}(x),\Phi_{\eta',\
\sigma'}(x')]=-i\pi\delta_{\eta,\eta'}\delta_{\sigma,\sigma'}{\rm sign}(x-x')$.
They in turn are combinations of the bosonic fields $\phi_c$ and $\phi_s$ 
and their conjugate momenta $\partial_x\theta_c$ and $\partial_x\theta_s$ 
\begin{equation}
\label{decomp}
\Phi_{\eta,\sigma}=\sqrt{\pi/2}\,[(\theta_c-\eta\phi_c)+\sigma(\theta_s-\eta
\phi_s)] \; .
\end{equation}
Physically, $\phi_c$ and $\phi_s$ are, respectively, the phases of the $2k_F$
charge density wave (CDW) and spin density wave (SDW) fluctuations, and 
$\theta_c$ is the superconducting phase. In terms of them the long wavelength 
component of the charge and spin densities are given by
\begin{eqnarray}
\label{chargedensity}
\rho(x)=&&
\sum_{\eta,\sigma}\psi^{\dagger}_{\eta,\sigma}\psi_{\eta,\sigma}
-\frac{2k_F}{\pi}=\sqrt{\frac{2}{\pi}}\partial_x \phi_c \; ,
\\ 
S_z(x)=&&
\frac{1}{2}\sum_{\eta,\sigma}\sigma\psi^{\dagger}_{\eta,\sigma}
\psi_{\eta,\sigma}=\sqrt{\frac{1}{2\pi}}\partial_x \phi_s \; .
\end{eqnarray}
\sindex{zzrho}{$\rho(x)$}{Charge density}
\sindex{sz}{$S_{z}(x)$}{Spin density}
The Klein factors $F_{\eta,\sigma}$ in Eq. (\ref{bidentity}) are responsible 
for reproducing the correct anticommutation relations between different 
fermionic species and $a$ is a short distance cutoff that is taken to zero 
at the end of the calculation.
\sindex{fzznzzs}{$F_{\eta,\sigma}$}{Klein factor}
\sindex{zzdelxzzthetac}{$\partial_{x}\theta_c$}{Conjugate momentum of $\phi_c$}
\sindex{zzphic}{$\phi_c$}{Bosonic charge field}
\sindex{zzphis}{$\phi_s$}{Bosonic spin field}
\sindex{zzdelxzzthetas}{$\partial_{x}\theta_c$}{Conjugate momentum of $\phi_s$}

The widely discussed separation of charge and spin in this problem is
formally
\marginpar{\em In 1D spin and charge separate.} 
a statement that the Hamiltonian density can be expressed as a sum of two 
pieces, each of the sine-Gordon variety, involving only charge or spin fields  
\begin{equation}
\label{bosH}
{\cal H}=\sum_{\alpha=c,s} \left\{ \frac {v_{\alpha}} 2\left [ 
K_{\alpha}(\partial_{x}\theta_{\alpha})^{2}
+\frac {(\partial_{x}\phi_{\alpha})^{2}} {K_{\alpha}}
\right ] + V_{\alpha} \cos(\sqrt{8\pi} \phi_{\alpha})\right\} \; .
\end{equation}
When the Hamiltonian is separable, wavefunctions, and therefore correlation
functions, factor.  (See Eqs.~(\ref{bCDW}) and (\ref{bSS}).)
In terms of the parameters of the fermionic formulation Eq. 
(\ref{hamiltonian1}) the charge and spin velocities are given by
\begin{eqnarray}
\label{velocities}
v_c&=&\frac{1}{2\pi}\sqrt{(2\pi v_F+g_4)^2-(g_{1\parallel}-2g_2)^2} \; , \\ 
v_s&=&\frac{1}{2\pi}\sqrt{(2\pi v_F-g_4)^2-g_{1\parallel}^2} \; ,
\end{eqnarray}
while the Luttinger parameters $K_\alpha$, which determine the power law 
behavior of the correlation functions, are 
\begin{eqnarray}
\label{lparam}
K_c&=&\sqrt{\frac{2\pi v_F+g_4-2g_2+g_{1\parallel}}
{2\pi v_F+g_4+2g_2-g_{1\parallel}}} \; , \\
K_s&=&\sqrt{\frac{2\pi v_F-g_4+g_{1\parallel}}
{2\pi v_F-g_4-g_{1\parallel}}} \; .
\end{eqnarray}
The cosine term in the spin sector of the bosonized version of the Hamiltonian 
(Eq.~(\ref{bosH})) originates from the back scattering term in Eq.~(\ref{hamiltonian1}) 
where the amplitudes are related according to 
\begin{equation}
V_s=\frac{g_{1\perp}}{2(\pi a)^2} \; .
\label{V_s}
\end{equation} 
The corresponding term in the charge sector describes umklapp 
processes and in view of our assumption will be set to zero $V_c=0$. 
Eqs. (\ref{velocities}-\ref{V_s}) complete the exact mapping between
the fermionic and bosonic field theories.
\sindex{vs}{$v_s$}{Spin velocity}
\sindex{vc}{$v_c$}{Charge velocity}
\sindex{kc}{$K_c$}{Charge Luttinger parameter}
\sindex{ks}{$K_s$}{Spin Luttinger parameter}
\sindex{vc}{$V_s$}{Strength of sine-Gordon potential}

In the absence of back scattering ($g_1=0$) this model is 
usually called the Tomonaga-Luttinger model. Since $\partial_x\theta_{c,s}$ 
and $\phi_{c,s}$ are canonically conjugate, it is clear from the form of 
the bosonized Hamiltonian (Eq.~(\ref{bosH})) that it describes a collection of 
independent charge and spin density waves with linear dispersion 
$\omega_{c,s}=v_{c,s}k$. The quadratic nature of the theory and the 
coherent representation (Eq.~(\ref{bidentity})) of the electronic operators in 
terms of the bosonic fields allow for a straightforward evaluation of 
various electronic correlation functions.

For $g_1\neq 0$ the spin sector of the theory turns into a sine-Gordon 
theory whose renormalization group flow is well known \cite{ktflow}. 
In particular, for repulsive interactions ($g_1>0$) the backscattering 
amplitude is renormalized to zero in the long wavelength low energy limit and 
consequently at the fixed point $K_s=1$. On the other hand, in the presence 
of attractive interactions ($g_1<0$) the model flows to strong (negative) 
coupling where the cosine term in Eq.~(\ref{bosH}) is relevant. As a result
$\phi_s$ is pinned in the sense that in the ground state, it executes only
small amplitude fluctuations about its classical ground state value ({\it
i.e.} one of the minima of the cosine). There is a spin gap to both
extended phonon-like small amplitude oscillations about this minimum and
large amplitude 
soliton excitations that are domain walls at which
$\phi_s$ changes between two adjacent minima.

The susceptibility of the interacting one dimensional electron gas to
various instabilities
can be investigated by calculating the correlation functions 
of the operators that describe its possible orders. They include, among 
others, the $2k_F$ CDW and SDW operators
\begin{eqnarray}
O_{CDW}(x)&=&e^{-i2k_Fx}\sum_{\tau}\psi_{1,\tau}^{\dagger}(x)
\psi_{-1,\tau}(x) \; , \\
\label{OCDW}
O_{SDW_\alpha}(x)&=&e^{-i2k_Fx}\sum_{\tau,\tau'}\psi_{1,\tau}^{\dagger}(x)
\sigma_{\tau,\tau'}^{\alpha}\psi_{-1,\tau'}(x) \; ,
\label{OSDW}
\end{eqnarray}
where \mbox{\boldmath $\sigma$} are the Pauli matrices, the $4k_F$ CDW (or
Wigner crystal) order
\begin{equation}
O_{4k_F}(x)=e^{-i4k_Fx}\sum_{\tau}\psi_{1,\tau}^{\dagger}(x)\psi_{1,-\tau}
^{\dagger}(x)\psi_{-1,-\tau}(x)\psi_{-1,\tau}(x) \; ,
\end{equation}
and the singlet (SS)
and triplet (TS) pair annihilation operators
\begin{eqnarray}
O_{SS}(x)&=&\sum_{\tau}\tau\psi_{1,\tau}(x)\psi_{-1,-\tau}(x) \; , \\
\label{SS}
O_{TS_\alpha}(x)&=&\sum_{\tau,\tau'}\tau\psi_{1,\tau}(x)
\sigma_{\tau,\tau'}^{\alpha}\psi_{-1,-\tau'}(x) \; .
\label{TS}
\end{eqnarray}
They can also be written in a suggestive bosonized form. For example the 
CDW and the singlet pairing operators are expressed as 
\footnote{For a discussion of some delicate points involving Klein factors 
in such expressions see Refs. \citen{vondelft98} and \citen{schulzreview}.}  
\begin{eqnarray}
O_{CDW}(x)&=&\frac{e^{-2i k_F x}}{\pi a}\cos[\sqrt{2\pi}\phi_s(x)]
e^{-i\sqrt{2\pi}\phi_c(x)} \; , \label{bCDW}\\
O_{SS}(x)&=&\frac{1}{\pi a}\cos[\sqrt{2\pi}\phi_s(x)]
e^{-i\sqrt{2\pi}\theta_c(x)} \; . \label{bSS}
\end{eqnarray}
\marginpar{\em 1D order parameters have ``spin'' amplitudes and 
``charge'' phases.}
The distinct roles of spin and charge are vividly apparent in these
expressions: the amplitude  of the order parameters is a function
of the spin fields while their phase is  determined by the charge
degrees of freedom. Similar relations are found  for the SDW and triplet
pairing operators.  However, the $4k_F$ CDW order is independent of the
spin fields.

If in the bare Hamiltonian, $g_1>0$ and $V_s$ is not too large, the system 
flows to the Gaussian fixed point with $K_s=1$ and no spin gap. The gapless 
fluctuations of the amplitude (spin) and phase (charge) of the various 
orders lead then to an algebraic decay of their zero temperature space-time 
correlation functions (with logarithmic corrections
which reflect the slow renormalization of marginally irrelevant
operators near the fixed point
\cite{giamarchi89}):   
\begin{eqnarray}
\label{corr}
\nonumber
\langle O_{CDW}^\dagger(x)O_{CDW}(0)\rangle&\propto& e^{2ik_F x}x^{-(1+K_c)}
\ln^{-3/2}(x) \; , \\
\nonumber
\langle O_{SDW_\alpha}^\dagger(x)O_{SDW_\alpha}(0)\rangle&\propto& e^{2ik_F x}
x^{-(1+K_c)}\ln^{1/2}(x) \; , \\
\nonumber
\langle O_{4k_F}^\dagger(x)O_{4k_F}(0)\rangle&\propto& e^{4ik_F x}
x^{-4K_c} \; , \\
\nonumber
\langle O_{SS}^\dagger(x)O_{SS}(0)\rangle&\propto& x^{-(1+1/K_c)}
\ln^{-3/2}(x) \; , \\
\langle O_{TS_\alpha}^\dagger(x)O_{TS_\alpha}(0)\rangle&\propto&  
x^{-(1+1/K_c)}\ln^{1/2}(x) \; , 
\end{eqnarray}
where the proportionality involves model dependent constants and where 
sub-leading terms have been omitted. In the presence of interactions 
that break spin rotation symmetry ($g_{1\parallel}\neq g_{1\perp}$) the 
model flows, for moderately repulsive bare $g_{1\parallel}$, to a point on 
a fixed line with $V_s=0$ and $K_s>1$. Correspondingly, the spin 
contribution to the decay exponent of the correlation functions 
(see Eq.~(\ref{corr})) changes from 1 to $K_s$ 
for the CDW, SS, and the z component of the SDW order, and from 1 to $1/K_s$
for TS and the x and y components of the SDW order.  (For $K_s\ne 1$, 
there are no logarithmic corrections and the leading behavior is that  of
a pure power law
\cite{giamarchi89}.)

The temporal dependence of the above correlation functions is easily 
obtained owing to the Lorentz invariance of the model (Eq.~(\ref{bosH})). By 
Fourier transforming them one obtains the related susceptibilities 
whose low temperature behavior for the spin rotationally invariant case 
is given according to
\begin{eqnarray}
\nonumber
\label{suscep}
\chi_{CDW}&\propto&T^{K_c-1}|\ln(T)|^{-3/2} \; , \\
\nonumber
\chi_{SDW}&\propto&T^{K_c-1}|\ln(T)|^{1/2} \; , \\
\nonumber
\chi_{4k_F}&\propto&T^{4K_c-2} \; , \\
\nonumber
\chi_{SS}&\propto&T^{1/K_c-1}|\ln(T)|^{-3/2} \; , \\
\chi_{TS}&\propto&T^{1/K_c-1}|\ln(T)|^{1/2} \; .
\end{eqnarray}
\sindex{zzxxi}{$\chi$}{Susceptibility}
\marginpar{\em Without a spin gap, SDW and triplet pairing fluctuations 
are most relevant.}
Therefore in the absence of a spin gap and for $1/3<K_c<1$, the $2k_F$ 
fluctuations are the most divergent, and the SDW is slightly more 
divergent than the CDW. In the presence of strong repulsive interactions
when
$K_c<1/3$,  the $4k_F$ correlations dominate. If $K_c>1$, the pairing
susceptibilities  diverge at low temperatures and triplet pairing is the
dominant channel.  
 
When $g_1<0$, a spin gap opens of magnitude
\begin{equation}
\label{spin gap}
\Delta_s\sim\frac{v_s}{a}\left(\frac{|g_1|}{2\pi v_s}\right)^{1/(2-2K_s)} 
\;. 
\end{equation}   
This can be explicitly demonstrated at the special Luther-Emery point
\cite{LE} $K_s=1/2$, where the spin sector is equivalent to a massive 
free Dirac theory. At this point,  a new set of spinless fermions can be
defined
\begin{equation}
\label{spinfermions}
\Psi_{\eta}\equiv \frac{1}{\sqrt{2\pi
a}}F_{\eta}\exp[i\sqrt{\pi/2}(\theta_s- 2\eta\phi_s)] \; ,
\end{equation}
in terms of which the spin part of the Hamiltonian can be refermionized 
\begin{equation}
\label{referH}
{\cal H}_s=-i v_s\sum_{\eta}\eta \Psi_{\eta}^{\dagger}\partial_x
\Psi_{\eta}+\Delta_s(\Psi_1^\dagger\Psi_{-1}+H.c.) \; ,
\end{equation}
and readily diagonalized to obtain the spin excitation spectrum 
\begin{equation}
\label{spinener}
E_s=\sqrt{v_s^2 k^2+\Delta_s^2} \; .
\end{equation} 
\sindex{es}{$E_s$}{Spin spectrum}

In the spin gapped phase, correlations involving spin 1 order parameters, 
such as SDW and triplet pairing, decay exponentially with correlation 
length $\xi_s=v_s/\Delta_s$. On the other hand the amplitude 
of the CDW and SS order parameters acquire a vacuum expectation 
\marginpar{\em With a spin gap, CDW or singlet pairing fluctuations 
are the most relevant.}
value.  Actual long range order, however, does not occur due to the
phase fluctuations associated with the still gapless charge modes.
Nevertheless,  the CDW and SS susceptibilities are enhanced compared to
the case  with no spin gap and in a spin rotationally invariant system
are given by
\begin{eqnarray}
\label{ensus}
\nonumber
\chi_{CDW}&\propto&\Delta_s T^{K_c-2} \; , \\
\chi_{SS}&\propto&\Delta_s T^{1/K_c-2} \; .
\label{chiofT}
\end{eqnarray}

\begin{figure}[!!!ht]
\begin{center}
\epsfig{figure=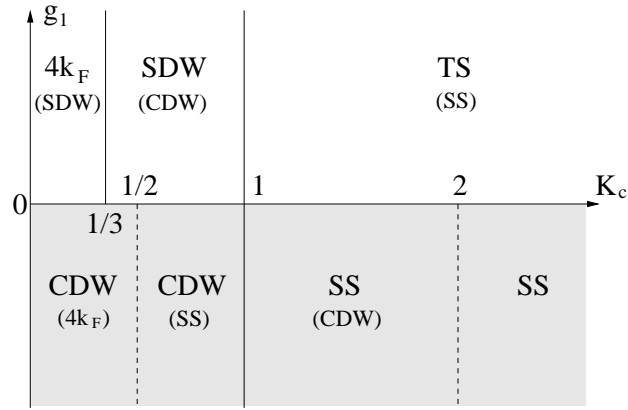,width=0.7\linewidth}
\end{center}
\caption[Phase diagram of the 1DEG]
{\label{fig1d1}Phase diagram for the one dimensional spin rotationally
invariant electron gas showing where various zero temperature
correlations diverge. Parentheses indicate subdivergent correlations
and the shaded region contains the spin gapped phases. The order parameters 
that appear in the figure are: singlet superconductivity (SS); triplet 
superconductivity (TS); $2k_F$ spin density wave (SDW); $2k_F$ charge density 
wave (CDW); and 
$4k_F$ charge density wave (4k$_{\rm F}$).}
\end{figure}

As long as $K_c>1/2$ the singlet pairing susceptibility is divergent but 
it becomes more divergent than the CDW susceptibility only when $K_c>1$. 
The latter diverges for $K_c<2$ and is the predominant channel provided 
$K_c<1$. Figure \ref{fig1d1} summarizes the situation for low temperatures 
showing where in parameter space each type of correlation diverges.

\begin{figure}[!!!ht]
\begin{center}
\epsfig{figure=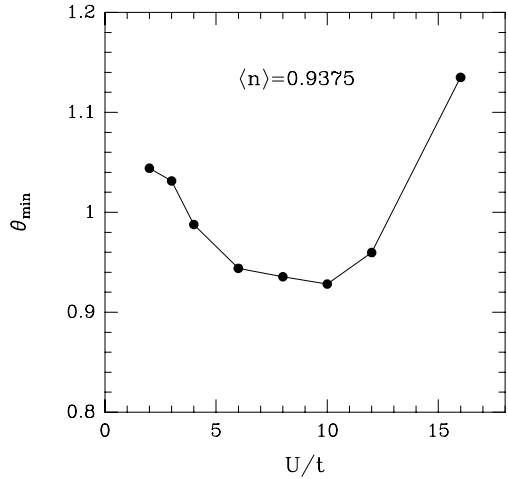,width=0.6\linewidth}
\end{center}
\caption
{\label{fig1d2}
Minimal value of the decay exponent, $\theta=1/K_c$, of the $d$-wave singlet
pairing  correlations in a two leg ladder with varying hopping ratio 
$t_\perp/t$ as a function of $U/t$. The electron filling 
is $\langle n\rangle=0.9375$. (From Noack et al.\cite{noackd})}
\end{figure}

We see that the low energy behavior of a system with a spin gap is basically
\marginpar{\em Concerning the sign of the effective interactions.} 
determined  by a single parameter $K_c$. For a Hubbard chain with
repulsive interactions, it is well known 
\cite{kawakami90} that $K_c<1$, but this is not a general physical bound. 
For instance, numerical experiments on two leg Hubbard ladders (which are 
spin gapped systems as we discuss in Sections~\ref{strong} and ~\ref{numerical})
have found a power law decay $r^{-\theta}$ of the singlet $d$-wave pairing 
correlations along the ladder. Fig.~\ref{fig1d2} presents the minimal value 
of the decay exponent $\theta$ obtained for ladders with varying ratio of 
inter- to intra-leg hopping $t_\perp/t$ as a function of the relative 
interaction strength $U/t$ \cite{noackd}. By comparing it with the 
corresponding 
exponent $\theta=1/K_c$ calculated for a spin gapped one dimensional
system, one can see that $K_c>1/2$ 
over the entire range of parameters
and that for some ranges $K_c>1$.
Our point is that in a multicomponent 1DEG,
it is possible to have $K_{c}>1$ (and thus singlet superconductivity
as the most divergent susceptibility) even for repulsive interactions.

\subsection{Spectral functions of the 1DEG---signatures of fractionalization}

The fact that one can obtain a strong (power law) divergence of the
superconducting susceptibility from repulsive interactions between electrons
is certainly reason enough to look to the 1DEG for clues
concerning the origins of high temperature superconductivity---we will
further pursue this in Sections \ref{strong} and \ref{numerical}, below. 
What we will do now is to continue to study the 1DEG as a solved model of a 
non-Fermi liquid.

In a Fermi liquid the elementary excitations have the quantum numbers of 
an electron and a nonvanishing overlap with the state created by the  
electronic creation operator acting on the ground state. As a result the
single  particle spectral function, $A(\vec k,\omega)$, 
\sindex{akzzomega}{$A^<(\vec k,\omega)$}{Single hole spectral function}
\sindex{zzomega}{$\omega$}{Frequency}
is peaked at 
$\omega=\epsilon(\vec k)=\vec v_F(\vec k_F)\cdot (\vec k-\vec k_F)$,
where $\epsilon(\vec k)$ is the quasiparticle dispersion relation.
This peak can be and has been \cite{Mo} directly observed using 
angle resolved photoemission spectroscopy (ARPES) which measures 
the single hole piece of the spectral function
\begin{equation}
\label{spectralhole}
A^<(\vec k,\omega)=\int_{-\infty}^{\infty}d\vec{r}\, dt \, 
e^{i(\vec{k}\cdot\vec{r}+\omega t)}
\langle\psi_\sigma^\dagger(\vec{r},t)\psi_\sigma(0,0)\rangle \; .
\end{equation}

The lifetime of the quasiparticle, $\tau(\vec k)$, can be determined from 
\sindex{zztau}{$\tau$}{Quasiparticle lifetime}
the width of the peak in the ``energy distribution curve'' (EDC) defined by
considering $A^<(\vec k,\omega)$ at fixed $\vec k$ as a function of $\omega$:
\begin{equation}
1/\tau =\Delta\omega \; .
\end{equation}
In a Fermi liquid, so long as the quasiparticle excitation is well defined
({\it i.e.} the decay rate is small compared to the binding energy)
this width is related via the Fermi velocity to the peak width $\Delta k$ 
in the ``momentum distribution curve'' (MDC).  
This curve is defined as a cross section of $A^<(\vec k,\omega)$
taken at constant binding energy,  $\omega$. 
Explicitly 
\begin{equation}
\Delta\omega=v_F\Delta k \; .
\label{widths}
\end{equation}

\begin{figure}[!!!ht]
\begin{center}
\epsfig{figure=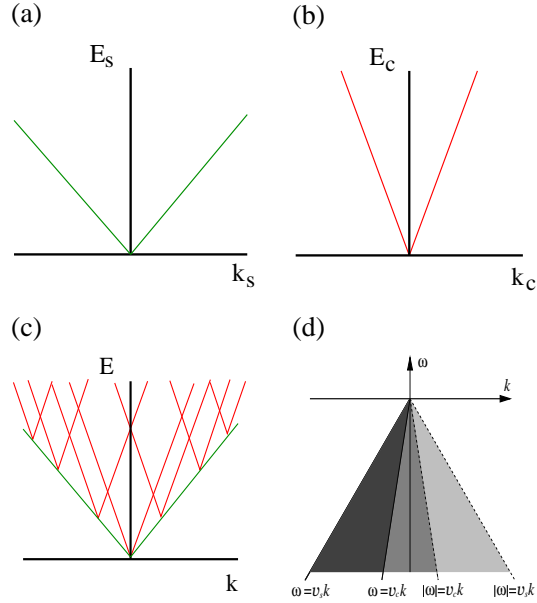,width=0.6\linewidth}
\end{center}
\caption{\label{fig1d5}  Kinematics of the 1DEG:
(a) Dispersion of the spin excitations. (b) Dispersion of the charge 
excitations. (c) The available electronic states. 
(d) Kinematic constraints on the spectral function: $A^<(k,\omega)$ 
for the 1DEG is nonzero at $T=0$ only in the shaded region of the 
$(k,\omega)$ plane. In the spin rotationally invariant case, $K_s=1$, 
$A^<(k,\omega)=0$ in the lightly shaded region, as well. If in addition, 
$K_c=1$, $A^<(k,\omega)=0$ outside of the darkest region. 
We have assumed $v_c>v_s$, which is usually the case in realistic systems.}
\end{figure}

A very different situation occurs in the theory of the 1DEG where 
\marginpar{\em There are no stable excitations of the 1DEG with quantum 
numbers of an electron.} 
the elementary excitations, charge and spin density waves, do not have the 
quantum numbers of a hole.   Despite the fact that the
elementary excitations are bosons, they give 
rise to a linear in $T$ specific heat that is not qualitatively 
different from that of a Fermi liquid.  However,
because of the separation of charge and spin, the creation of 
a hole (or an electron) necessarily implies the creation of 
two or more elementary excitations, of which one or more carries 
its spin and one or more carries its charge. Consequently, $A^<(k,\omega)$ 
does not have a pole contribution, but rather consists of a multiparticle 
continuum which is distributed over a wide region in the $(k,\omega)$ plane. 
The shape of this region is determined predominantly by the kinematics. 
The energy and momentum of an added electron are distributed between the 
constituent charge and spin pieces. In the case where both of them 
are gapless [see Figs.~\ref{fig1d5}(a) and \ref{fig1d5}(b)] this means
\begin{eqnarray}
\label{kinematics}
\nonumber
E&=&v_c|k_c|+v_s|k_s| \; , \\ 
k&=&k_c+k_s \; ,
\end{eqnarray}

\noindent
where energy and momentum are measured relative to $E_F$ and $k_F$ 
respectively. Consequently any point above the dispersion curve of 
the slower excitation (taken here to be the spin) may be reached by 
placing an appropriate amount of energy and momentum into the spin 
degrees of freedom, and the remaining energy and momentum into the 
charge degrees of freedom, as shown in Fig.~\ref{fig1d5}(c). The 
addition of a hole is similarly constrained kinematically, and the 
corresponding zero temperature ARPES response has weight only 
within the shaded regions of Fig.~\ref{fig1d5}(d).  

Further constraints on the distribution of spectral weight may arise from 
symmetries. In the spin rotationally invariant case, at the fixed point 
$K_s=1$, the spin correlators do not mix left and right moving pieces. 
As a consequence, $A^<(T=0)$ for a right moving hole vanishes  
when $\omega$ is in the range 
$v_s k\le |\omega|\le v_c k$ 
(assuming $v_s< v_c$ and $k>0$),
even if the kinematic conditions are satisfied;  
See Fig.~\ref{fig1d5}(d).\footnote{While the kinematic constraints are 
symmetric under $k\to -k$, the dynamical considerations are not, since 
although we have shifted the origin of $k$, we are in fact considering a 
right moving electron, 
{\it i.e.} one with momentum near $+k_F$.} If in addition $K_c=1$, so that 
the charge piece also does not mix left and right movers, $A^<(T=0)$ vanishes 
unless $k<0$ and $v_s |k|\le|\omega|\le v_c |k|$, (the darkest region in 
Fig.~\ref{fig1d5}(d)). While $K_s=1$ is fixed by symmetry, there is no reason 
why $K_c$ should be precisely equal to 1. However, if the interactions are 
weak, ({\it i.e.} if $K_c$ is near 1) most of the spectral weight is still
concentrated in this region. It spreads throughout the rest of the triangle 
with increasing interaction strength. 

Clearly, the total width of the MDC is bounded by kinematics and is at most
\marginpar{\em A dichotomy between sharp MDC's and broad EDC's is a telltale sign 
of electron fractionalization.}   
$\Delta k_{max}=2|\omega|/{\rm min}(v_c,v_s)$. Any peak in the MDC will have 
a width which  equals a fraction of this, depending on the interactions 
and symmetries of 
the problem, but in any case will vanish as the Fermi energy is approached. 
By contrast, at $k=0$, the shape of the EDC is not given by the kinematics 
at all, but is rather determined by the details of the matrix elements 
linking the one hole state to the various multi particle-hole states 
which form the continuum. In this case, the spectrum has a nonuniversal 
power law behavior with exponents determined by the interactions in the 1DEG.
Whenever such a dichotomy between the MDC and EDC is present, it can be
taken as evidence of  electron fractionalization \cite{frac}.

These general considerations can be substantiated by examining the explicit 
expression for the spectral function of the Tomonaga-Luttinger model 
\cite{Luther74,Meden92,Voit93,spectral}. The quantum criticality and the 
spin-charge separation of the model imply a scaling form for its correlation 
functions 
\begin{equation}
\label{Aform}
A^<(k,\omega) \propto T^{2(\gamma_c+\gamma_s)+1}\int dq\, d\nu \, 
G_c(q,\nu) \, G_s(\tilde{k}-rq,\tilde{\omega}-\nu) \; ,
\end{equation}
\sindex{gs}{$G_s$}{Spin piece of the one hole spectral function}
\sindex{gc}{$G_c$}{Charge piece of the one hole spectral function}
where we introduce the velocity ratio $r=v_s/v_c$ and define the 
scaling variables
\begin{equation}
\label{scalevar}
\tilde k=\frac{v_s k}{\pi T} \;\;\;\; , \;\;\;\; 
\tilde\omega=\frac{\omega}{\pi T} \; .
\end{equation}

\begin{figure}[!!!ht]
\begin{center}
\epsfig{figure=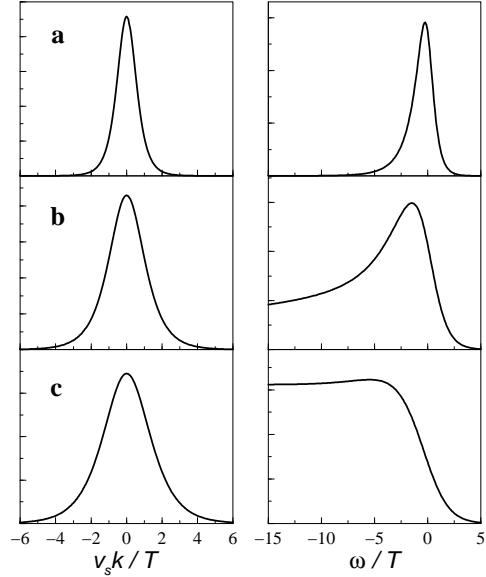,width=0.55\linewidth}
\end{center}
\caption{\label{fig1d6}
MDC's at $\omega=0$ (left) and EDC's at $k=0$ (right), for a spin 
rotationally invariant Tomonaga-Luttinger liquid, with $v_c/v_s=3$ and 
a) $\gamma_c=0$, 
b) $\gamma_c=0.25$, 
and c) $\gamma_c=0.5$.}
\end{figure}
\sindex{zzgammac}{$\gamma_c$}{Charge Luttinger exponent}
\sindex{zzgammas}{$\gamma_s$}{Spin Luttinger exponent}

\begin{figure}[ht!!!]
\begin{center}
\epsfig{figure=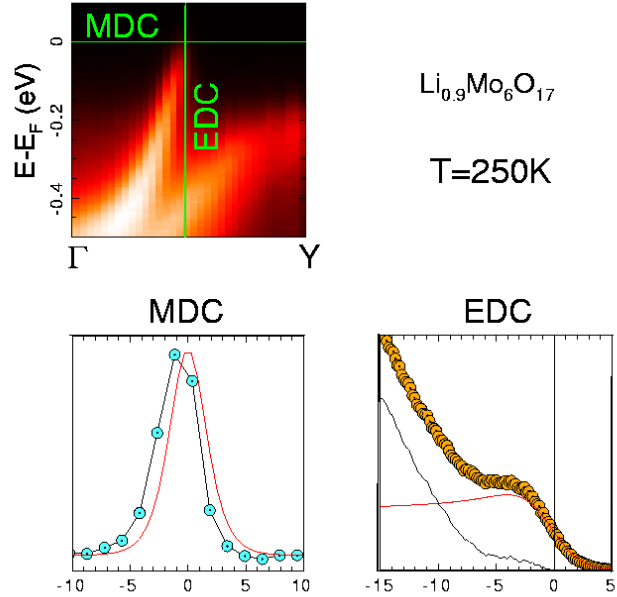,width=0.8\linewidth}
\end{center}
\caption{\label{allen}
ARPES intensity map for the purple bronze Li$_{0.9}$Mo$_9$O$_{17}$. The lower 
left panel depicts the MDC at the Fermi energy together with a 
Tomonaga-Luttinger theoretical curve. The lower right panel contains the EDC 
at the Fermi wavevector. The red line corresponds to the Tomonaga-Luttinger 
result and the black curve is its deviation from the experimental data. 
(From Ref.~\citen{allennature}.)}
\end{figure}

\noindent
Since the spin and charge sectors are formally invariant under separate 
Lorentz transformations, the functions $G_\alpha, (\alpha=c,s)$ 
also split into right  and left moving parts
\begin{equation}
\label{gk}
G_\alpha(k,\omega)=\frac{1}{2}h_{\gamma_\alpha+\frac{1}{2}}
\left(\frac{\omega+k}
{2}\right)h_{\gamma_\alpha}\left(\frac{\omega-k}{2}\right) \; ,
\end{equation}
where $h_\gamma$ is expressed via the beta function
\begin{equation}
\label{hk}
h_\gamma(k)={\rm Re}\left[(2i)^{\gamma}B\left(\frac{\gamma-ik}{2},1-\gamma
\right)\right] \; ,
\end{equation}
and the exponents 
\begin{equation}
\label{gammadef}
\gamma_\alpha=\frac{1}{8}(K_\alpha+K_\alpha^{-1}-2) \;,
\end{equation}
are defined so that $\gamma_\alpha=0$ for noninteracting fermions.

Fig.~\ref{fig1d6} depicts MDC's at the Fermi energy ($\omega=0$) and EDC's at 
the Fermi wavevector ($k=0$) for a spin rotationally invariant ($\gamma_s=0$) 
Tomonaga-Luttinger model for various values of the parameter $\gamma_c$. 
While the MDC's broaden somewhat with increasing interaction strength they 
remain relatively sharp with a well defined peak structure. On the other 
hand any corresponding structure in the EDC's is completely wiped out in the 
presence of strong interactions. Such behavior has been observed in ARPES 
studies of quasi-one dimensional compounds as depicted in Fig.~\ref{allen} 
as well as in the cuprate high temperature superconductors \cite{frac}. 

Away from the Fermi energy and Fermi 
wavevector and for not too strong interactions the peaks in the MDC and EDC 
split into a double peak structure, one dispersing with $v_s$ and the other 
with $v_c$. If observed this can be taken as further evidence for spin-charge
separation. 

We now turn to the interesting case in which the superconducting
susceptibility is enhanced due to the opening of a spin gap, $\Delta_s$. At
temperatures large compared to $\Delta_s$, the spin gap can be ignored, and the
spectral function is well approximated by that of the Tomonaga-Luttinger
liquid.  However, even below the spin gap scale, many of the characteristics
of  the Tomonaga-Luttinger spectral function are retained. Spin-charge 
separation still holds in the spin gapped Luther-Emery liquids and there  are
no stable ``electron-like'' excitations. The charge excitations are still 
the gapless charge density waves of the Tomonaga-Luttinger liquid but the
spin  excitations now consist of massive spin solitons with dispersion 
$E_s(k)=\sqrt{v_s^2 k^2+\Delta_s^2}$. As a result the spin piece of the 
spectral function is modified and from kinematics it follows that it 
consists of a coherent one spin soliton piece and an incoherent 
multisoliton part
\begin{equation}
\label{eq:Gs}
G_{s}(k,\omega)=  Z_s(k) \delta[\omega+E_{s}(k)]
+G_{s}^{(multi)}(k,\omega) \; ,
\end{equation}
\sindex{zs}{$Z_s$}{Coherent spin soliton weight}
where the multisoliton piece is proportional (at $T=0$) to 
$\Theta[-\omega-3E_{s}(k/3)]$. (For $K_s<1/2$ formation of spin 
soliton-antisoliton bound states, ``breathers'', may shift the threshold 
energy for multisoliton excitations somewhat). 
The form of $Z_s(k)$ has been calculated explicitly \cite{tsvelikcond}, 
but a simple scaling argument gleans the essential physics \cite{dimxover}. 
It follows from the fact that the Luther-Emery 
liquid is asymptotically free that at high energies and short 
distances compared to the spin gap, the physics looks the same as in the 
gapless state. Therefore the dependence of the correlation functions on 
high energy physics, such as the short distance cutoff $a$, cannot 
change with the opening of the gap. Since in the gapless system $G_s$ 
is proportional to $a^{2\gamma_s-1/2}$, it is a matter of dimensional 
analysis to see that 
\begin{equation}
\label{zsgeneral}
Z_s(k)=(\xi_s/a)^{\frac{1}{2}-2\gamma_s}f_s(k\xi_s) \;,
\end{equation}
where $f_s$ is a scaling function and $\xi_s=v_s/\Delta_s$ is the spin 
correlation length. 
\sindex{zzxis}{$\xi_s$}{Spin correlation length}

\marginpar{\em The Luther-Emery liquid is a pseudogap state.}
Despite the appearance of a coherent piece in the spin sector, 
the spectral function (Eq.~\ref{Aform}) still exhibits an overall incoherent 
response owing to the convolution with the incoherent charge part. The result 
is grossly similar to the gapless case, aside from the fact that the Fermi 
edge (the tip of the triangular support of $A^<$ in Fig.~\ref{fig1d5}d) 
is pushed back from the Fermi energy by the magnitude of the spin gap 
(thus rounding the tip of the triangle).  If,
as suggested in Section~\ref{general}, the Luther-Emery liquid is the
paradigmatic example of a pseudogap state, clearly the above spectral
function gives us an impression of what to expect the signature of the
pseudogap to be in the one electron properties.

\subsection{Dimensional crossover in a quasi-1D superconductor}
\label{dimensional}

Continuous global symmetries cannot be spontaneously broken in one 
dimension, even at $T=0$. Since the one dimensional Hamiltonian
(Eq.~(\ref{hamiltonian1}))  is invariant under translations and spin $SU(2)$ and
charge $U(1)$  transformations, no CDW, SDW, or superconducting long range
order can exist in its ground state. 
Therefore, in a
quasi-one dimensional  system made out of an array of coupled 1DEG's, a
transition into an ordered  state necessarily signifies a dimensional
crossover at which, owing to  relevant interchain couplings, phases of
individual chains lock together\cite{dimxover,guinea}. The ultimate low 
temperature fate of the system is fixed by the identity of the first phase 
to do so. This, in turn, is determined by the relative strength of the 
various couplings and the nature of the low energy correlations in the 1DEG.  

In the spin gapped phase, which we consider in the rest of this section, both
the CDW and  the superconducting susceptibilities are enhanced.  To begin
with, we will analyze the simplest model of a quasi-one dimensional
superconductor.  We defer until the following  section any serious
discussion of the competition between CDW and superconducting order.  We will
also defer until then any discussion of the richer possibilities which arise
when the quasi-one dimensional physics arises from a self-organized
structure, {\it i.e.} stripes, with their own additional degrees of
freedom.  

\subsubsection{Interchain coupling and the onset of order}

The simplest and most widely studied model of a quasi-one dimensional
spin gapped fluid is
\begin{eqnarray}
{\cal H} = &&  \sum_j {\cal H}_j + {\cal J} \sum_{<i,j>}
[O_{SS}^{\dagger}(i,x)O_{SS}(j,x) + {\rm H.C.} ] \nonumber  \\ 
&& + {\cal V} \sum_{<i,j>}
[O_{CDW}^{\dagger}(i,x)O_{CDW}(j,x) + {\rm H.C.} ] \; ,
\label{2dhamiltonian}
\end{eqnarray}
where ${\cal H}_j$ describes the Luther-Emery liquid on chain,
pairs of nearest neighbor chains are denoted $<i,j>$, and
$O_{\alpha}(j,x)$ is the appropriate order parameter field on chain $j$. The
bosonized form of these operators is given in Eqs. (\ref{bCDW}) and
(\ref{bSS}), above.  It is assumed that the interchain couplings,
${\cal J}$ and
${\cal V}$, are small compared to all intrachain energies.  
There are two more or less complementary ways of approaching this problem: 
\sindex{cdw}{CDW}{Charge density wave}
\sindex{sdw}{SDW}{Spin density wave}
\sindex{ts}{TS}{Triplet superconductivity}
\sindex{ss}{SS}{Singlet superconductivity}

\paragraph{\bf 1)}  The first is to perform a perturbative renormalization 
group (RG) analysis about the decoupled fixed point, {\it i.e.} compute the 
beta function perturbatively in powers
of the interchain couplings. To lowest order, the beta function is simply 
determined by the
scaling dimension,
$D_{\alpha}$, of each operator -- if $D_{\alpha} <2$, it means that
\sindex{dzzalpha}{$D_{\alpha}$}{Scaling dimension of $O_{\alpha}$}
$O_{\alpha}$ is perturbatively relevant, and otherwise it is irrelevant. It
turns out that the CDW and SC orders are dual to each other, so that
\begin{equation}
D_{SS} = 1/K_c \ \ \ \ , \ \ \ \ D_{CDW}=K_c \; .
\end{equation}
This has the implication that one, or the other, or both of the interchain
couplings is always relevant.  From this, we conclude with a high level of
confidence that at low temperature, even if the interchain couplings are
arbitrarily weak, the system eventually undergoes a phase transition to a
higher dimensional ordered state. An estimate of $T_c$ can be derived from
these equations in the standard way, by identifying the transition
temperature with the scale at which an initially weak interchain coupling
grows to be of order 1.  In this way, for $D_{SS}<2$, one obtains an
estimate of the superconducting transition temperature
\begin{equation}
T_c \sim E_F [{\cal J}/E_F]^{1/(2-D_{SS})} = {\cal J}[{\cal J}/E_F]^{(D_{SS}-
1)/(2-D_{SS})},
\end{equation} 
and similarly for the CDW ordering temperature.  Note that as
$D_{SS}\to 2^-$, $T_c\to 0$, and that $T_c \gg {\cal
J}$ for $D_{SS}<1$.  Clearly, the power law dependence of $T_c$ on
coupling constant offers the promise of a high $T_c$  when compared
with the exponential dependence in BCS theory.  

\paragraph{\bf 2)}   The other way is to use interchain
mean field theory\cite{scalapino}. Here, one treats the
one dimensional fluctuations exactly, but the interchain couplings in
mean field theory.  For instance, in the case of interchain SS ordering, one
considers each chain in the presence of an external field
\begin{equation}
{\cal H}^{eff} = {\cal H}_j + [ \Delta_{SS}^* O_{SS}(j,x) + {\rm H.C.}] \; ,
\label{mfhamiltonian}
\end{equation}
where $\Delta_{SS}$ is determined self-consistently:
\begin{equation}
\Delta_{SS}=z{\cal J} \langle O_{SS}(j,x)\rangle \; ,
\end{equation}
where $z$ is the number of nearest neighbor chains and the expectation value 
is taken with respect to the effective Hamiltonian. 
This mean field theory is exact\cite{dimxover,arrigoni1} 
in the limit of large $z$, and is expected
to be reliable so long as the interchain coupling is weak. It can be shown to 
give exact results in the limit of extreme anisotropy for the Ising model, 
even in two dimensions (where $z=2$) \cite{scalapino}. 
More generally, it is a well controlled approximation at least for 
temperatures $T\gg
{\cal J}$ (which includes temperatures in the neighborhood of $T_c$ as long as 
$D_{SS}<1$). 
This approach gives an estimate of
$T_c$ which is related to the susceptibility of the single chain, 
\begin{equation}
1=z{\cal J} \chi_{SS}(T_c) \; ,
\end{equation}
which, from the expression in Eq. (\ref{chiofT}), can be seen to produce 
qualitatively
the same estimate for $T_c$ as the perturbative RG treatment.  The advantage
of the mean field treatment is not only that it gives an explicit, and very
physical expression for $T_c$, but that it permits us to compute explicitly
the effect of ordering on various response functions, including the one
particle spectral function.  The case of CDW ordering is a straightforward extension.

\subsubsection{Emergence of the quasiparticle in the ordered state}

The excitation spectrum changes dramatically below $T_c$ when
\marginpar{\em Superconducting order 
binds fractionalized excitations into 
``ordinary'' quasiparticles.}
the interchain ``Josephson'' coupling ${\cal J}$  triggers long range
order\cite{dimxover}. The fractionalized excitations of the 
Tomonaga-Luttinger and the Luther-Emery liquids are replaced by new 
excitations with familiar ``BCS'' quantum numbers. Formally, 
superconducting order leads to a confinement phenomenon. While the  
spin gap in the Luther-Emery state already implies suppressed fluctuations 
of $\phi_s$ on each chain, and correspondingly a finite amplitude 
$\cos(\sqrt{2\pi}\phi_s)$ of the superconducting order parameter, 
it is the interchain Josephson coupling
that tends to lock its phase $\theta_c$ from one chain to the next.

Operating with the hole operator, Eq. (\ref{bidentity}), on the ground state 
at the position of the $j$th chain creates a pair of kinks (solitons) of 
magnitude $\sqrt{\pi/2}$ in both the charge and spin fields $\theta_c$ 
and $\phi_s$ of this chain. As a result the phase of the order parameter 
changes by $\pi$ upon passing either the spin or the charge soliton. 
This introduces a negative Josephson coupling between the affected chain 
and its neighbors along the entire distance between the charge 
and spin solitons. The energy penalty due to this frustration 
grows linearly with the separation between the solitons and 
causes a bound pair to form. In fact, all solitonic excitations 
are confined into pairs, including charge-charge and spin-spin pairs. The 
bound state between the charge and the spin pieces restores the electron, 
or more precisely the Bogoliubov quasiparticle, as an elementary excitation, 
causing a coherent (delta function) peak to appear in the single particle  
spectral function.  

\begin{figure}[!!!ht]
\begin{center}
\epsfig{figure=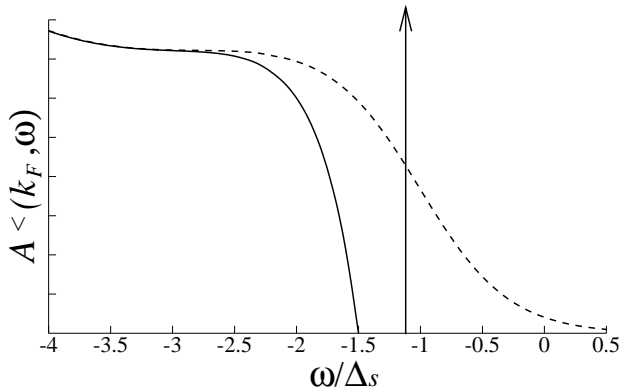,width=0.7\linewidth}
\end{center}
\caption{\label{fig1d7}
The temperature evolution of the spectral function. The dashed line
depicts 
$A^<$ at intermediate temperatures below the spin gap $\Delta_s$ but above 
$T_c$. The solid line represents the spectral function at zero temperature. 
A coherent delta function peak onsets near $T_c$ at energy 
$\Delta_0=\Delta_s+\Delta_c(0)/2$. The multiparticle piece starts at a 
threshold $2\Delta_c(0)$ away from the coherent peak.} 
\end{figure}

An explicit expression for the spectral function in the superconducting 
state can be obtained in the context of the effective Hamiltonian in Eq. 
(\ref{mfhamiltonian}): 
\begin{equation}
\label{ASC}
A^<(k,\omega)=Z(k)\delta[\omega-E(k)]+A^{(incoherent)}(k,\omega) \; ,
\end{equation}
where $E(k)=\sqrt{v_s^2 k^2+\Delta_0^2}$. Here $\Delta_0=\Delta_s+\Delta_c/2$ 
is the creation energy of the bound state where $\Delta_c\propto \Delta_{SS}$ 
is  the mean field gap ($\Delta_c \ll\Delta_s$) that opens in the 
charge sector below $T_c$ \cite{dimxover}.  
The multiparticle incoherent piece has a threshold slightly above the 
single hole threshold at $\omega=E(k)+2\Delta_c$. The shape of $A^<(k,\omega)$
at $T=0$ is presented schematically in Fig.~\ref{fig1d7}.

Once again, we may employ the asymptotic freedom of the system to construct a 
scaling argument. In this case, high energy physics dependent upon either
the cutoff 
or the spin gap (which is by assumption much larger than $T_c$) cannot change 
upon entering the superconducting state. Comparing the form of the spectral 
response in the normal spin gapped state with that of the superconductor 
reveals the weight of the coherent peak
\begin{equation}
\label{Z}
Z(k)=Z_s(0)(\xi_c/a)^{-\frac{1}{2}-2\gamma_c}f(k\xi_c) \;,
\end{equation}
\sindex{z}{$Z$}{Coherent quasiparticle weight}
where $f$ is a scaling function and $\xi_c=v_c/\Delta_c$ is the charge 
\sindex{zzxic}{$\xi_c$}{Charge correlation length}
correlation length. Physically, the dependence of the weight on $\Delta_c$, 
which also equals the (local) superfluid density \cite{dimxover}, reflects the 
fact that the superfluid stiffness between chains controls the strength of 
the bound state forming the quasiparticle.

Since the superfluid density is a rapid function of temperature upon 
entering the superconducting state, the weight of the coherent peak 
will also rapidly increase as the temperature is lowered. Because the 
Josephson coupling is weak, the energy of the bound state is largely 
set by the spin gap, so that the energy of the coherent peak will not be 
a strong function of temperature in the neighborhood of $T_c$. 
Likewise, since the gap is not
changing  rapidly, the scattering rate and therefore the lifetime of the new 
quasiparticle will not have strong temperature dependence either. All of the 
above signatures have been observed in ARPES measurements of the coherent 
peak in \BSCCO  \cite{fedorov,shenprb,feng,arpesding} and \YBCO  
\cite{ybcoanis}.

\marginpar{\em The temperature evolution of the spectral function is in marked 
contrast with that in a BCS superconductor} 
The behavior we have just described is in sharp 
contrast to that of a conventional superconductor, where the gap opens 
precisely at $T_c$. Since in that case the gap is a rapid
function of temperature, so is the energy of the conventional quasiparticle
peak. Moreover, scattering processes are rapidly gapped out upon entering the
BCS superconducting state, so that the quasiparticle often sharpens 
substantially as the temperature is lowered below $T_c$. Most importantly, 
in the conventional case, quasiparticles exist above the transition 
temperature, so the intensity ($Z$ factor) of
the peak does not change much upon entering the superconducting state. 
By contrast, in a quasi-one dimensional superconductor, there are no
quasiparticle excitations in the normal state. The existence of the
quasiparticle is due to the dimensional crossover to the three dimensional
state, and is an entirely collective effect!

\begin{figure}[ht!!!]
\begin{center}
\epsfig{figure=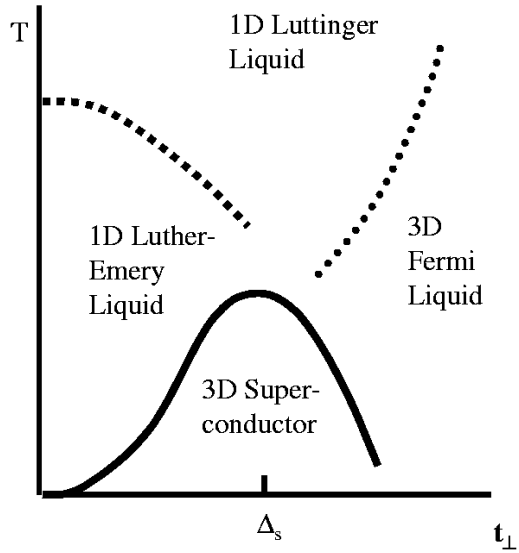,width=0.6\linewidth}
\end{center}
\caption[Two Routes to Dimensional Crossover]{\label{crossover}
{\em Two Routes to Dimensional Crossover.} 
In an array of multicomponent 1DEG's, 
for temperatures large compared to
the transverse single particle tunnelling, $t_\perp$, 
the system behaves as a collection of independent (1D) Luttinger Liquids.
For weak $t_{\perp}$, the dimensional crossover may proceed as described in 
\sindex{tperp}{$t_{\perp}$}{Interchain single particle tunnelling}
Section~\ref{dimensional}, with a crossover first to a (1D) Luther-Emery
Liquid, and a lower temperature dimensional crossover to a (3D)
superconductor.  For large $t_\perp$, there may be a dimensional crossover
into a (3D) Fermi liquid, before the system becomes a (3D) superconductor.
}
\end{figure}

\subsection{Alternative routes to dimensional crossover}

Until now, we have assumed that the spin gap is large compared to the
interchain couplings, and this assumption leads inevitably to the existence
of a quasi-1D pseudogap regime above $T_c$ and a dimensional
crossover associated with the phase ordering at $T_c$.  Since under some
circumstances, the spin gap in 1D can be zero or exponentially small compared
to $E_F$, it is possible for a system to be quasi-1D, in the sense that the
interchain couplings are small compared to the intrachain interactions, and
yet have the dimensional crossover occur above any putative spin gap scale. 
In this case, most likely the dimensional crossover is triggered by the
relevance of the interchain, single particle hopping operator---since any
spin gap is negligible, the previous argument for its irrelevance is
invalidated. What this means is that there is a dimensional crossover, $T^{*}_{3D}$,
at which the system transforms from a Luttinger liquid at high temperatures
to a Fermi liquid at lower temperatures.  (See Fig.~\ref{fig:ourview}.)
If there
are residual effective attractive interactions, the system will ultimately
become a superconductor at still lower temperatures.  However, in this case,
the transition will be more or less of the BCS type---a Fermi surface
instability (albeit on a highly anisotropic Fermi surface) with well defined
quasiparticles existing both above and below $T_c$.

\marginpar{\em The case where dimensional crossover to a Fermi liquid occurs 
well above $T_c$ may serve as a model for the overdoped cuprates.}
The crossover from a Luttinger liquid to a Fermi liquid is not as
well characterized, theoretically, as the crossover to a superconductor.  
The reason is that no simple form of interchain mean field theory can be 
employed to study it. Various energy scales associated with the crossover 
can be readily obtained from a scaling analysis. A recent interesting 
advance\cite{arrigoni1,giamarchidmft,orgdimxover} has been made on this 
problem using ``dynamical mean field theory,'' again based on the idea of 
using $1/z$ (where $z$ is the number of neighboring chains) as a small
parameter, which gives some justification for a widely used RPA-like 
approximation for the spectral function \cite{tsvelikbook}. 
However, there are still serious shortcomings with this
approximation\cite{arrigoni1,yin}. Clearly more interesting 
work remains to be done to sort out the physics in this limit, 
which may be a caricature of the physics of the overdoped cuprates. 
More complicated routes to dimensional crossover can also
be studied\cite{matsprl}, relevant to systems with more than 
one flavor of chain. For instance, it has recently been found 
that it is possible for a two component quasi-1D system to produce 
a superconducting state which supports gapless ``nodal quasiparticles,'' 
even in the limit of extreme anisotropy \cite{matsprl}.

\section{Quasi-1D Physics in a Dynamical Stripe Array}
\label{smectic}

\addtocontents{toc}{

\noindent{\em An interesting generalization of the quasi-1D system occurs
when the background geometry on which the constituent 1DEG's reside is,
itself, dynamically fluctuating.  This situation arises in conducting
stripe phases. } }

As mentioned before, in the simplest microscopic realizations of  
\marginpar{\em Competition between CDW and SS  is  key  in quasi-1D 
systems.}
the 1DEG with repulsive interactions, $0<K_c<1$ and hence the CDW 
susceptibility is the most divergent as $T\rightarrow 0$ 
(See Eq. (\ref{ensus}).) This seemingly implies that the typical 
fate of a quasi-one dimensional system with a spin gap is to wind 
up a CDW insulator in  which CDW modulations on neighboring 
chains phase lock to each other.  And, indeed, many quasi-one dimensional
metals in nature suffer precisely this fate---the competition between CDW
and SS order is a real feature of quasi-1D systems. Of course, as shown 
in Fig. \ref{fig1d2}, above, the $K_c$ inequality need not be satisfied 
in more complicated realizations of the 1DEG.  

What we will examine in this section is another way
in which the balance between CDW and SS ordering can 
be affected.\cite{kfe,prl,recoil} 
Specifically, we will show below that transverse fluctuations of the
backbone on which the quasi-1D system lives significantly enhance the
tendency to SS while suppressing CDW ordering. Such fluctuations 
are unimportant in conventional quasi-one dimensional solids, where 
the constituent molecules, upon which the electrons move, have a large 
mass and a rigid structure. But 
when the 1DEG's live along highly quantum electronic textures,
or ``stripes,'' transverse stripe fluctuations are probably always large.

\subsection{Ordering in the presence of quasi-static stripe fluctuations}
Consider a two dimensional array of stripes that run along the $x$ direction, 
and imagine that there is a 1DEG which lives on each stripe.
To begin with, we will consider the case in which the stripe fluctuations are
sufficiently slow that they can be treated as static---in other words, we
consider an array of imperfectly ordered stripes, over whose meanderings we
will eventually take an equilibrium (annealed) average. We will use a
coordinate system in which points on the stripes are labeled by the
coordinate $x$, the stripe number $j$, and in which transverse displacements
of the stripe in the $y$ direction are labeled by $h_j(x)$. We therefore 
ignore the  possibility of overhangs which is a safe assumption in the 
ordered state.  
\sindex{hjx}{$h_{j}(x)$}{Transverse stripe displacement}

We now consider the effect that stripe geometry fluctuations have on the 
inter-stripe couplings. Because the CDW order (and any other 
$2k_F$ or $4k_F$ orders) occurs at a large wave vector, the geometric 
fluctuations profoundly affect its phase: 
\begin{equation}
\label{newcdw}
O_{CDW}(j,x)=\frac{e^{-2i k_F L_j(x)}}{\pi a}\cos[\sqrt{2\pi}\phi_s(j,x)]
e^{-i\sqrt{2\pi}\phi_c(j,x)} \; ,  
\end{equation}
where
\begin{equation}
\label{arclength}
L_j(x)=\int_0^x dx'\sqrt{1+(\partial_{x'} h_j)^2} \; , 
\end{equation}
is the arc length, {\it i.e.} the distance measured along stripe $j$ to point 
$x$. At the same time $O_{SS}$  is unchanged, as are
other $k=0$ orders. 
This results in a fundamental difference in the way CDW and Josephson 
inter-stripe couplings evolve with growing stripe fluctuations. 
\sindex{ljx}{$L_{j}(x)$}{Arc length}

The CDW and Josephson couplings between neighboring stripes are of the form 
\begin{eqnarray}
\nonumber
H_{{\cal V}}&=&\frac{1}{(2\pi a)^2}
\sum_j\int dx {\cal V}(\Delta_j h)\cos[\sqrt{2\pi}\phi_s(j,x)]
\cos[\sqrt{2\pi}\phi_s(j+1,x)] \\
\label{cdwcoupling}
&&\hspace{2cm} \times
\cos[\sqrt{2\pi}\Delta_j\phi_c+2k_f\Delta_j L] \; ,\\
\nonumber
H_{\cal J}&=&-\frac{1}{(2\pi a)^2}
\sum_j\int dx {\cal J}(\Delta_j h)\cos[\sqrt{2\pi}\phi_s(j,x)]
\cos[\sqrt{2\pi}\phi_s(j+1,x)] \\
\label{josephson}
&&\hspace{2cm} \times\cos[\sqrt{2\pi}\Delta_j\theta_c] \; ,
\end{eqnarray}  
where $\Delta_j h\equiv h(j+1,x)-h(j,x)$ etc. The coupling constants 
${\cal V}(\Delta_j h)$ and ${\cal J}(\Delta_j h)$, depend on the local 
spacing between adjacent stripes, since they are more strongly coupled when 
they are close together than when they are far apart. This is particularly 
important for the Josephson coupling which depends on the pair tunnelling 
amplitude and therefore roughly exponentially on the local spacing between 
the stripes
\begin{equation}
\label{joscoupl}
{\cal J}(\Delta_j h)\approx {\cal J}_0 e^{-\alpha\Delta_j h} \; .
\end{equation}

By integrating out the stripe fluctuations $h$ one obtains the effective 
Hamiltonian of an equivalent rigid system of stripes. To first order 
in ${\cal V}$ the CDW coupling is similar to Eq. (\ref{cdwcoupling}) but 
with $\Delta_j L$ set equal to 0 in the last term and 
${\cal V}(\Delta_j h)$ 
replaced by 
\begin{equation}
\label{newV}
\langle {\cal V}(\Delta_j h)\rangle\exp[-2k_F^2\langle(\Delta_j L)^2\rangle ] \; ,
\end{equation}
\marginpar{\em Stripe fluctuations dephase CDW order...}
where $\langle\ \rangle$ signifies averaging over transverse stripe 
fluctuations. Since $\Delta_j L= L_{j+1}(x)-L_j(x)$ is a sum of contributions 
with random signs, which are more or less independently distributed along the 
distance $|x|$, we expect it to grow roughly as in a random walk, {\it i.e.} 
$\langle(\Delta_j L)^2\rangle\sim D|x|$, where $D$ is a constant.
Indeed one can show that at finite temperature 
$\langle(\Delta_j L)^2\rangle\sim T|x|$ while at $T=0$ 
$\langle(\Delta_j L)^2\rangle\sim \hbar\bar{\omega}\log|x|$, where 
$\hbar\bar{\omega}$ is a
suitable measure of the transverse stripe zero point energy. As a result of 
this dephasing effect, coupling between CDW's vanishes rapidly except in
a  narrow region near the ends of the stripes and hence can be ignored in
the  thermodynamic limit. 
In short, transverse stripe fluctuations cause
destructive interference of $k \ne 0$ order on neighboring chains,
strongly suppressing those orders.

The effects of stripe fluctuations on the Josephson coupling can be 
analyzed in the same way. To first order in the inter-stripe coupling, 
${\cal J}(\Delta_j h)$ is simply replaced by its average value, 
$\bar {\cal J}\equiv<{\cal J}(\Delta_j h)>$.  
\marginpar{\em ... but they enhance SS order.}
In other words, once quasi-static stripe fluctuations are integrated out, 
the result is once again the Hamiltonian we studied in 
Eq. (\ref{2dhamiltonian}), above, but with ${\cal V}=0$ and 
${\cal J}=\bar{\cal J}$.  Moreover, due to the exponential dependence of 
${\cal J}(\Delta_j h)$ on $(\Delta_j h)$, it is clear that 
$\bar {\cal J} > {\cal J}(0)$, {\it i.e.}  transverse stripe
fluctuations strongly enhance the Josephson coupling between stripes. 
(There is a similar enhancement of the CDW coupling but it is overwhelmed 
by the dephasing effect.) Physically, this enhancement reflects the fact 
that the mean value of ${\cal J}$ is dominated by regions
where  neighboring stripes come close together.  
In the case of small amplitude fluctuations,
this enhancemnt can be viewed as an inverse Debye-Waller factor,  
\begin{equation}
\label{newJ}
\langle {\cal J}\rangle\approx {\cal J}_0 
e^{\frac{\alpha^2}{2}\langle(\Delta_j h)^2 \rangle} \; .
\end{equation}
Where the transverse stripe fluctuations are comparable
in  magnitude to the inter-stripe spacing, the mean Josephson coupling is 
geometrically determined by the mean density of points at which neighboring 
stripes actually ``bump'' (i.e. separated by one lattice constant $a$). 
In this limit, treating the stripe fluctuations as a random walk yields 
the estimate
\begin{equation}
{\cal J} \sim \left(\frac{a}{R}\right)^2 {\cal J}_0 \; , 
\end{equation}
where $R$ is the mean distance between stripes.

\subsection{The general smectic fixed point}
  
The quasi-static limit discussed above is presumably inadequate at low enough
temperatures, where the quantum dynamics of stripe fluctuations must always
be relevant. The complete problem, in which both the stripe dynamics and the 
dynamics of the 1DEG's are treated on an equal footing remains unsolved. 
However, since in a crystalline background, the stripe fluctuations are 
typically not gapless, we expect that at low enough temperatures, 
the stripe fluctuations can be treated as fast, and be integrated 
out to produce new effective interactions. So long as the stripes are 
reasonably smooth, these induced interactions will consist of long 
wavelength (around $k=0$) density-density and current-current interactions 
between the neighboring Luttinger liquids---interactions that we have ignored 
until now. These interactions should undoubtedly be present in the bare model, 
as well, even in the absence of stripe fluctuations. They are marginal 
operators and should be included in the fixed point action
\cite{prl,carpentier}. We are still interested in the spin gapped case so in
the following analysis consider the charge sector only. Consequently we drop
the subscript $c$  from the various quantities.

Using Eq. (\ref{chargedensity}) and the bosonization formula
for the  current density along the chain, $-\sqrt{\frac{2}{\pi}}v K \partial_x\theta$,
the phase-space Lagrangian density for $N$ coupled chains is
\begin{equation}
\label{smecticlag}
{\cal L}=\sum_j\partial_x\theta_j\partial_t\phi_j -
\frac{1}{2}\sum_{j,j'=1}^N [\partial_x\phi_j
\tilde W_0(j-j')\partial_x\phi_{j'} + \partial_x\theta_j
\tilde W_1(j-j')\partial_x\theta_{j'} ] \; . 
\end{equation}
The diagonal terms $(j=j')$ in Eq. (\ref{smecticlag}) describe the 
decoupled system with $\tilde W_0(0)=v/K$ and $\tilde W_1(0)=v K$. The off
diagonal  terms preserve the smectic symmetry
$\phi_j(x)\rightarrow\phi_j(x)+\alpha_j$  and
$\theta_j(x)\rightarrow\theta_j(x)+\beta_j$  (where $\alpha_j$ and
$\beta_j$ are constant on each stripe) of the  decoupled Luttinger
fluids. Whenever this symmetry is unbroken, the $2k_F$  charge density
profiles and the superconducting order parameters on each  stripe can
slide relative to each other without an energy cost. 
\marginpar{\em The fixed point is an ``electron smectic''.}
This Hamiltonian thus describes a general ``smectic metal
phase.'' It is smectic in the sense that it can flow and has no resistance to 
shear, but it has a broken translational symmetry in the direction 
transverse to the stripes--- broken by the stripe array itself. 
Similar ``sliding'' phases of coupled  classical two dimensional
$XY$ models have also been discussed \cite{lubensky22}.

The Lagrangian density in Eq. (\ref{smecticlag}) can be simplified by
integrating out the dual fields, and expressing the result  in terms of the
Fourier transform of $\phi_a$ with respect to the chain index, 
$\phi_a=\frac{1}{\sqrt{N}}\sum_{k_\perp}e^{i k_\perp a}
\phi(k_\perp)$: 
\begin{equation}
\label{FTsmectic}
\nonumber
{\cal L}=\sum_{k_\perp}\frac{1}{2} \kappa(k_\perp)\left[\frac{1}{v(k_\perp)}
|\partial_t\phi(k_\perp)|^2-v(k_\perp)|\partial_x\phi(k_\perp)|^2\right] \; ,
\end{equation} 
where $\tilde W(a)=\frac{1}{N}\sum_{k_\perp}e^{i k_\perp    a} 
W(k_\perp)$ so that  
the smectic fixed point is characterized by the  $k_{\perp}$
dependent velocities  and inverse Luttinger parameters
\begin{eqnarray}
\label{effectvel}
v(k_\perp)&=&\sqrt{W_0(k_\perp)W_1(k_\perp)} \; , \\
\label{effectK}
\kappa(k_\perp)&=&\sqrt{W_0(k_\perp)/W_1(k_\perp)} \; .
\end{eqnarray}
Alternatively, in terms of the dual fields,
\begin{equation}
\label{FTsmecticdual}
\nonumber
{\cal L}=\sum_{k_\perp}\frac{1}{2\kappa(k_\perp)}\left[\frac{1}{v(k_\perp)}
|\partial_t\theta(k_\perp)|^2-v(k_\perp)|\partial_x\theta(k_\perp)|^2\right] 
\; .
\end{equation} 
\sindex{vkperp}{$v(k_\perp)$}{Velocity at the smectic fixed point}
\sindex{zzkappakperp}{$\kappa(k_\perp)$}{Luttinger parameter at the smectic fixed point}

\begin{figure}[ht!!!]
\begin{center}
\epsfig{figure=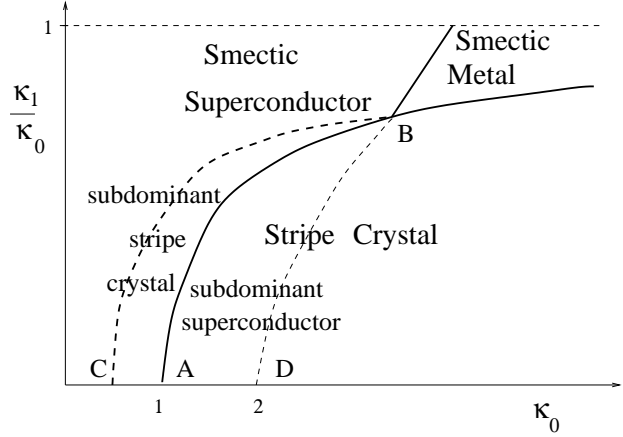,width=0.7\linewidth}
\end{center}
\caption{\label{fig1d4}
Phase diagram of a spin gapped stripe array with model interactions as 
discussed in the text.}
\end{figure}

In the presence of a spin gap, single electron tunnelling is irrelevant, 
and the only potentially relevant interactions involving pairs of stripes 
are singlet tunnelling and the coupling between the CDW order 
parameters, {\it i.e.}, Eqs. (\ref{josephson}, \ref{cdwcoupling}) with the 
cosine terms involving the spin fields replaced by their vacuum expectation 
values and with $\Delta_j L$ and $\Delta_j h$ set equal to 0. The scaling 
dimensions of these
perturbations can be readily  evaluated \cite{prl,carpentier}:
\begin{eqnarray}
\label{dimSC}
D_{SC}&=&\int_{-\pi}^{\pi}\frac{d k_\perp}{2\pi}\kappa(k_\perp)
(1-\cos k_\perp)=\kappa_0-\frac{\kappa_1}{2} \; , \\
\label{dimCDW}
D_{CDW}&=&\int_{-\pi}^{\pi}\frac{d k_\perp}{2\pi}\frac{1}{\kappa(k_\perp)}
(1-\cos k_\perp)=\frac{2}{\kappa_0-\kappa_1+\sqrt{\kappa_0^2-\kappa_1^2}} \; .
\end{eqnarray}
To be explicit, in the above, we have 
\marginpar{\em Long wavelength couplings suppress CDW even more.}
(for purposes of illustration) evaluated the integrals
for the simple model in which 
$\kappa(k_\perp)=\kappa_0+\kappa_1\cos k_\perp$.  Here
$\kappa_0$ can be thought of as the intra-stripe inverse Luttinger 
parameter and $\kappa_1$ is a measure of the nearest neighbor inter-stripe 
coupling. For stability, $\kappa_0>\kappa_1$ is required.  
Comparing the scaling dimensions in Eqs. (\ref{dimSC}) and (\ref{dimCDW}) one 
obtains the phase diagram which is presented in Fig.~\ref{fig1d4}. 
The line AB is a line of first order transitions between the smectic 
superconductor and the electronic crystal. It terminates at a 
bicritical point from which two continuous transition lines emanate.   
They separate the smectic superconductor and the crystal from a 
strong coupling regime where both Josephson tunnelling and CDW coupling are 
irrelevant at low energies. In this regime the smectic metal is stable.

An important lesson from this model is that inter-stripe long wavelength 
interactions rapidly increase the scaling dimension of the inter-stripe 
CDW coupling while the scaling dimension of the Josephson coupling is less 
strongly affected (in this model it is actually reduced). Indeed one 
can see from Fig.~\ref{fig1d4} that there is a region of $\kappa_0\geq 1$  
and large enough $\kappa_1$ where the global order is superconducting 
although in the absence of inter-stripe interactions ($\kappa_1\sim 0$) 
the superconducting fluctuations are sub-dominant. 

Extensions of this model to a three dimensional array of chains 
\cite{slidingkane} and the inclusion of a  magnetic field 
\cite{kun} have been considered as well.  In particular, it is found that
the magnetic field supresses the region of superconducting order in the
phase diagram in Fig. (\ref{fig1d4}), thus expanding the regime in which
the smectic metal is stable.  Similar considerations lead one  naturally to
consider other states obtained when the stripe fluctuations  become still more
violent. Assuming that the long range stripe order is  destroyed by such
fluctuations, while the short distance physics remains that of quasi-1DEG's
living along the locally  defined stripes, one is led to investigate the
physics of electron  nematic and stripe liquid phases. We shall return to this
point in the final section.

\section{Electron Fractionalization in $D>1$ as a Mechanism of High
Temperature Superconductivity}

\addtocontents{toc}{
\noindent{\em Spin-charge separation seems to offer an attractive route to 
high temperature superconductivity. It occurs robustly in 1D, but is now 
known to occur in higher dimensions as well, although seemingly only under 
very special circumstances.  } }
\label{fraction}

We briefly discuss here a remarkable set of ideas for a novel mechanism of 
high temperature superconductivity based on higher dimensional generalizations 
of the 1D notion of spin-charge separation. Boasting a high pairing scale as 
well as crisp experimental predictions, these theories have many attractive 
features. They also bear a strong family resemblance to the 
``spin gap proximity effect mechanism,'' which we develop in 
some detail in Section~\ref{sgpfx}.  
These appealing ideas, while valid, require the proximity of a spin 
liquid phase which in turn appears to be a fragile state of matter; 
for this reason, and others which will be made clear  below, 
it is our opinion that these ideas are probably not applicable 
to the cuprate superconductors. The discussion in this section is 
therefore somewhat disconnected from the development in the rest of the paper. 
We merely sketch the central ideas, without providing any derivations.
There are a number of  recent papers dealing with this 
subject to which the interested reader can refer; see 
Refs.~\citen{myreview,senthilsreview,senthilfisher,sondhifradkin,fradkinbook}.

\subsection{RVB and spin-charge separation in two dimensions}

Immediately following the discovery of high temperature
superconductivity\cite{mueller}, Anderson proposed\cite{pwa87} that the key
to the problem lay in the occurrence of a never before documented state of
matter (in $D>1$), a spin liquid or ``resonating valence bond'' (RVB) state, 
related to a state he originally proposed\cite{fazekas} for quantum 
antiferromagnets on a triangular (or similarly frustrating) lattice. 
In this context\cite{sondhifradkin}, a spin liquid is defined to be an 
insulating state with an odd  number of electrons per unit cell 
(and a charge gap) which breaks neither spin rotational nor translational 
symmetry.  Building on this proposal, Kivelson,
Rokhsar, and Sethna\cite{krs} showed that a consequence of the existence
of such a spin liquid state is that there exist quasiparticles with reversed
charge spin relations, just like the solitons in the 1DEG discussed in Section 
\ref{1D}, above. Specifically, there exist charge 0 spin
1/2  ``spinons'' and charge e spin 0 ``holons.''  Indeed, these quasiparticles
were recognized as having a topological character\cite{krs,laughscience}
analogous to that of the Laughlin quasiparticles in the quantum Hall effect. 

There was a debate at the time concerning the proper exchange statistics, with
proposals presented identifying the holon as a boson\cite{krs,bza}, a
fermion\cite{read}, and a semion\cite{kalmeyer}.  It is now
clear that all sides of this debate were correct, in the sense that there is
no universal answer to the question. The statistics of the fractionalized 
quasiparticles is dynamically determined, and  is sensitive to a form of 
``topological order''\cite{wentopo,senthil,holonstatistics,read,myreview} 
which differentiates various spin liquids.  
There are even transitions between states in which the holon has different 
statistics\cite{holonstatistics,kimnayak}. 

Two features of this proposal are particularly attractive:

{\bf 1)} It is possible to envisage a high pairing scale in the Mott 
insulating parent state, since the strong repulsive interactions
between electrons, which result in the insulating behavior, are insensitive 
to any subtler correlations between electrons. Thus, the ``$\mu^*$ issue'' 
does not arise:  the spin liquid can be viewed as an insulating liquid of
preformed cooper pairs\cite{pwa87,krs,rk}, or equivalently a superconductor 
with zero superfluid density.\footnote{An oxymoron since in this case
$T_{\theta}=T_{c}=0$, but the intuitive notion is clear:
we refer to a state which is derived from a superconductor by taking
the limit of zero superfluid density while holding the pairing scale fixed.}
If this pairing scale is somehow preserved upon doping, then the transition 
temperature of the doped system is determined by superfluid stiffness and is 
not limited by a low pairing scale, as it would be in a BCS superconductor. 
Indeed, as in the case of the 1D Luther-Emery liquid
discussed in Section~\ref{1D}, pairing becomes primarily 
a property of the spin degrees of freedom, and involves
little or no pairing of actual charge.

{\bf 2)} When the holons are bosonic, their density directly determines the
superfluid density. Thus the superconducting $T_c$ can be crudely viewed as 
the bose condensation temperature of the holons. The result is that for small 
concentration of doped holes $x$\cite{pwa87}, the transition temperature is 
proportional to a positive power of $x$ (presumably\cite{krs}
$T_c \sim x$ in 2D), in contrast to the exponential dependence on parameters 
in a BCS superconductor.  

In short, many of the same features that would make a quasi-1D system 
attractive from the point of view of high temperature superconductivity 
(see Sections~\ref{1D} and \ref{strong}) would make a doped spin liquid 
even more attractive. However, there are both theoretical and
phenomenological reasons for discounting this idea in the context of 
the cuprates.

\subsection{Is an insulating spin liquid ground state possible in $D>1$?}

\marginpar{\em Is this simply angels dancing on the head of a pin?}
The most basic theoretical issue concerning the applicability of 
the fractionalization idea is whether a  spin liquid state occurs
at all in $D>1$. The typical consequence of the Mott physics is an
antiferromagnetically ordered (``spin crystalline'') state, especially the
N\'{e}el state, which indeed occurs at $x=0$ in the cuprates. 
Moreover, the most straightforward quantum disordering of an
antiferromagnet will lead to a spin Peierls state, rather than 
a spin liquid, as was elegantly demonstrated by Haldane\cite{haldane4} and Read and
Sachdev\cite{rs}.  Indeed, despite many heroic efforts, the theoretical ``proof of
principle,''  {\it i.e.} a theoretically tractable microscopic model with plausible
short  range interactions which exhibits a spin liquid ground state phase, was
difficult to achieve. A liquid is an intermediate phase, between solid 
and gas, and so cannot readily be understood in a
strong or weak coupling limit\cite{sondhifradkin}.  

Very recently, Moessner and Sondhi\cite{shivaji} have managed to demonstrate 
just this point of principle! They have considered a model\cite{rk} on a 
triangular lattice (thus returning very closely to the
original proposal of Anderson) which is a bit of a
caricature in the sense that the constituents are not single electrons, but
rather valence bonds (hard core dimers), much in the spirit pioneered by
Pauling.\footnote{Indeed, it is tempting to interpret the dimer model as 
the strong coupling, high density limit  of a fluid of Cooper pairs\cite{rk}.} 
The model is sufficiently well motivated microscopically,
and the spin liquid character robust enough, that it is
reasonable to declare the  spin liquid a theoretical possibility.   
The spin liquid state of Moessner and Sondhi does not break any 
obvious symmetry.
\footnote{This work was, to some extent, anticipated in studies of  large $N$
generalizations of the Heisenberg antiferromagnet.\cite{rs}}

That said, the difficulty in finding such a spin liquid ground state in
model calculations is still a telling point. A time reversal 
\marginpar{\em Spin liquids are fragile.}
invariant insulating state cannot be adiabatically connected to a
problem of noninteracting quasiparticles with an effective 
band structure\footnote{In a time reversal symmetry broken state, 
the band structure need not
exhibit the Kramer's degeneracy,  so that a weak coupling state with
an odd number of electrons per unit cell is possible.}---band 
insulators always have an even number of electrons per unit cell. Thus, an 
insulating spin liquid is actually quite an exotic state of matter.  
Presumably, it only occurs when all more obvious types of ordered 
states are frustrated, {\it i.e.} those which break spin rotational symmetry, 
translational symmetry, or both.  The best indications at present
are that this occurs in an exceedingly small corner of model space,
and that consequently  spin liquids are likely to be rather delicate 
phenomena, if they occur at all in nature. This, in our
opinion, is the basic theoretical reason for discarding this 
appealing idea in the cuprates, where high temperature 
superconductivity is an amazingly robust phenomenon.

\marginpar{\em The cuprates appear to be doped spin crystals, not doped spin 
liquids.}
One could still imagine that the insulating state is magnetically ordered, as 
indeed it is in the cuprates, but that upon doping, once the magnetic order 
is suppressed, the system looks more like a doped spin liquid than a doped 
antiferromagnet. In this context, there are a number of phenomenological 
points about the cuprates that strongly discourage this
viewpoint. In the first place, the undoped system is not only an ordered 
antiferromagnet, it is a nearly classical one: 
its ground state and elementary excitation
spectrum\cite{chn1,chn2,aepplihighenergy,birgeneau} are quantitatively 
understood using lowest order spin wave theory. This state is as far from a 
spin liquid as can be imagined! Moreover, even in the doped system, 
spin glass and other types of magnetic order are seen to persist up to 
(and even into) the superconducting state, often with frozen moments with 
magnitude comparable to the ordered moments in the undoped
system\cite{birgeneau,budnick,neidermeyer,panagopolis}. 
These and other indications show that the doped system ``remembers'' 
that it is a doped antiferromagnet, rather a doped spin liquid.

Regardless of applicability to the cuprates,
\marginpar{\em Where to look for spin liquids}
it would be worthwhile to search for materials that do exhibit spin 
liquid states, and even more so to look for superconductivity when they 
are doped. Numerical studies\cite{tsvelikkagome,spinliq1,spinliq2,spinliq3} 
indicate that good candidates for this are electrons on a triangular lattice 
with substantial longer range ring exchange interactions, such as may
occur in a 2D Wigner crystal near to its quantum melting point\cite{cknv}, 
and the Kagom\'{e} lattice.  It is also possible, as discussed in 
Section~\ref{numerical}, to look for superconductivity in systems 
that exhibit some form of spin-charge separation at
intermediate length scales.  (See also Ref.~\citen{C60}.)  

\subsection{Topological order and electron fractionalization}
\label{topological}

Finally, we address the problem of
classifying phases in which true electron fractionalization occurs, 
{\it e.g.} in which spinons are deconfined. It is now clear from the 
work of  Wen\cite{wentopo} and Senthil and Fisher\cite{senthil} that 
the best macroscopic characterization of fractionalized phases in
two or more dimensions is topological, since  they frequently possess 
no local order parameter. Specifically, a fractionalized phase 
exhibits certain predictable ground state degeneracies on various 
closed surfaces---degeneracies which Senthil and Fisher have given
a physical interpretation in terms of ``vison expulsion.'' Unlike the 
degeneracies associated with conventional broken symmetries, these 
degeneracies are not lifted by small external fields which break either 
translational or spin rotational symmetry.  It has even
been shown\cite{holonstatistics,sachdevsenthil,senthil} 
(as funny as this may sound) that topological order is amenable 
to experimental detection. Once topological classification is accepted, 
the one to one relation between spin liquids and electron
fractionalization, implied in our previous discussion, is eliminated. 
Indeed, it is possible to imagine\cite{senthil,nayakfisherdual} ordered 
(broken symmetry) states, proximate to a spin liquid phase, which will 
preserve the ground state degeneracies of the nearby spin
liquid, and hence will exhibit spin-charge separation.

\section{ Superconductors with Small Superfluid Density}
\label{phase}

\addtocontents{toc}{
\noindent{\em In contrast to the case of conventional superconductors, 
in superconductors with small
superfluid density, fluctuations of the phase of the superconducting
order parameter affect the properties of the system in profound ways.}
}

A hallmark of BCS theory is that pairing precipitates order.
But it is possible for the two phenomena to happen separately: 
pairing can occur at a higher temperature than superconductivity.  
In this case, there is an intermediate temperature range described 
by electron pairs which have not condensed. In the order parameter language, 
this corresponds to a well developed amplitude of the order parameter, 
but with a phase which varies throughout the sample.  
Superconductivity then occurs with the onset of long range phase coherence.
(This is how ordering occurs in a quasi-1D superconductor, as
discussed in Section \ref{1D}, above.)  Such superconductors, while they
may have a large pairing scale, have a small stiffness to phase
fluctuations, or equivalently a small superfluid density.

\subsection{What ground state properties predict $T_c$?}

When the normal state is understood, it is reasonable to describe 
superconductivity as an instability of the normal state as temperature 
is  lowered, which BCS theory does quite successfully in simple metals.
Another approach, useful especially when the normal state is not well 
understood, is to consider which thermal fluctuations degrade the 
superconducting order as the temperature is raised.
Put another way, we address the question, ``What measurable ground state 
($T=0$) properties permit us to predict $T_c$?''

Two classes of thermal excitations are responsible for disordering
the ground state of a superconductor: amplitude fluctuations of the
complex order parameter (associated with pair breaking), 
and fluctuations of the phase (associated with pair currents).

The strength of the pairing at $T=0$ is quantifiable as a typical gap value,
\marginpar{\em Pairing is one energy scale...}
$\Delta_0$, 
where 
\be
T_p \equiv \Delta_0/2 \; ,
\label{Tp}
\ee
is the characteristic temperature at which the pairs fall apart.  
In a BCS superconductor, it is possible to estimate that $T_c \approx T_p$.
(The factor 1/2 in this definition approximates the weak coupling BCS
expression, $T_c=\Delta_0/1.78$.) Certainly, more generally, $T_p$ marks
a loose upper bound to $T_c$, since if there is no pairing, there
is probably no superconductivity.

We can construct another ground state energy scale as follows:
Divide the sample into blocks of linear dimension, $L$, and ask 
how much energy it costs to flip the sign of the superconducting
order parameter at the center of one such region. So long as $L$
is larger than the coherence length, $\xi_0$, the cheapest way to 
do this is by winding the phase of the order parameter, so the
energy is determined by the superfluid phase stiffness   
\marginpar{\em ...the superfluid phase stiffness sets another.}
\be
T_{\theta}= \frac{1}{2}A \gamma L^{d-2} \; ,
\label{Ttheta}
\ee
where $d$ is the number of spatial dimensions, $A$ is a 
geometry dependent dimensionless number of order 1 and 
the ``helicity modulus'', $\gamma$, is traditionally expressed 
in terms of the ratio of the superfluid density, $n_{s}$, to the 
pair effective mass, $m^{*}$:
\be
\gamma \equiv {\hbar^2 n_{s} \over  m^*} \; .
\label{gamma}
\ee  
(We will discuss the quantitative aspects of this relation 
in Subsection \ref{XY}.) Note that for $d=2$, this
energy is independent of $L$, while for $d=3$ it is minimized for the
smallest
allowable value of $L\sim\xi_0$.  Clearly, when the temperature is
comparable to $T_{\theta}$, thermal agitation  will produce random phase
changes from block to block, and hence destroy any long range order. Again, a
rough upper bound to $T_c$ is obtained in this way.
\sindex{zzgamma}{$\gamma$}{Helicity modulus}
\sindex{ns}{$n_s$}{Superfluid density}
\sindex{ma}{$m^*$}{Pair effective mass}

\begin{table}
\begin{center}
\caption{}
\begin{tabular}{|c||c|c|c|c|c||c|}
\hline 
Material&
\( L \) {[}\AA{]}&
\( \lambda _{L} \){[}\AA{]} &
\( T_{p} \){[}K{]} &
\( T_{c} \){[}K{]} &
\( T_{\theta } \){[}K{]} &
Ref.\\
\hline
\hline 
Pb&
830&
390&
7.9&
7.2&
6\( \times 10^{5} \)&
\citen{meserveyinparks1969,nature17}\\
\hline 
Nb\( _{3} \)Sn &
60&
640&
18.7&
17.8&
2\( \times 10^{4} \)&
\citen{orlando1979}\\
\hline 
UBe\( _{13} \) &
140&
10,000&
0.8&
0.9&
\( 10^{2} \)&
\citen{naidyuk1998,nature19,nature20}\\
\hline 
Ba\( _{0.6} \)K\( _{0.4} \)BiO\( _{3} \)  &
40&
3000&
17.4&
26&
5\( \times 10^{2} \)&
\citen{sharifi1991,uemura91}\\
\hline 
K\( _{3} \)C\( _{60} \)  &
30&
4800&
26&
20&
\( 10^{2} \)&
\citen{gunnarsson1997,nature22,nature23}\\
\hline 
MgB\( _{2} \) &
50&
1400&
15&
39&
1.4\( \times 10^{3} \)&
\citen{alexandrov2001,schmidt2002,sologubenko2002}\\
\hline
\hline 
ET&
15.2&
8000&
17.4&
10.4&
15&
\citen{arai2000,uemuranature}\\
\hline 
NCCO&
6.0&
1600&
10&
21-24&
130&
\citen{nugroho1999,armitage2001}\\
\hline 
PCCO&
6.2&
2800&
23&
23&
86&
\citen{prozorov2000,biswas2001,pccodist}\\
\hline 
Tl-2201 (op)&
11.6&
&
122&
91&
&
\citen{ozyuzer1998}\\
\hline 
Tl-2201 (od)&
11.6&
2000&
&
80&
160&
\citen{niedermayer1993,uemuranature}\\
\hline 
Tl-2201 (od)&
11.6&
2200&
&
48&
130&
\citen{niedermayer1993,uemuranature}\\
\hline 
Tl-2201 (od)&
11.6&
&
26&
25&
&
\citen{kang1996} \\
\hline 
Tl-2201 (od)&
11.6&
4000&
&
13&
40&
\citen{uemuranature,niedermayer1993}\\
\hline 
Bi-2212 (ud) x=.11&
7.5&
&
275&
83&
&
\citen{miyakawa1998,renner}\\
\hline 
Bi-2212 (op)&
7.5&
&
220&
95&
&
\citen{miyakawa1998}\\
\hline 
Bi-2212 (op) &
7.5&
2700&
&
90-93&
60&
\citen{prozorov2000,niderost1998}\\
\hline 
Bi-2212 (op)&
7.5&
1800&
&
84&
130&
\citen{sllee1993,weber1993}\\
\hline 
Bi-2212 (od) x=.19&
7.5&
&
143&
82&
&
\citen{miyakawa1998}\\
\hline 
Bi-2212 (od) x=.225&
7.5&
&
104&
62&
&
\citen{miyakawa1998}\\
\hline 
Y-123 (ud) x=.075&
5.9&
2800&
&
38&
42&
\citen{bernhard2001}\\
\hline 
Y-123 (ud) x= .1&
5.9&
1900&
&
64&
90&
\citen{bernhard2001}\\
\hline 
Y-123 (op) x=.16&
5.9&
1500&
&
85.5&
140&
\citen{panagopoulos1998,bernhard2001}\\
\hline 
Y-123 (op) &
5.9&
&
116&
91-92&
&
\citen{maggio1995}\\
\hline 
Y-123 (od) x=.19&
5.9&
1300&
&
79&
180&
\citen{bernhard2001}\\
\hline 
Y-123 (od) x=.23&
5.9&
1500&
&
55&
140&
\citen{bernhard2001}\\
\hline 
Y-248&
6.8&
1600&
&
82&
150&
\citen{zimmermann1995}\\
\hline 
Hg-1201 (op)&
9.5&
1700&
192&
95-97&
180&
\citen{panagopoulos1998,wei1998}\\
\hline 
Hg-1212 (op)&
6.4&
1700&
290&
108&
130&
\citen{fabrega1999,wei1998}\\
\hline 
Hg-1223 (op)&
5.3 &
1500&
435&
132-135&
130&
\citen{fabrega1999,wei1998,panagopoulos1998}\\
\hline 
Hg-1223 (op)&
7.9&
1500&
&
135&
190&
\citen{fabrega1999,panagopoulos1998}\\
\hline 
LSCO (ud) x=.1&
6.6&
2800&
75&
30&
47&
\citen{fujimori2000,panagopoulos,lscodist}\\
\hline 
LSCO (op) x=.15&
6.6&
2600&
58&
38&
54&
\citen{fujimori2000,panagopoulos}\\
\hline
LSCO (od) x=.20&
6.6&
1950&
&
34&
96&
\citen{panagopoulos}\\
\hline
LSCO (od) x=.22&
6.6&
1900&
&
27&
100&
\citen{panagopoulos}\\
\hline
LSCO (od) x=.24&
6.6&
1900&
&
20&
100&
\citen{panagopoulos}\\
\hline
\end{tabular}
\\
(See next page for caption.)
\end{center}
\label{phasetable}
\end{table}

\begin{table}
{\small Caption for Table~1: {\em Zero temperature properties of the superconducting state as
predictors of $T_{c}$.}  
Here,  $T_p$ is computed from Eq.~(\ref{Tp}) using
values of $\Delta_0$ obtained from
either tunnelling or  ARPES, except for overdoped Tl-2201, for
which we have used Raman data.
In computing $T_{\theta}$ from Eq.~(\ref{Ttheta}) 
for nearly isotropic materials (those above the double line), we have
taken $d=3$, $A=2.2$, $L=\sqrt{\pi}\xi_0$, and $n_s/2m^{*}= 
(8\pi)^{-1}(c/e)^2\lambda_{L}^{-2}$ where $\lambda_{L}$ and $\xi_{0}$  are
the zero  temperature London penetration depth and
coherence length, respectively.
For layered materials, we have taken $d=2$, 
$A=0.9$, and the areal superfluid density
$n_s/2m^{*}= (8\pi)^{-1}(c/e)^2L\lambda_{L}^{-2}$  where $L$ is now
the mean spacing  between layers and $\lambda_{L}$ is the
in-plane London penetration depth.   The precise 
numerical values of $A$ and the factor of $\sqrt{\pi}$ should not 
be taken seriously---they depend on microscopic details, which can 
vary from material to material as discussed in Section {\protect 
\ref{XY}}.  
Penetration depth measurements on Y-123 refer to polycrystalline
Y$_{0.8}$Ca$_{0.2}$Ba$_2$Cu$_3$O$_{7-\delta}$, and report $\lambda_{ab}$.
The two entries for Hg-1223 assume that the superfluid density
resides in all three planes (L=5.3\AA), or the outer two planes
only (L=7.9\AA).
In the 
case of the high temperature superconductors, the notations `ud', `op', and 
`od' refer to under, optimally, and overdoped materials, respectively.} 
\end{table}

In short, it is possible to conclude on very general grounds that
\be
T_c \le min[T_p,T_{\theta}] \; .
\ee
When $T_p\ll T_{\theta}$,  
phase fluctuations can be completely
neglected except in
the immediate neighborhood of $T_c$---this is the case 
in BCS superconductors.  If $T_p \gg T_{\theta}$,
quasiparticle excitations, {\it i.e.} the broken Cooper pairs, play no
significant thermodynamic role up to $T_c$.  In this case
a considerable amount of local pairing, and consequently a pseudogap, must
persist to temperatures well above $T_c$.  When both $T_p$ and $T_{\theta}$
are comparable to $T_c$, as is the case in most optimally doped high
temperature superconductors, neither class of thermal excitation can be safely
neglected.

\marginpar{\em Of this there is no possible doubt
whatever.}
In Table 1, following Ref. \citen{nature}, we tabulate
$T_{\theta}$, $T_p$, and $T_c$ for various superconducting materials. 
Clearly, in bulk Pb, phase fluctuations are not
terribly important, while in the cuprate superconductors (and the ET
superconductors), phase fluctuations are an order 1 effect.  Of this there is
no possible doubt!  Looking more closely at the table, one sees that the ratio
of $T_{\theta}/T_c$ is generally smaller for the underdoped materials, and
larger for  overdoped, which implies that phase fluctuations are progressively
less dominant with increasing doping.  
The ratio of $T_p/T_c$ varies in the opposite manner with doping.  

The obvious implication 
of the trends exhibited in Table 1 is that optimal doping marks a 
gradual crossover from an underdoped regime, where $T_c$ is predominantly 
a phase ordering transition, to an overdoped regime in which it is 
predominantly a pairing transition. This also implies that both
pairing and phase fluctuation physics play a nonnegligible role, 
except in the regimes of extreme underdoping or overdoping where 
$T_c\rightarrow 0$.

\subsection{An illustrative example: granular superconductors}

We now turn to a beautiful set of experiments
carried out by Merchant {\it et al.}\cite{merchant} on granular
Pb films with a thin coating of Ag. This is a system in which the microscopic
physics is well understood. The $T_c$ of bulk Pb is 7.2K while Ag
remains normal down to the lowest accessible temperatures, so that
$T_{\theta}$ can be varied with respect to $T_p$ by changing the 
thickness of Ag. In this way, the system can be tuned from an ``underdoped''
regime, where $T_c$ is a phase ordering transition and pairing persists to much
higher temperatures, to an ``overdoped'' regime, where the transition is very
BCS-like.

\begin{figure}[ht!!!]
\begin{center}
\epsfig{figure=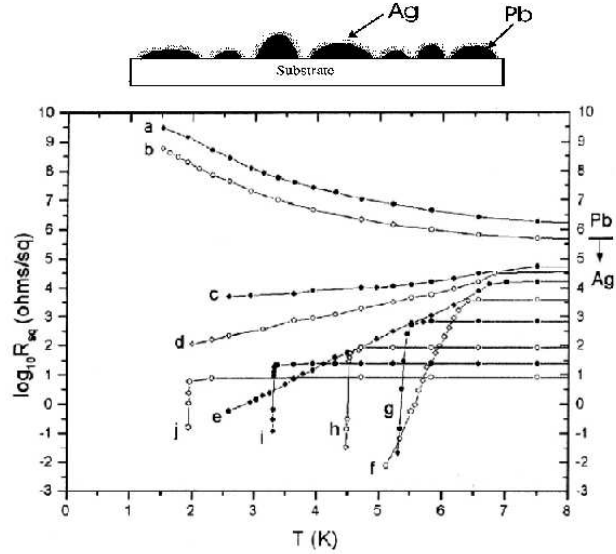,width=0.8\linewidth}
\end{center}
\caption[]
{\label{dynes5}
The logarithm of the resistance {\em vs.} temperature for a sequence of films, starting with
a granular Pb film (a) to which is added successively larger coverage
of Ag.  From Fig. 5 of Merchant {\it et al.}\protect{\cite{merchant}}.}
\end{figure}

\begin{figure}[ht!!!]
\begin{center}
\epsfig{figure=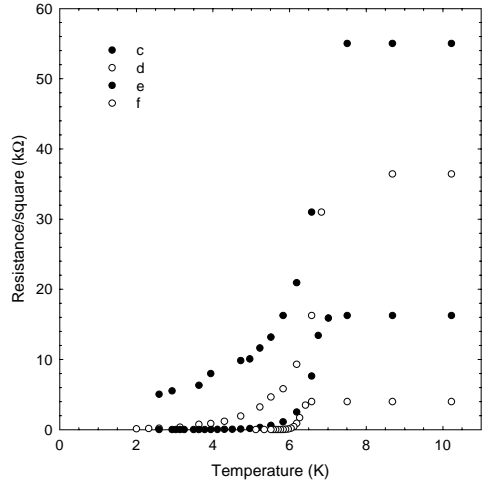,width=0.6\linewidth}
\end{center}
\caption[]
{\label{dyneslinear}
The same data as in Fig. \protect{(\ref{dynes5})}, but on a linear, as
opposed to a logarithmic, scale of resistivity.}
\end{figure}

Figure \ref{dynes5} shows the log of the resistance {\em vs.} temperature 
for a sequence of films (a-j) obtained by adding successive layers of Ag to a
granular Pb substrate. Films a and b are seen to be globally
insulating, despite being locally superconducting below $7.2K$.  
Films g-j are clearly superconductors.
Films c-f are anomalous metals of some still not understood variety.
It is important to note that Fig.~\ref{dynes5} is plotted on a 
log-linear scale, so that although it is unclear whether films
c-f will ever become truly superconducting, films e and f, for example,
have low temperature resistances which are $5$ or $6$ orders of magnitude
lower than their normal state values, due to significant 
superconducting fluctuations; see Fig.~\ref{dyneslinear}.

\begin{figure}[ht!!!]
\begin{center}
\epsfig{figure=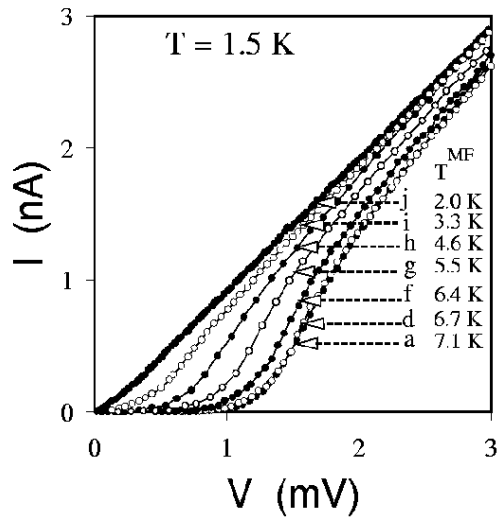,width=0.6\linewidth}
\end{center}
\caption[]
{\label{dynes6}
I-V curves from planar tunnelling into the same sequence of films
shown in Fig. \protect{(\ref{dynes5})}.  From Fig. 6 of Merchant {\it et
al}\protect{\cite{merchant}}.}
\end{figure}

Figure \ref{dynes6} shows I-V curves obtained
from planar tunnelling in the direction perpendicular to the same set of films.
As $dI/dV$ is proportional to the single particle density of states at energy
$V$, this can be interpreted as the analogue of an ARPES or tunnelling
experiment in the high temperature superconductors. 
Among other things, the gap seen in films a-d is roughly independent of Ag
coverage, and looks precisely like the gap that is seen upon tunnelling into
thick Pb films. In these films, the gap seen in tunnelling
is clearly a superconducting pseudogap. 

The analogy between the behavior of these films as a function of Ag coverage,
and the cuprate high temperature superconductors as a function of hole
concentration is immediately apparent:

With little or no Ag, the typical Josephson coupling, $J$, between far 
separated
\marginpar{\em ``$T_c$'' increases with increasing Ag...}
grains of Pb is small;  thermal phase fluctuations preclude any possibility of
long range phase order for $T > J$. Clearly, increasing Ag coverage increases
the coupling between grains, or more correctly, since the granular character
of the films is gradually obscured with increasing Ag coverage, it increases 
the phase stiffness or superfluid density. This causes the phase ordering
temperature to rise, much like the underdoped regime of the cuprates.

However, the pairing
scale, or equivalently the mean field $T_c$, is a decreasing function of Ag
\marginpar{\em ... and then $T_c$ decreases.}
coverage due to the proximity effect. Since the Pb grains are small compared
to the bulk coherence length, $\xi_0$, the granularity of
the films has little effect on the BCS gap equation. 
The pairing scale is equivalent to that of a homogeneous system with an 
effective pairing interaction, 
\be
\lambda^{eff}=\lambda_{Pb}\times f_{Pb}  + 0\times f_{Ag} \; ,
\ee
where $f_{Pb}$ and $f_{Ag}$ are, respectively, the volume fraction of Pb 
and Ag. Consequently, the pairing gap,
\be
\Delta_0\sim \exp[-1/(\lambda^{eff}-\mu^*)]
\ee
is a decreasing function of Ag coverage. So long as $f_{Ag}\ll 1$ (films a-f)
this effect is rather slight, as can be seen directly from the figures,
but then the gap value can be seen to plummet with increasing Ag
coverage.  In films g-j, this leads to a decrease of $T_c$,
reminiscent of overdoped cuprates.

Of course, it is clear that there is more going on in the experiment than 
this simple theoretical discussion implies:

1) {\it Disorder:} The effects of disorder 
\marginpar{\em Things we swept under the rug.}
are neglected in this discussion.
A priori these should be strong, especially at low Ag coverage.

2)  {\it Coulomb Blockade:}  As best one
can tell from the existing data, films a-f are not
superconductors with a reduced $T_c$---in fact films a and b appear 
to be headed toward an insulating
ground state, presumably due to quantum phase fluctuations
induced by the charging energy of the grains. The energy to
transfer a Cooper pair (charge 2e) between grains is
\be
V_{C}=4 \alpha e^2/L \; ,
\ee
where $L$ is the grain size and $\alpha$ is a dimensionless constant which
takes into account the grain shape and screening. 
When $V_C > J$, the number of pairs per grain becomes fixed 
at low temperature and the ground state is a type of paired Mott
insulator.
Since the number of pairs and the phase are
quantum mechanically conjugate on each grain,  
 when number fluctuations are suppressed by the
charging energy, quantum phase fluctuations flourish, and
prevent superconducting order.
The screening of the Coulomb interaction can mitigate this effect.
Screening clearly improves with increasing Ag coverage, so coverage
dependent effects of quantum phase fluctuations contribute to the
evolution observed in the experiments, as well.  

3)  {\it Dissipation:}  
There is even more to this story than the $\omega=0$
charging energies. In contrast with classical statistical mechanics,
the dynamics and the thermodynamics are inexorably
linked in quantum statistical mechanics, and
finite frequency physics becomes important.  
This issue has been addressed experimentally by Rimberg {\it et al.}
\cite{rimberg} While there has been considerable
progress in understanding the theory of quantum phase fluctuations (See, for
example, Ref.~\citen{kapitulnik} for a recent review), there are still many basic
issues that are unresolved. For instance, films c-f show
no sign of becoming truly superconducting or insulating as $T\rightarrow 0$!
What is the nature of this intermediate state?  
\marginpar{\em A mysterious ground state}
This is a widely observed phenomenon in systems which are expected to be
undergoing a superconductor to insulator transition\cite{mason,kapitulnik}. 
The physics of this anomalous metallic state is not understood {\em at all},
even in systems, such as the present one, where the microscopic physics is
believed to be understood.  (See Section~\ref{quantum} 
for a taste of the theoretical subtleties involved.)

\subsection{Classical phase fluctuations}
\label{XY}

We now undertake a critical analysis of thermal phase fluctuations. 
We will for now ignore the effects of thermal
quasiparticle excitations, as well as the quantum 
dynamics which certainly dominate the phase mode physics at 
temperatures low compared to its effective Debye temperature.
These important omissions will be addressed in Section~\ref{quantum}.

\subsubsection{Superconductors and classical XY models}

When $T_{\theta} \ll T_{p}$, the superconducting transition temperature
\marginpar{\em The superfluid density sets
the phase stiffness.}
$T_c \approx T_{\theta}$, and the transition can be well described by a 
phase only model.  
On general symmetry grounds,
the free energy associated with time independent deformations of the
phase must be of the form
\be
V_{phase} = (\gamma/2)\int d{\vec r} (\nabla \theta)^2 \; ,
\label{Fsc}
\ee
where the helicity modulus, $\gamma$, is given by the superfluid 
density, $n_s$, and the effective pair mass, $m^*$, 
according to Eq.~(\ref{gamma}). 
Since  $\vec v_s = {\hbar \over  m^*} \nabla \theta$ is the 
superfluid velocity, $V_{phase}$ is easily seen to have an 
interpretation as the kinetic energy of the superfluid, 
$V_{phase}=\int d\vec r n_{s}m^{*} v_{s}^{2}/2$,
so that classical phase fluctuations correspond to thermally 
induced pair currents.
Eqs.~(\ref{Fsc}) and~(\ref{gamma}) 
establish the sense in which the superfluid
density controls the stiffness to phase fluctuations.

Eq.~(\ref{Fsc}) is the continuum form of the classical XY model.
Both in a superconductor and in the XY model,
$\theta$ is a periodic variable (defined modulo $2\pi$).
Thus, we must handle the short distance physics with some care
to permit the vortex excitations which are the expression of that 
periodicity.  When this is done, typically by defining the model on
a lattice, it captures the essential physics of the 
transition between a low temperature ordered and a high temperature 
disordered state.  

To be concrete, let us consider an XY model on a $d$ dimensional 
hypercubic lattice
\be
H_{XY}= -\sum_{<i,j>}{\cal V}(\theta_{i}-\theta_{j}) \; ,
\label{Hxy}
\ee
where $<i,j>$ are nearest neighbor sites and ${\cal V}$ is an even, periodic 
function ${\cal V}(\theta)={\cal V}(\theta+2\pi)={\cal V}(-\theta)$, 
with a maximum at $\theta=0$ such that the Hamiltonian 
is minimized by the uniform state.  
The lattice constant, $a$, in this model has a physical interpretation---it 
defines the size of the vortex core.  
To generalize this model to the case of an
anisotropic ({\it e.g.} layered) superconductor, we let both the 
lattice constant, $a_{\mu}$, and the potential, ${\cal V}_{\mu}(\theta)$, 
depend on the direction, $\mu$.  

At zero temperature, the helicity modulus 
can be simply computed:
\be
\gamma_{\mu}(T=0)=2[a_{\mu}^{2}/\nu]
{\cal V}_{\mu}^{\prime\prime}(0) \; ,
\ee
where $\nu=(\prod_{\nu}a_{\nu})$ is the unit cell volume.
Thus, the relation between $\gamma(0)$ and $T_{\theta}$, the ordering
temperature of the model, depends both on
the detailed form of ${\cal V}$ and on the lattice cutoff.
In constructing Table 1 above, we have taken ${\cal V}=V\cos(\theta)$,
and identified the area of the vortex core, $\pi \xi_{0}^{2}$, with the 
plaquette area, $a^{2}$ - this is the origin of the somewhat 
arbitrary $\sqrt{\pi}$ which appears in the three dimensional 
expression for $T_{\theta}$. Fortuitously,
for layered materials, $\gamma_{x} = \gamma_{y}\equiv \gamma_{xy}$ 
depends only on the spacing 
between planes, $a_{z}$, and not on the in-plane lattice constant.

One can, in principle, handle the short 
distance physics in a more systematic way by solving the microscopic 
problem (probably numerically) on large systems 
(large compared to $\xi_{0}$), and then
matching the results with  the short distance behavior of the XY model.  
In this way, one could, in principle, derive explicit expressions for 
${\cal V}$ and $a_{\mu}$ in terms of the microscopic properties of
a given material.  However, no one (to the best of our knowledge) has 
carried through such an analysis for any relevant microscopic model.

What we\cite{classphs} have done, instead, is to keep at most the first 
2 terms in a 
\marginpar{\em How much does the detailed shape of ${\cal V}$ matter?}
Fourier cosine series of Eq.~(\ref{Hxy}).  With the cuprates in mind,
we have studied planar systems:
\ba
H= && -J_{\parallel} \sum_{<ij>_{\parallel}} \left\{ \cos(\theta_{ij}) 
+ \delta\cos(2\theta_{ij})\right \} \nonumber \\ 
&& -J_{\perp}\sum_{<ij>_{\perp}} \left\{
\cos(\theta_{ij}) \right \} \; ,
\label{hamiltonian}
\ea
where $<ij>_{\parallel}$ denotes
nearest neighbors within a plane, and 
$<ij>_{\perp}$ denotes
nearest neighbors between planes.
It is assumed that $J_{\parallel}$, $J_{\perp}$, and $\delta$ are 
positive, since there is no reason to expect any frustration
in the problem,\cite{spivak}
and that $\delta \le 0.25$, since for $\delta > 0.25$ there is 
a secondary minimum in the potential for $\theta_{ij} = \pi$,
which is probably unphysical. 
Since dimensional analysis arguments of 
the sort made above are essentially 
independent of $\delta$, varying $\delta$ permits us to obtain some 
feeling for how quantitatively robust the results are with regard
to ``microscopic details.''

\subsubsection{Properties of classical XY models}

The XY model is one of the most studied models in physics\cite{lubenskybook}.
We\cite{classphs} have recently carried
out a series of quantitative analytic and numerical studies of XY
models (using Eq.~(\ref{hamiltonian})).
In particular, we have focused on the thermal
evolution of the superfluid density and the relation between the
superfluid density and the ordering temperature.   

As long as $J_{\perp}$ is nonzero, this model is in the
universality class of the 3D XY model, and near enough to $T_c$,
$\gamma(T) \sim |T_c-T|^{\nu}$, where $\nu$ is the correlation
length exponent of the 3DXY model, $\nu \approx .67$.  
For sufficient anisotropy, there may be a crossover from 2D critical
behavior close (but not too close) to $T_c$, to 3D critical behavior
very near $T_c$.  
In practice, this crossover is very hard to see due to the
special character of the critical phenomena of the 2D XY model;
even a very weak $J_{\perp}$ significantly increases the transition
temperature.  

To see this, consider the case in which
$J_{\parallel} \gg J_{\perp}$;  
in this limit, one can study the physics of the
system using an asymptotically exact interplane mean field theory
\cite{scalapino}. We define the order parameter, 
$m(T) \equiv \langle \cos[\theta_j]\rangle$, and
consider the behavior of a single decoupled planar $XY$ model in the
presence of an external field, $h(T)=2J_{\perp} m(T)$ due to the
mean field of the neighboring two planes. The self-consistency
condition thus reads
\be
m(T)=m_{2D}(T,h) \; ,
\ee
where $m_{2D}(T,h)$ is computed for the 2D model.  
A simple estimate for $T_c$ can be obtained by linearizing this
equation:
\be
1=2J_{\perp}\chi_{2D}(T_c) \; .
\ee
\marginpar{\em 2D critical behavior may be hard to see.}
Here the 2D susceptibility is
\be
\chi_{2D}\sim T_{2D}^{-1}exp\left\{A_{\chi}\sqrt{T_{2D}/(T_c-T_{2D})}\right\} \; ,
\ee
where $T_{2D}$ is the Kosterlitz-Thouless transition temperature and
$A_{\chi}$ is a nonuniversal number of order $1$.
A consequence of this is that even a very small interlayer coupling
leads to a very large fractional increase in $T_c$
\be
T_c-T_{2D}\sim T_{2D}A_{\chi}^2/\log^2[J_{\parallel}/J_{\perp}] \; .
\ee 
Only if $(T_c-T_{2D})/ T_{2D}\ll 1$ will there be clear 2D critical
behavior observed in the thermodynamics.

To make contact with a range of experiments it is necessary that we
focus attention not only on universal critical properties, but also on
other properties which are at least relatively robust to changes in
microscopic details.  One such property is the width of the critical
region, but we are not aware of any systematic studies of the factors
that influence this.  For the simple ($\delta=0$) isotropic 3D XY model,
the critical region certainly does not extend further than $10\%$ away
from $T_c$.

\begin{figure}[ht!!!]
\begin{center}
\epsfig{figure=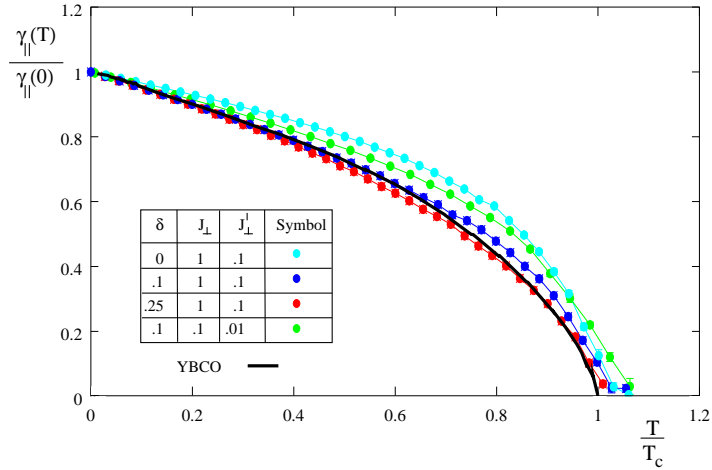,width=0.8\linewidth}
\end{center}
\caption[]
{\label{sfdens}
Superfluid density {\it vs.} temperature, scaled
by the zero temperature superfluid density and by $T_c$, respectively,
from Ref.~\citen{classphs}.
Experimental data on $YBCO$ is depicted by the black line, and
is taken from Kamal {\it et al}.\protect \cite{3dxy2}  
(The data are essentially the same for a range of doping
concentration.)  Our Monte Carlo results 
for system size $16\times16\times16$ are the filled symbols.
Calculations are for two planes per unit cell, with coupling
$J_{\parallel}=1$ within each plane, and $J_{\perp}$ and $J_{\perp}^{\prime}$ between alternate planes.  
Monte Carlo points above $T_{c}$ are nonzero due to finite size effects.
Except where explicitly shown, error bars are smaller than symbol size.}
\end{figure}

Another such property is the low temperature slope of 
superfluid density curves as a function of temperature.
Using linear spin wave theory\cite{stroud,coffey}, one can
\marginpar{\em The superfluid density is linear at low T.}
obtain a low temperature expansion of the in-plane helicity modulus,
\be
{\gamma_{\parallel}(T) \over a_{\perp}} = J_{\parallel}(1+4\delta)
-{\alpha(1-16\delta) \over 4(1+\delta)}T + {\cal{O}}(T^{2}) \; ,
\ee
where we have used $a_{x}=a_{y}\equiv a_{\perp}$ 
and $\gamma_{x}=\gamma_{y}\equiv \gamma_{\parallel}$ 
for a planar system and $\alpha$ is a nonuniversal number which depends on
$J_{\perp}/J_{\parallel}$.  It is easy to show\cite{classphs} that  
$\alpha =1$ in the two dimensional limit 
($J_{\perp}/J_{\parallel}=0$),  and that  $\alpha=2/3$ in the 
three dimensional limit ($J_{\perp}=J_{\parallel}(1+4\delta)$).
The $T$-linear term is independent of $J_{\parallel}$, 
so that we expect the slope of scaled superfluid density curves, 
$\gamma_{\parallel}(T)/\gamma_{\parallel}(0)$ {\it vs.} $T/T_c$, 
to be much less sensitive to microscopic parameters 
({\it i.e.} material dependent properties such as 
doping in the cuprates) than is $\gamma_{\parallel}(0)$. That 
this expectation is realized can be seen from our Monte Carlo 
simulation results presented in Fig.~\ref{sfdens}.

In addition, we find that there is a
{\it characteristic shape} to the superfluid density vs. temperature
curves in XY models.  
We have used Monte Carlo simulations to focus on two other
dimensionless nonuniversal parameters:  
$A_1=T_{c}/\gamma_{\parallel}(0)$ and
$A_2=T_{c} \gamma^{'}_{\parallel}(0)/\gamma_{\parallel}(0)$,
where $\gamma^{'}_{\parallel}(0) = d\gamma_{\parallel}(0)/dT$.  
$A_1$ is a measure of how well the ground state property 
$\gamma_{\parallel}(0)$ (measurable through the superfluid density)
predicts $T_c$, which is equivalent to $T_{\theta}$ in 
this model.  $A_2$ can be expressed in the more intuitive
form $A_2=T_{c}/T_{ex}$, where
$T_{ex} \equiv \gamma_{\parallel}(0)/\gamma_{\parallel}^{/}(0)$,
is the estimate of $T_c$ one would obtain by extrapolating from  
the low temperature slope of $\gamma_{\parallel}(T)$ to the point 
at which the superfluid stiffness would vanish. 
\marginpar{\em The shape of $\gamma(T)$is robust!}
Over {\it orders of magnitude} of couplings
($0\le J_{\perp}/J_{\parallel}\le .1$), and throughout the
range $0 \le \delta \le .25$, $A_1$ and $A_2$ are
remarkably robust:  $A_1 \sim .6 - 1.7$,
and $A_2 \sim .2 - .5$.

\subsection{Quantum considerations}
\label{quantum}

In quantum systems, the dynamics affects the thermodynamics.  
However, the role of quantum effects on the phase dynamics is a large
topic, and one in which many uncertainties remain.  
We will briefly discuss the simplest case here, mostly to
illustrate the complexity of the problem.

Let us consider  a simple two fluid model~\cite{badmetals} in
which a phase fluctuating superconductor is capacitively coupled to a
normal fluid. The continuum limit of the effective action obtained upon
integrating out the normal fluid can be derived from simple hydrodynamic
considerations. From the Josephson relation, it follows that the electric field
\be
\vec E = -(\hbar/2e) \vec \nabla \dot \theta \; .
\ee
The Euclidean effective action is obtained by augmenting the classical
action, Eq.~(\ref{Fsc}), with the Maxwell term, and analytically continuing to
imaginary time:
\be
S[\theta]=\int _0^{\beta} d\tau \left\{\int d\vec r{\cal L}_{quantum} +
V_{phase}\right\}, \ \ \ {\cal L}_{quantum}=\vec E\cdot \vec D/8\pi \; ,
\ee
where $\beta=1/T$,  
$\vec D(\vec k,\omega)=\epsilon_0(\vec k,\omega)\vec E(\vec k,\omega)$, and 
$\epsilon_0$ is the normal fluid dielectric function
(analytically continued to imaginary time).    
Again, this effective action must be cutoff at short distances in such a
way as to preserve the periodicity of $\theta$ by allowing vortex
excitations. 

An analysis of the Maxwell term, 
\marginpar{\em The order of limits matters.}
$S_{quantum}$, allows us to illustrate 
some of the complexity of this
problem.  At $\vec k=\vec 0$ and small $\omega$, 
$\epsilon_0\approx 4\pi \sigma_0/i\omega$, where $\sigma_{0}$ is the D.C.
conductivity of the normal fluid.  Thus, if we first consider the 
spatial continuum limit before going to low frequencies, 
$S_{quantum}\sim \sum_{\omega_n}\int d\vec
 r\sigma_0|\omega_n| |\vec
\nabla\theta|^2$, where $\omega_n=2\pi n T$ are the Matsubara frequencies.
We recognize the resulting action as the continuum limit of an array 
of resistively shunted Josephson junctions\cite{cikl,wagenblast} (RSJ).  
Here, the normal fluid plays the role of an ``Ohmic heat bath.''  

On the other
hand,  if we first take $\omega=0$, and then $k$  small,
$\epsilon\approx (k_{TF}/k)^2$ where $k_{TF}$ is the Thomas-Fermi screening
length.  In this limit, the Maxwell term has the form of a
phase kinetic energy, ${\cal L}_{quantum}\sim 
(M_{\theta}/2)|\dot\theta|^2$, with an effective mass,
$M_{\theta}\propto [e^{2}/k_{TF}^{2}]^{-1}$ inversely proportional to
an appropriately defined local charging energy. The resulting 
effective action is the continuum limit of the ``lattice quantum
rotor''  (QR) model, also a widely studied problem\cite{subirbook}.  
The RSJ and QR models have 
quite different behavior at low temperatures.  Without a rather 
complete understanding of the physics of the normal fluid, it is 
impossible in general to determine which, if either, of these limits 
captures the essential quantum physics.  

There is nonetheless one important issue which can be addressed in 
a theoretically straightforward fashion:  
the temperature scale below which quantum effects dominate.  
\marginpar{\em The classical to quantum crossover temperature is estimated.}  
The classical 
physics we studied in the previous section is readily obtained from 
the quantum model by suppressing all fluctuations with nonzero 
Matsubara frequency.  We thus estimate a classical to quantum 
crossover temperature, $T_{cl}$, by comparing the classical 
($\omega=0$)
and first finite frequency ($\omega=\omega_{1}=2\pi T$) contributions to 
$S[\theta]$.  This leads to the implicit equation for $T_{cl}$:
\be
T_{cl}=\sqrt{e^{2}n_{s}/\epsilon_0 m^{*}} \; ,
\ee
where $\epsilon_0$ is evaluated at temperature $T=T_{cl}$, 
frequency $\omega \sim 2\pi T_{cl}$, and a 
typical momentum, $k\sim 1/a$.
So long as $T\gg T_{cl}$, the imaginary time independent (classical) 
field configurations dominate the thermodynamics.  Clearly, depending 
on how good the screening is, $T_{cl}$ can be much smaller or much 
larger than $T_{c}$.  If we approximate $\epsilon_0$ by its finite 
frequency, $k\rightarrow 0$ form, this estimate can be recast in an intuitively 
appealing form\cite{classphs}:
\be
T_{cl}\sim \left(\frac{\sigma_{Q}}{\sigma_{0}}\right) T_{\theta} \; ,
\ee
where $\sigma_{Q}=e^{2}/(h a)$ is the quantum of conductance in which 
the vortex core radius enters as the quantum of length.

Recent theoretical developments have uncovered yet more
subtleties. Although the low energy physics involves only
phase fluctuations, phase slips 
(short imaginary time events where the
phase spontaneously ``slips" by $2\pi$) involve amplitude fluctuations.  
In the presence of an ohmic heat bath, there are subtle,
long time consequences of these amplitude fluctuations 
\cite{feigelman,spivak2,demler}.
Another interesting possibility is electron fractionalization.
Under some circumstances, it has been proposed\cite{senthil}
that $hc/e$ vortices may be energetically preferred to 
the usual $hc/2e$ vortices, leading to a fractionalized state.  

Combine this exciting 
\marginpar{\em This is an important unsolved problem!} 
but incomplete jumble of theoretical ideas 
with the remarkably simple but entirely unexplained behavior 
observed experimentally in granular superconducting films as they 
crossover from superconducting to insulating behavior, and one is 
forced to concede that the theory of quantum phase fluctuations is 
seriously incomplete.

\subsection{Applicability to the cuprates}

 Both phase and pair breaking fluctuations
are more prevalent at low $T$ in the cuprate superconductors than in 
conventional BCS superconductors.
The low superfluid density provides only a weak stiffness
to thermal phase fluctuations of the order parameter.  In addition,
the nodes in the gap mean that  there are low energy 
quasiparticle excitations 
down to arbitrarily low temperature.
However, it is important to remember that nodal quasiparticles occupy only a small fraction of the Brillouin zone so long as $\Delta_{o}\gg T$. 

\subsubsection{$T_c$ is unrelated to the gap in underdoped cuprates}

As mentioned in Section \ref{general}, in underdoped cuprates, 
many probes detect a pseudogap in the normal state, such as NMR, 
STM, junction tunnelling, and ARPES.  
Whereas BCS theory would predict $T_c \sim \Delta_o / 2$, 
where $\Delta_o$ is the superconducting gap maximum at zero temperature, 
the low temperature magnitude of the single particle gap as measured by 
ARPES or tunnelling experiments does not follow this relation, qualitatively 
or quantitatively. On the underdoped side, $T_c$ increases with increasing 
doping, whereas $\Delta_{o}$ moves in the opposite direction in all cases 
studied to date. Even at optimal doping, $T_c$  is always considerably 
smaller than
the BCS value of $\Delta_{o}/2$. In optimally doped BSCCO, for example, 
$T_c \sim \Delta_{o}/5$, where $\Delta_{o}$ is the peak energy observed 
in low temperature tunnelling experiments.\cite{bsccotun,doniach,levi} 
(See also Table 1.)

The ARPES experiments provide $k$-space information demonstrating 
\marginpar{\em There is no signature of the transition in the single particle gap.}
that the gap, above and below $T_c$, has 
an anisotropy consistent with a $d$-wave order parameter.
Furthermore, $\Delta_{o}(T)$ is largely undiminished in going 
from $T=0$ to $T=T_c$ in underdoped samples, and the size and shape of 
the gap are basically unchanged through the transition.    
Add to this the contravariance of $T_c$ with the low temperature magnitude 
of the gap as the doping is changed, and it appears the gap and 
$T_c$ are simply independent energies \cite{harris79,fischerpseudo}.  
The gap decreases with overdoping, which may be 
responsible for the depression of $T_c$ in that region, 
so that the transition may be more conventional on the overdoped side.   

\subsubsection{$T_c$ is set by the superfluid density in underdoped cuprates}

As emphasized above, the superfluid density in 
cuprates is orders of magnitude smaller than in conventional superconductors.\cite{nature}  
In addition, when the superfluid density is converted to an energy scale, 
it is comparable to $T_c$, whereas in conventional superconductors this 
phase stiffness energy scale is far above the transition temperature. 
In those conventional cases, BCS theory works quite well, 
but in the cuprates, the phase stiffness energy scale 
should also be considered.

This is further emphasized by the Uemura plot\cite{uemura}, 
which  compares the transition temperature to the superfluid density.  
For underdoped systems, the relationship is linear within experimental 
errors. This is strong evidence that $T_c$ is determined by the superfluid 
density, and therefore set by phase ordering.

\subsubsection{Experimental signatures of phase fluctuations}

In YBCO, $3DXY$ critical fluctuations have been observed
in the superfluid density within $10\%$ of $T_c$\cite{3dxy,xypasler}, 
implying that the temperature dependence of the superfluid density below and 
near $T_c$ is governed by phase fluctuations. It needs to be stressed that in 
conventional superconductors, such fluctuations that are seen are Gaussian 
in character---that is they involve fluctuations of both the amplitude and the
phase of the order parameter 
\footnote{An interesting way to identify separate Gaussian and phase
fluctuation regimes in YBCO is presented in Ref.~\citen{ando2}. 
See also Ref.~\citen{shivaji}.}. 
The purely critical phase fluctuations observed in YBCO are
entirely different.  
At low temperature (as low as $T=1$K~\cite{hardynodes}), 
the superfluid density is a linearly decreasing function of 
temperature\cite{bonnnodal}. While this linear behavior is generally believed
to be the result of amplitude fluctuations of an order parameter with nodes, 
it is difficult\cite{wenlee,wenleeqp,dhlphase,millisnodes} from this perspective to 
understand why the slope is nearly independent of $x$ and of $\Delta_0/T_c$. 
This feature of the data is naturally explained if it is assumed that 
the linear temperature dependence, too, arises from classical phase 
fluctuations, but then it is hard to understand\cite{classphs} why 
quantum effects would not quench these fluctuations at such low temperatures.

\section{Lessons From Weak Coupling}
\label{weak}

\addtocontents{toc}{
\noindent{\em The weak coupling renormalization group approach to the
Fermi liquid and the 1DEG is presented. The role of retardation, the
physics of the Coulomb pseudopotential, and the nonrenormalization of
the electron phonon coupling in a BCS superconductor are systematically 
derived. The strong renormalization of the electron-phonon
interaction in the 1DEG is contrasted with this---it is suggested that
this may be a more general feature of non-Fermi liquids.} }

\subsection{Perturbative RG approach in $D > 1$}

In recent years, Fermi liquid theory, and with it the characterization of the
BCS instability, has been recast
in the language of 
a perturbative renormalization group (RG) treatment. We will adopt this 
approach as we reconsider the conventional BCS-Eliashberg theory of the
phonon mediated mechanism of superconductivity in simple metals.  In 
particular, we are interested in exploring the interplay between a 
short ranged instantaneous electron-electron repulsion of strength $\mu$ 
and a retarded attraction (which we can think of as being mediated by the 
exchange of phonons) of strength $\lambda$, which operates only below 
a frequency scale $\omega_{D}$. Although we will make use of a 
perturbative expression for the beta function which is valid only 
for $\mu$ and $\lambda$ small compared to 1, the results are 
nonperturbative in the sense that we will recover the nonanalytic 
behavior of the pairing scale, $T_{p}$, expected from BCS mean field 
theory. The results are valid for any relative strength of 
$\mu/\lambda$ and, moreover, the corrections due to higher order 
terms in the beta function are generally smooth, and so are not 
expected to have large qualitative effects on the results so long as
$\mu$ and $\lambda$ are not large compared to 1.  

All the results obtained in this section have been well 
understood by experts since the golden age of many-body theory, 
along with some of the most important higher order corrections which occur
for $\lambda$ of order 1 (which will be entirely neglected here). 
Our principal purpose in including this section is to provide a simple 
derivation of these results in a language that may be more accessible 
to the modern reader. A most insightful exposition of this approach is 
available in the articles by Polchinski, Ref.~\citen{polchinski}, and Shankar, 
Ref.~\citen{shankar}, which can be consulted wherever the reader is curious 
about parts of the analysis we have skipped over.
The one technical modification we adopt here is to employ an energy shell 
RG  transformation, rather than the momentum shell approach adopted in
Ref.~\citen{shankar}; this method allows us to handle the retarded 
and instantaneous interactions on an equal footing. It can also be 
viewed as an extension of the analogous treatment of the 1D problem 
adopted in Ref.~\citen{zkl}, as discussed in the next subsection.

We start by defining a scale invariant (fixed point) Euclidean
action for a noninteracting Fermi gas
\ba
S_{fp}[\Psi_{\uparrow},\Psi_{\downarrow}]&=&(2\pi)^{-(d+1)}k_{F}^{d-1}
\sum_{\sigma}\int d\omega d\hat k dk
{\cal L}_{0}[\Psi_{\sigma}]  \; , \\
\label{fixedpoint}
{\cal L}_{0}[\Psi_{\sigma}]&=&\bar 
\Psi_{\sigma}[i\omega+v_{F}(\hat k) k]\Psi_{\sigma} \; , \nonumber
\ea
where $d\vec k=k_{F}^{d-1}d\hat k dk$, the unit vector 
$\hat k$ is the direction of $\vec k$ and $k$ is the displacement 
from the Fermi surface; we have assumed a simple spherical Fermi 
surface. The treatment that we present here breaks down when the 
Fermi surface is nested or contains Van Hove singularities. 
To regularize the theory, it is necessary to cut off the integrals;  
whereas Shankar confines $k$ to a narrow shell about 
the Fermi surface, $|k|< \Lambda \ll k_{F}$, we allow $k$ to vary 
from $-\infty$ to $+\infty$, but confine the $\omega$ integral to a 
narrow shell $|\omega| < \Omega \ll E_{F}$.  
\sindex{zzomega}{$\Omega$}{Frequency}

We now introduce electron-electron interactions. Naive power 
counting leads to the conclusion that the four fermion terms are 
marginal, and all higher order terms are irrelevant, so we take 
\ba
S_{int}=&&\sum_{\sigma,\sigma'}
\int \prod_{j=1}^{3}\frac {d\vec k_{j}d\omega_{j}}{(2\pi)^{d+1}}
\bar \Psi_{\sigma}(\vec k_{1},\omega_{1})
\bar \Psi_{\sigma'}(\vec k_{2},\omega_{2}) \nonumber
\\
&& \times [g(\vec k_{2}-\vec k_{3}) 
 + \Theta(\omega_{D}-|\omega_{2}-\omega_{3}|)\tilde
g(\vec 
k_{2}-\vec k_{3}) ] \nonumber \\
&& \times \Psi_{\sigma'}(\vec k_{3},\omega_{3})
 \Psi_{\sigma}(\vec k_{1}+\vec k_{2}-\vec 
 k_{3},\omega_{1}+\omega_{2}-\omega_{3}) \; ,
\label{Sint}
\ea
where $\Theta$ is the Heavyside function, and $g$ and $\tilde g$ are, 
respectively, the instantaneous and retarded interactions. Signs are 
such that positive $g$ corresponds to repulsive interactions. The 
distinction between retarded and instantaneous interactions is 
important so long as $\Omega \gg \omega_{D}$. 
We have invoked spin rotation invariance in order to ignore
the dependence of $g$  and $\tilde g$ on the spin indices.

It should be stressed, as already mentioned in Section \ref{1D}, that
this should already be interpreted as an effective field theory, in
which the microscopic properties that depend on the band structure away
from the Fermi surface such as mixing with other bands, more complicated 
three and four-body interactions, etc. have already fed into the parameters 
that appear in the model. What we do now is to address the question of what
further changes in the effective interactions are produced when we
integrate out electronic modes in a narrow shell between $\Omega$ and
$\Omega  e^{-\ell}$, ($\ell>0$ and small), and then rescale all frequencies 
according to
\be
\omega\rightarrow e^{\ell}\omega, \ \ k\rightarrow e^{\ell} k
\ \ {\rm and } \ \ \Psi\rightarrow e^{-(3/2 + \eta_{F})\ell}\Psi \; ,
\ee
to restore the cutoff to its original form and where, as usual, 
$\eta_{F}$ is a critical exponent that is determined by the the 
properties of the interacting fixed point. We will carry this 
procedure out perturbatively in powers of $g$ and $\tilde g$---to 
the one loop order we (and everyone else) analyzes, $\eta_{F}=0$.

To first order in perturbation theory, simple power counting insures 
that the entire effective action is invariant under the RG 
transformation, other than the parameter $\omega_{D}$ which changes 
according to
\be
d\omega_{D}/d\ell = \omega_{D} \; .
\ee

\begin{figure}[htb!!!]
\begin{center}
\epsfig{figure=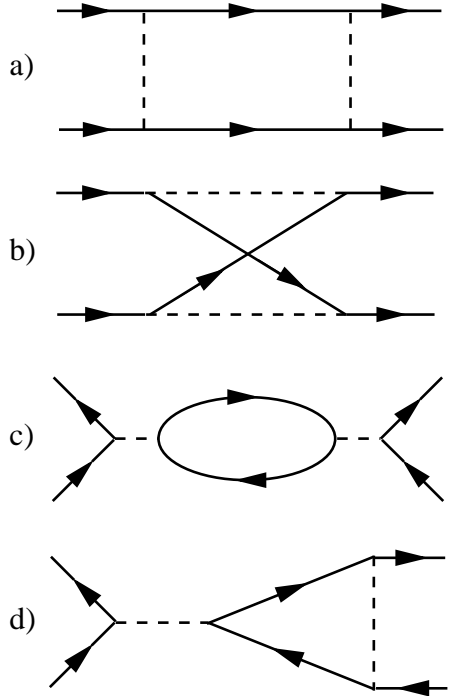,width=0.5\linewidth}
\end{center}
\caption{\label{loops}
The one loop diagrams that are invoked in the discussion of the 
renormalization of the effective interactions. a) and b) are referred to
as the ``Cooper channel'' and c) and d) as ``particle-hole channels''.
The loop  is made out of electronic propagators with frequencies in the
shell  which is being integrated. The dashed lines represent
interactions.}
\end{figure} 

To second (one loop) order, the forward 
scattering interactions are still unchanged;
they produce the Fermi liquid parameters, and should
actually be included as part of the fixed point action and treated 
non-peturbatively. This can be done straightforwardly, but for 
simplicity will be ignored here. The one loop diagrams which 
potentially produce contributions to the beta function are shown in 
Fig.~\ref{loops}. All internal legs of the diagrams refer to electron 
propagators at arbitrary momenta but with their frequencies constrained 
to lie in the shell which is being integrated out, 
$\Omega > |\omega|\ge \Omega e^{-\ell}$. The dashed 
lines represent interactions. All external legs are taken to lie on 
or near the Fermi surface. Clearly, the energy transfer along the 
interaction lines in the Cooper channel, Figs. \ref{loops}a and \ref{loops}b,
is of order $\Omega$, and so for 
$\Omega \gg \omega_{D}$, 
$\tilde g$ does not contribute, while in 
Figs.~\ref{loops}c and \ref{loops}d there is zero frequency transfer 
along the interaction lines, and so $g$ and $\tilde g$ contribute equally.  

Since $\Omega\ll E_{F}$, we can 
classify the magnitude of each diagram in powers of $\Omega$;  any 
term of order $|\Omega|^{-1}$ makes a logarithmically divergent
contribution to the effective interaction upon integration over 
frequency, while any terms that are proportional to $E_F^{-1}$ are much 
smaller and make only finite contributions which can be ignored for the present 
purposes. When the Cooper diagrams, shown in Figs.~\ref{loops}a and 
\ref{loops}b, are evaluated for zero center of mass momentum, ({\it i.e.} 
if the momenta on the external legs are $\vec k_F$ and
$-\vec k_F$), the bubble is easily seen to be proportional to 
$\Omega^{-1}$. However, if the center of mass momentum is nonzero 
({\it i.e.} if the external momenta are $\vec k_F+\vec q$ and $-\vec k_F$), 
the same bubble is proportional to $1/v_F|\vec q|$, and hence 
is negligible. 
The particle-hole diagrams in Figs.~\ref{loops}c and \ref{loops}d are
a bit more complicated.  The bubble is zero  for total momentum 0, 
and proportional to $1/v_Fk_F$ for  momentum transfer near $2k_{F}$. 
Thus, in more than one dimension, the particle hole bubbles can be neglected 
entirely.  (We will treat the 1d case separately, below.)  
Putting all this together in the usual manner, we are left with 
the one-loop RG equations for the interactions between
electrons on opposing sides of the Fermi surface,
 \be
 \frac {d g_{l}} {d\ell} = -\frac 1 {\pi v_F} g_{l}^{2}\; , \ \ 
\ \ \frac {d \tilde g_{l}} {d\ell} =0 \; ,
 \ee
where $l$ refers to the 
appropriate Fermi surface harmonic;  for the case of a circular Fermi 
surface in two dimensions, $l$ is simply angular momentum.
(Implicit in this is the fact that odd $l$ are associated with
interactions in the triplet channel while even $l$ are in the singlet channel.)
 
These equations describe the changes in the effective interactions 
upon an infinitesimal RG transformation.    They can be easily  
integrated to obtain expressions for the scale dependent 
interactions.  However, these equations are only valid so long as all 
the interactions are weak (to justify perturbation theory) and so 
long as $\Omega\gg \omega_{D}$.  Assuming that it is the second 
\marginpar{\em Note the nonrenormalization of $\lambda$ for $\Omega > \omega_D$.} 
condition that is violated first, we can obtain expression for the 
effective interactions at this scale by integrating to the point at 
which $\Omega=\omega_{D}$;  the result is
\be
\mu(\omega_{D}) = \frac{\mu_{0}}{1+\mu_{0}\log(\Omega_{0}/\omega_{D})}\; , \ \ 
\lambda(\omega_{D})=\lambda_{0} \; ,
\ee
where $\mu=g/\pi v_{F}$, $\lambda=\tilde g/\pi v_{F}$, the symmetry 
labels on $g$ and $\tilde g$ are left implicit, and 
the subscript ``0'' refers to the initial 
values of the couplings at a microscopic scale, $\Omega_{0}\sim E_{F}$.

The fact that the retarded interactions do not renormalize is 
certainly as noteworthy as the famous renormalization of $\mu$.  
This means that it is possible to estimate $\lambda$ from 
microscopic calculations or from high temperature measurements, such 
as resistivity measurements in the quasi-classical regime where
$\rho \propto \lambda T$.  

Once the scale $\Omega=\omega_{D}$ is 
reached, a new RG procedure must be adopted.  At this point, the 
retarded and instantaneous interactions are not distinguishable, so we 
must simply add them to obtain a new, effective interaction,
$g^{eff}(\omega_{D})= g(\omega_{D})+\tilde g$,  which upon further 
reduction of $\Omega$ renormalizes as a nonretarded interaction.  
If $g^{eff}(\omega_{D})$ is repulsive, it will be further reduced 
with decreasing $\Omega$.  However, if it is attractive in any channel, 
the RG flows 
carry the system to stronger couplings, and eventually the 
\marginpar{\em Fermi liquid behavior breaks down at the pairing scale.}
perturbation theory breaks down.  We can estimate the characteristic 
energy scale at which this breakdown occurs by integrating the one loop 
equations until the running coupling constant reaches a certain finite 
value $ -1/\alpha$:
\begin{equation}
\Omega_{1}=\omega_{D}e^{\alpha}\exp[-1/|g^{eff}(\omega_{D})|] \; .
\end{equation}
Of course, the RG approach does not tell us how to interpret this 
energy scale, other than that it is the scale at which Fermi liquid 
behavior breaks down.  However, we know on other grounds that this 
scale is the pairing scale, and that the breakdown of Fermi liquid 
behavior is associated with the onset of superconducting behavior. 

\subsection{Perturbative RG approach in $D = 1$}
\label{rg1}

\subsubsection{The one loop beta function}

In one dimension, the structure of the perturbative beta function is very
different from in higher dimensions.  In addition to the familiar logarithmic
divergences in the particle-particle (or Cooper) channel, there appear
similar logarithms in the particle-hole channel.  That these lead to a
serious breakdown of Fermi liquid theory can be deduced directly from the
perturbation theory, although it is only through the magic of 
bosonization (discussed in Section \ref{1D}) that it is possible 
to understand what these divergences lead to.

To highlight the differences with the higher dimensional case, we 
will treat the 1d case using the perturbative RG approach, 
but now taking into account the dimension specific interference between 
the Cooper and particle-hole channels.
However, having belabored the derivation of the perturbative beta function for 
the higher dimensional case, we will simply write down the result for 
the 1d case;  the reader interested in the details of the derivation 
is referred to Refs.~\citen{zkl} and \citen{KZ}.

In 1d,  there are only two 
potentially important momentum transfers which scatter electrons at 
the Fermi surface, as contrasted with the continuum of possibilities 
in high dimension.  It is conventional to indicate by $g_{1}$
 the interaction with 
momentum transfer $2k_{F}$, and by $g_{2}$ that with zero momentum 
transfer.  If we are interested in the case of a nearly half filled 
band, we also need to keep track of the umklapp scattering, $g_{3}$, 
which involves a momentum transfer $2\pi$ to the lattice 
(see Section~\ref{1D}).  
Consequently, we must introduce a chemical potential, $\mu$, defined 
such that $\mu=0$ corresponds to the half filled band.  Finally, we 
consider the retarded interactions, $\tilde g_{1}$, $\tilde g_{2}$, 
and $\tilde g_{3}$ which operate at frequencies less than $\omega_{D}$.
For simplicity, we consider only the case of spin rotationally 
invariant interactions.

The one loop RG equations (obtained by evaluating precisely the 
diagrams in Fig. \ref{loops}), under conditions $\Omega \gg 
\omega_{D}, \mu$, are
\ba
&&\frac {d g_{1}} {d\ell} = - \frac {g_{1}^{2}} {\pi v_{F}} \; ,
\ \ \ \
\frac {d g_{c}} {d\ell} = - \frac {g_{3}^2} {\pi v_{F}}\; , \ \ \ \ 
\frac {d g_{3}} {d\ell} = - \frac {g_{3}g_c} {\pi v_{F}} \; ,\nonumber \\
&&\frac {d \tilde g_{\pm}} {d\ell} = - \frac {g_{\pm}} {\pi v_F}
[\frac 3 2 g_1\pm g_3+\frac 1 2g_c + \tilde g_{\pm}] \; ,
\ \ \ \ \frac {d \tilde g_{2}} {d\ell} = 0 \; ,\nonumber\\
&&\frac {d \mu} {d\ell} = \mu \; , \ \ \ \ \frac {d \omega_{D}} {d\ell}
=[1+\frac {\tilde g_+} {\pi v_F}]\omega_{D}  \; ,
\label{1drg}
\ea
where $g_{c}\equiv g_{1}-2g_{2}$ and $\tilde g_{\pm}=\tilde g_{1} \pm 
\tilde g_{3}$.  For $\mu \gg \Omega \gg \omega_{D}$, the same 
equations apply, except now we must set $g_{3}=\tilde g_{3}=0$.  And, 
of course, if $\omega_{D}> \Omega$, we simply drop the notion of 
retarded interactions, altogether.

\marginpar{\em The electron-phonon interaction in a non-Fermi liquid can be
strongly renormalized.} 
There are many remarkable qualitative aspects to these
equations, many of which differ markedly from the analogous equations in 
higher dimensions. The most obvious feature is that the retarded interactions
are strongly renormalized, even when the states being eliminated have energies
large compared to $\omega_D$. What this means is that in one dimension, the
effective electron-phonon interaction at low energies is {\em not} simply
related to the microscopic interaction strength. Some of the effects of 
this strong coupling on the spectral properties of quasi-one dimensional 
systems can be found in Refs.~\citen{KZ,finkelstein,finkelstein2}.

\subsubsection{Away from half filling}

To see how this works out, let us consider the typical case in which the
nonretarded interactions are repulsive ($g_1$, and $g_2 >0$) and the retarded
interactions are attractive ($\tilde g_{\pm} <0$) and strongly
retarded, $\omega_D/E_F \ll 1$.  
Far from half filling, we can also set $g_3=\tilde g_3=0$.  The presence or
absence of a spin gap is determined by the sign of $g_1$.  Thus, just as in the
3d case, in order to derive the effective theory with nonretarded interactions
which is appropriate to study the low energy physics at scales small compared
to $\omega_D$, we integrate out the fermionic degrees of freedom  at scales
between
$E_F$ and $\omega_D$, and then compute the effective backscattering
interaction,
\be
g_1^{eff}=g_1(\omega_D)+ \tilde g_1(\omega_D) \; .
\ee
If $g_1^{eff}>0$ ({\it i.e.} if $g_1(\omega_D) > |\tilde g_1(\omega_D)|$), then
the Luttinger liquid is a stable fixed point, and in particular no spin gap
develops.  If $g_1^{eff}<0$, however, the Luttinger liquid fixed point is
unstable;  now, the system flows to a Luther-Emery fixed point with a spin gap
which can be determined in the familiar way to be
\be
\Delta_s\sim \omega_D \exp[-\pi v_F/g_1^{eff} ] \; .
\ee
This looks very much like the BCS result from high dimensions.
The parallel with BCS theory goes even a bit further, since under the RG
transformation, a repulsive
$g_1$ scales to weaker values in just the same way as the Coulomb
pseudopotential in higher dimensions:
\be
g_1(\omega_D)=\frac{g_1^0}{1+(g_1^0/\pi v_F) \log{(E_F/\omega_D)}} \; ,
\label{g1omega}
\ee
where $g_1^0\equiv g_1(E_F)$.
However, in contrast to the higher dimensional case, $\tilde g_1$ is strongly
renormalized;  integrating the one-loop equations, it is easy to show that
\ba
\tilde g_1(\omega_D)&=&\left(\frac {\tilde g_1^0 }{1+ \tilde g_1^0 L}\right)
 \left( \frac {g_1(\omega_D)} {g_1^0} \right)^{3/2} \left(\frac {E_F}
{\omega_D}\right)^{-g_c/2\pi v_F} \; , \\
L&=& \int_0^{\log(E_F/\omega_D)} \frac {dx} {\pi v_F} \frac{\exp[-g_c x/2\pi
v_F]} {[1+(g_1^0/\pi v_F) x]^{3/2}} \; .
\label{tildeg1omega}
\ea

Various limits of this expression can easily be analyzed---we will not give an
exhaustive analysis here.  For $g_1=g_c=0$, Eq.~(\ref{tildeg1omega}) reduces to
the same logarithmic expression, Eq.~(\ref{g1omega}), as for $g_1$, although
because $\tilde g_1$ has the opposite sign, the result is a logarithmic 
{\em increase} of the effective interaction; 
this is simply the familiar Peierls renormalization of
the electron-phonon interaction. 
For $g_c <0$, this renormalization is substantially amplified. Thus, in marked
contrast to the higher dimensional case, strong repulsive interactions 
actually  enhance the effects of weak retarded attractions!
\marginpar{\em Repulsive interactions 
enhance the effects of weak retarded attractions.} 

Finally, there is bad news as well as good news. As discussed in Section \ref{1D},
the behavior of the charge modes is largely determined by 
the ``charge Luttinger exponent, $K_c$, which is in
turn determined by the effective interaction
\be
g_c^{eff}= g_c +\tilde g_1^{eff}-2\tilde g_2 \; ,
\label{gceff}
\ee
according to the relation 
(See Eq. (\ref{lparam}).)
\be
K_c=\sqrt{\frac{1+(g_c^{eff}/\pi v_F)}{1-(g_c^{eff}/\pi v_F)}} \; . 
\ee
In particular, the relative strength of the superconducting and CDW
fluctuations are determined by $K_c$;  the smaller $K_c$, {\it i.e.} the more
negative
$g_c^{eff}$, the more dominant are the CDW fluctuations. It therefore 
follows from Eq. (\ref{gceff}) that a large negative value of 
$\tilde g_1^{eff}$ due to the renormalization of
the electron-phonon interaction only throws the balance more strongly in favor
of the CDW order.  
For this reason, most quasi 1D systems with a
spin gap  are CDW insulators, rather than superconductors.

\subsubsection{Half filling}

Near half filling, the interference between the retarded and instantaneous
interactions becomes even stronger.  In the presence of Umklapp scattering, an
initially negative $g_c$ renormalizes to stronger coupling, as does $g_3$
itself.  Without loss of generality, we can take $g_3>0$ since
its sign can be reversed by a change of basis.  Then we can see that
both $g_3$ and $g_c$ contribute to an inflationary growth of $\tilde g_-$. 
The RG equations have been integrated in Ref.~\citen{salkolaandme}, and  we
will not repeat the analysis here.  The point is that all the effects 
discussed above apply still more strongly near half filling.  
In addition, we now encounter an
entirely novel phenomenon---we find that the effective electron-phonon
interaction strength at energy scale $\omega_D$ is strongly doping dependent,
as well.  It is possible\cite{salkolaandme}, as indeed seems to be the
case in the model conducting polymer polyacetylene, for the
electron-phonon coupling to be sufficiently strong to open a Peierls gap
of magnitude 2eV (roughly, 1/5 of the $\pi$-band width) at half filling,
and yet be so weak at a microscopic scale that for doping concentrations 
greater than 5\%, no
sign of a Peierls gap is seen down to temperatures of order 1K! 
\marginpar{\em The effective electron-phonon coupling can even be strongly
doping dependent.}   

How many of the features seen from this study of the 1DEG are specific to one
dimensional systems is not presently clear.  Conversely, these
results prove by example that familiar properties of Fermi liquids cannot be
taken as generic.  In particular, strongly energy and doping dependent
electron phonon interactions are certainly possibilities that should be taken
seriously in systems that are not Fermi liquids.

\section{ Lessons from Strong Coupling}
\label{strong}

\addtocontents{toc}{
\noindent{\em In certain special cases, well controlled analytic results
can be obtained in the limit in which the bare electron-electron and/or
electron-phonon interactions are strong.  We discuss several such
cases, and in particular we demonstrate a
theoretically well established mechanism in one dimension, 
the ``spin gap proximity
effect,'' by which strong repulsive interactions between electrons can
result in a large and robust spin gap and strongly enhanced local
superconducting correlations.  We propose this as the paradigmatic
mechanism of high temperature superconductivity.} }

In certain special cases, well controlled analytic results
can be obtained in the limit in which the bare electron-electron 
interactions are nonperturbative. We discuss several such models.

\subsection{The Holstein model of interacting electrons and phonons}

The simplest model of strong electron-phonon coupling is the Holstein model 
of an optic phonon, treated as an Einstein oscillator, coupled to a single 
tight binding electron band,
\be
H_{Hol}= -t\sum_{<i,j>,\sigma}[c_{i,\sigma}^{\dagger}c_{j,\sigma} + 
{\rm H.C.}] +\alpha \sum_j x_j\hat n_j +\sum_j\left [\frac {P_j^2} {2M} + 
\frac {K x_j^2} 2\right] \; ,
\ee
where $\hat n_j=\sum_{\sigma}c_{j,\sigma}^{\dagger}c_{j,\sigma}$ is the 
electron density operator and $P_j$ is the momentum conjugate to $x_j$.  

In treating the interesting strong coupling physics of this problem, 
it is sometimes
useful to transform this model so that the phonon displacements are
defined relative to their instantaneous ground state configuration. 
This is done by means of the unitary transformation,
\be
U=\prod_j\exp[ i(\alpha/K) P_j \hat n_j] \; ,
\ee
which shifts the origin of oscillation as 
$U^{\dagger}x_jU=x_j - (\alpha/K) \hat n_j$. 
Consequently, the transformed Hamiltonian has the form
\be
U^{\dagger} H_{Hol} U =-t\sum_{<i,j>,\sigma}
[\hat S_{ij}c_{i,\sigma}^{\dagger}c_{j,\sigma} + {\rm H.C.}] - 
\frac {U_{eff}} 2 \sum_j [\hat n_j]^2 +\sum_j\left [\frac {P_j^2} {2M} + 
\frac {K x_j^2} 2\right] \; ,
\ee
where $\hat S_{i,j}=\exp[-i(\alpha/K)(P_i-P_j)]$ and $U_{eff}= \alpha^2/K$.

There are several limits in which this model can be readily analyzed:

\subsubsection{Adiabatic limit: $E_F \gg \omega_D$} In the limit 
$t\gg \omega_D$, where $\omega_D=\sqrt{K/M}$ is the phonon frequency and
for $\alpha$ not too large, this is just the sort of model considered in the weak
coupling section, or any other conventional treatment of the electron-phonon 
problem. Here, Migdal's theorem provides us with guidance, and at
least for not too strong coupling, the BCS-Eliashberg treatment discussed 
in Section \ref{weak} can be applied. While $U_{eff}$ is, indeed, the effective 
interaction which enters the BCS expression for the superconducting $T_c$,  
because the fluctuations of $P_i$ are large if $M$ is large, it is not useful 
to work with the transformed version of the Hamiltonian.
  
\subsubsection{Inverse adiabatic limit; negative $U$ Hubbard model} In the 
inverse adiabatic limit, $M\rightarrow 0$, fluctuations of $P_j$ are 
negligible, so that $\hat S_{ij} \to 1$. Hence, in this limit, 
the Holstein model is precisely equivalent to the Hubbard model, 
but with an effective negative $U$. If $U_{eff}\ll t$, this 
is again a weak coupling model, and will yield a superconducting $T_c$ 
given by the usual BCS expression, although in this case with a prefactor 
proportional to $t$ rather than $\omega_D$.  

In contrast, if $U_{eff} \gg t$, a strong coupling expansion is
required.  Here, we first find the (degenerate) ground states of the 
unperturbed model with $t=0$, and then perform perturbation theory in 
small $t/U_{eff}$.  In the zeroth order ground states, each site is 
either unoccupied, or is occupied by a singlet pair of electrons. 
The energy of this state is $-U_{eff} N^{el}$, where $N^{el}$ is 
the number of electrons. These states can be thought of as the states of
infinite mass, hard core charge 2e bosons on the lattice. There is a gap to 
the first excited state of magnitude $U_{eff}$. Second order perturbation 
theory in the ground state manifold straightforwardly yields an effective 
Hamiltonian which is equivalent\footnote{Clearly, $\tilde
b_j\equiv c_{j\uparrow}c_{j\downarrow}$ does not satisfy the same-site 
piece of the bosonic commutation relation, but the hard core constraint 
on the $b_j$ bosons corrects any
errors introduced by neglecting this.} to a model of hard core bosons
($[b_i^{\dagger},b_j]=\delta_{i,j}$)
\be
H_{boson}=-t_{eff}\sum_{<i,j>} [b_i^{\dagger} b_j + {\rm H.C.}]  
+ V_{eff}\sum_{<i,j>} b_i^{\dagger} b_i b_j^{\dagger} b_j
+ [\infty]\sum_j b_j^{\dagger}b_j[b_j^{\dagger}b_j-1] \; ,
\label{bosons}
\ee 
with nearest neighbor hopping $t_{eff}=2t^2/U_{eff}$ and nearest neighbor 
repulsion $V_{eff}=2t_{eff}$. This effective model is applicable for 
energies and temperatures small compared to $U_{eff}$.  

The properties of this bosonic Hamiltonian, and closely related
models where additional interactions between bosons are included, 
have been widely studied\cite{zimanyigang,mpafbosons}.  
It has a large number of possible phases, including
superconducting, crystalline, and striped or liquid crystalline phases. 
\marginpar{\em Strong attractions impede coherent motion, and enhance charge ordering.}
The equivalence between hard core bosons and spin-1/2 operators can be 
used to relate this model to various spin models that have been studied in
their own right. However, for the present purposes, there are two clear 
lessons we wish to draw from this exercise. The first is that there are 
ordered states, in particular insulating charge ordered states, which can 
compete very successfully with the superconducting state in strong coupling.  
The second is that, even if the system does manage to achieve a 
superconducting ground state, the characteristic superconducting $T_c$ 
will be proportional to $t_{eff}$, and hence to the small parameter,
$t/U_{eff}$.

\subsubsection{Large $U_{eff}$:  bipolarons} More generally, in the 
strong coupling limit, $U_{eff}\gg t$, a perturbative approach in powers of 
$t/U_{eff}$ can be undertaken, regardless of the value of $M$. Once again, 
the zeroth order ground states are those of charge 2$e$ hard core bosons, 
as in Eq. (\ref{bosons}).  However, now the  phonons make a contribution 
to the ground state---the ground state energy is $-U_{eff} N^{el}+(1/2)
\omega_D N$ where $N$ is the number of sites, and the gap to the first 
excited state is the smaller of $U_{eff}$ and $\omega_D$. Still, we can 
study the properties of the model at energies and temperatures small 
compared to the gap in terms of the hard core bosonic model. Now, however,
\ba
t_{eff}&=&  2 \frac{t^2}{U_{eff}} F_+\left(X\right) 
 \; ,  \nonumber \\
V_{eff}&=&4 \frac{t^2}{U_{eff}} F_-\left(X\right) \; ,
\ea 
where $X\equiv\frac{U_{eff}}{\omega_D}$ and
\be
F_{\pm}(X)=\int_0^{\infty} d t \exp\{ -t -X[1\pm \exp(-t/X)]\} \; .
\ee
This is often referred to as
a model of bipolarons. 
In the inverse adiabatic limit, $F_{\pm}(X)\to 1$ as $X\to 0$, and hence these
expressions reduce to those of the previous subsection. 
However, in the adiabatic limit, $X \gg 1$, $F_+(X)\sim e^{-2X}$, 
so $t_{eff}$ is exponentially reduced by a Frank-Condon factor! 
However, $F_-(X)\to 1$ as $X\to \infty$, so $V_{eff}$ remains substantial. 
Clearly, the lessons concerning the difficulty of obtaining high
temperature superconductivity from strong coupling drawn from the 
negative $U$ Hubbard model apply even more strongly to the case in 
which the phonon frequency is small. A bipolaron mechanism of 
superconductivity is simply impossible unless the phonon frequency
is greater than or comparable to $U_{eff}$; in the opposite limit, 
the exponential suppression of $t_{eff}$ relative to the effective 
interactions, $V_{eff}$, strongly suppresses the coherent Bose-condensed 
state, and favors various types of insulating, charge ordered states.

\subsection{Insulating quantum antiferromagnets}

We now turn to models with repulsive interactions. To begin with, we discuss 
the ``Mott limit'' of the antiferromagnetic insulating state. Here, we 
imagine that there is one electron per site, and such strong interactions 
between them that charge fluctuations can be treated petrubatively.  
In this limit, as is well known, the only low energy degrees of freedom 
involve the electron spins, and hence the problem reduces to that of
an effective quantum Heisenberg antiferromagnet.  

\subsubsection{Quantum antiferromagnets in more than one dimension}

In recent years, there has been considerable interest 
\cite{myreview,shivaji,kirillchetan,nayakfisherdual,sondhifradkin,
coldea,spinliq1,spinliq2,spinliq3} 
in the many remarkable quantum states that can occur in quantum spin models 
with sufficiently strong frustration---these studies are beyond 
the scope of the present review. 
\marginpar{\em In more than one dimension, it is a solved problem.} 
On a hypercubic lattice (probably on any simple, bipartite lattice) and
in dimension 2 or greater, there is by now no doubt that even the spin 1/2 
model (in which quantum fluctuations are the most severe) has a N\`{e}el 
ordered ground state\cite{chn1}. Consequently, the properties of such systems 
at temperatures and energies low compared to the antiferromagnetic exchange 
energy, $J$, are determined by the properties of interacting spin waves. 
This physics, in turn, is  well described in terms of a simple field theory, 
known as the $O(3)$ nonlinear sigma model. While interesting work is still 
ongoing on this problem, it is in essence a solved problem, and excellent 
modern reviews exist\cite{chakreview}.

In its ordered phase, the antiferromagnet has:  i) gapless spin wave 
excitations, 
\marginpar{\em Antiferromagnetic order is bad for superconductivity.}
and ii)  reduced tendency to phase ordering due to the frustration of
charge motion. Since the superconducting state 
possesses a spin gap (or, for d-wave, a partial gap) and is characterized 
by the extreme coherence of charge motion, it is clear that both these 
features of the antiferromagnet are disadvantageous for superconductivity.
\footnote{There is a very
interesting line of reasoning\cite{SO5} which takes the opposite viewpoint:
it is argued that the important point to focus on is that both 
the superconductor and the antiferromagnet have gapless Goldstone modes, 
not whether those modes are spinless or spinful. In this line of thought 
there is a near symmetry, which turns out to be SO(5), between the $d$-wave 
superconducting and the N\'{e}el ordered antiferromagnetic states.  This is an 
attractive notion, but it is not clear to us precisely how this line of 
reasoning relates to the more microscopic considerations discussed here.}   

There is a body of thought\cite{pinessfe,levinsfe,bulutscal,scalspin} 
that holds that it is
possible,  at sufficiently strong doping of an antiferromagnet to reach a
state in which  the antiferromagnetic order and the consequent low
energy spin fluctuations  are eliminated and electron itineracy is
restored, which yet has vestiges of  the high energy spin wave
excitations of the parent ordered state that can  serve to induce a
sufficiently strong effective attraction between electrons  for high
temperature superconductivity. 
Various strong critiques of this approach have also been articulated 
\cite{andersonadvances}.  
\marginpar{\em A spin fluctuation exchange mechanism in a nearly antiferromagnetic
electron fluid is critiqued.}
We feel that the theoretical viability of this ``spin fluctuation exchange'' 
idea has yet to be firmly established. As an example of how this could be 
done, one could imagine studying a two component system consisting of a 
planar, Heisenberg antiferromagnet coupled to a planar Fermi liquid. 
One would like to see that, as some well articulated measure of the 
strength of the antiferromagnetism is increased, 
the superconducting pairing scale 
likewise increases. If such a system could be shown to be a high temperature 
superconductor, it would establish the point of principal. However, it has 
been shown by Schrieffer\cite{schrieffer} that Ward identities, 
which are ultimately 
related to Goldstone's theorem, imply that long wavelength spin waves cannot 
produce any pairing interaction at all. A model of this sort that has 
been analyzed in 
detail is the one dimensional Kondo-Heisenberg model, which is the 1D 
analogue of this system\cite{sikkema,zek,zacharkondo}. This system does not 
exhibit significant superconducting fluctuations of any conventional kind.  
While there  certainly does not exist a ``no-go'' theorem, it does not seem 
likely to us that an exchange of spin waves in a nearly
anitferromagnetic system can ever give rise to high temperature
superconductivity.\footnote{Under 
circumstances in which antiferromagnetic correlations are
very short ranged, it may still be possible to think of an effective 
attraction between electrons mediated by the exchange of very local 
spin excitations\cite{scalspin}. This escapes most of the critiques 
discussed above---neither Ward identities nor the general
incompatibility between antiferromagnetism and easy electron itineracy 
have any crisp meaning at  short distances. By the same token, however, 
it is not easy to  unambiguously show that such short range magnetic 
correlations are the  origin of strong superconducting correlations 
in any system, despite  some recent progress along these
lines\cite{whiteandscalapino}.}

\subsubsection{Spin gap in even leg Heisenberg ladders}  

The physics of quantum antiferromagnets in one dimension is quite different 
from that in higher dimension, since the ground state is not magnetically 
ordered. However, its general features have been well understood for many 
years. In particular, for spin-1/2 Heisenberg ladders or cylinders with an 
even number of sites on a rung, quantum fluctuations result in a state with 
a spin gap. This is a special case of a general result \cite{haldane}, 
known as ``Haldane's conjecture,'' that any 1D spin system with an even 
integer number of electrons per unit cell has a spin rotationally invariant 
ground state and a finite spin gap in the excitation spectrum.  
This conjecture has not been proven, but  has been validated in many 
limits and there are no known exceptions 
\footnote{One can hardly fail to notice that the Haldane conjecture
is closely related to the conventional band structure view that insulators 
are systems with a gap to both charge and spin excitations due to the fact 
that there are an even number of electrons per unit cell and all bands are 
either full or empty.}.  

\marginpar{\em Insulating ladders are good parents for high temperature superconductors.}
The physics of interacting electrons on ladders---{\it i.e.} ``fat'' 1D 
systems, will be discussed at length below. We believe this is an important, 
paradigmatic system for understanding the physics of high temperature 
superconductivity. The fact that even the undoped (insulating) ladder 
has a spin gap can be interpreted as a form of incipient superconducting 
pairing.  Where that gap is large, {\it i.e.} a substantial
fraction of the exchange energy, $J$, it is reasonable to hope that 
doping it will lead to a conducting state which inherits  from the parent 
insulating state this large gap, now directly interpretable as a pairing gap.

Let us start by considering an $N$ leg spin-1/2 Heisenberg model
\be
H=\sum_{<i,j>}J_{ij}\vec S_{i}\cdot \vec S_{j} \; ,
\ee
where $\vec S_{i}$ is the spin operator on site $i$, so for $a, b, c = 
\{x, y, z$\}, $[S_{i}^{a},S_{j}^{b}] = i\delta_{ij}\epsilon^{abc}S_{i}^{c}$ 
and $\vec S_{i}\cdot \vec S_{i}= 3/4$. Here, we still take the lattice to 
be infinite in one (``parallel'') direction  but of width $N$ sites in 
the other.  At times, we will distinguish between a ladder, with open 
boundary conditions in the ``perpendicular'' direction, and a cylinder, 
with periodic boundary conditions in this direction. We will typically 
consider isotropic antiferromagnetic couplings, $J_{ij}=J >0$.

\paragraph{Ladders with many legs:}
In the limit of large $N$, it is clear that the model can be viewed as 
a two dimensional antiferromagnet up to a crossover scale, beyond 
which the asymptotic one dimensional behavior is manifest. This 
viewpoint was exploited by Chakravarty\cite{chak96} to obtain a remarkably 
accurate analytic estimate of the crossover scale. His approach was to first 
employ the equivalence between the Heisenberg model and the quantum 
nonlinear sigma model.
\marginpar{\em The spin gap falls exponentially with $N$.}  
One feature of this mapping is that the thermodynamic properties of the $d$ 
dimensional Heisenberg model are related to a $d+1$ 
dimensional sigma model, with an imaginary time direction which, by 
suitable rescaling, is {\em precisely} equivalent to any of the spatial 
directions.  The properties of the Heisenberg model 
at finite temperatures are then related to the sigma model on a 
generalized cylinder, which is periodic in the imaginary time 
direction with circumference $\hbar v_{s}/T$ where $v_{s}$ is the 
spin wave velocity.  What Chakravarty pointed out is that, through 
this mapping,  there is an equivalence between the Heisenberg cylinder 
with circumference $L=Na$ at zero temperature and the infinite planar 
Heisenberg magnet at temperature, $T = v_{s}/L$.  From the well known 
exponential divergence of the correlation length with decreasing 
temperature in the 2d system, he obtained the
asymptotic expression for the dimensional crossover length in the cylinder,
\be
\xi_{dim} \sim a \exp[0.682 N] \; .
\label{xicylinder}
\ee 
As this estimate is obtained from the continuum theory, it is only 
well justified in the large $N$ limit. However, comparison with 
numerical experiments described in Section \ref{numerical} 
(some of which predated the analytic theory \cite{2legwhite}) 
reveal that it is amazingly accurate, even for $N=2$, and that 
the distinction between ladders and cylinders is not very significant, either.

This result is worth contemplating.  It implies that the special 
physics of one dimensional magnets is only manifest at exponentially 
long distances in fat systems.  Correspondingly, it means that these 
effects are confined to energies (or temperatures) smaller than the
characteristic scale
\be
\Delta_{dim} =  v_{s}/\xi_{dim} \; .
\ee
As a practical matter, it means that 
only the very narrowest systems, with $N$ no bigger than 3 or 4, will  
exhibit the peculiarities of one dimensional magnetism at any 
reasonable temperature.

To understand more physically what these crossover scales mean, one 
needs to know something about the behavior of one dimensional 
magnets.  Since even leg ladders and cylinders have a spin gap, it
is intuitively clear (and correct) that $\Delta_{dim}$ is nothing but
the spin gap and $\xi_{dim}$ the correlation length associated with 
the exponential fall of magnetic correlations at $T=0$.  For odd leg 
ladders, $\xi_{dim}$ is analogous to a Josephson length, where 
correlations crossover from the two dimensional power law behavior 
associated with the existence of Goldstone modes, to the peculiar 
quantum critical behavior of the one dimensional spin 1/2 Heisenberg chain.

\paragraph{The two leg ladder:}  It is often useful in developing 
intuition to consider limiting cases in which the mathematics 
becomes trivial, although one must always be sensitive to the danger of 
being overly influenced by the naive intuitions that result.  

In the case of the two leg ladder, there exists such a limit,
$J_{\perp}\gg J_{\parallel}$,  where $J_{\perp}$ and $J_{\parallel}$ are, 
respectively, the exchange couplings across the rungs, and along the sides 
of the ladder. Here the zeroth order ground state is a  direct product of 
singlet pairs (valence bonds) on the rungs of the ladder. 
Perturbative corrections to the ground state cause these valence  bonds to
resonant, locally, but do not fundamentally affect the  
character of the ground state. The ground state energy per site is
\be
E_{0}=-(3/8)J_{\perp}[1 + (J_{\parallel}/J_{\perp})^{2}+\ldots] \; .
\ee
Since each valence bond is nothing but 
a singlet pair of electrons, this makes it clear that there is a very 
direct sense in which the two leg ladder can be thought of as a 
paired insulator.  The lowest lying spin-1 excited states are a 
superposition of bond triplets on different rungs, and have a 
dispersion relation which can easily be derived in perturbation 
theory:
\be
E_{triplet}= J_{\perp}+J_{\parallel}\cos(k) + {\cal 
O}(J_{\parallel}^{2}/J_{\parallel}) \; .
\ee
This, too, reveals some features that are more general, such as a minimal spin gap of 
magnitude $\Delta_s= J_{\perp}[1 -(J_{\parallel}/J_{\perp})+{\cal 
O}(J_{\parallel}^{2}/J_{\parallel}^{2})]$ at what would be the 
antiferromagnetic ordering wavevector $k=\pi$.

\subsection{The isolated square}
\label{square}
While we are considering mathematically trivial problems, it is worth taking a 
minute to discuss the solution of the $t-J$ model (defined in Eq. (\ref{tJ}), 
below) on an isolated 4-site square. The pedagogic value of this problem, 
which is exactly diagonalizable, was first stressed 
by Trugman and Scalapino \cite{trugscal}.   This idea was recently
carried further by Auerbach and collaborators\cite{assa3,2000}, who have
attempted to build a theory of the 2D $t-J$ model by linking together
fundamental squares. 
The main properties of the lowest energy states of this system are given 
in Table 2 for any number of doped holes. 

\begin{table}
\label{squaretable}
\begin{center}
\begin{tabular}{|c|}
\hline
0 holes \\
\hline
\hline
\end{tabular}
\end{center}
\vspace{-0.75cm}
\begin{center}
\begin{tabular}{|c||c|c|c|}
\hline
& Energy & Spin & Momentum \\
\hline
\hline
g.s. & $E=-3J$ & $S=0$ & $P=\pi$ \\
\hline
$1^{\rm st}$ e.s. & $E=-2J$ & $S=1$ & $P=0$ \\
\hline
\end{tabular}
\end{center}
\begin{center}
\begin{tabular}{|c|}
\hline
1 hole \\
\hline
\hline
\end{tabular}
\end{center}
\vspace{-0.75cm}
\begin{center}
\begin{tabular}{|c||c|c|c|}
\hline
& Energy & Spin & Momentum \\
\hline
\hline
$0<J/t<0.263$ & & & \\
\hline
g.s. & $E=-2t$ & $S=3/2$ & $P=0$ \\
\hline
$1^{\rm st}$ e.s. & $E=-J-\sqrt{J^2/4+3t^2}$ & $S=1/2$ & $P=\pm\pi/2$ \\
\hline
\hline
$0.263<J/t<2/3$ & & & \\
\hline
g.s. & $E=-J-\sqrt{J^2/4+3t^2}$ & $S=1/2$ & $P=\pm\pi/2$ \\
\hline
$1^{\rm st}$ e.s. & $E=-2t$ & $S=3/2$ & $P=0$ \\
\hline
\hline
$2/3<J/t<2$ & & & \\
\hline
g.s. & $E=-J-\sqrt{J^2/4+3t^2}$ & $S=1/2$ & $P=\pm\pi/2$ \\
\hline
$1^{\rm st}$ e.s. & $E=-3J/2-t$ & $S=1/2$ & $P=\pi$ \\
\hline
\hline
$2<J/t$ & & & \\
\hline 
g.s. & $E=-3J/2-t$ & $S=1/2$ & $P=\pi$ \\
\hline
$1^{\rm st}$ e.s. & $E=-J-\sqrt{J^2/4+3t^2}$ & $S=1/2$ & $P=\pm\pi/2$ \\
\hline
\end{tabular}
\end{center}
\begin{center}
\begin{tabular}{|c|}
\hline
2 holes \\
\hline
\hline
\end{tabular}
\end{center}
\vspace{-0.75cm}
\begin{center}
\begin{tabular}{|c||c|c|c|}
\hline
& Energy & Spin & Momentum \\
\hline
\hline
$0<J/t<2$ & & & \\
\hline
g.s. & $E=-J/2-\sqrt{J^2/4+8t^2}$ & $S=0$ & $P=0$ \\
\hline
$1^{\rm st}$ e.s. & $E=-2t$ & $S=1$ & $P=\pm\pi/2$ \\
\hline
\hline
$2<J/t$ & & & \\
\hline
g.s. & $E=-J/2-\sqrt{J^2/4+8t^2}$ & $S=0$ & $P=0$ \\
\hline
$1^{\rm st}$ e.s. & $E=-J$ & $S=0$ & $P=\pi,\pm\pi/2$ \\
\hline
\end{tabular}
\end{center}
\caption{The low energy spectrum of the 4-site $t-J$ square for 0 holes
(4 electrons), 1 hole (3 electrons), and 2 holes (2 electrons).  The 3 and
4 hole problems are left as an exercise for the reader.}
\end{table}

The ``undoped'' state of this system ({\it i.e.} with 4  
electrons) is a singlet with ground state energy $E_0=-3J$. However, 
interestingly, it is not in the identity representation of the symmetry 
group of the problem---it is odd under 90$^{o}$ rotation. If we number 
the sites of the square sequentially from 1 to 4, then the ground state 
wavefunction is
\be
|4-electron\rangle = [\hat P^{\dagger}_{1,2}\hat P^{\dagger}_{3,4}
-\hat P^{\dagger}_{1,4}\hat P^{\dagger}_{2,3}]|0\rangle
\ee
where
$\hat P^{\dagger}_{i,j}=\hat P^{\dagger}_{j,i}=[c_{i,\uparrow}^\dagger 
c_{j,\downarrow}^\dagger+c_{j,\uparrow}^\dagger c_{i,\downarrow}^\dagger]
/\sqrt{2}$ 
creates a singlet pair of electrons on the bond between sites $i$ and $j$.  
Manifestly, $|4-electron\rangle$ has the form of an odd superposition of 
nearest neighbor valence bond states---in this sense, it is the quintessential 
resonating valence bond state. The lowest lying excitation is a spin-1 state 
with energy $-2J$, so the spin gap is $J$.

There are level crossings as a function of $J/t$ in the ``one hole'' (3
electron) spectrum. For $0<J/t<(8-\sqrt{52})/3\approx 0.263$ the
ground state is a spin  3/2 multiplet with energy $E_1=-2t$. It is
orbitally nondegenerate with zero  momentum (we consider the square as a
4-site chain with periodic boundary  conditions and refer to the momentum
along the chain.)  For $(8-\sqrt{52})/3<J/t<2$ the ground state has spin
1/2,  is two-fold degenerate with crystal momentum $\pm \pi/2$, and has
energy 
$E_1=-[2J+\sqrt{J^2+12t^2}]/2$. For $2<J/t$, the 
ground state has spin 1/2, zero momentum, and energy $E_1=-3J/2-t$.

The two hole (2 electron) ground state has energy 
$E_2=-[J+\sqrt{J^2+32t^2}]/2$, and spin 0. It lies in the identity 
representation of the symmetry group. The lowest excitation  
is a spin 1 state. For $0<J/t<2$ it has crystal momentum $k=\pm \pi/2$ 
({\it i.e.} it has a two-fold orbital and 3-fold spin degeneracy) and has 
energy $E_2(S=1)=-2t$. For $2<J/t$ it is orbitally nondegenerate with 
energy $E_2(S=1)=-J$.

One important 
\marginpar{\em Pair field correlations have $d_{x^2-y^2}$ symmetry.}
consequence of this, which follows directly from the Wigner-Eckhart 
theorem, is that the pair annihilation operator that connects the 
zero hole and the two hole ground states must transform as $d_{x^2-y^{2}}$. 
This is, perhaps, the most important result of this exercise. It 
shows the robustness of the $d$ wave character of the pairing in a 
broad class of highly correlated systems.  
The dominant component of this operator is of the form 
\be
\phi_{1}= \hat P_{12}-\hat P_{23}+\hat P_{34}-\hat P_{41}. 
\ee
It also includes terms that create holes on next nearest neighbor 
diagonal sites \cite{whitepair,poilblancpair}.

There are a few other interesting aspects of this solution. In the 
single hole sector, the ground state is maximally polarized, in 
agreement with Nagaoka's theorem, for sufficiently large $t/J$, but 
there is a level crossing to a state with smaller spin when $t/J$ is still
moderately large.
Moreover, even when the single hole state 
is maximally polarized, the two hole state, like the zero hole state, 
is always a spin singlet. Both of these features have been observed
in numerical studies on larger $t-J$ clusters\cite{nagaokanumerics}.  

If we look still more closely at the $J/t\to 0$ limit, there is
another interesting aspect of the physics: It is intuitively clear 
that in this limit, the holes should behave as spinless fermions. 
This statement requires no apology in the maximum spin state. 
Thus, the lowest energy spin-1 state with two holes has energy 
$E_2(S=1)=-2t$ in this limit.  It corresponds to a state in which 
one spinless fermion has crystal momentum $k=0$ and energy $-2t$, 
and the other has crystal momentum $\pm \pi/2$ and energy 0. However, 
what is more interesting is that there is also a simple interpretation of
the two hole ground state in the same representation. The antisymmetry 
of the spins in their singlet state means that they affect the hole 
dynamics through a Berry's phase as if half a magnetic flux quantum 
were threaded through the square. This Berry's phase implies
that the spinless fermions satisfy antiperiodic boundary conditions. 
The ground state is thus formed by occupying the single particle states 
with $k=\pm \pi/4$ for a total ground state energy of $E_2=-2\sqrt{2}t$, 
precisely the $J/t\to 0$ limit of the expression obtained above. 
The interesting thing is that, in this case, it is the hole kinetic 
energy, and {\em not} the exchange energy, which favors the singlet 
over the triplet state. This simple exercise provides an intuitive 
motivation for the existence of various forms of ``flux phase''
in strongly interacting systems \cite{rokhsar}.

Finally, it is worth noting that pair binding occurs, in the sense that 
$2E_{1}-E_{0}-E_{2}>0$, so long as $J/t>\sqrt{(39-\sqrt{491})/\sqrt{3}}
\approx 0.2068$.  
We will return to the 
issue of pair binding in Section \ref{numerical} where we will show a 
similar behavior in Hubbard and $t-J$ ladders.

\subsection{The spin gap proximity effect mechanism}
\label{sgpfx}

The final strong coupling model we will consider 
consists of two inequivalent 1DEG's weakly coupled together---a generalization 
of a two leg ladder.  Each 1DEG is represented by an appropriate 
bosonized field theory---either a Luttinger liquid or a Luther-Emery 
liquid.  Most importantly, the two systems are assumed to have 
substantially different values of the Fermi momentum, $k_{F}$ and 
$\tilde k_{F}$.  We consider the case in which the interactions 
between the two systems are weak, but the interactions within each 
1DEG may be arbitrarily large.  The issue we address is what changes 
in the properties of the coupled system are induced by these 
interactions.  (For all technical details, see Refs.~\citen{spin gap} and \citen{spin gap2}.)

There is an important intuitive reason to expect this system to 
\marginpar{\em Intuitive description of the spin gap proximity effect \ldots}
exhibit a novel form of kinetic energy driven superconducting 
pairing.  Because $k_{F}\ne \tilde k_{F}$, single particle tunnelling 
between the two 1DEG's is not a low energy process---it is irrelevant 
in the renormalization group sense, and can be ignored as anything but 
a high energy virtual fluctuation.  The same conclusion holds for any 
weak coupling between the $2k_{F}$ or $4k_{F}$ density wave 
fluctuations.  There are only two types of coupling that are 
potentially important at low energies:  pair tunnelling, since the 
relevant pairs have 0 momentum, and coupling between 
long wavelength spin fluctuations.  

The magnetic
interactions are marginal to leading order in a perturbative RG 
analysis---they turn out to be marginally relevant if 
the interactions are antiferromagnetic and marginally irrelevant if 
ferromagnetic\cite{sikkema,zacharkondo}. The effect of purely magnetic
interactions  has been widely studied in the context of Kondo-Heisenberg
chains,  but will not be discussed here.  The effect of triplet pair
tunnelling  has only been superficially analyzed in the literature
\cite{VZ,linbalents,spin gap2}---it would be
worthwhile  extending this analysis, as it may provide some insight into the 
origin of the triplet superconductivity that has been observed 
recently in certain highly correlated materials.  However, in the 
interest of brevity, we will  ignore these interactions.

Singlet pair tunnelling interactions between the two 1DEG's have a scaling 
\marginpar{\em  \ldots as a
kinetic energy driven mechanism of pairing.} 
dimension which depends on the nature of the correlations in the decoupled 
system.  Under appropriate 
circumstances, they can be relevant.  When this is the case, the 
coupled system scales to a new strong coupling fixed point which 
exhibits a total spin gap and strong global superconducting 
fluctuations.  This is what we refer to as the spin gap proximity 
effect, because the underlying physics is analogous to the proximity 
effect in conventional superconductors.  The point is that even if it 
is energetically costly to form pairs in one or both of the 1DEGs, 
once the pairs are formed they can coherently tunnel between the two 
systems, thereby lowering their zero point kinetic energy.  Under 
appropriate circumstances, the kinetic energy gain outweighs the cost 
of pairing.  This mechanism is quite distinct from any relative of the 
BCS mechanism---it does not involve an induced attraction.

The explicit model which is analyzed here is expressed in terms of four 
bosonic fields: $\phi_{c}$ and $\phi_{s}$ represent the charge and spin 
degrees of freedom of the first 1DEG, and $\tilde \phi_{c}$ and 
$\tilde \phi_{s}$ of  the other, as is discussed in Section \ref{1D}, above.  
The Hamiltonian of the decoupled system is the general bosonized Hamiltonian 
described in that section, with appropriate velocities and charge 
Luttinger exponents, $v_{s}$, $v_{c}$, $\tilde v_{s}$, $\tilde v_{c}$, 
$K_{c}$, and $\tilde K_{c}$ if both are Luttinger liquids, and values 
of the spin gap, $\Delta_{s}$ and $\tilde \Delta_{s}$ in the case of 
 Luther-Emery liquids ({\it i.e.} if the cosine potential in the 
sine-Gordon theory for the spin degrees of freedom is relevant).  
If we ignore the long wavelength magnetic couplings and triplet pair 
tunnelling between the two systems, the remaining possibly important 
interactions at low energy,
\be
H_{inter}=\int dx [{\cal H}_{for}+{\cal H}_{pair}] \; ,
\ee
are the forward scattering (density-density and current-current) interactions
in the charge sector
\be
{\cal H}_{for}= V_{1}\partial_{x}\phi_{c}\partial_{x}\tilde \phi_{c} 
+V_{2}\partial_{x}\theta_{c}\partial_{x}\tilde \theta_{c} \; ,
\ee
where $\theta$ designates the field dual to $\phi$ 
(see Section \ref{1D}), and the singlet 
pair tunnelling
\be
{\cal H}_{pair}={\cal J} \cos[\sqrt{2\pi}\phi_{s}]\cos[\sqrt{2\pi}
\tilde\phi_{s}] \cos[\sqrt{2\pi}(\theta_{c}-\tilde\theta_{c})] \; .
\ee
As discussed previously, the singlet pair creation operator 
involves both the spin and the charge fields.

The forward scattering interactions are precisely marginal, 
and should properly be incorporated in the definition of 
the fixed point Hamiltonian.  
${\cal H}_{pair}$ is a nonlinear interaction;  the coupled problem 
with nonzero ${\cal J}$ has not been
exactly solved.  However, it is relatively straightforward to asses the  
perturbative relevance of this interaction, and to deduce the 
properties of the most likely strong coupling fixed point (large 
${\cal J}$) which governs the low energy physics when it is relevant.  

The general expression for the scaling dimension of ${\cal H}_{pair}$ 
is a complicated analytic combination of the parameters of the 
decoupled problem
\be
\delta_{pair}=\frac{1}{2} \left[ \frac A {K_c} + \frac {B} {\tilde K_c} 
+ K_s + \tilde K_s \right] \; ,
\ee
where $A=1$ and  $B = 1$ in
\marginpar{\em The scaling dimension of the pair tunnelling interaction is introduced.}
the absence of intersystem forward scattering interactions, but more 
generally $A$ and $B$ are complicated functions of the coupling constants.  
For illustrative purposes, one can consider the explicit expression 
for these functions under the special circumstances
$V_{2}=-(\tilde v_c/v_c)(K_c\tilde K_c)V_{1}$;  
then $A=\sqrt{1-(V_1^2 K_c\tilde{K_c}/v_c\tilde{v_c})}$ and 
$B=(v_c-V_1 K_c)^2/ \sqrt{v_c^4-V_1^2 v_c\tilde{v_c}K_c\tilde{K_c}}$.
Here, if both 1DEG's are Luttinger liquids, spin rotational  
invariance implies that $K_{s}=\tilde K_{s}=1$.  
If one or the other  1DEG is a Luther-Emery liquid, one should substitute
$K_{s}=0$ or $\tilde K_{s}=0$ in the above expression.  

Pair tunnelling is 
perturbatively relevant if $\delta_{pair}< 2$, and irrelevant 
otherwise.  Clearly, having a preexisting spin gap in either of the 
1DEG's dramatically decreases $\delta_{pair}$---if there is already 
pairing in one subsystem, then it stands to reason that pair 
tunnelling will more easily produce pairing in the other. However, 
\marginpar{\em The physical effects which make pair tunnelling relevant are described.}
even if neither system has a preexisting spin gap, there are a wide 
set of physical circumstances for which $\delta_{pair}<2$. Notice, in 
particular, that repulsive intersystem interactions, $V_{1}>0$, 
produce a  reduction of $\delta_{pair}$.  Again, the physics of 
this is intuitive---an induced anticorrelation between regions of 
higher than average electron density in the two 1DEG's means that 
where there is a pair in one system, there tends to be a low density 
region on the other which is just waiting for a pair to tunnel into it.  
(See, also, Section~\ref{smectic}.)

In the limit that ${\cal J}$ is large, the spin fields in both 1DEG's 
are locked, which implies a total spin gap, and the out-of-phase fluctuations
\marginpar{\em The implications of strong pair tunnelling are discussed.}
of the dual
charge phases are gapped as well.  This means that the 
only possible gapless modes of the system involve the total charge 
phase, $\phi\equiv [\phi_{c}+\tilde \phi_{c}]/\sqrt{2}$, and its dual,
$\theta\equiv [\theta_{c}+\tilde \theta_{c}]/\sqrt{2}$.  $\theta$ is 
simply the total superconducting phase of the coupled system, and 
$\phi$ the total CDW phase.  At the end of the day, this strong coupling fixed point of the 
coupled system is a Luther-Emery liquid, and 
consequently has a strong tendency to superconductivity.  
In general,
there will be substantial renormalization of the effective parameters 
as the system scales from the weak to the strong coupling fixed point.  
Thus, it is difficult to estimate the effective Luttinger
parameters which govern the charge modes of the resulting Luther-Emery liquid.
A naive estimate, which may well be unreliable, can be be made by simply 
setting ${\cal J}\rightarrow \infty$.  In this case, all the induced gaps are
infinite, and the  velocity and Luttinger exponent that govern the dynamics 
of the remaining mode are 
\ba
K_c^{total}&=&\sqrt{v_cK_c+\tilde v_c\tilde K_c +2 V_2\over
v_c/K_c+\tilde {v_c}/\tilde {K_c} + 2 V_1} \; ,\\
v_c^{total}&=&\frac{1}{4}\sqrt{[v_cK_c+\tilde{v_c}\tilde{K_c} + 2 V_2]
[v_c/K_c+ \tilde{v_c}/\tilde{K_c} + 2 V_1]} \; . \nonumber
\ea

\section{Lessons from Numerical Studies of Hubbard and Related
Models}
\label{numerical}

\addtocontents{toc}{ 
\noindent{\em The careful use of numerical studies to understand the
physics on scales relevant to the mechanism of high temperature
superconductivity is advocated.}
}

High temperature superconductivity is a result of strong electronic correlations.
\marginpar{\em Numerical studies are motivated...}
Couple this prevailing thesis with the lack of controlled analytic methods for most 
relevant models, and the strong motivation for numerical approaches becomes evident.
Such numerical studies are limited to relatively small systems, due to a rapid growth in 
complexity with system size. However, many of the interesting aspects of the high 
temperature superconductors, especially those which relate to the ``mechanism'' of 
pairing, are moderately local, involving physics on the length scale of the 
superconducting coherence length $\xi$. Since $\xi$  is typically a few lattice spacings in 
the high $T_c$ compounds, one expects that numerical solutions of model problems on 
clusters with as few as 50-100 sites should be able to reveal the salient features of high 
temperature superconductivity, if it exists in these models.  Moreover, numerical studies 
can guide our mesoscale intuition, and serve as important tests of analytic predictions.

Notwithstanding these merits,
\marginpar{\em ... with caution.}
a few words of caution are in order. Even the largest systems that
have been studied so far\footnote{The largest are about 250 sites
\cite{whitescalapino,whitestripesallx2d} using the density matrix 
renormalization group method (DMRG) and up to approximately 800
sites  in Green function Monte Carlo simulations.
\cite{hm}}  are still relatively small. Therefore, the results are 
manifestly  sensitive to the shape and size of the cluster and other
finite  size effects. Some features, especially with regard to stripes, 
appear particularly sensitive to small changes in the model such 
as the presence of second neighbor hopping, \cite{2legt',stripeSCtt'}, 
the type of boundary conditions\cite{hellman99},
\nocite{hellman99-2,hellman99-3}  
{\em etc.} Less subtle 
modifications seem to have important 
consequences, too \cite{arrigoniandme},
most notably the inclusion of long range 
Coulomb forces (although this has been much less studied). 
This sensitivity has resulted in
considerable controversy in the field concerning the true
ground state phase diagrams of the stated models in the thermodynamic
limit; see Refs. \citen{hellman99,hellman99-2,hellman99-3} and \citen{whitecompare},
among others. 

The best numerical data, especially in terms of system size, exists for 
narrow Hubbard and $t-J$ ladders. We therefore begin by considering 
them. Apart from their intrinsic appeal, these systems also offer 
several lessons which we believe are pertinent to the two dimensional
models.
The second
part of this section provides a brief review of the conflicting  
results and views which have emerged from attempts to extrapolate from   
fat ladders and small periodic clusters to the entire plane.  

We feel that numerical studies are {\em essential} in order to explore the
important
\marginpar{\em What do we learn from numerical studies?}
mesoscale physics of highly correlated systems, but except in the few
cases where a careful finite size scaling analysis has been possible over a
wide range of system sizes, conclusions concerning the long distance
physics should be viewed as speculative.  Even where the extrapolation to
the thermodynamic limit has been convincingly established for a given
model, the established fact that there are so many closely competing
phases in the strong correlation limit carries with it the corollary 
that small changes in the Hamiltonian can sometimes tip the
balance one way or the other.  Thus, there are significant
limitations concerning the conclusions that can be drawn from numerical
studies.  In the present section we  
focus on the reproducible features of the local correlations that
follow robustly from the physics of strong, short ranged repulsions
between electrons, paying somewhat less attention to the various
controversies concerning the actual phase diagram of this or that model. 

We entirely omit any discussion of the technical details
of the  numerical calculations.  Methods that have been used
include exact diagonalization by Lanczos techniques, Monte Carlo 
simulations of various sorts, numerical renormalization group
approaches, and variational ansatz. The reader who is  interested in such aspects is
invited to consult Refs. 
\citen{dagreview,hellps00,sorella98qmc,chen95pl,dmrgreview,improvedDMRG}.

\subsection{Properties of doped ladders}

Ladder systems, that is, quasi-one dimensional systems obtained by assembling 
chains one next to the other, constitute a bridge between the essentially 
understood behavior of strictly one dimensional models and the incompletely
understood behavior in two dimensions.
Such systems are not merely a theoretical creation but
are realized  in nature \cite{ladder,dagotto5}. For example, two leg $S=1/2$ ladders 
(two coupled spin-1/2 chains)  are found in vanadyl pyrophosphate 
$(VO)_2P_2O_7$. Similarly, the cuprate 
compounds $SrCu_2O_3$ and $Sr_2Cu_3O_5$ consist of weakly coupled arrays of 
2-leg and 3-leg ladders, respectively. It is likely that ladder physics is 
also relevant to the high temperature superconductors, at least in the 
underdoped regime, where ample experimental evidence exists for the 
formation of self-organized stripes. 

In this section we review some of the most prominent features of 
\marginpar{\em Synopsis of findings}
Hubbard and especially $t-J$ ladders. As we shall see the data offers 
extensive support in favor of the contention that a purely electronic 
mechanism of superconductivity requires mesoscale structure\cite{C60}.  
Specifically, we will find that 
spin gap formation and 
pairing correlations, with robust
$d$-wave-like character,  are intimately connected. Both of these signatures
of local superconductivity
appear as distinct and  universal features in the physics of doped ladders. 
Nevertheless, they tend to diminish, in some cases very rapidly, with the 
lateral extent of the ladder, thus strongly suggesting that such structures 
are {\em essential} for the attainment of high temperature superconductivity. 
In addition we shall demonstrate the tendency of 
these systems to develop charge density wave correlations upon doping; 
it is natural to imagine that as the transverse width of the ladder tends to 
infinity, these density wave correlations will evolve into true 
two dimensional stripe order.

\subsubsection{Spin gap and pairing correlations} 

\paragraph{\bf Hubbard chains:} The
purely one dimensional Hubbard model can  be solved exactly using  
Bethe ansatz \cite{lw,Schulz92} and thus may seem out of place in this 
section. 
However, like other models in this section, it is a lattice fermion model.
In analyzing it, we will 
encounter many of the concepts that will figure  prominently in our 
discussion of the other models treated here, most 
notably the importance of intermediate scales. Anyway, 
in many cases, the Bethe ansatz equations themselves must be solved  
numerically, so we can view this as simply a more efficient numerical 
algorithm which permits us to study larger systems (up to 1000 sites 
\cite{C60} or more). 

The Hubbard Hamiltonian is   
\begin{equation}
\label{hubbard}
H_U=-t\sum_{\langle i,j\rangle,s}(c_{i,s}^{\dagger}c_{j,s}+h.c.)
+U\sum_i n_{i,\uparrow}n_{i,\downarrow} \;,
\end{equation}
where $\langle\ \ \rangle$ denotes nearest neighbors on a ring with an even
number  of sites $N$ and $N+Q$ electrons. We define $E(Q,S)$ to be the 
lowest lying energy eigenvalue with total spin $S$ and ``charge'' $Q$.   
Whenever the ground state is a spin singlet we can define the spin gap 
$\Delta_s$ as the energy gap to the lowest $S=1$ excitation
\begin{equation}
\label{spin gapdef}
\Delta_s(Q)=E(Q,1)-E(Q,0) \; .
\end{equation} 
The pair binding energy is defined as
\begin{equation}
\label{pbdef}
E_{pb}(Q)=2E(Q+1)-E(Q+2)-E(Q) \; ,
\end{equation}
where $E(Q)$ has been minimized with respect to $S$. 
A positive pair binding 
energy means that given $2(N+Q+1)$ electrons and two clusters, it is 
energetically more favorable to place $N+Q+2$ electrons on one cluster and 
$N+Q$ on the other than it is to put $N+Q+1$ electrons on each cluster. 
In this sense, a positive $E_{pb}$ signifies an effective {\em attraction} 
between electrons. The exact particle-hole symmetry of the Hubbard model 
on a bipartite lattice implies that electron doping $Q>0$
is equivalent to hole doping $Q<0$.

\begin{figure}[h!!!]
\begin{center}
\epsfig{figure=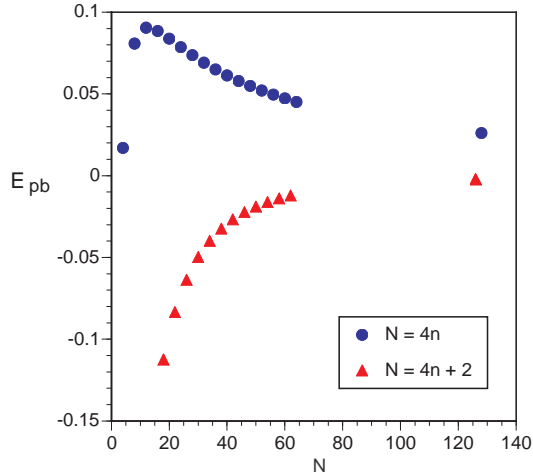,width=0.6\linewidth}
\end{center}
\caption{\label{fignum1}
Pair binding energy, $E_{pb}$, of $N=4n$ and $N=4n+2$ site Hubbard rings 
with $t=1$ and $U=4$. (From Chakravarty and Kivelson. \cite{C60})}
\end{figure}

\begin{figure}[ht!!!]
\begin{center}
\epsfig{figure=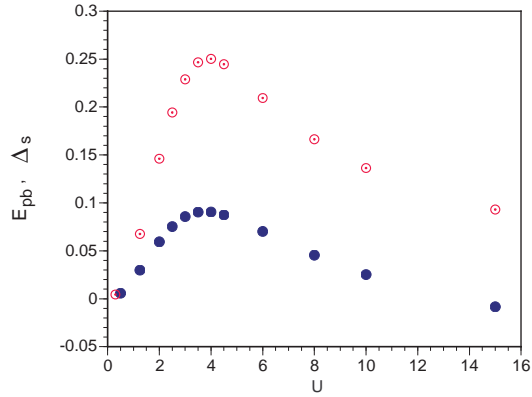,width=0.6\linewidth}
\end{center}
\caption{\label{fignum2}
Pair binding energy, $E_{pb}$ (solid symbols), and spin gap, $\Delta_s$
(open symbols), of a 12 site $Q=0$ Hubbard ring as a function of $U$ in 
units of $t=1$. (From Chakravarty and Kivelson. \cite{C60})}
\end{figure}

Fig. \ref{fignum1}
\marginpar{\em Intermediate scales play an important role.}
displays the pair binding energy for electrons added to $Q=0$ rings. 
The role of intermediate scales is apparent:
$E_{pb}$ vanishes for large $N$ and is maximal at an intermediate value 
of $N$. (The fact that pair binding occurs for $N=4n$ rings but not when 
$N=4n+2$ is readily understood from low order perturbation theory in $U/t$ 
\cite{C60}). Moreover, the spin gap $\Delta_s$ reaches a maximum 
at intermediate interaction strength, and then decreases for large 
values of $U$, as expected from its proportionality to the exchange 
constant $J=4t^2/U$ in this limit. The pair binding energy $E_{pb}$ follows 
suit with a similar dependence, as seen from Fig. \ref{fignum2}.

We have already seen the intimate relation between the spin gap and 
the superconducting susceptibility in the context of quasi-one
dimensional superconductors (see Section \ref{1D}). 
Further  understanding
of the relation between
pair binding and the spin gap can be gained 
by using bosonization to study the Hubbard model in the large $N$ limit 
\cite{Schulz92,C60}. The result for $N=4n\gg 1$ is
\begin{eqnarray}
\Delta_s&=&\frac{v_s}{N}\left[B_1\ln^{1/2}(N)+B_2\right] +
\ldots\label{eq:ds}\\
E_{pb}&=&\Delta_s +B_3 \frac {v_s} N- \frac{B_4}{N^2}
\left[\frac{{v_c}^2}{\Delta_c}\right]+
\ldots\label{eq:epair}
\end{eqnarray}
Here, $v_{s}$ and $v_{c}$ are the spin and charge velocities, respectively 
\marginpar{\em The spin gap and pairing are related.} 
(in units in which the lattice constant is unity), and
$\Delta_{c}$ is the charge gap in the $N\to \infty$ limit.
The constants, $B_j$, are numbers of order unity. The important 
lesson of this analysis is that pair binding is closely related to the 
phenomenon of spin gap formation. Indeed, for large $N$, 
$E_{pb}\approx\Delta_s$. 

\paragraph{\bf Hubbard and $t-J$ ladders:} In the thermodynamic limit, 
where the number of sites $N\to\infty$, Hubbard chains, and their strong coupling descendants the 
$t-J$ chains, have no spin gap
and a small
superconducting susceptibility, irrespective  of the doping level. In contrast,
ladder systems can exhibit both a spin gap  and a strong tendency towards
superconducting order even in the thermodynamic limit. While these systems are
infinite in extent, the mesoscopic physics comes in through the finiteness 
of the transverse dimension. 

\begin{figure}[ht!!!]
\begin{center}
\epsfig{figure=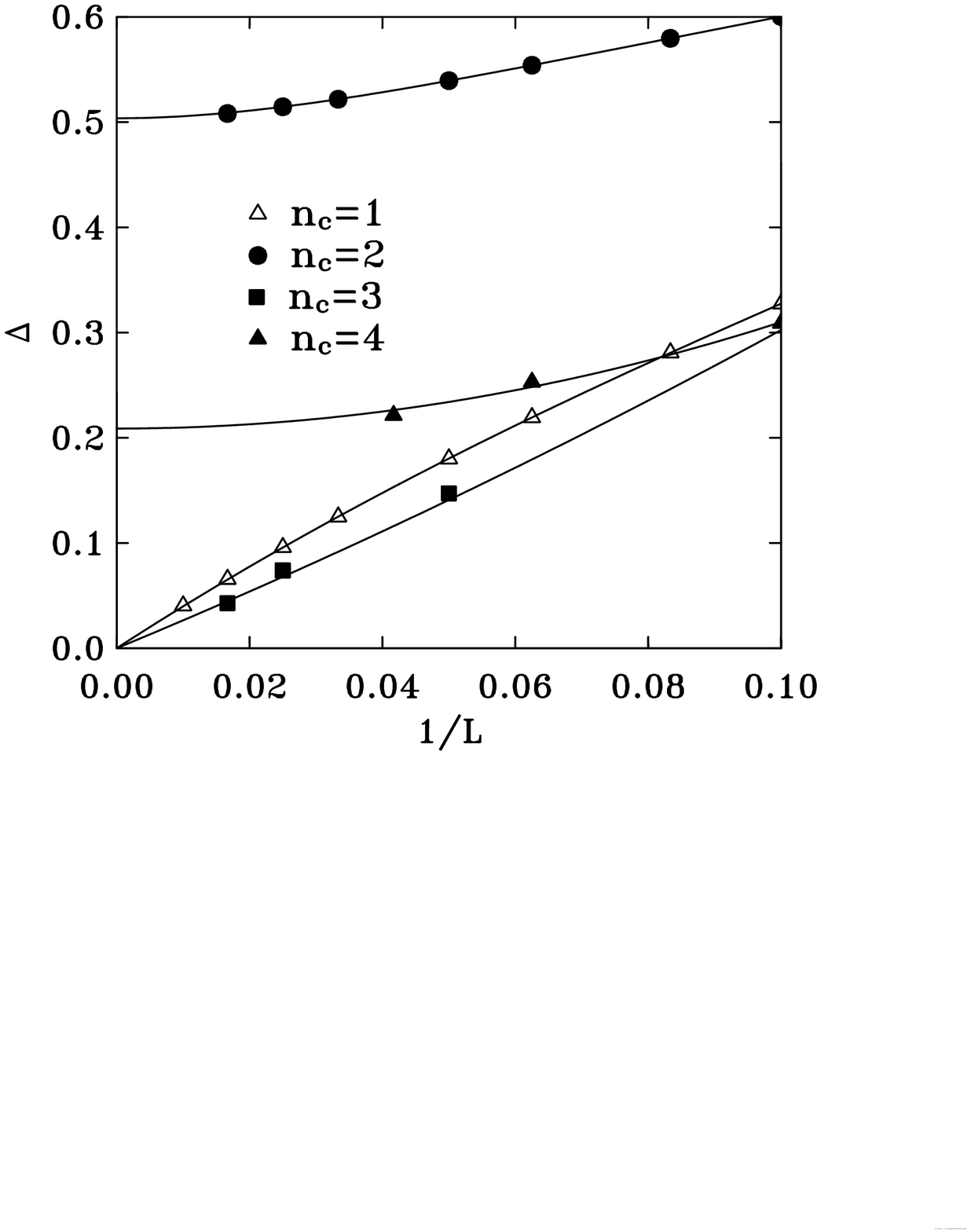,width=0.6\linewidth}
\end{center}
\caption{\label{fignum3}
Spin gaps as a function of system size $L$ for open $L\times n_c$  
Heisenberg ladders. (From White {\em et al.} \cite{2legwhite})}
\end{figure}

In the large $U$ limit and at half filling (one electron per site) the Hubbard 
ladder is equivalent to the spin-1/2 Heisenberg ladder
\begin{equation}
\label{heisenberg}
H_J=J \sum_{\langle i,j\rangle}{\bf S}_i\cdot{\bf S}_j \; ,
\end{equation}
where ${\bf S}_i$ is a spin 1/2 operator, $J=4t^2/U\ll t$ is the 
antiferromagnetic exchange interaction, and $\langle i,j\rangle$ now 
signifies nearest neighbor sites of spacing $a$ on the ladder.
As discussed in Section \ref{strong}, 
there is a marked difference between the behavior of ladders with 
even and odd numbers of chains or ``legs''. 
\marginpar{\em The number of legs matters!}
While even leg ladders 
are spin gapped with exponentially decaying spin-spin correlations,
odd leg ladders are gapless and exhibit power law 
falloff of these correlations (up to logarithmic 
corrections). This difference is clearly 
demonstrated in Fig. \ref{fignum3}. The spin gaps for the first few 
even leg ladders are known numerically \cite{2legwhite,sgtroyer}. 

For the two, four, and six leg ladders, $\Delta_s=0.51(1)J$, 
$\Delta_s=0.17(1)J$, and $\Delta_s=0.05(1)J$, respectively. 
This gap appears to vanish exponentially with the width $W$ 
of the system, in accordance with the theoretical estimate \cite{chak96} 
$\Delta_s\sim3.35J\exp[-0.682(W/a)]$, as discussed in Section \ref{strong}.
\marginpar{\em Widening the ladder closes the gap.}
Although odd leg 
Heisenberg ladders are gapless, they are characterized by an
energy scale
which has the same functional dependence on $W$ as 
$\Delta_s$. Below this energy, the excitations are gapless
spinons analogous to those in the Heisenberg chain \cite{chak96},
while above it they are weakly interacting
spin waves. 
Based  on our experience with
the Hubbard rings we expect that spin gap formation  is related to
superconductivity. As we shall see below this is indeed the  case once
the ladders are doped with holes. On the face of it, this implies that
only  rather narrow ladders are good candidates for the mesoscopic
building  blocks of a high temperature superconductors.

When the Hubbard ladder is doped with holes away from half filling, its strong 
coupling description is modified from the Heisenberg model (Eq.~(\ref{heisenberg})) 
to the $t-J$ model
\begin{equation}
\label{tJ}
H_{t-J}=-t\sum_{\langle i,j\rangle,s}(c_{i,s}^{\dagger}c_{j,s}+h.c.)
+J\sum_{\langle i,j\rangle}\left[{\bf S}_i\cdot{\bf S}_j-\frac{1}{4}n_i n_j
\right] \; ,
\end{equation}
which is defined with the supplementary constraint of no doubly occupied 
sites. This is the version which has been most extensively studied 
numerically. 
Unless otherwise stated, we will quote results for representative values 
of $J/t$ in the range $J/t=0.35$ to $0.5$.
\sindex{ta}{$t$}{Nearest neighbor hopping}
\sindex{ja}{$J$}{Nearest neighbor exchange coupling}

\begin{figure}[ht!!!]
\begin{center}
\epsfig{figure=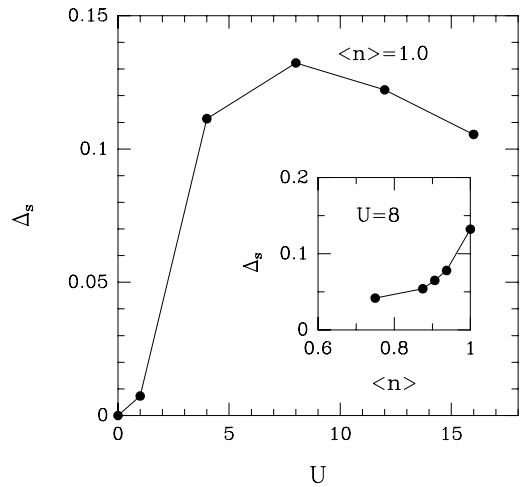,width=0.6\linewidth}
\end{center}
\caption{\label{fignum4}
The spin gap as a function of $U$ for a half filled $2\times 32$ Hubbard 
ladder. The inset shows $\Delta_s$ as a function of filling 
$\langle n\rangle$ for $U=8$. Energies are measured in units of 
$t=1$. (From Noack {\em et al.}\cite{noackphys})}
\end{figure}

Numerical studies of the two leg Hubbard model \cite{noackprl,noackphys} 
have demonstrated that doping tends to decrease the spin gap 
{\em continuously} from its 
value in the undoped system but it persists down to at least an average 
filling of $\langle n\rangle=0.75$, as can be seen from the inset in Fig. 
\ref{fignum4}. A similar behavior is observed in the $t-J$ ladder although 
the precise evolution of the spin gap upon doping depends on details of 
the model\cite{2legt'}.

\marginpar{\em Holes like to $d$-pair.} 
Pairs of holes in two leg Hubbard or $t-J$ ladders 
form bound pairs as can be seen both from the fact that the pair binding
energy is positive, and from the fact that positional correlations between
holes are indicative of a bound state. 
The pairs have 
a predominant $d_{x^2-y^2}$ symmetry as is revealed by
the relative minus sign between the ground state to
ground state amplitudes for adding a singlet pair on neighboring sites
along and across the legs\cite{noackprl,noackd}. It seems that the
dominance of the $d_{x^2-y^2}$ channel is universally shared by 
all models over the entire range of doping that has been studied.  
(See Section \ref{square} for a discussion of this phenomenon in the 
$2\times 2$ plaquette.)

\begin{figure}[ht!!!]
\begin{center}
\epsfig{figure=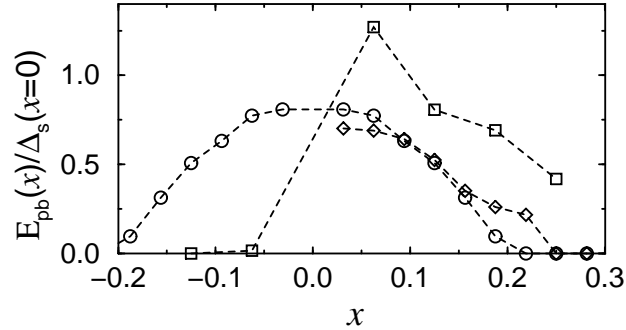,width=0.8\linewidth}
\end{center}
\caption{\label{fignum5}
The ratio of the pair binding energy to the undoped spin gap as a function 
of hole doping $x=1-\langle n\rangle$. The diamonds are for a $32\times 2$ 
$t-J$ ladder with $J/t=0.3$. The circles are for a one band $32\times 2$ 
Hubbard ladder with $U/t=12$. The squares are for a three band Hubbard 
model of a two leg Cu-O ladder, i.e. a ladder made of Cu sites where  
nearest neighbor sites are connected by a link containing an O atom.
Here $U_d/t_{pd}=8$, where $U_d$ is the on-site Cu Coulomb interaction and 
$t_{pd}$ is the hopping matrix element between the O and Cu sites. The 
energy difference between the O and Cu sites is 
$(\epsilon_p-\epsilon_d)/t_{pd}=2$, and the calculation is done on a 
$16\times 2$ ladder. (From Jeckelmann {\em et al.} \cite{jeck})}
\end{figure}

The doping dependence of the pair binding energy roughly follows the spin 
gap in various versions of the two leg ladder as shown in Fig. \ref{fignum5}. 
The correlation function 
$D(l)$ of the pair field 
\begin{equation}
\label{pairf}
\Delta_i^\dagger=(c^\dagger_{i1\uparrow} 
c^\dagger_{i2\downarrow}-c^\dagger_{i1\downarrow}c^\dagger_{i2\uparrow}) \; ,
\end{equation} 
exhibits behavior consistent with a power law decay 
\cite{dag92,noackprl,noackd,kimura}
\begin{equation}
\label{pfpfcor}
D(l)=\langle\Delta_{i+l}\Delta_i^\dagger\rangle\sim\l^{-\theta} \; .
\end{equation}
There exists less data concerning its doping dependence, but from the 
relevant studies \cite{noackprl,noackd} we can conclude that the 
pair correlations increase from the undoped system to a maximum at 
$x\sim 0.0625$ and then decrease when more holes are added to the system.  
\sindex{dl}{$D(l)$}{Correlation function of the pair field}

Both the spin gap and the pairing correlations in 
\marginpar{\em Details and their importance}  
doped Hubbard and $t-J$ ladders can be appreciably enhanced by slight 
generalizations of the models. For example, the exponent $\theta$ in 
Eq.~(\ref{pfpfcor}), which depends on the coupling strengths $U/t$ or $J/t$
and the doping level $x$, is also sensitive to the ratio of the hopping 
amplitudes between neighboring sites on a rung and within a chain 
$t_\perp/t$. By varying this parameter, the exponent $\theta$ can be tuned 
over the range $0.9\le\theta\le 2.1$. In particular, for $x=0.0625$ and 
intermediate values of the (repulsive) interaction $5\le U/t\le 15$, it can be made 
smaller than 1 \cite{noackd}; see Fig. \ref{fig1d2}. This is
significant since, as we saw in Section \ref{1dcompete}, whenever $\theta<1$
the  superconducting susceptibility is the most divergent among the
various  susceptibilities of the ladder. Adding a nearest neighbor 
exchange coupling, $J$, to $H_U$ 
also leads to 
stronger superconducting signatures owing to an increase in the pair 
mobility and binding energy \cite{tuj}. 
\marginpar{\em Another lesson in humility} 
The moral here is that details are
important as far as they reveal the nonuniversal properties of the 
Hamiltonians that we study, and indicate relevant directions in model space.
It should also imprint on us a sense of humility when attempting to fit 
real world data with such theoretical results. 

\begin{figure}[ht!!!]
\begin{center}
\epsfig{figure=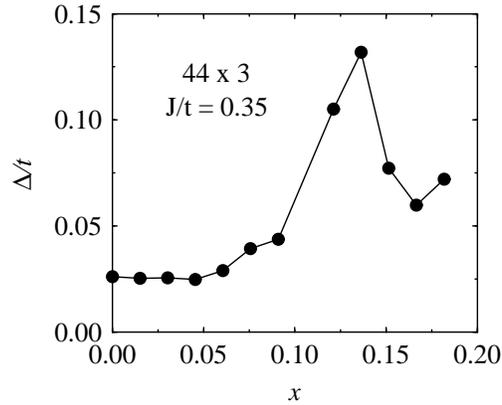,width=0.6\linewidth}
\end{center}
\caption{\label{fignum6}
Spin gap for a $44\times 3$ ladder with open boundary conditions and 
$J/t=0.35$ as a function of doping. (From White and Scalapino. \cite{3leg})}
\end{figure}

We already noted that, in contrast to the two leg ladder, the three leg
\marginpar{\em Odd and want a gap? --Dope!}
system does not possess a spin gap at half filling. This situation persists 
up to hole doping of about $x=1-\langle n\rangle=0.05$, as can be seen in Fig. 
\ref{fignum6}.\footnote{The nonvanishing spin gap in this region is presumably 
a finite size effect; see Fig. \ref{fignum3}.}
However, with moderate doping a spin gap 
is formed which reaches a maximum value at a doping level of $x=0.125$. For 
the system shown here, with $J/t=0.35$, the gap is only 20 
percent smaller than that of the undoped two leg Heisenberg ladder. 
Upon further doping, the spin gap 
decreases and possibly vanishes as $x$ gets to be 0.2 or larger. 

\begin{figure}[ht!!!]
\begin{center}
\epsfig{figure=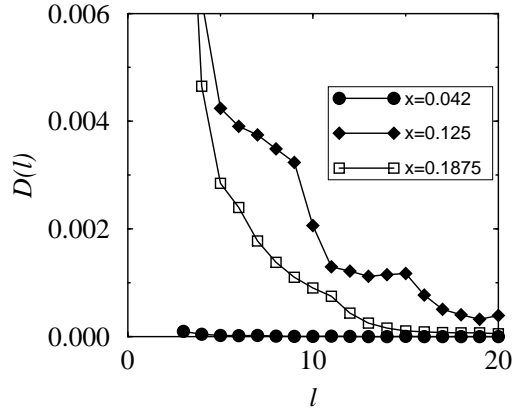,width=0.6\linewidth}
\end{center}
\caption{\label{fignum7}
The $d_{x^2-y^2}$ pair field correlations $D(l)$ for three different 
densities, calculated on $32\times 3$ ($x=0.1875$) and $48\times 3$ 
($x=0.042$, $x=0.125$) open $t-J$ ladders with $J/t=0.35$. 
(From White and Scalapino. \cite{3leg})}
\end{figure}

The establishment of a spin gap is concurrent with the onset of pairing 
\marginpar{\em The same goes for pairing.}
correlations in the system. While two holes introduced into a long, 
half filled three chain ladder do not bind \cite{whitepair}, 
indications of pairing 
emerge as soon as the spin gap builds up \cite{rice97,3leg}. 
As an example, Fig. \ref{fignum7} plots the pair field--pair field
correlation function of Eq.~(\ref{pfpfcor}) for various values of the hole doping, 
defined with
\begin{equation}
\label{delta}
\Delta_i^\dagger=c_{i,2\uparrow}^\dagger(c_{i+1,2\downarrow}^\dagger+
c_{i-1,2\downarrow}^\dagger-c_{i,1\downarrow}^\dagger-
c_{i,3\downarrow}^\dagger)-(\uparrow\leftrightarrow\downarrow) 
\end{equation}
which creates a $d_{x^2-y^2}$ pair around the $i$th site of the middle leg 
(the leg index runs from 1 to 3).\footnote{There also exists a small $s$-wave component in the pair field 
due to the one dimensional nature of the cluster.} 
In the regime of low doping $x\leq 0.05$, the pair field correlations are 
negligible. However, clear pair field correlations are present at $x=0.125$,
where they are comparable to those in a two leg ladder under similar 
conditions.  The pair field correlations are less strong at $x=0.1875$;  they
follow an approximate power law  decay as a function of the distance.
\cite{kimura,3leg}  (The oscillations in $D(l)$ are produced 
by the open boundary conditions used in this calculation.) This behavior 
can be understood from strong coupling bosonization considerations 
\cite{spingap} in which the two even modes (with respect to reflection 
about the center leg) form a spin gapped two leg ladder and for small doping 
the holes enter the odd mode giving rise to a gapless one dimensional 
electron gas. As the doping increases, pair hopping between the two 
subsystems may induce a gap in the gapless channel via the spin gap 
proximity effect \cite{spingap}.

\begin{figure}[ht!!!]
\begin{center}
\epsfig{figure=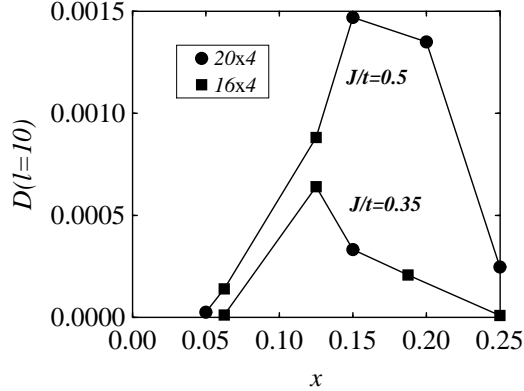,width=0.6\linewidth}
\end{center}
\caption{\label{fignum8}
The $d_{x^2-y^2}$ pair field correlation $D(l)$ at a separation of $l=10$ 
rungs as a function of doping $x$, for $20\times 4$ and $16\times 4$ open 
ladders with $J/t=0.35$ and 0.5. (From White and Scalapino. \cite{4leg})}
\end{figure}

Increasing the number of legs from three
to four 
leads to behavior similar to that
exhibited by the two leg ladder. The
system is  spin gapped and two holes in a half filled four leg ladder tend to
bind.  The pair exhibits features common to all pairs in an antiferromagnetic 
environment, including 
a $d$-wave-like symmetry \cite{whitepair}. 
Further similarity with the two leg ladder is seen in the 
$d$-wave pair field correlations $D(l)$. Fig. \ref{fignum8} shows $D(l=10)$ 
for a $t-J$ four leg ladder as a function of doping (extended $s$-wave 
correlations are much smaller in magnitude). The pairing correlations 
for $J/t=0.5$ increase with doping, reaching a maximum between $x=0.15$ and 
$x=0.2$, and then decrease. 
\marginpar{\em Four legs are good; two legs are better.}
The magnitude of the correlations near the 
maximum is similar to that of a two leg Hubbard ladder with $U=8t$ 
(corresponding to $J\sim 4t^2/U=0.5$) with the same doping, but smaller 
than the maximum in the two leg ladder which occurs at smaller doping
\cite{noackprl,noackd}. For $J/t=0.35$ the peak is reduced in magnitude and 
occurs at lower doping. The behavior of $D(l)$ near the maximum is consistent
with power law decay for short to moderate distances but seems to fall more 
rapidly at long distances (perhaps even exponentially. \cite{dougprivate})

\begin{figure}[ht!!!]
\begin{center}
\epsfig{figure=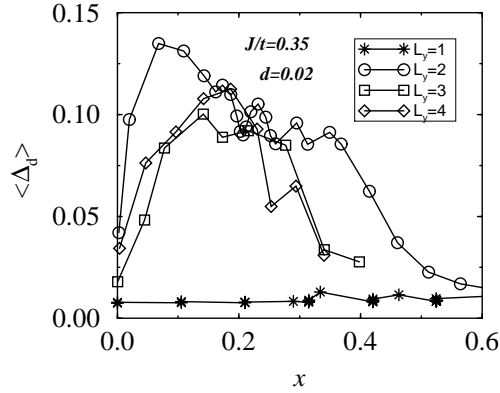,width=0.6\linewidth}
\end{center}
\caption{\label{fignum9}
The $d_{x^2-y^2}$ pairing response to a proximity pair field operator 
as a function of doping for a single chain and two, three, and four leg 
ladders. For the single chain, near neighbor pairing is measured. 
(From White and Scalapino. \cite{3leg})}
\end{figure}

Lastly, we present in Fig. \ref{fignum9} the response of a few ladder
systems to a proximity pairing field
\begin{equation}
\label{proxfield}
H_1=d\sum_i(c^\dagger_{i,\uparrow}c^\dagger_{i+\hat{y},\downarrow}
-c^\dagger_{i,\downarrow}c^\dagger_{i+\hat{y},\uparrow}+h.c.) \; ,
\end{equation}
which adds and destroys a singlet electron pair along the ladder. The 
response is given by the average $d_{x^2-y^2}$ pair field
\begin{equation}
\label{deltad}
\langle\Delta_d\rangle=\frac{1}{N}\sum_i\langle\Delta_i\rangle \; ,
\end{equation}
with $\Delta_i$ defined in Eq.~(\ref{pairf}). We see that the pair field 
response tends to decrease somewhat with the width of the system but is overall
similar for the two, three and four leg ladders. We suspect it gets rapidly 
smaller for wider ladders.

\subsubsection{Phase separation and stripe formation in ladders} 

We now address the issue of whether there is any apparent tendency to 
form charge density and/or spin density wave order in ladder systems, 
and whether there is a tendency of the doped holes to phase separate. Since
incommensurate density wave long range order, like superconducting order, is 
destroyed by quantum fluctuations in one dimension, we will again be looking
primarily at local correlations, rather than actual ordered states.  
Of course, we have in mind that local correlations and enhanced 
susceptibilities in a one dimensional context can be interpreted as 
indications that in two dimensions true superconductivity, stripe order, 
or phase separation may occur.

\paragraph{\bf Phase Separation:}  Phase separation was first found in the
one dimensional chain \cite{ogata91, hellberg93} and subsequently in the 
two leg ladder \cite{troyer96,hayward96, sierra98}. As a rule, the phase 
separation line has been determined by calculating  the coupling $J$ at 
which the compressibility diverges. (See, however, Ref. \citen{hm}.)  
This is in principle an incorrect criterion. The compressibility only 
diverges at the consolute point. Thermodynamically appropriate criteria 
for identifying regimes of phase separation from finite size studies 
include the Maxwell construction (discussed explicitly in Section \ref{doped}, 
below), and measurements of the surface tension in the presence of boundary 
conditions that force phase coexistence. The divergent compressibility is 
most directly related to the spinodal line, which is  not even strictly well 
defined beyond mean field theory. Thus, while in many cases the phase 
diagrams obtained in this way may be qualitatively correct, they are always 
subject to some uncertainty.  

\begin{figure}[ht!!!]
\begin{center}
\epsfig{figure=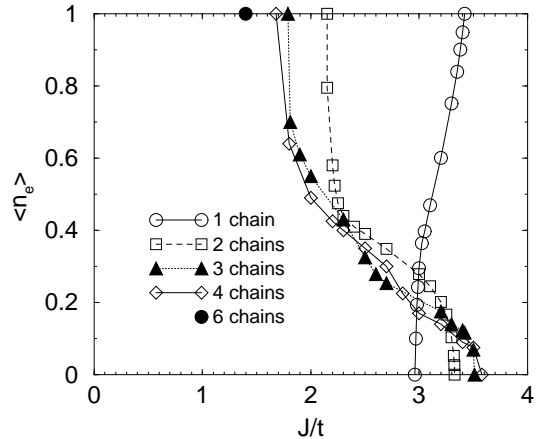,width=0.6\linewidth}
\end{center}
\caption{\label{fignum10}
Boundary to phase separated region in $t-J$ ladders. Open boundary 
conditions were used in both the leg and rung directions except for the 
six leg ladder where periodic boundary conditions were imposed along the 
rung. Phase separation is realized to the right of the curves. 
$\langle n_e\rangle$ is the total electron density in the system. 
(From Rommer {\em et al.} \cite{rommer00})}
\end{figure}

More recently Rommer, White, and Scalapino 
\cite{rommer00} have used DMRG methods to extend the study to ladders 
of up to six legs. Since these calculations are carried out with open boundary
conditions, which break the translational symmetry of the system, they have 
used as their criterion the appearence of an inhomogeneous state with a 
hole rich region at one edge of the ladder and hole free regions near the 
other, which is a thermodynamically correct criterion for phase separation.
However, where the hole rich phase has relatively low hole density, and in all
cases for the six leg ladder, they were forced to use a different criterion 
which is not thermodynamic in character, but is at least intuitively 
appealing.  From earlier studies (which we discuss below) it appears that 
the ``uniform density'' phase, which replaces the phase separated state 
for $J/t$ less than the critical value for phase separation, is a ``striped'' 
state, in which the holes congregate into puddles (identified as stripes) 
with fixed number of holes, but with the density of stripes determined by 
the mean hole density on the ladder. With this in mind, Rommer {\em et al.}  
computed the interaction energy between two stripes, and estimated the 
phase separation boundary as the point at which this interaction
turns from repulsive to attractive.  
The results, summarized in Fig.~\ref{fignum10},
agree with the thermodynamically determined phase boundary
where they can be compared.

\sindex{ne}{$n_e$}{Electron density}

\marginpar{\em Ladders phase separate for large enough $J/t$.} 
For large enough values of $J/t$, both the single chain and the ladders are 
fully phase separated into a Heisenberg phase ($\langle n_e\rangle=1$) and 
an empty phase ($\langle n_e\rangle=0$). However, the evolution of this state 
as $J/t$ is reduced is apparently different for the two cases. For the chain, 
the Heisenberg phase is destroyed first by holes that diffuse into it; this
presumably reflects the fact that hole motion is not significantly frustrated 
in the single chain system. In the ladders, on the other hand, the empty phase
is the one that becomes unstable due to the sublimation of electron pairs from
the Heisenberg region.  This difference is evident in Fig. \ref{fignum10} 
where the phase separation  boundary occurs first at high electron density in 
the chain and high hole  density in the ladders. It is also clear from looking
at this figure that  the value of $J/t$ at which phase separation first
occurs for small electron densities is hardly sensitive to the width of 
the ladder. However, as more electrons are added to the system 
(removing holes), 
phase separation is realized for smaller values of $J/t$ in wider ladders. 
Whether this is an indication that phase separation takes place at arbitrarily
small $J/t$ for small enough hole densities in the two dimensional system is 
currently under debate, as we discuss in Section~\ref{2dphasesep}.    
 
\paragraph{\bf ``Stripes'' in ladders:}
\marginpar{\em Stripes appear at smaller $J/t$.}    
At intermediate values of $J/t$, and not too close to half filling, the 
doped holes tend to segregate into puddles which straddle the
ladders, as is apparent from the spatial modulation of the mean charge density
along the ladder.  Intuitively, we can think of this state as consisting of an
array of stripes with a spacing which is determined by the doped hole density.
From this perspective, the total number of doped holes associated with
each puddle, $N_{puddle}=\varrho L$, is interpreted as arising from a stripe
with a mean linear density of holes, $\varrho$, times the length of the stripe,
$L$.\footnote{For instance, on a long, $N$ site, 4 leg ladder with $4n$ holes, 
where $n \ll N$, one typically observes $n$ or $2n$ distinct peaks in the 
rung-averaged charge density, which is then interpreted as indicating a 
stripe array with $\varrho=1$ or $\varrho=1/2$, respectively.}  
($L$ is also the width of the ladder.) 
In the thermodynamic limit, long wavelength quantum fluctuations of the
stripe array would presumably result in a uniform charge density, but the 
ladder ends, even in the longest systems studied to date, are a sufficiently 
strong perturbation that they 
pin the stripe array\cite{affleckwhitescalapino}.  
In two and three leg ladders, the observed stripes apparently always have
$\varrho=1$. For the four leg ladder, typically $\varrho=1$, but under 
appropriate circumstances (especially for $x=1/8$), $\varrho=1/2$ stripes 
are observed. In six and eight leg ladders, the charge density oscillations 
are particularly strong, and correspond to stripes with $\varrho = 2/3$ and 
$1/2$, respectively.
Various arguments have been presented to identify certain of these stripe 
arrays as being ``vertical'' ({\it i.e.} preferentially oriented along the 
rungs of the ladder) or ``diagonal'' ({\it i.e.} preferentially oriented 
at 45$^o$ to the rung), but these arguments, while intuitively appealing, 
do not have a rigorous basis.
\sindex{zzrhozzz}{$\varrho$}{Hole density on a stripe}

We will return to the results on the wider ladders, below, where we discuss
attempts to extrapolate these results to two dimensions.

\subsection{Properties of the two dimensional $t-J$ model}

It is a subtle affair to draw conclusions about the properties of the two dimensional Hubbard and 
$t-J$ models from numerical studies of finite systems.
The present numerical capabilities do not generally permit a systematic  
finite size scaling analysis. As a result, extrapolating results from 
small clusters with periodic boundary conditions, typically used when 
utilizing Monte Carlo or Lanczos techniques,  or from strips with open 
boundary conditions as used in DMRG studies, is 
susceptible to criticism \cite{hellman99,whitecompare}. 
It comes as no surprise 
then that several key issues concerning the ground state properties of the 
two dimensional models are under dispute. In the following we present 
a brief account of some of the conflicting results and views. 
However, at least two things do not seem to be in dispute:
1) there is a strong tendency for doped holes 
in an antiferromagnet to clump in order relieve the
frustration of hole motion\cite{kelosalamos}, and 2) where it occurs, 
hole pairing has a $d_{x^2-y^2}$ character. Thus, in one way or another, 
the local correlations that lead to stripe formation and $d$-wave 
superconductivity are clearly present in $t-J$-like models!

\subsubsection{Phase separation and stripe formation}
\label{2dphasesep}

There have been relatively few numerical studies 
of large two dimensional Hubbard model clusters.
Monte Carlo  simulations on square
systems with sizes up to
$12\times 12$ 
and temperatures down to roughly $t/8$ have been
carried out, 
typically with $U/t=4$\cite{dagreview}. A signature of phase separation in
the form of a discontinuity 
in the chemical  potential as a function of doping was looked for and not 
found. No evidence of stripe formation was found, either. Given the limited 
size and temperature range of these studies, and the absence of results that 
would permit a Maxwell construction to determine the boundary of phase 
separation, it is difficult to reach a firm conclusion on the basis of 
these studies. 
Certainly at relatively elevated temperatures, holes in the Hubbard model
do not show a strong tendency to cluster,
but it is difficult to draw conclusions concerning lower temperature, 
or more subtle tendencies. (Variational ``fixed node'' studies by Cosentini 
{\it et al}\cite{cosentini} are suggestive of phase separation at small $x$, 
but more recent studies by Becca {\it et al.}\cite{beccaps} reached the 
opposite conclusion.)
 
\marginpar{\em Everybody agrees on the phase separation boundary for 
$x\sim 1$.}
There are many more studies of phase separation in the $t-J$ model.
Most of them agree on the behavior in the regime of very low electron 
density $n_e=1-x \ll 1$. The critical $J/t$ value for phase separation at
vanishingly  small $n_e$ was calculated very accurately by Hellberg and 
Manousakis \cite{hm2} and was found to be $J/t=3.4367$. 
However, there are conflicting results for systems close to half filling ($n_e\sim 1$) and with 
small $t-J$. This is the most delicate region where 
high numerical accuracy is hard to obtain. Consequently, there is no 
agreement on 
whether the two dimensional 
$t-J$ model phase separates for all values of $J/t$ at sufficiently 
low hole doping $x$.    

Emery {\it et al}. \cite{ekl,kel} presented a variational argument
(recently extended and substantially improved by Eisenberg {\it et
al.}\cite{eisenberg}) that for $J/t \ll 1$ and for $x$ less than a critical
\marginpar{\em The situation for $x\sim 0$ is murkier, but...}
concentration, $x_c\sim\sqrt{J/t}$, phase separation occurs between a
 hole free antiferromagnetic and a metallic ferromagnetic  state. Since for
large $J/t$ there is clearly phase separation for all $x$, they proposed that 
for sufficiently small $x$, phase separation is likely to occur for all $J/t$. 
To test this, they computed the ground state energy by exact diagonalization of
$4\times 4$ doped $t-J$ clusters.  If taken at face value and interpreted
via a Maxwell construction, these results imply that for any $x < 1/8$, 
phase separation occurs at least for all $J/t> 0.2$.  Hellberg and Manousakis
\cite{hm,hellps00}  calculated the ground state energy on larger clusters
of up to $28\times 28$  sites using Green function Monte Carlo methods.
By implementing a Maxwell  construction, they reached the similar conclusion that the
$t-J$ model phase  separates for all values of $J/t$ in the low hole doping
regime.   

On the other hand, Putikka {\it et al}. \cite{putikka92} studied this problem 
using a high temperature series expansion extrapolated to $T=0$ and concluded 
that phase separation only occurs above a line extending from $J/t=3.8$ at 
zero filling to $J/t=1.2$ at half filling. In other words, they concluded that 
there is no phase separation for any $x$ so long as $J/t<1.2$.
Exact diaganolization results for the compressibility and  
the binding energy of $n$-hole clusters in systems of up to 26 sites by  
Poilblanc \cite{poilblanc95ps} were interpreted as suggesting that the 
ground state is phase  separated close to half filling only if
$J/t>1$. Quantum Monte Carlo  simulations of up to 242 sites using stochastic
reconfiguration by  Calandra {\it et al.}\cite{calandra98ps} have found a phase
separation instability  for $J/t\sim 0.5$ at similar doping levels, but no
phase separation for $J/t <0.5$, while earlier variational  Monte Carlo
calculations \cite{yokoyama96ps} reported a critical value of
$J/t=1.5$.  
Using Lanczos techniques to calculate the ground state energy on lattices of up
to 122 sites,
Shin {\it et al}. \cite{tklee,shin01ps} estimate the lower critical value for
phase separation as $J/t=0.3-0.5$, a somewhat lower bound than previously 
found using similar numerical methods \cite{khono97ps}. 
Finally, DMRG 
calculations on wide ladders with open boundary conditions in one direction 
by White and Scalapino \cite{whitescalapino,whitestripesallx2d} found striped 
ground states for $J/t=0.35$ and $0<x<0.3$, but no indication of phase 
separation.  

\begin{figure}[ht!!!]
\begin{center}
\epsfig{figure=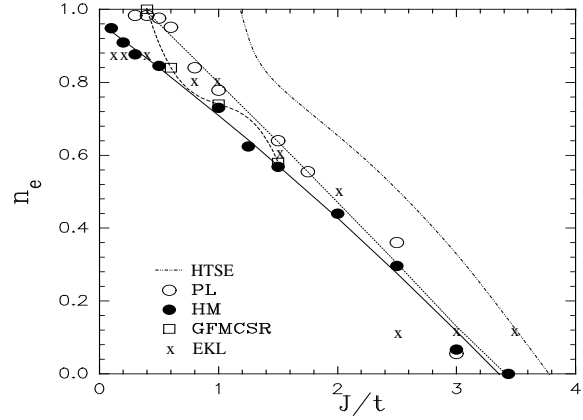,angle=-90,width=0.65\linewidth}
\end{center}
\caption{\label{fignum11}
Phase separation boundary of the two dimensional $t-J$ model according 
to various numerical studies. The dashed-dotted line represents the 
high temperature series expansion results by Puttika {\em et al.} \cite{putikka92}. 
Also shown are results from calculations using the Power-Lanczos method by 
Shin {\em et al.} \cite{shin01ps} (open circles), Greens function Monte Carlo 
simulations by Hellberg and Manousakis \cite{hellps00} (closed circles) and by 
Calandra {\em et al.} \cite{calandra98ps} (open squares), and exact diagonalization 
of $4\times 4$ clusters by Emery {\em et al.} \cite{ekl} (x's). (Adapted from  
Shin {\em et al.} \cite{shin01ps})}
\end{figure}

\marginpar{\em ... it seems that the model is either phase separated,
or very close to it.}
For comparison, we have gathered a few of the results mentioned above in 
Fig.~\ref{fignum11}. The scatter of the data at the upper left corner of the 
$n_e-J/t$ plane is a reflection of the near linearity of the 
the ground state energy as a function of doping in this region 
\cite{whitecompare}. High numerical accuracy is needed in order to 
establish 
a true linear  behavior which would be indicative of phase separation. 
While there is currently no definitive answer concerning phase separation at small doping,
it seems  clear that in this region the two dimensional $t-J$ 
model is in delicate balance,  either in or close to a phase separation 
instability.     

The nature of the ground state for moderately small $J/t$ beyond any 
phase separated regime
is also in dispute. While DMRG calculations on fat ladders 
\cite{whitescalapino,whitestripesallx2d} find striped ground states 
for $J/t=0.35$ and $x=1/8$, Monte Carlo simulations on a torus\cite{hellman99} 
exhibit stripes only as excited states.
Whether this discrepancy is due to finite size  effects or the type of 
boundary conditions used is still not settled. 
(The fixed node Monte Carlo studies 
of  Becca {\it et al.}\cite{sorella2} likewise conclude that stripes do not 
occur in the ground state, although they can be induced by the addition of 
rather modest anisotropy into the $t-J$ model, suggesting that they are at 
\marginpar{\em Stripes are important low energy configurations of the $t-J$ 
model.}
least energetically competitive.) While these conflicting conclusions may 
be difficult to resolve, it seems inescapable to us  that 
stripes are important low
energy configurations of the two dimensional $t-J$ model for small  doping
and moderatly small $J/t$.

The most reliable results concerning the internal structure of the stripes
\marginpar{\em Typically stripes are quarter-filled antiphase domain walls.}
themselves come from studies of fat $t-J$ ladders, 
where stripes are certainly a prominent part of the electronic structure.
In all studies of ladders, the doped holes aggregate into ``stripes'' which
are oriented either perpendicular or 
parallel to the extended direction of the ladder, 
depending on boundary conditions. 
In many cases the spin correlations in
the hole poor regions between stripes locally resemble those in the undoped
antiferromagnet but suffer a $\pi$-phase shift across the hole rich stripe.  
This magnetic structure is 
vividly apparent in studies for which the 
low energy orientational fluctuations of the spins are suppressed by the 
application of staggered magnetic fields on certain boundary sites 
of the ladders---then, these magnetic correlations are 
directly seen
in the expectation values of the spins \cite{whydo00}. However, such findings 
are not universal:  in the case of the four leg ladder, with stripes along the 
ladder rungs, 
Arrigoni {\it et al.} 
\cite{arrigoniandme} recently showed that in long systems (up to $4\times 27$),
these antiphase magnetic correlations are weak or nonexistent, 
despite strong evidence of charge stripe correlations. 
Ladder studies have also demonstrated 
that stripes tend to 
favor a linear charge density of $\varrho=1/2$ along each stripe.\footnote{At 
about the same time, Nayak and Wilczek\cite{nayakwilczek}
presented  an interesting analytic argument which leads to the same bottom line.}
Specifically, by applying boundary conditions which force a single stripe to 
lie along the long axis of the ladder, White {\it et al}
\cite{whitestripesallx2d} were able to study the energy of 
a stripe as a function of $\varrho$. They found an energy which is apparently 
a smooth function of $\varrho$ ({\it i.e.} with no evidence of a 
nonanalyticity which would lock $\varrho$ to a specific value), but with 
a pronounced minimum at $\varrho=1/2$.
Moreover, with boundary conditions favoring stripes perpendicular to the
ladder axis, they found that for $x\leq 1/8$ stripes tend to form with
$\varrho=1/2$ so that the spacing between neighboring stripes is approximately
$1/2x$, while at larger $x$, 
a first order transition occurs to ``empty domain walls'' with $\varrho=1$
and an inter-stripe spacing of $1/x$.  
In the region $0.125<x<0.17$ the two types of stripes can coexist.

It is worth noting that the original indications of stripe order came from
Hartree-Fock treatments\cite{zaanengunnarsson,poilblanc89,schulzstripes,
machida89}.  Hartree-Fock stripes are primarily spin textures. 
In comparison to the DMRG results on ladders, they correspond to ``empty'' 
($\varrho=1$) antiphase ($\pi$-phase shifted) domain walls, and so are 
insulating and overemphasize the spin component of the stripe order, 
but otherwise capture much of the physics of stripe formation remarkably 
accurately. 

\begin{figure}[ht!!!]
\begin{center}
\epsfig{figure=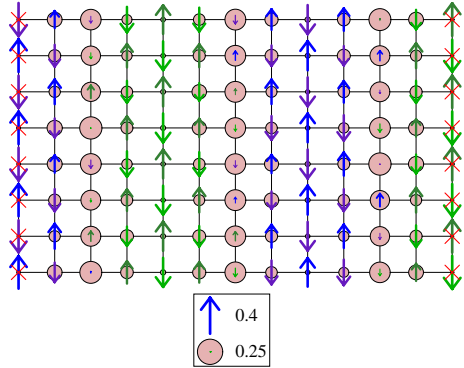,width=0.54\linewidth}
\end{center}
\caption{\label{fignum12}
Hole density and spin moments on a $13\times8$ cylinder 
with 12 holes, $J/t=0.35$, periodic boundary conditions along the $y$ 
direction and $\pi$-shifted staggered magnetic field of magnitude $0.1t$ 
on the open edges. The diameter of the circles is proportional to the hole 
density $1-\langle n_i\rangle$ and the length of the arrows is proportional to 
$\langle S_i^z\rangle$.  
(From White and Scalapino. \cite{whydo00})}
\end{figure}

\begin{figure}[ht!!!]
\begin{center}
\epsfig{figure=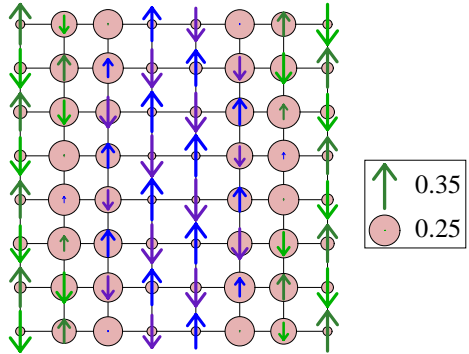,width=0.55\linewidth}
\end{center}
\caption{\label{fignum13}
Hole density and spin moments on a central section of a $16\times 8$ 
cylinder with 16 holes, $J/t=0.35$, with 
periodic boundary conditions along the $y$ direction and staggered magnetic 
field of magnitude $0.1t$ on the open  edges. The notation is similar to 
Fig.~\ref{fignum12}.
(From White and Scalapino. \cite{whydo00})}
\end{figure}

Further insight into the physics that generates the domain walls 
can be gained by 
\marginpar{\em They can be site- or bond-centered.}
looking more closely at their hole density and spin structures. Both 
site-centered and bond-centered stripes are observed. They are close in energy 
and each type can be stabilized by adjusting the boundary conditions 
\cite{whitescalapino}. Fig.~\ref{fignum12} depicts three site-centered 
stripes in a $13\times 8$ system with 12 holes, periodic boundary conditions 
along the $y$ direction and a $\pi$-shifted staggered magnetic field on the 
open ends of magnitude $0.1t$. These stripes are quarter-filled antiphase
domain walls.
Fig.~\ref{fignum13} shows a central section of a $16\times 8$ cluster 
containing two bond-centered domain walls. This system is similar to the 
one considered above except that the magnetic field on the open ends 
is not $\pi$-shifted. 
Like their site-centered counterparts,
the bond-centered stripes are antiphase domain walls, but with one hole per 
two domain wall unit cells. 

The $\pi$-phase 
\marginpar{\em The topological character of spin stripes can be inferred 
from local considerations.} 
shift in the exchange field across the stripe can probably be traced, in both
the bond- and  site-centered cases, to a gain in the transverse kinetic energy 
of the holes. To demonstrate this point consider a pair of holes in a 
$2\times 2$ $t-J$ plaquette, as was done in Section~\ref{square}. One can 
simulate the effect of the exchange field running on both sides of the 
plaquette through a mean field $h$ which couples to the spins on the square 
\cite{whydo00}. For the in-phase domain wall such a coupling introduces 
a perturbation $h(S_1^z-S_2^z-S_3^z+S_4^z)$ which, to lowest order in $h$, 
lowers the ground state energy by $-h^2/\sqrt{J^2+32t^2}$. For the 
$\pi$-shifted stripe the perturbation is $h(S_1^z+S_2^z-S_3^z-S_4^z)$ 
with a gain of $-4h^2/\sqrt{J^2+32t^2}$ in energy, thereby being more 
advantageous for the pair.  Indeed, this physics has been confirmed by several
serious studies, which combine analyatic and numerical work, 
by Zachar\cite{zachar}, Liu and Fradkin\cite{liufradkin}, and
Chernyshev {\it et al.}\cite{netowhite}  These studies indicate that there is a
transition from a tendency for in-phase magnetic order across a stripe for 
small $\varrho$, when the direct magnetic interactions are dominant, 
to antiphase magnetic order for $\varrho >0.3$, when the transverse 
hole kinetic energy is dominant. 

\subsubsection{Superconductivity and stripes}

\marginpar{\em There is no evidence for superconductivity in the Hubbard 
model.}
There is no unambiguous evidence for superconductivity in the Hubbard model.  
The original finite temperature Monte Carlo simulations on small 
periodic clusters with $U/t=4$ and $x=0.15$ \cite{moreo92sc,dagreview} 
found only short range pair-pair correlations. The same conclusion 
was reached by a later zero temperature constrained path Monte Carlo 
calculation \cite{zhang97sc}.

There are conflicting results concerning the question of superconductivity  
in the $t-J$ model. 

In the unphysical region of large $J/t$, solid conclusions
can  be reached:  Emery {\it et al}\cite{ekl} showed that  proximate to the
phase separation boundary at $J/t \leq 3.8$, the hole rich phase
(which is actually a dilute electron phase with $x\sim 1$) has an $s$-wave
superconducting ground state.  This result was confirmed and extended by
Hellberg and Manousakis\cite{hm}, who further argued that in the dilute
electron limit, $x\to 1^-$, there is a transition from an $s$-wave state for 
$2<J/t<3.5$ to a $p$-wave superconducting state for $J/t<2$, possibly with a
$d$-wave state at intermediate $J/t$.  Early Lanczos calculations were carried 
out by Dagotto and Riera 
\marginpar{\em There is conflicting evidence for superconductivity in the 
$t-J$ model.}
\cite{dagotto93sc,dagreview,dagotto4} in which various quantitites,
such as the pair field correlation function and the 
superfluid density, were computed
to search for signs of superconductivity in
$4\times 4$ $t-J$ clusters.  In agreement with the analytic results, these
studies gave strong evidence of superconductivity for large
$J/t$. Interestingly, the strongest signatures of superconductivity were 
found for $J/t=3$ and $x=0.5$ and decayed rapidly for larger $J/t$. 
This was interpreted as due to a transition into the phase separation region. 
(Note, however, that {\em all} the studies summarized in Fig.~\ref{fignum11}  
suggest that $x=0.5$ is already inside the region that, in the thermodynamic 
limit, would be unstable to phase separation.) 

\begin{figure}[ht!!!]
\begin{center}
\epsfig{figure=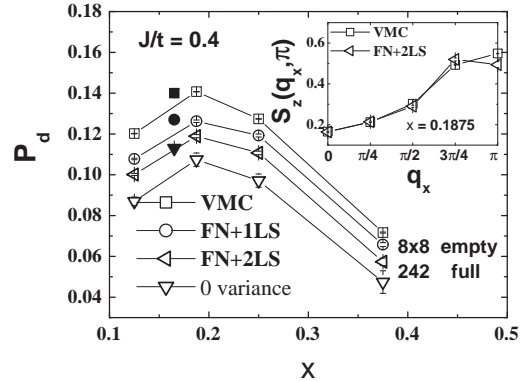,width=0.66\linewidth}
\end{center}
\caption{\label{fignum14}
The superconducting order parameter $P_d=2\lim_{l\rightarrow\infty}
\sqrt{D(l)}$ calculated for the largest distance on a $8\times 8$, $J/t=0.4$ 
cluster as function of hole doping $x$. Results for $x=0.17$ on a 242 site 
cluster are also shown. The different sets of data correspond to various 
Monte Carlo techniques. The inset shows the spin structure factor at 
$x=0.1875$. (From Sorella {\em et al.} \cite{sorella01sc})}
\end{figure}

More recent Monte Carlo simulations by Sorella {\em et al.}
\cite{calandra00sc,sorella01sc} showed evidence for long range
superconducting order in $J/t=0.4$ clusters  of up to 242 sites with periodic
boundary conditions and for a range of $x>0.1$, 
as shown in Fig.~\ref{fignum14}.
No signs of static stripes have been found in the parameter region that 
was investigated in these studies. A slight tendency towards 
incommensurability appears in the spin structure factor 
at (and sometimes above)
optimal doping, suggesting perhaps very weak dynamical stripe 
correlations. This finding is in sharp contrast to DMRG 
\cite{whitescalapino,whitestripesallx2d} and other \cite{tohyama99st} 
calculations that find striped ground states for the same parameters.  

\marginpar{\em Static stripes hamper superconductivity, but dynamic stripes 
may enhance it.}
Notwithstanding this controversy, these results seem to add to the general 
consensus that {\em static} stripe order and superconductivity compete. 
This is not to say that stripes and superconductivity cannot coexist. 
As we saw, evidence for both stripes and pairing have been found in three 
and four leg $t-J$ ladders \cite{3leg,4leg}. 
In fact pairing is enhanced in both of these systems when stripes are formed 
compared to the unstriped states found at small doping levels. 
Because of the open boundary conditions that were used in these studies 
the stripes were open ended and more dynamic. Imposing periodic 
boundary conditions in wider ladders (and also the four leg ladder) 
results in stripes that wrap around the periodic direction. 
These stripes appear to be more static,
and pairing correlations are suppressed. A similar behavior is observed 
when the stripes are pinned by external potentials. 

\begin{figure}[ht!!!]
\begin{center}
\epsfig{figure=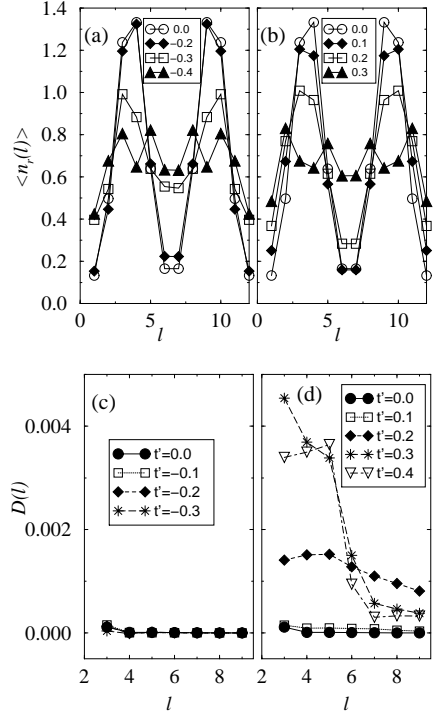,width=0.5\linewidth}
\end{center}
\caption{\label{fignum15}
Hole density per rung for a $12\times 6$ ladder with periodic boundary 
conditions along the rungs, 8 holes, $J/t=0.35$ and a) $t'\leq 0$ and 
b) $t'\geq 0$. c) and d) depict the $d$-wave pairing correlations for the 
same systems. (From White and Scalapino. \cite{stripeSCtt'})}
\end{figure}
Further evidence for the delicate interplay between stripes and 
pairing comes from studies of the $t-t'-J$ model in which a 
diagonal, single particle, next nearest neighbor hopping $t'$ is 
added to the basic $t-J$ model \cite{stripeSCtt',tohyama99st}. Stripes 
destabilize for either sign of $t'$. This is probably 
due to the enhanced mobility of the holes that can now hop on the same 
sublattice without interfering with the antiferromagnetic background. 
Pairing is suppressed for $t'<0$ and enhanced for $t'>0$. 
\footnote{This is surprising since $T_c$ is generally higher for hole doped 
cuprates (believed to have $t'<0$)  than it is for electron doped cuprates 
(which have  $t'>0$).} It is not clear whether the complete elimination of stripes or 
only a slight destabilization is more favorable to pairing correlations.
Fig.~\ref{fignum15} suggests that optimal pairing occurs in between 
the strongly modulated ladder and the homogeneous system. 

Finally, allowing for extra hopping terms in the Hamiltonian is not the 
only way tip the balance between static charge order and superconductivity. 
So far we have not mentioned the effects of long range Coulomb interactions 
on the properties of Hubbard related systems. This is not a coincidence since 
the treatment of such interactions in any standard numerical method is 
difficult. Nevertheless, a recent DMRG study of four leg ladders with open 
and periodic boundary conditions which takes into account the Coulomb 
potential in a self-consistent Hartree way \cite{arrigoniandme}, gives 
interesting results. It suggests that the inclusion of Coulomb interactions 
suppresses the charge modulations associated with stripes while enhancing 
the long range superconducting pairing correlations. At the same time the 
local superconducting pairing is not suppressed. Taken together, these facts 
support the notion that enhanced correlations come from 
long range phase ordering between stripes with well-established pairing. 
This enhanced phase stiffness is presumably due to  pair tunnelling 
between stripes produced by increased stripe fluctuations.

\section{Doped Antiferromagnets}
\label{doped}
\addtocontents{toc}{

\noindent{\em There are many indications that the cuprate superconductors
should be viewed as doped antiferromagnetic insulators. The motion of dilute
holes in an aniterromagnet is highly frustrated, and attempts to understand
the implications of this problem correspondingly frustrating. However, one
generic solution is macroscopic or microscopic phase separation into 
hole poor antiferromagnetic and hole rich metallic regions 
(where the hole motion is unfrustrated). 
} }

The undoped state of the cuprate superconductors is a strongly 
insulating antiferromagnet. It is now widely believed that the
existence of such a parent correlated insulator is 
an essential feature of high temperature superconductivity,
as was emphasized in some of the earliest studies of this problem 
\cite{pwa87,cuo}.
However, the doped antiferromagnet is a complicated theoretical problem---to 
even cursorily review what is known about it would more than double the size 
of this document. In this section we very briefly discuss the aspects 
of this problem which we consider most germane to the cuprates, 
and in particular to the physics of stripes. 
More extensive reviews of the subject can be found in
\cite{pnas,kelosalamos,frustrated,nagaevbook}. 

\subsection{Frustration of the motion of dilute holes in an antiferromagnet}

The most important local interactions in a doped antiferromagnet 
are well represented by the large U Hubbard model, the $t-J$ model, 
and their various relatives. To be concrete, we will focus on the 
$t-J-V$ model\cite{kel} (a slight generalization of the $t-J$ model, 
Eq. (\ref{tJ}), to which it reduces for for $V=-J/4$.)
\ba
H= -t  \sum_{<i,j>,\sigma} 
\left\{c^{\dagger}_{i,\sigma}c_{j,\sigma} + {\rm h.c.} \right \} 
+  \sum_{<i,j>}  \left \{ J \vec S_i \cdot \vec S_j 
+  V n_i n_j\right \} \; ,
\label{eq:tj} 
\ea
where $\vec S_i=\sum_{\sigma,\sigma^{\prime}} c^{\dagger}_{i,\sigma}
\vec \sigma_{\sigma,\sigma^{\prime}}c_{i,\sigma^{\prime}}$ is the spin  
of an electron on site $i$. Here $\vec \sigma$ are the Pauli matrices and 
$<i,j>$ signifies nearest neighbor sites on a hypercubic lattice in $d$ 
dimensions. There is a constraint of no double occupancy on any site,
\be
n_i = \Sigma_{\sigma} c_{i,\sigma}^{\dagger}c_{i,\sigma}=0, 1 \; .
\ee
The concentration of doped holes, $x$, is taken to be much smaller 
than 1, and is defined as 
\be
x=N^{-1}\sum_j n_j \; ,
\ee
where $N$ is the number of sites. 
\sindex{ta}{$t$}{Hey!}
\sindex{ja}{$J$}{Hey!}
\sindex{va}{$V$}{Nearest neighbor interaction}

The essential feature of this model is that it embodies a strong, short range 
repulsion between electrons, manifest in the constraint of no double 
occupancy. 
The exchange integral $J$  arises 
through virtual processes wherein the intermediate state has a doubly 
occupied site, producing an antiferromagnetic coupling. 
Doping is assumed to remove electrons thereby producing a ``hole" 
or missing spin which is mobile because neighboring electrons can hop 
into its place with amplitude $t$.

Like a good game, the rules are simple: antialign adjacent spins, and let 
holes hop. And like any good game, the winning strategy is complex.
The ground state of this model must simultaneously minimize the zero point 
kinetic energy of the doped holes and the exchange energy, 
but the two terms compete. 
\marginpar{\em For $t>J>xt$, the problem is highly frustrated.} 
The spatially confined wavefunction of a localized hole has a high 
kinetic energy; the $t$ term accounts for the tendency of a doped 
hole to delocalize by hopping from site to site. 
However, as holes move through an antiferromagnet they 
scramble the spins: each time a hole hops from one site to its 
nearest neighbor, a spin is also moved one register in the lattice, 
{\it onto the wrong sublattice}. So it is impossible to minimize both 
energies simultaneously in $d>1$. Moreover, in the physically relevant 
range of parameters, $ t> J > tx$, neither energy is dominant. 
On the one hand, because $t>J$, one cannot simply perturb about the 
$t=0$ state which minimizes the exchange energy. On the other hand, 
because $J > tx$ one cannot simply perturb about the ground state of 
the kinetic energy.

A number of strategies, usually involving further generalizations of the 
model, have been applied to the study of this problem, including: large 
$n$\cite{AssaN}, large $S$\cite{AssaS,assa4}, large $d$\cite{larged}, 
small $t/J$\cite{ekl}, large $t/J$\cite{ekl,visscher,ioffe}, and 
various numerical studies of finite size clusters. 
(Some of the latter are reviewed in Section \ref{numerical}.) 
For pedagogic purposes, we will frame aspects of the ensuing 
discussion in terms of the large $d$ behavior of the model since 
it is tractable, and involves no additional theoretical technology, 
but similar conclusions can be drawn from a study of any of the 
analytically tractable limits listed above\footnote{In some ways, the
large $S$ limit is the most physically transparent of all these
approaches---see Ref. \citen{assa4} for further discussion.}. One common
feature
\footnote{It is still controversial whether or not  phase separation 
is universal in $d=2$ and $3$ at small enough $x$---
see
Refs.~\citen{kelosalamos,pryadkoprl,hm,whitecompare,tklee,freerickslieb,nagaev96}.} 
of these solutions is a tendency of the doped holes to phase separate  at
small $x$. The reason for this is intuitive:
in a phase separated state, the holes are expelled from the 
pure antiferromagnetic fraction of the system, where the exchange energy is 
minimized and the hole kinetic energy is not an issue, 
while in the hole rich regions, the kinetic energy of the holes is minimized, 
and the exchange energy can be neglected to zeroth order since 
$J < t x_{rich}$, where $x_{rich}$ is the concentration of 
doped holes in the hole rich regions.

We employ the following large dimension strategy. 
\marginpar{\em A large dimension expansion} 
We take as the unperturbed Hamiltonian the Ising piece of the interaction: 
\be
H_{o} =  \sum_{<i,j>} \big\{J_{z}  S_{i}^z  S_{j}^z +  
V n_{i} n_{j} \big\} \; , 
\ee 
and treat as perturbations the XY piece of the interaction and the hopping: 
\ba
&&H_{1}=  \frac{J_{\perp}}{2} \sum_{<i,j>}   \left\{  S_{i}^+  S_{j}^- 
+ {\rm h.c.} \right\} \; ,
\label{eq:H1} \\ 
&&H_{2}= -t  \sum_{<i,j>,\sigma} \left\{
c^{\dagger}_{i,\sigma}c_{j,\sigma} + {\rm h.c.} \right \} \; .
\label{eq:H2}
\ea
Expansions derived in powers of $J_{\perp}/J_{z}$ and $t/J_{z}$ can be 
reorganized in powers of $1/d$,\cite{larged} at which point we will again set 
$J_{\perp}=J_{z} \equiv J$ as in the original model (Eq.~(\ref{eq:tj})), 
and allow the ratio $t/J$ to assume physical values.

\subsubsection{One hole in an antiferromagnet}

\begin{figure}[ht!!!]
\begin{center}
\epsfig{figure=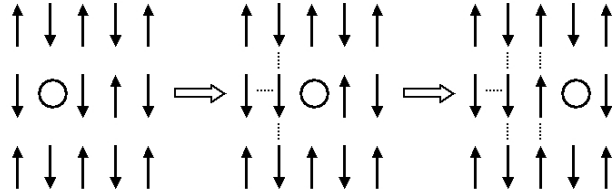,width=0.7\linewidth}
\end{center}
\caption{\label{fig:tJstring}
{\em Frustration of one hole's motion in an antiferromagnet.} 
As the hole hops, it leaves behind a string of frustrated bonds 
designated here by dashed lines.} 
\end{figure}

\begin{figure}[ht!!!]
\begin{center}
\epsfig{figure=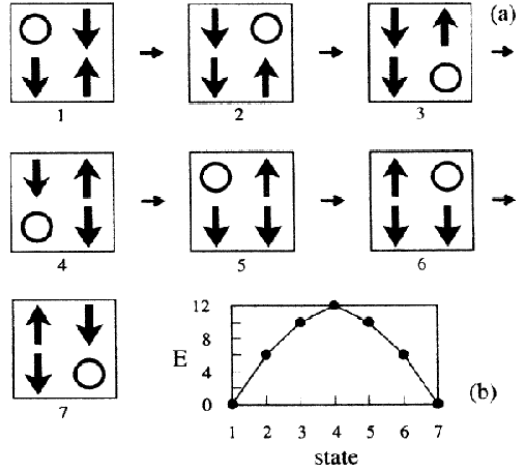,width=0.6\linewidth}
\end{center}
\caption{\label{fig:trugman}
{\em Trugman terms.} (a) A hole moving one and a half times around a plaquette translates a 
degenerate ground state without leaving a frustrated string of spins behind. 
(b) The energy of the intermediate states in units of $J$. The hole has to 
tunnel through this barrier as it moves. From Ref.~\citen{trugman}.} 
\end{figure}

It is universally recognized that a key principle governing the physics 
of doped antiferromagnets is that the motion of a single hole is highly 
frustrated. To illustrate this point, it is convenient to examine it from 
the perspective of a large dimension treatment in which 
the motion of one hole in an 
\marginpar{\em The motion of one hole in an antiferromagnet is frustrated.} 
antiferromagnet is seen to be frustrated by a ``string'' left in 
its wake (see Fig.~\ref{fig:tJstring}), which costs an energy of order 
$(d-1)J$ times the length of the string. The unperturbed ground state of 
one hole on, say, the ``black'' sublattice, is $N/2$-fold degenerate 
(equal to the number of black sublattice sites), once a direction for the 
N\'eel order is chosen (the other $N/2$ degenerate ground states describing 
a hole on the ``red'' sublattice form a disjoint Hilbert space under the 
operation of $H_1$ and $H_2$). These ground states are only connected in 
degenerate perturbation theory of third or higher order, via, {\it e.g.} 
two operations of $H_2$ and one of $H_1$.   
They are connected in perturbation theory of sixth or higher order 
by operations solely of the hopping term $H_1$ via the Trugman\cite{trugman} 
terms, in which a hole traces any closed, nonintersecting path two steps 
less than two full circuits; see Fig.~\ref{fig:trugman} for an example 
(such paths become important when $J\ll t$). 
In this manner a hole can ``eat its own string''. Owing to such processes 
a hole can propagate through an antiferromagnet. However, the high order in  
the perturbation series and the energetic barriers involved render the 
effective hopping matrix elements significantly smaller than their 
unperturbed values. 

\subsubsection{Two holes in an antiferromagnet} 

In early work on high temperature superconductivity, it was often claimed 
that, whereas the motion of a single hole is inhibited by antiferromagnetic 
order, pair motion appears to be entirely unfrustrated. It was even 
suggested \cite{hirsch} that this might be the basis of a novel, 
kinetic energy driven mechanism of pairing---perhaps the first such 
suggestion. However, a flaw with this argument was revealed in the 
work of Trugman\cite{trugman}, who showed that this mode of propagation 
of the hole pair is frustrated by a quantum effect which originates from 
the fermionic character of the background spins. While Trugman's original 
argument was based on a careful analysis of numerical studies in $d=2$, 
the same essential effect can be seen analytically in the context of a 
large $d$ expansion. The effective Hamiltonian of two holes can be written 
as follows \cite{larged}: 
\be
H_{2}^{eff} = U^{eff}\sum_{<i,j>}
c_{i}^{\dagger}c_{j}^{\dagger}c_{j}c_{i}  - T^{eff}\sum_{<i,j,k>}
c_{j}^{\dagger}c_{i}^{\dagger}c_{j}c_{k}  + {\cal O}(1/d^{2}) \; ,
\ee
where $<i,j,k>$ signifies a set of sites such that $i$ and $k$ are both 
nearest neighbors of $j$, 
and the $c_{i}^{\dagger}$ creates a hole at site $i$.
To lowest order in $(1/d)$, $U^{eff}=V-J/4$ and 
$T^{eff}=t^{2}/Jd$. For states with the two holes as nearest neighbors, 
$H_{2}^{eff}$ can be block diagonalized by Fourier transform, yielding 
$d$ bands of eigenstates labeled by a band index and 
a Bloch wavevector $\vec k$. The result is that $d-1$ of these 
bands have energy $U^{eff}$ and do not disperse. 
The remaining band has energy 
$U^{eff}+4T^{eff}\sum_{a=1}^{d}sin^{2}(k_{a}/2)$, 
where $k_{a}$ is the component of $\vec k$ along $a$.
This final band, which feels the effects of pair propagation, 
has the largest energy. 
\begin{figure}[ht!!!]
\begin{center}
\epsfig{figure=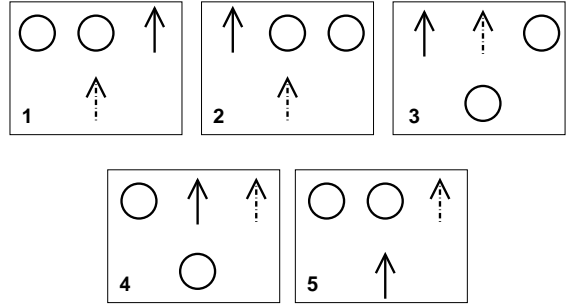,width=0.7\linewidth}
\end{center}
\caption{\label{fig:interference}
{\em Frustration of a hole pair's motion in an antiferromagnet.} 
The figure shows a sequence of snapshots in a process that takes 
a pair of holes back to their original position, but with a pair 
of spins switched. The sequence is as follows: 
1) Initial two hole state. 2) A spin has moved two sites to the left. 
3) The other spin has moved one site up. 4) A hole has moved two sites to the 
left. 5)  A hole has moved up. Due to the fermionic nature of the spins, 
the above process leads to an increase in the pair energy, so that pair 
propagation is not an effective mechanism of pair binding.}
\end{figure} 
This counterintuitive result follows from the fermionic nature 
\marginpar{\em Two holes are no less frustrated.} 
of the background spins. A similar calculation for bosons 
would differ by a minus sign:  in that case, the final band 
has energy  
$U^{eff}-4T^{eff}\sum_{a=1}^{d}sin^{2}(k_{a}/2)$, 
which is much closer to what one might have 
expected.\footnote{This corrects similar expressions in Ref.~\citen{larged}.}
The interference effect for the fermionic problem  
is illustrated in Fig.~\ref{fig:interference}. 
Different paths that carry the system from one hole pair configuration to 
another generally interfere with each other, and when two such paths 
differ by the exchange of two electrons, they interfere destructively in 
the fermionic case and constructively in the bosonic. It 
follows from this argument that pair motion, too, is frustrated---it 
actually results in an effective kinetic repulsion between holes, rather 
than in pair binding\footnote{It is apparent that  second neighbor
hopping terms, $t^{\prime}$, produce less frustration of the
single particle motion, and ``pair hopping'' terms, which arise naturally
in the $t/U$ expansion of the Hubbard model, lead to unfrustrated
pair motion\cite{assa2}.  However, $t^{\prime}$ is generally
substantially smaller than $t$, and if pair hopping is derived from the
Hubbard model, it is of order $J$, and hence relatively small.}.

\subsubsection{Many holes: phase separation}

In large $d$, the frustration of the kinetic energy of doped holes 
in an antiferromagnet 
leads to a miscibility gap\cite{larged}. 
Perhaps this should not be surprising, since phase separation is the generic 
fate of mixtures at low temperatures. 
At any finite temperature, two-phase coexistence occurs whenever the 
chemical potentials of the two phases are equal. In the present 
case, one of the phases, the undoped antiferromagnet, is 
incompressible, which means that at $T=0$ its chemical potential lies 
at an indeterminate point within the Mott gap. Under these circumstances, 
phase coexistence is instead established by considering the total energy of 
the system: 
\ba 
E_{tot}&=&N_{AF}e_{AF} + N_{h}e_{h} \nonumber \\ 
&=&Ne_{AF} + N_{h}(e_{h}-e_{AF}) \; ,
\ea
where $N_{AF}$ and $N_{h}$ are the number of sites occupied by the 
undoped antiferromagnet and by the hole rich phase, respectively; 
$N=N_{AF}+N_{h}$; $e_{AF}$ is the energy per site of the antiferromagnet 
and $e_{h}$ is the energy per site of the hole rich phase, in which 
the concentration of doped holes is $x_{rich}=x(N/N_h) \ge x$. 
If $E_{tot}$ has a minimum with respect to $N_{h}$ at a value 
$N_{h}<N$, there is phase coexistence. This minimization leads 
to the equation 
\be 
\mu = \frac{e_{AF}-e_{h}(\mu)}{1-n(\mu)} \; ,
\label{mueq} 
\ee 
where $\mu$ is the chemical potential of the hole rich phase, and 
$n=1-x_c$ is the electron density in the hole rich phase. 

As we shall see, in the limit of large dimension, $n(\mu)$ 
(and hence $e_h$ as well) is either 0 or exponentially small, 
so Eq.~(\ref{mueq}) reduces to 
\be
\mu \approx e_{AF}  \; .
\ee
We can see already how phase separation can transpire. As the electron 
\marginpar{\em Phase separation occurs below a critical concentration 
of doped holes.} 
density is raised from zero ({\it i.e.} starting from 
$x=1$ and lowering $x$), the chemical potential of the electron gas 
increases. Once $\mu$ reaches $e_{AF}$, the added electrons must 
go into the antiferromagnetic phase, and the density of the electron gas 
stops increasing. We can employ a small $k$ expansion of the electronic 
dispersion, ${\epsilon}(\vec k)= -2td+tk^{2}+\ldots$, to determine
that 
$\mu \approx-2td+tk_{F}^{2}$. Thus if $e_{AF} < -2td$, the electron 
gas is completely unstable, and there is phase separation into the pure 
antiferromagnet, and an insulating hole rich phase with $n=0$. In this case, 
$x_c=1$. Otherwise, the density of the electron gas is 
\be
n={2A_{d} \over d}\bigg({k_{F} \over 2 \pi} \bigg)^{d}= 
{2A_{d} \over d}\bigg({\sqrt{(\mu+2td)/t} 
\over 2 \pi}\bigg)^{d} \; .
\label{nc}
\ee
Here $A_d$ is the hypersurface area of a $d$ dimensional unit sphere. 
In large $d$, the energy per site of the pure antiferromagnet 
approaches that of the classical N\'eel state: 
\be 
e_{AF}=-d\left(\frac{J}{4}-V\right)[ 1 + {\cal O}(1/d^2)] \; .
\ee 
{F}rom this, it follows that the hole rich phase is insulating ({\it i.e.} 
it has no electrons) if $J-4V > 8t$ and it is metallic ($x_c <1$) if 
$J-4V < 8t$. However, even when the hole rich phase is metallic, 
its electron density  
is exponentially small (as promised): 
\be 
n=1-x_c 
={2 \over \sqrt{\pi d}}\
\left[{e \over \pi}\left(1-\left[{J-4V \over 8t}\right]\right)\right]^{d/2} 
[1+{\cal O}(1/d)] \; ,
\label{xcofJ} 
\ee 
where we have used the asymptotic large $d$ expression\cite{larged} 
$A_d \approx \sqrt{\frac{d}{\pi}} ({2 \pi e \over d})^{d/2}$. 
\begin{figure}[ht!!!]
\begin{center}
\epsfig{figure=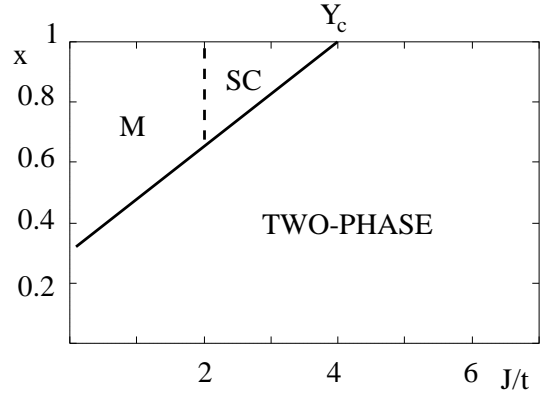,width=0.6\linewidth}
\end{center}
\caption{\label{fig:larged} 
Phase diagram of the $t-J$ model deduced from large the $d$ expansion. 
In the figure, we have set $d=2$. ``Two-phase'' 
labels the region in which phase separation occurs between the pure 
antiferromagnet and a hole rich phase, ``SC'' labels a region of $s$-wave 
superconductivity, and ``M'' labels a region of metallic behavior. 
At parametrically small $J/t \propto 1/\sqrt{d}$, a ferromagnetic phase 
intervenes at small doping. From Ref.~\citen{larged}. } 
\end{figure}  
As illustrated in Fig.~\ref{fig:larged} in large $d$, so long as $0<x<x_c$, 
the ground state of the $t-J-V$ model is phase separated, with an undoped 
antiferromagnetic region and a hole rich region which, if $8t>J-4V$, 
is a Fermi liquid of dilute electrons, or if $8t<J-4V$, is an insulator. 
(Under these same circumstance, if $x_c<x<1$, the ground state is a uniform, 
Fermi liquid metal\footnote{This statement neglects a possible subtlety due 
to the Kohn-Luttinger theorem.}.)

\begin{figure}[ht!!!]
\begin{center}
\epsfig{figure=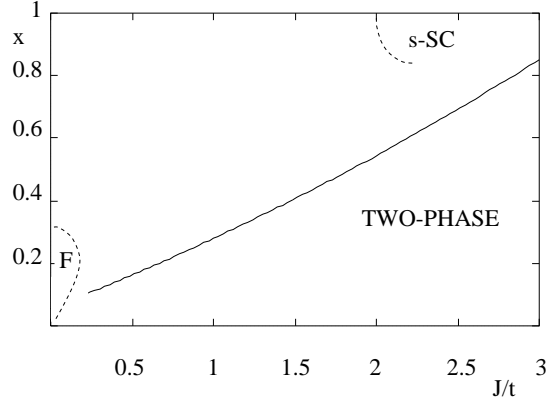,width=0.6\linewidth}
\end{center}
\caption{\label{fig:manousakis} 
Phase diagram of the $t-J$ model in two dimensions at zero temperature, 
deduced from numerical studies with up to 60 electrons. 
``Two-phase'' labels the region of phase separation,  
``s-SC'' labels a region of $s$-wave superconductivity, and 
``F'' labels a region of ferromagnetism. This figure 
is abstracted from Hellberg and Manousakis\cite{hm}. } 
\end{figure}

In the low dimensions of physical interest, such as $d=2$ and 
$d=3$, the quantitative accuracy of a large dimension expansion is certainly 
suspect. Nonetheless, we expect the qualitative physics of $d=2$ and $d=3$ 
to be captured in a large dimension treatment, since the lower critical 
dimension of most long range $T=0$ ordered states is $d=1$. 
For comparison, in Fig.~\ref{fig:manousakis} we reproduce the 
 phase diagram of the 2D $t-J$ model which was proposed by
Hellberg and Manousakis\cite{hm} on the basis of Monte Carlo 
studies of systems with up to 60 electrons. There is clearly 
substantial similarity between this and the large $D$ result 
in Fig.~\ref{fig:larged}. 

In one sense phase separation certainly can be thought of as a 
strong attractive interaction between holes, although in 
reality the mechanism is more properly regarded 
as the ejection of holes from the 
antiferromagnet.\footnote{Like salt crystallizing from a 
solution of salt water, the spin crystal is pure.}
The characteristic energy scale of this interaction 
is set by magnetic energies, so one expects to see phase separation 
only at temperatures that are small compared to the antiferromagnetic 
exchange energy $J$. 

\subsection{Coulomb frustrated phase separation and stripes}
\label{coulfrust}

Were holes neutral, 
phase separation 
would be a physically reasonable solution to the problem of frustrated hole 
motion in an antiferromagnet.
But there is another 
competition if the holes carry charge. 
In this case, full phase separation is impossible 
because of  the infinite Coulomb energy density it would entail. 
Thus, there is a second competition between the short range tendency 
to phase separation embodied in the $t-J$ model, and the long range piece 
of the Coulomb interaction. 
The compromise solution to this second level of frustration results in an 
emergent length scale\cite{antonioprb}---a crossover between phase separation 
on short length scales, and the required homogeneity on long length scales. 
\marginpar{\em Stripes are a unidirectional density wave.}
Depending upon microscopic details, many solutions are possible \cite{seul} 
which are inhomogeneous on intermediate length scales, such as checkerboard 
patterns, stripes, bubbles, or others. 

Of these, the stripe solution
is remarkably stable in simple 
models\cite{ekl,larged,stojkovic1999}, and moreover is widely observed in the 
cuprates\cite{pnas}. 
A stripe state is a unidirectional density wave state---we 
think of such a state, at an intuitive level, as consisting of 
alternating strips of hole rich and hole poor phase. 
A fully ordered stripe phase has charge density wave and spin density wave
order interleaved.  

Certain aspects of stripe states can be made precise 
on the basis of long distance considerations.  
If we consider the Landau theory\cite{zkelandau} of coupled order parameters for a
spin density wave $\vec S$ with ordering vector $\vec k$ and a charge 
density wave $\rho$ with ordering vector $\vec q$, then if $2\vec k \equiv 
\vec q$ (where $\equiv$, in this case, means equal modulo a reciprocal 
lattice vector), then there is a cubic
term in the Landau free energy allowed by symmetry,
\begin{equation}
F_{coupling} = \gamma_{stripe} \left[ \rho_{-\vec q}\vec S_{\vec k}\cdot 
\vec S_{\vec k} +
 {\rm C.C.} \right ].
\end{equation}
There are two important consequences of this term.  
Firstly, the system can lower its energy by locking the 
ordering vectors of the spin and charge density wave components
of the order, such that the period of the spin order is twice that 
of the charge order. At order parameter level, is the origin of 
the antiphase character of the stripe
order\footnote{In the context of Landau-Ginzberg theory, the situation 
is somewhat more complex, and whether the spin and charge order have 
this relation, or have the same period turns out to depend on short 
distance physics, see footnote \ref{inphaseft} and \cite{pryadkolg}.}.  
Secondly, because
this term is linear in $\rho$, it means that if there is spin order, $<\vec 
S_{\vec k}>\ne 0$, there must {\em necessarily}\footnote{Here, we exclude the 
possibility of perfectly circular spiral spin order, 
in which $Re\{<\vec S>\}\cdot Im\{<\vec S>\}=0$ 
and $[Re\{<\vec S>\}]^2=[Im\{<\vec S>\}]^2\ne 0$.} 
be charge order, $<\rho_{2\vec k}>\ne 0$,
although the converse is not true.

The Landau theory also allows us to 
distinguish three macroscopically distinct scenarios for the onset 
of stripe order. If charge order onsets at a higher critical temperature, 
and spin order either does not occur, or onsets at a lower critical 
temperature, the stripe order can be called ``charge driven.''  
If spin and charge order onset at the same critical temperature, 
but the charge order is parasitic, in the sense that 
$<\rho_{2\vec k}>\sim <\vec S_{\vec k}>^2$, the stripe order 
is ``spin driven.'' Finally, if charge and spin order onset simultaneously
by a first order transition, the stripe order is driven by the symbiosis 
between charge and spin order. This is discussed in more detail in 
Ref. \citen{zkelandau}.

The antiphase nature of the stripes 
was first predicted by the Hartree-Fock theory and 
has been confirmed as being the most probable outcome in various later, 
more detailed studies of the 
problem\cite{whitescalapino,zachar,liufradkin,netoandco}. 
In this case, the spin texture undergoes a $\pi$ phase shift across every
charge stripe, 
so that every other spin stripe has the opposite 
N{\'e}el vector,
cancelling out any 
magnetic intensity at the commensurate 
wavevector, $<\pi,\pi>$.
This situation\cite{topo,zaanennussinov} has been 
called ``topological doping.'' And, indeed, the predicted factor of two 
ratio between the spin and charge periodicities has been observed in all well 
established experimental realizations of stripe order in doped 
antiferromagnets.\cite{tranquadareview} Still, it is important to 
remember that nontopological stripes are also a logical possibility
\cite{zachar,liufradkin,hanandlee,pryadkolg,pryadkoprl,netoz}, and we should 
keep our eyes open for this form of order, as well.\footnote{For example, 
an analogous Landau theory of stripes near the N{\'e}el state must include 
the order parameter $\vec S_{\vec \pi}$, which favors in-phase domain walls 
\cite{pryadkolg}.\label{inphaseft}}

In the context of frustrated phase separation,
the formation of inhomogeneous structures is predominantly a 
\marginpar{\em The Coulomb interaction sets the stripe spacing.}
statement about the charge density, and its scale is set by the Coulomb 
interaction. This has several implications. Firstly, this means that 
charge stripes may begin to self-organize (at least locally) at relatively
high temperatures, {\it i.e.} they are charge driven in the sense
described above.\footnote{In Hartree-Fock theory, stripes are spin driven.}
Secondly, charge density wave order 
always couples linearly to lattice distortions, so we should expect dramatic 
signatures of stripe formation 
to show up in the phonon spectrum. 
Indeed, phonons may significantly affect the energetics of stripe 
formation\cite{zaanenlittlewood}. Thirdly, although we are used to 
thinking of density wave states as insulating, or at least as having a 
dramatically reduced density of states at the Fermi energy, this is not 
necessarily true. 
If the average hole 
concentration on each stripe
\marginpar{\em Competition sets the hole concentration on a stripe.} 
is determined primarily  by the competition between the Coulomb interaction 
and the local tendency to phase separation, 
the linear hole density per site along each stripe can vary as a function
of $x$ 
and consequently there is no reason to expect 
the Fermi energy to lie in a gap or pseudogap. 
In essence, stripes may be intrinsically metallic, or even superconducting.
Moreover, such compressible stripes are highly prone to
lattice commensurability effects which  tend to pin the inter-stripe
spacing at commensurate values. Conversely, if the stripes are a
consequence of some sort of Fermi surface nesting, as is the case in 
the Hartree-Fock studies\cite{zaanengunnarsson,schulz,machida89} of stripe
formation,  the stripe period always adjusts precisely so as to maintain
a gap or  pseudogap at the Fermi surface: there is always one doped hole
per site along each charge stripe. This insulating behavior is likely a
generic feature of  all local models of stripe
formation\cite{topo}, although more sophisticated treatments can lead to
other preferred linear hole densities along a
stripe\cite{nayakwilczek,whitescalapino}.

In short, stripe order is theoretically expected to be a common form 
of self-organized charge ordering in doped antiferromagnets. 
In a $d$-dimensional striped state, the doped holes are concentrated 
in an ordered array of parallel $(d-1)$ dimensional hypersurfaces: 
solitons in $d=1$, ``rivers of charge'' in $d=2$, and sheets of charge 
in $d=3$. This ``charge stripe order'' can either coexist with 
antiferromagnetism with twice the period (topological doping) or
with the same period as the charge order, or the magnetic order 
can be destroyed by quantum or thermal fluctuations of the spins. 
\marginpar{\em ``Stripe glasses'' and ``stripe liquids'' are also possible.} 
Moreover, the stripes can be insulating, conducting, or even superconducting. 
It is important to recall that for $d<4$ quenched disorder is always a 
relevant perturbation for charge density waves,\cite{larkin} so rather 
than stripe ordered states, real experiments may often require interpretation 
in terms of a ``stripe glass''\cite{kestripes,ichikawa,schmalianglass,dagotto1}. 
Finally, for many purposes, it is useful to think of systems that are not 
quite ordered, but have substantial short range stripe order as low 
frequency fluctuations, as a ``fluctuating stripe liquid''.  
We will present an example of such a state in the next subsection.

\subsection{Avoided critical phenomena}

Let us examine a simple model of Coulomb frustrated phase separation. 
We seek to embody a system with two coexisting phases, 
which are forced to interleave due to the charged nature 
of one of the phases. To account for the short range tendency 
to phase separation, we include a short range ``ferromagnetic'' 
interaction which encourages nearest neighbor regions to be of the same phase, 
and also a long range ``antiferromagnetic'' interaction which prevents any 
domain from growing too large: 
\begin{equation}
\label{sphericalH}
H=-L\sum_{<i,j>}\vec S_{i}\cdot \vec S_{j} + \frac {Qa^{d-2}} 2\sum_{i\ne j} 
{\vec S_{i}\cdot \vec S_{j} \over |\vec R_{i}-\vec R_{j}|^{d-2}} \; .
\end{equation} 
Here $\vec S_{j}$ is an $N$ component unit vector, 
$\vec S_{i}\cdot \vec S_{i}=1$, $L$ is a nearest neighbor ferromagnetic 
interaction, $Q$ is an antiferromagnetic ``Coulomb'' term which represents 
the frustration (and is always assumed small, $Q \ll L$), $d$ is the spatial 
dimension, $<i,j>$ signifies nearest neighbor sites, $a$ is the lattice 
constant, and $\vec R_{j}$ is the location of lattice site $j$. 
The Ising ($N=1$) version of this model is the simplest 
coarse grained model\cite{kelosalamos,ute} of Coulomb frustrated 
phase separation, in which $S_j=1$ represents a hole rich, 
and $S_j=-1$ a hole poor region. In this case, $L>0$ is the surface 
tension of an interface between the two phases, and $Q$ is the strength 
of the Coulomb frustration. While the phase diagram of this model has been 
analyzed\cite{ute} at $T=0$, it is fairly complicated, and its extension to 
finite temperature has only been attempted numerically\cite{tarjus}. 
However, all the thermodynamic properties of this model can be 
obtained\cite{spherical,zohar} exactly in the large $N$ limit.
\begin{figure}[ht!!!]
\begin{center}
\epsfig{figure=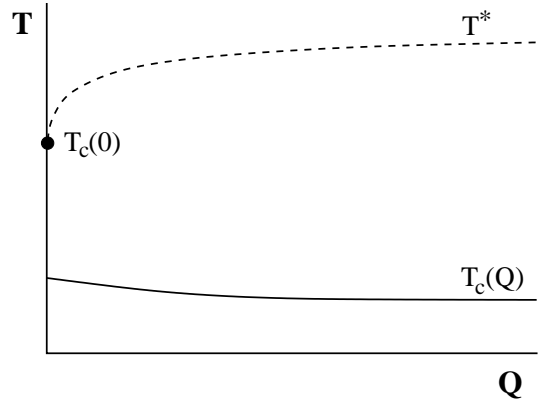,width=0.6\linewidth}
\end{center}
\caption{\label{fig:spherical} 
Schematic phase diagram of the model in Eq. (\ref{sphericalH})
of avoided critical phenomena. 
The thick black dot marks $T_{c}(Q=0)$, 
the ordering temperature in the absence of frustration; 
this is ``the avoided critical point''.  
Notice that $T_{c}(Q\rightarrow 0) < T_{c}(Q=0)$.
From Ref.~\citen{zohar}. }
\end{figure} 

Fig.~\ref{fig:spherical} shows the phase diagram for this model. 
Both for $Q=0$ and $Q\ne 0$, there is a low temperature ordered state,
but  the ordered state is fundamentally different for the two cases. 
For the unfrustrated case, the ordered state is homogeneous, 
whereas with frustration, there is an emergent length scale 
in the ordered state which governs the modulation of the order parameter. 
To be specific, in dimensions $d >2$ and for $N > 2$, 
there is a low temperature ordered unidirectional spiral phase, which 
one can think of as a sort of stripe ordered phase\cite{zohar}. 
Clearly, as $Q\rightarrow 0$, the modulation length scale must diverge, 
so that the homogeneous ordered state is recovered. 
However, like an antiferromagnet doped with neutral holes, 
there is a discontinuous change in the physics from $Q=0$ to 
any finite $Q$: for $d \le 3$, 
$\lim_{Q\rightarrow 0} T_{c}(Q)\equiv T_{c}(0^{+})$ is strictly less than 
$T_{c}(0)$. 
In other words, an infinitesimal amount of frustration 
depresses the ordering temperature discontinuously.

Although for any finite $Q$ the system does not experience a 
\marginpar{\em This model exhibits a ``fluctuating stripe'' phase.} 
phase transition as the temperature is lowered through $T_{c}(0)$, 
the avoided critical point heavily influences the short range physics. 
For temperatures in the range $T_{c}(0)> T > T_{c}(0^{+})$, substantial 
local order develops. An explicit expression for the spin-spin correlator 
can be obtained in this temperature range: At distances less than 
the correlation length $\xi_{0}(T)$ of the unfrustrated magnet, 
$R_{ij}<\xi_{0}(T)$, the correlator is critical, 
\begin{equation}
\langle\vec S_{i}\cdot \vec S_{j}\rangle \sim \left(a/R_{ij}\right)
^{d-2-\eta} \; ,
\end{equation}
but for longer distances, $R_{ij}>\xi_0(T)$, it exhibits a damped version of 
the Goldstone behavior of a fluctuating stripe phase, 
\begin{equation} 
\langle\vec S_{i}\cdot \vec S_{j}\rangle \sim 
\left(a/R_{ij}\right)^{\frac{d-1}{2}}\cos[K R_{ij}]
\exp[-\kappa R_{ij}] \; .
\end{equation}
At $T_{c}(Q)$, the wavevector $K$ is equal to the stripe 
ordering wavevector of the low temperature ordered state, 
$K(T_{c})=(Q/L)^{1/4}$.  As the temperature is raised, 
$K$ decreases until it vanishes at a disorder line marked 
$T^{*}$ in the figure. 
The inverse domain size is given by 
\begin{equation} 
\kappa(T)=\sqrt{ (Q/L)^{1/4}-K^{2}(T)} \; .
\end{equation} 
For a broad range of temperatures (which does not narrow as $Q\rightarrow 0$), 
this model is in a fluctuating stripe phase in a sense that can be 
made arbitrarily precise for small enough $Q$.

\subsection{The cuprates as doped antiferromagnets}

\subsubsection{General considerations}

\marginpar{\em Our theoretical understanding of the undoped antiferromagnets 
is extolled.}
There is no question that the undoped parents of the high temperature 
superconductors are Mott insulators, in which the strong short range 
repulsion between electrons is responsible for the insulating behavior, and 
the residual effects of the electron kinetic energy (superexchange) lead 
to the observed antiferromagnetism. Indeed, one of the great theoretical 
triumphs of the field is the complete description, based on interacting 
spin waves and the resulting nonlinear sigma model, of the magnetism 
in these materials.\cite{chn2,chakreview,aepplihighenergy} 

However, it is certainly less clear that one should inevitably view the 
superconducting materials as doped antiferromagnets, 
especially given that we have presented strong reasons to expect 
a first order phase transition between $x=0$ and $x>0$. 
Nonetheless, many experiments on the cuprates are suggestive of a doped 
antiferromagnetic character. In the first place, various measurements of 
the density of mobile charge, including the superfluid 
density\cite{uemura,uemura91}, 
the ``Drude weight'' measured in optical conductivity\cite{tanner}, and
the Hall number\cite{onghall,onghall2}, 
are all consistent with a density proportional
to the {\it doped hole}  density, $x$, rather than the total hole density,
$1+x$, expected from a  band structure approach. Moreover, over a broad
range of doping, the  cuprates retain a clear memory of the
antiferromagnetism of the parent  correlated insulator. Local magnetism
abounds. NMR, $\mu$SR, and neutron  scattering find evidence 
(some of which is summarized in Section~\ref{1241}) of static, or slowly
fluctuating, spin patterns,  including stripes, spin glasses, and perhaps
staggered orbital currents. 
\marginpar{\em Why the cuprates should be viewed as doped antiferromagnets} 
Static magnetic moments, or slowly fluctuating ones, are hard to 
reconcile with a Fermi liquid picture. There is also some evidence from STM 
of local electronic inhomogeneity\cite{pan,davis,kap} in BSCCO, indicative
of the  short range tendency to phase separate. The Fermi liquid state in
a simple metal 
is highly structured  in $k$-space, and so is highly homogeneous (rigid)
in real space.   This is certainly in contrast with  experiments on the 
cuprates which indicate significant real space structure.

\subsubsection{Stripes} 

There is increasingly strong evidence that 
stripe correlations, as a specific feature of doped antiferromagnets, occur 
in {\em at least some} high temperature superconducting materials. 
The occurrence of stripe phases in the high temperature superconductors in 
\marginpar{\em Another triumph of theory! \\  
(Look, there are painfully few of them.)} 
particular, and in doped antiferromagnets more generally, 
was successfully predicted
\footnote{The theoretical 
predictions  predated any clear body of well accepted 
experimental facts, although in all fairness it must be admitted that there 
was some empirical evidence of stripe-like structures which predated all 
of the theoretical inquiry: Even at the time of the first Hartree-Fock 
studies, there was already dramatic experimental 
evidence\cite{thursten,japstripes} of incommensurate magnetic structure 
in {\LSCO}.
}  
by theory\cite{zaanengunnarsson,schulz,machida89}. 
Indeed, it is clear that a fair fraction of the theoretical inferences 
discussed in Section~\ref{coulfrust} 
are, at 
least in broad outline, applicable 
to a large number of materials, including at least some 
high temperature superconductors\cite{pnas}. 
In particular, 
the seminal discovery\cite{tranqnature} that in {\LNSCO}, 
first charge stripe order, then spin stripe order, and then superconductivity 
onset at successively lower critical temperatures
is consistent with Coulomb frustrated phase separation. 
(See Fig.~\ref{fig:ichikawa} in Section~\ref{1241}.) 
Somewhat earlier work on the closely related 
nickelates\cite{jtniprl} established that the charge 
stripes are, indeed, antiphase domain walls in the spin order. 

Controversy remains as to how universal stripe phases are in the 
cuprate superconducting materials, and even how the observed phases 
should be precisely characterized. This is also an exciting topic, 
on which there is considerable ongoing theoretical and experimental study. 
We will defer further discussion of this topic to Section \ref{finalchapter}.

\subsection{Additional considerations and alternative perspectives}

There are a number of additional aspects of this problem which we 
have not discussed here, but which we feel warrant a mention. 
In each case, clear discussions exist in the literature to which 
the interested reader is directed for a fuller exposition.

\subsubsection{Phonons}
\label{phonons}

There is no doubt that strong electron-phonon coupling can drive a system
to phase separate.  Strong correlation effects necessarily enhance such 
tendencies, since
they reduce the rigidity of the electron wavefunction to spatial modulation.  
(See, {\it e.g.}, the 1D example in Section~\ref{rg1}.)
In particular, when there is already a tendency to some form of charge
ordering, 
on very general grounds we expect it to be strongly
enhanced by electron-phonon interactions.  

This observation makes us very leery of any attempt at a {\em quantitative} 
comparison between results on phase separation or stripe formation in the 
$t-J$ or Hubbard models with experiments in the cuprates, where
the electron-phonon interaction is manifestly strong\cite{lanzara}. 
Conversely, there should generally be substantial signatures of various 
stripe-related phenomena in the phonon dynamics, and this can be used to 
obtain an experimental handle on these behaviors\cite{mook}. 
Indeed, there exists a parallel development of stripe-related 
theories of high temperature superconductivity based on Coulomb frustration
of a phase separation instability which is driven by strong electron-phonon
interactions\cite{gorkov,castro1,castro2}. The similarity between 
many of the  notions that have emerged from these studies, and those 
that have grown out of studies of doped antiferromagnets illustrates 
both how robust the consequences of frustrated phase separation are 
in highly correlated systems, and how difficult it is to unambiguously
identify a ``mechanism'' for it.  For a recent discussion of many of the 
same phenomena discussed here from this alternative viewpoint, 
see Ref.~\citen{castro5}.

\subsubsection{Spin-Peierls order}
Another approach to this problem, which emerges naturally from 
an analysis of the large $N$ limit\cite{rs}, is to view the doped system 
as a ``spin-Peierls'' insulator, by which we mean  a quantum disordered 
magnet in which the unit cell size is doubled but spin rotational invariance 
is preserved.\footnote{Alternatively, this state can be viewed as a 
bond-centered charge density wave\cite{sshrmp,campbell}.}  
While the undoped system is certainly antiferromagnetically ordered, 
it is argued that when the doping exceeds the critical value at which 
spin rotational symmetry is restored, the doped Mott insulating features 
of the resulting state are better viewed as if they arose from a doped 
spin-Peierls state.  Moreover, since the spin-Peierls state has a spin gap, it
can profitably be treated as a crystal of Cooper pairs, which makes the 
connection to superconductivity very natural. Finally, as mentioned in 
Section \ref{fraction}, this approach has a natural connection with 
various spin liquid ideas.  

Interestingly, it turns out that the doped spin-Peierls state also 
generically phase separates\cite{IL,marder,sachdev1,AssaN}. When the 
effect of long range Coulomb interactions are included, the result is a 
staircase of commensurate stripe phases\cite{sachdev1}. Again, the 
convergence of the pictures emerging from diverse starting points 
convinces us of the generality of stripey physics in correlated systems. 
For a recent discussion of the physics of stripe phases, and their 
connection to the cuprate high temperature superconductors approached 
from the large N/spin-Peierls perspective, see Ref.~\citen{sachdevzhang}.

\subsubsection{Stripes in other systems}

It is not only the robustness of stripes in various theories that 
warrants mention, but also the fact that they are observed, in one 
way or another, in diverse physical realizations of correlated electrons. 
Stripes, and even a tendency to electronic phase
separation, are by now well documented in the manganites---the colossal 
magnetoresistance materials. (For recent discussions, which review some of 
the literature, see Refs.~\citen{khomskiistripes,dagotto2} and \citen{dagotto3}.) 
This system, like the nickelates and cuprates, is a doped
antiferromagnet, so the  analogy is quite precise.  

Although the  microscopic physics of
quantum Hall systems is quite different from that of doped antiferromagnets, 
it has been realized for some time\cite{foglershklovskii,chalker} that
in higher Landau levels, a similar drama occurs due to the interplay 
between a short ranged attraction and a long range repulsion between 
electrons which gives rise to stripe and bubble phases.  Evidence of 
these, as well as quantum Hall nematic
phases,\cite{eisenstein,eisenstein1999}
has become increasingly 
compelling in recent years.  (For a recent review, see \cite{foglerreview}.)  
On a more speculative note, it has been noticed that such behavior may be 
expected in the neighborhood of many first order transitions in electronic 
systems, and it has been suggested that various charge inhomogeneous states
may play a role in the apparent metal-insulator transition observed 
in the two dimensional electron gas\cite{spivaknew}.

\section{Stripes and High Temperature Superconductivity}
\label{finalchapter}

\addtocontents{toc}{
\noindent{\em We present a coherent view---our view---of high
temperature superconductivity in the cuprate superconductors.
This section is
more broadly phenomenological than is the rest of this paper.}  }

In this article, we have analyzed the problem of high temperature  
superconductivity 
\marginpar{\em We boast, and yet yearn for the unified understanding of 
BCS theory.}
in a highly correlated electron liquid, with particular emphasis on doped
antiferromagnets. We have identified theoretical issues, and even some 
solutions. We have also discussed aspects of the physics that elude a 
BCS description. This is progress.  

However, 
we have not presented a single, unified 
solution to the problem. 
Contrast this with BCS, a theory so elegant it may 
captured in haiku:
\bigskip

\centerline{Instability}
\centerline{Of a tranquil Fermi sea --}
\centerline{Broken symmetry.}

\bigskip
{\noindent Of} course, to obtain a more  quantitative
understanding of particular materials 
would require a few more verses---we might need to study the Eliashberg equations to treat the 
phonon dynamics in a more realistic fashion, and we may need to include 
Fermi liquid corrections, and we may also have to wave our hands a bit about 
$\mu^*$, {\em etc}.  But basically, in the context of a single 
approximate solution of a very simple model problem, we obtain a remarkably 
detailed and satisfactory understanding of the physics. And while we may 
not be able to compute $T_c$ very accurately---it does, after all, depend 
exponentially on parameters---we can understand what sort of metals  
will tend to be good superconductors: metals with strong electron-phonon 
coupling, and consequently high room temperature resistances, are good 
candidates, as are metals with large density of states at the Fermi 
energy. We can also compute various dimensionless ratios of physical 
quantities, predict dramatic coherence effects (which {\em do} 
depend on microscopic details), and
understand the qualitative effects of disorder.

The theory of high temperature superconductivity presented here
\marginpar{\em We outline a less ambitious goal for theory.} 
reads more like a Russian novel, with exciting chapters and
fascinating characters, but there are many intricate subplots, 
and the pages are awash in
familiars, diminutives, and patronymics.  
To some extent, this is probably unavoidable.  Fluctuation effects 
matter in the superconducting state: the phase  ordering
temperature,
$T_{\theta}$, is approximately equal to 
$T_c$, and the zero temperature coherence length, $\xi_0$, 
is a couple of lattice constants. In addition, the existence 
of one or more physical pseudogap scales (the $T^{*}$'s) in addition 
to $T_c$ means that there are multiple distinct qualitative changes 
in the physics in going from high temperature to $T=0$. 
Moreover,
various other types of ordered states are seen in close proximity to or 
in coexistence with the superconducting state. Thus, it is more plausible 
that we will weave together a qualitative understanding of the basic physics 
in terms of a number of effective field theories, each capturing the 
important physics in some range of energy and length scales. 
Ideally, these different theories will be nested,
with each effective Hamiltonian derived as the low energy limit of the 
preceding one.

While not as satisfying as the unified description 
of BCS-Eliashberg-Migdal theory, there
is certainly ample precedent for the validity of this kind of 
multiscale approach. The number of quantitative predictions 
may be limited, but 
we should expect the approach to provide a simple understanding of a large 
number of qualitative observations.
In fact, we may never be able to predict $T_c$ reliably, or even whether a 
particular material, if made, will be a good superconductor, but 
a successful theory should certainly give us some guidance concerning 
what {\em types} of new materials are good candidate high temperature 
superconductors\cite{skfisk,geballe}.  

Before we continue, we wish to state a major change of emphasis. 
\marginpar{\em We now consider applicability.}
Up until this point, we have presented only results that we 
consider to be on secure theoretical footing. That is, we have
presented a valid theory.\footnote{High temperature superconductivity 
being a contentious field, it will not surprise the reader to learn that 
there is controversy over how important each of the issues discussed  above 
is to the physics of the cuprates. As the field progresses, and especially
as new data are brought to light, it may be that in a future version of
this  article we, too, might change matters of {\em emphasis}, but we are 
confident that no new understanding will challenge the {\em validity} 
of the theoretical constructs discussed  until now.} We now allow ourselves 
free rein to discuss the {\em applicability} of these ideas to the real
world. In particular, we discuss the cuprate high temperature
superconductors, and whether the salient physics therein finds a natural
explanation in terms of stripes in doped antiferromagnets.
Various open issues are laid out, as well as some general strategies 
for addressing them.

\subsection{Experimental signatures of stripes}

At the simplest level, stripes refer to a broken symmetry 
state in which the discrete translational symmetry of the crystal is 
broken in one direction: stripes is a term for a 
{\em unidirectional density wave}.  
``Charge stripes" refer to a unidirectional charge density wave (CDW).
``Spin stripes" are unidirectional colinear spin density waves (SDW).
\footnote{Spiral SDW order has somewhat different character, even when 
unidirectional, and is not generally included in the class of striped states.}
More subtle local forms of stripes, such as stripe liquids, nematics
and glasses are addressed in Section~\ref{sec:liqcryst}.

\subsubsection{Where do stripes occur in the phase diagram?}
\label{1241}

As discussed in Sections~\ref{numerical} and \ref{doped},
holes doped into 
an antiferromagnet have a tendency to self-organize into rivers of charge,
and these charge stripes
tend to associate with antiphase domain walls in the spin texture.
As shown in Section \ref{coulfrust}, stripe order is typically either 
``charge driven,''  in which case spin order onsets (if at all) at a
temperature less than  the charge ordering temperature, or
``spin driven,'' if the charge order  onsets as a weak parasitic order at
the same temperature as the spin order.
To the extent that stripes are indeed a consequence of Coulomb
frustrated  phase separation, we expect them to be charge driven, in this
sense.

Neutron scattering has proven the most useful probe for unambiguously 
\marginpar{\em  Experimental evidence of stripes has been detected in:}
detecting stripe order. Neutrons can scatter directly from the electron 
spins. However, neutrons (and, for practical reasons, X-rays as well) 
can only detect charge stripes indirectly by imaging the induced 
lattice distortions.  Alternatively, (as discussed in
Section
\ref{coulfrust}) since spin stripe order
implies charge order, the magnetic neutron scattering 
itself can be viewed as an indirect measure of charge order.  Since
stripe order is unidirectional, it should ideally show up in a 
diffraction experiment as pairs of new Bragg peaks at positions
$\vec k_{\pm}=\vec Q\pm 2\pi \hat e/\lambda$ where $\hat e$ is 
the unit vector perpendicular to the stripe direction,
$\lambda$ is the stripe period, and $\vec Q $ is an appropriate 
fiduciary point. For charge stripes, $\vec Q$ is any reciprocal lattice 
vector of the underlying crystal, while for spin stripes, $\vec Q$ is offset 
from this by the N\'{e}el ordering vector, $<\pi,\pi>$. Where both spin and 
charge order are present, the fact that the charge stripes are associated 
with magnetic antiphase domain walls is reflected in the fact that 
$\lambda_{spin}=2\lambda_{charge}$, or equivalently 
$\vec k_{charge} = 2\vec k_{spin}$.

\begin{figure}[ht!!!]
\begin{center}
\epsfig{figure=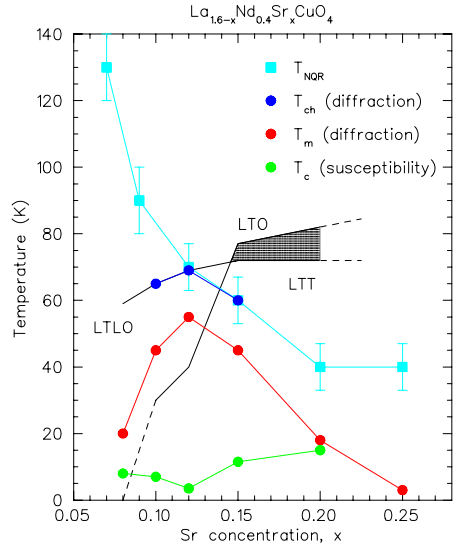,width=0.5\linewidth}
\end{center}
\caption[Diffraction data on LNSCO]{\label{fig:ichikawa}
Blue data points refer to the onset of charge inhomogeneity.
Red data points denote the onset of incommensurate magnetic peaks.
Green data points are the superconducting $T_c$.
{F}rom Ichikawa {\em et al.}\cite{ichikawa}}
\end{figure}

La$_{1.6-x}$Nd$_{0.4}$Sr$_x$CuO$_{4}$ (LNSCO) is stripe ordered, 
\marginpar{\em LNSCO}
and the onset of stripe ordering with temperature is clear. 
Fig.~\ref{fig:ichikawa} shows data from neutron scattering, NQR, 
and susceptibility measurements\cite{ichikawa}. In this material, 
charge stripes form at a higher temperature than spin stripes. Note 
also that static charge and spin stripes coexist with superconductivity 
throughout the superconducting dome. In fact
experiments reveal quartets of new Bragg peaks, 
at $\vec Q \pm 2\pi \hat x/\lambda$ and $\vec Q \pm 2\pi \hat y/\lambda$. 
In this material, the reason for this is understood to be a
bilayer effect---there is a crystallographically imposed tendency 
for the stripes on neighboring planes to be oriented at right 
angles to each other, giving rise to two equivalent pairs of peaks.
\marginpar{\em LBCO} 
Charge and spin peaks have also been detected\cite{fujita1} in neutron
scattering studies of La$_{1.875}$Ba$_{0.125-x}$Sr$_x$CuO$_4$.

Spin stripe order  has also been observed
from elastic neutron scattering in 
\marginpar{\em LSCO}
La$_{2-x}$Sr$_x$CuO$_4$ (LSCO) for dopings between $x=.02$ and $x=.05$ 
where the material is not superconducting at any $T$; these  stripes are 
called diagonal, because they lie along a direction rotated $45^{o}$ 
to the Cu-O bond direction\cite{lsco1d}.  
Above $x=.05$\cite{yamada}, 
the stripes are vertical\footnote{We should say {\em mostly} vertical.
Careful neutron scattering work\cite{yshift1,yshift2} on LSCO and LCO has
shown that the incommensurate peaks are slightly rotated from the 
$Cu-O$ bond direction, corresponding to the orthorhombicity of the
crystal.\label{foot:yshift}},
{\it i.e.} along the Cu-O bond direction, and
the samples are superconducting at low temperature.  
For dopings between $x=.05$ and $x=.13$, the stripes 
have an ordered (static) component.  
In the region $x=.13$ to $x=.25$, incommensurate 
magnetic peaks have been detected with inelastic neutron scattering. 
Because of the close resemblance between these peaks and the static 
order observed at lower doping, this can be unambiguously interpreted 
as being due to slowly fluctuating stripes.

Neutron scattering has also detected spin stripes in 
\marginpar{\em LCO}
La$_2$CuO$_{4+\delta}$ (LCO) with $\delta=$
$.12$ \cite{yshift1}. In this material, static stripes coexist 
with superconductivity even at optimal doping. 
In the $T_c = 42K$ samples (the highest $T_c$ for this
family thus far), superconductivity and spin stripe order onset 
simultaneously \cite{yshift1,wells1}.
Application of a magnetic field suppresses the superconducting transition 
temperature, but has little effect on the ordering temperature of the 
spins\cite{khaykovich}.  

In very underdoped nonsuperconducting LSCO, because the stripes 
lie along one of the orthorhombic axes, it has been possible to confirm
\cite{kimura1,fujita2} 
that stripe order leads, as expected, to pairs of equivalent Bragg spots, 
indicating unidirectional density wave order. In both superconducting LSCO 
and LCO, quartets of equivalent Bragg peaks are observed whenever stripe 
order occurs. This could be due to a bilayer effect, as in LNSCO, 
or due to a large distance domain structure of the stripes within 
a given plane, such that different domains contribute weight to 
one or the other of the two pairs of peaks. 
However, because the stripe character in these materials so closely 
resembles that in LNSCO, there is no real doubt that the observed 
ordering peaks are associated with stripe order, as opposed to some 
form of checkerboard order. 

In YBaCu$_{2}$O$_{6+y}$ (YBCO), incommensurate spin fluctuations have been 
\marginpar{\em YBCO}
identified throughout the superconducting doping range. 
\cite{mook,mook2,ybco6.7,mookchg} By themselves, these peaks 
(which are only observed at frequencies above a rather substantial spin gap) 
are subject to more than one possible interpretation\cite{bourges1},
 although their 
similarity\cite{aeppli22} to the stripe signals seen in LSCO is 
strong circumstantial evidence that they are associated with stripe 
fluctuations.
Recently, this interpretation has been strongly reinforced by several 
additional observations. Neutron scattering evidence\cite{mookchg} 
has been found of static charge stripe order in underdoped 
YBCO with $y=.35$ and $T_c=39$K.
The charge peaks persist to at least $300K$. The presence of a static stripe 
phase in YBCO means that inelastic peaks seen at higher doping are very
likely fluctuations of this ordered phase.  
In addition, phonon anomalies have been linked to the 
static charge stripes at $y=.35$, and used to 
detect charge fluctuations at $y=.6$ \cite{mook}. 
By studying a partially detwinned sample with $y=.6$, 
with a $2:1$ ratio of domains of crystallographic orientation, 
Mook and collaborators were able to show that the quartet of 
incommensurate magnetic peaks consists of two inequivalent pairs, 
also with a 2:1
ratio of intensities in the two directions\cite{ybco1d}. 
This confirms that in YBCO, as well, the 
signal arises from unidirectional spin and charge modulations (stripes), 
and not from a checkerboard-like pattern.

Empirically, charge stripe formation precedes 
spin stripe formation as the temperature is lowered, and charge 
stripes also form at higher temperatures than $T_c$. Both types of stripe 
formation may be a phase transition, or may simply be a crossover of local 
stripe ordering, depending upon the material and doping.  
Where it can be detected, charge stripe formation occurs at a higher 
temperature than the formation of the pairing gap,\footnote{See
our discussion of the pseudogap(s) in Section~\ref{sec:pseudogap}.} 
consistent with the spin gap proximity effect (see Section~\ref{sgpfx}).

Although some neutron scattering has been done on 
\marginpar{\em BSCCO}
Bi$_2$Sr$_2$CaCu$_2$O$_{8+\delta}$ (BSCCO), 
the probe has only produced weak evidence of significant 
incommensurate structure \cite{mookbscco}. The weak coupling of planes
in BSCCO makes neutron scattering difficult, 
as it is difficult to grow the requisite large crystals. 
However, BSCCO is very well suited to surface probes
such as ARPES and STM. Recent STM data, both with\cite{hoffman} and 
without\cite{aharon} an external magnetic field have revealed a 
{\em static} modulation in the local density of states 
that is very reminiscent of the incommensurate peaks observed with neutron  
scattering. Indeed, in both cases, the Fourier transform of the STM image 
exhibits a clear quartet of incommensurate peaks, just like those seen in 
neutron scattering in LSCO and YBCO. Here, however, unlike in the 
neutron scattering data, phase information is available in that 
Fourier transform. Using standard image enhancement methods, 
this phase information can be exploited\cite{aharon} to directly 
confirm that the quartet of intensity peaks is a consequence 
of a domain structure, in which the observed density of states 
modulation is locally one dimensional, but with an orientation 
that switches from domain to domain.  The use of STM as a probe of 
charge order is new, and there is much about the method that needs 
to be better understood\cite{leeoscillations}
before definitive conclusions can be reached, but the 
results to date certainly look very promising.

\marginpar{\em Preliminary evidence of 
nematic order has been detected, as well.}
Finally, striking evidence of electronic anisotropy has been seen in 
untwinned crystals of La$_{2-x}$Sr$_x$CuO$_4$ ($x=0.02-0.04$) 
and YBa$_{2}$CuO$_{6+y}$ ($y=0.35-1.0$) by Ando and collaborators\cite{ando1d}.
The resistivity differs in the two in-plane directions
in a way that cannot be readily accounted for by crystalline anisotropy alone.
It is notable that in YBCO, the anisotropy increases as $y$ is decreased.  
That is, the electrical anisotropy  {\em increases} 
as the orthorhombicity is {\em reduced}. In some cases, substantial 
anisotropy persists up to temperatures as high as 300K. Furthermore, 
for $y<0.6$, the anisotropy increases with decreasing temperature, 
much as would be expected\cite{qhnematic} for  an electron nematic.  These
observations from transport correlate well with the evidence from neutron
scattering\cite{ybco1d}, discussed above, of substantial orientational order of the
stripe correlations in YBCO, and with the substantial, and largely temperature
independent anisotropy of the superfluid density observed in the same
material.\cite{anis248}  Together, these observations constitute important, but still
preliminary evidence of a nematic stripe phase in the cuprates.

\subsection{Stripe crystals, fluids, and electronic liquid crystals}
\label{sec:liqcryst}

Stripe ordered phases are precisely defined in terms of broken symmetry. 
A charge stripe phase spontaneously breaks the discrete 
translational 
symmetry and typically also the point group symmetry 
(e.g. four-fold rotational symmetry) of the host crystal. 
A spin stripe phase breaks spin rotational 
symmetry as well.
While experiments to detect these orders 
in one or another specific material
may be difficult to implement 
for practical reasons and because of the complicating effects of 
quenched disorder, 
the issues are unambiguous. 
Where these broken symmetries occur, it is certainly 
reasonable to conclude that the existence of stripe order is an 
established fact. That this can be said to be the case
in a number of superconducting cuprates is responsible for the upsurge of 
interest in stripe physics.

\begin{figure}[ht!!!]
\begin{center}
\epsfig{figure=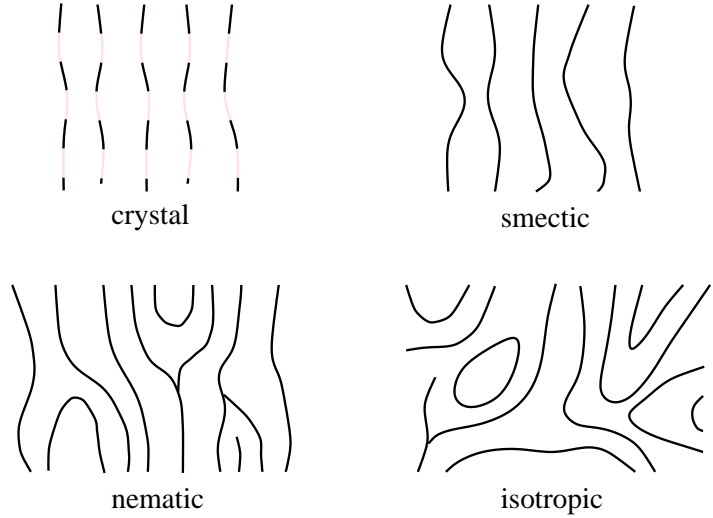,width=0.8\linewidth}
\end{center}
\caption[Schematic representation of stripe phases in 2D]
{\label{stripephases}
Schematic representation of various stripe phases in two dimensions. 
The broken lines represent density modulations along the stripes. In 
the electronic crystal, density waves on neighboring stripes are locked 
in phase and pinned. The resulting state is insulating and breaks 
translation symmetry in all directions. Solid lines represent metallic 
stripes along which electrons can flow. They execute increasingly violent transverse 
fluctuations as the system is driven towards the transition into the 
nematic phase. The transition itself is associated with unbinding of 
dislocations that are seen in the snapshot of the nematic state.
The isotropic stripe fluid breaks no spatial
symmetries
of the host crystal, but retains a local vestige of stripe order.}
\end{figure}

It is much more complicated to 
define precisely the intuitive notion of a
``stripe fluid''.\footnote{For the present purposes, 
the term ``fluctuating stripes'' is
taken to be synonymous with a stripe fluid. See, 
for example, Refs.~\citen{topo} and~\citen{zaanenfluid}.}
Operationally, it means there is sufficient
short ranged stripe order that, for the 
purposes of understanding the mesoscale 
physics, it is possible to treat the system {\em as if} it were stripe 
ordered, even though translational symmetry is not actually broken. 
It is possible to imagine intermediate stripe liquid phases which are 
translationally invariant, but which still break some symmetries which
directly 
reflect the existence of local stripe order.  
The simplest example of this is an ``electron nematic'' phase.  
\marginpar{\em Some stripe liquids break rotational symmetry.}
In classical liquid crystals, the nematic phase occurs when the 
constituent molecules are more or less cigar shaped. It can be thought 
of as a phase in which the cigars are preferentially aligned
in one direction, so that the rotational symmetry of free space is broken
(leaving only rotation by $\pi$ intact) but translational symmetry
is unbroken. In a very direct sense,
this pattern of macroscopic symmetry 
breaking is thus encoding information about the 
microscopic constituents of the liquid. In a similar fashion, we can envisage 
an electron nematic as consisting of a melted stripe ordered phase in which 
the stripes meander, and even break into finite segments, but maintain some 
degree of orientational order---for instance, the stripes are
more likely to lie in the x rather than the y direction; 
see Fig.~\ref{stripephases}.   

One way to think about different types of stripe order is to imagine 
\marginpar{\em Melting stripes}
starting with an initial ``classical'' ordered state, with coexisting 
unidirectional SDW and CDW order.  As quantum fluctuations are 
increased (metaphorically, by increasing $\hbar$), one can envisage 
that the soft orientational fluctuations of the spins will first cause 
the spin order to quantum melt, while the charge order remains.
If the charge order, too, is to quantum melt in a continuous phase 
transition, the resulting state will still have the stripes generally 
oriented in the same direction as in the ordered state, but with 
unbound dislocations which restore 
translational symmetry. 
\footnote{It is also possible to view the electron nematic from a 
weak coupling perspective, where it occurs as a Fermi surface instability
\cite{vadim}, sometimes referred
to as a Pomeranchuk instability.\cite{metzner,wegner} 
This instability is ``natural'' 
when the Fermi surface lies near a  Van Hove singularity. 
The relation between the weak coupling and the stripe fluid pictures 
is currently a subject of ongoing investigation\cite{vadimandeduardo}.}
If the underlying crystal is tetragonal \cite{isingnematic},
this state still spontaneously breaks the crystal point group symmetry.  In
analogy with  the corresponding classical state, it has been called an
electron  nematic, but it could also be viewed as an electronically
driven  orthorhombicity.  This is still a state with broken symmetry, so 
in principle its existence should be unambiguously identifiable from
experiment.\footnote{It is probable that when nematic order is lost, 
the resulting stripe liquid phase is not thermodynamically distinct 
from a conventional metallic phase, although the local order is sufficiently 
different that one might expect them to be separated by a first order 
transition.  However, it is also possible that some more subtle form of order
could distinguish a stripe liquid from other electron liquid phases---for 
instance, it has been proposed by Zaanen and collaborators\cite{zaanentopo} 
that a stripe liquid might posses an interesting, discrete topological order 
which is a vestige of the antiphase character of the magnetic correlations 
across a stripe.} The order parameter can be identified with the matrix 
elements of any traceless symmetric tensor quantity, for instance the 
traceless piece of the dielectric or conductivity tensors.

\begin{figure}[ht!!!]
\begin{center}
\epsfig{figure=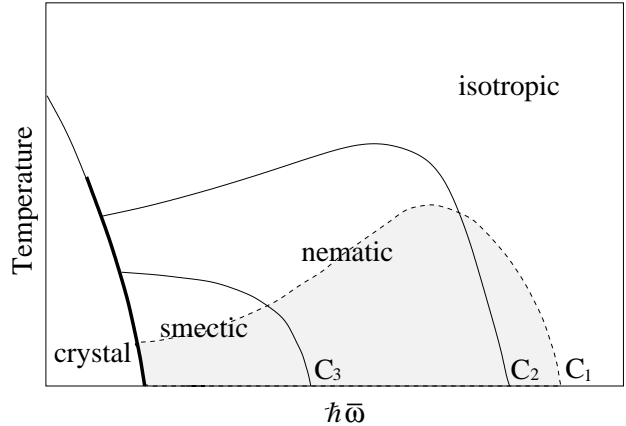,width=0.7\linewidth}
\end{center}
\caption[Schematic phase diagram of a fluctuating stripe array]{\label{fig1d3}
Schematic phase diagram of a fluctuating stripe array in a (tetragonal) 
system with four-fold rotational symmetry in $D=2$. Here 
$\hbar\bar{\omega}$ is a measure of the magnitude of the transverse 
zero point stripe fluctuations. Thin lines represent continuous 
transitions and the thick line a first order transition. We have 
assumed that the superconducting susceptibility on an isolated stripe 
diverges as $T\to 0$, so that at finite stripe density,
there will be a transition to a globally superconducting state 
below a finite transition temperature. On the basis of qualitative arguments, 
discussed in the text, we have sketched a boundary of the superconducting 
phase, indicated by  the shaded region. Depending on microscopic details the 
positions  of the quantum critical points $C_1$ and
$C_2$ could be interchanged. Distinctions between various 
possible commensurate and incommensurate stripe crystalline and smectic 
phases are not indicated in the figure. Similarly, all forms of spin order 
are neglected in the interest of simplicity.}
\end{figure}

With this physics in mind, we have sketched a qualitative phase diagram,
shown in Fig.~\ref{fig1d3}, which provides a physical picture of the 
consequences of melting a stripe ordered phase. As a function of increasing 
quantum and thermal transverse stripe fluctuations one expects the insulating 
electronic crystal, which exists at low temperatures and small
$\hbar$, to evolve eventually into an isotropic disordered phase. 
At zero temperature this melting occurs in a sequence of quantum 
transitions\cite{kfe}. 
The first is a first order transition into a smectic phase, then by 
dislocation unbinding a continuous transition leads to a nematic phase 
that eventually evolves (by a transition that can be continuous in $D=2$, 
but is first order in a cubic system) into the isotropic phase. 
Similar transitions exist at finite temperature as indicated in 
Fig.~\ref{fig1d3}. 

We have also sketched a superconducting phase boundary in the same figure. 
Provided that there is a spin gap on each stripe, and that the charge 
\marginpar{\em Superconducting electronic liquid crystals}
Luttinger exponent $K_c>1/2$, then (as discussed in Section \ref{1D}) there 
is a divergent superconducting susceptibility on an isolated stripe. 
In this case, the superconducting $T_c$ is determined by the Josephson 
coupling between stripes. Since, as discussed in Section~\ref{smectic}, the mean
Josephson coupling increases with increasing stripe fluctuations, 
$T_c$ also rises with increasing $\hbar$ throughout the smectic phase. 
While there is currently no well developed theory of the superconducting 
properties of the nematic phase,\footnote{Some very promising recent progress 
toward developing a microscopic theory of the electron nematic phase has been 
reported in Refs.~\citen{vadim} and~\citen{keenematic}.} 
to the extent that we can think of the nematic as being locally smectic,
it is reasonable to expect a continued increase in $T_c$ across much, 
or all of the nematic phase, as shown in the figure.
However, as the stripes lose their local integrity toward the 
transition to the isotropic phase we expect, assuming that stripes 
are essential to the mechanism of pairing, that $T_c$ will decrease, as shown.

The study of electronic liquid crystalline phases is in its infancy.  
Increasingly unambiguous experimental evidence of the existence of 
nematic phases has been  recently reported in quantum Hall systems 
\cite{eisenstein,qhnematic,foglerliqcryst,eisenstein2,eisenstein1999} 
in addition 
to the preliminary evidence of such phases in highly underdoped 
cuprates discussed above.   
Other more exotic electronic  liquid crystalline phases are being studied 
theoretically. This is a very  promising area for obtaining precise answers 
to well posed questions that may yield critical information concerning the 
important mesoscale physics of the high temperature superconductors.

\subsection{Our view of the phase diagram---Reprise}

Since the motion of dilute holes in a doped antiferromagnet is frustrated,
\marginpar{\em It's all about kinetic energy.}
the minimization of their kinetic energy is 
a complicated, multistage process.
We have argued that this is 
accomplished in three stages:  
(a)  the formation of static or dynamical charge inhomogeneity (stripes) at
$T^{*}_{stripe}$, 
(b) the creation of local spin pairs at $T^{*}_{pair}$, 
which creates a spin gap, and 
(c)  the establishment of a phase-coherent 
superconducting state at $T_c$.
The zero point kinetic energy is lowered along a stripe in the first stage, 
and perpendicular to the stripe in the second and third stages.  
Steps (a), (b), and (c) above are clearcut only if the 
energy scales are well separated, that is, 
if $T^{*}_{stripe} >> T^{*}_{pair} >> T_{c}$.  
On the underdoped side at least, if we identify $T^{*}_{stripe}$ 
and $T^{*}_{pair}$ with the appropriate observed pseudogap
phenomena (see Section~\ref{sec:pseudogap}) there is a substantial 
(if not enormous) separation of these temperature scales.  

\subsubsection{Pseudogap scales}
At high temperatures, the system must be disordered.
As temperature is lowered, the antiferromagnet ejects holes, and 
charge stripe correlations develop. This may be either a phase transition
or a crossover. We have called this temperature $T^{*}_{stripe}$ 
in Fig.~\ref{fig:ourview}. Even if it is a phase transition, 
for instance a transition to a stripe nematic state, local order may 
develop above the ordering temperature, and probes on various time scales 
may yield different answers for $T^{*}_{stripe}$. As the antiferromagnet 
ejects holes, local antiferromagnet correlations are allowed to develop.  
Probes bearing on this temperature include the Knight shift, NQR, and 
diffraction. At a lower temperature, through communication with the locally 
antiferromagnetic environment, a spin gap develops on stripes.  
We identify this spin gap with the pairing gap, and have labeled this 
temperature (which is always a crossover) $T^{*}_{pair}$.  
Probes bearing on this temperature measure the single particle gap,
and include ARPES, tunnelling, and NMR.

\subsubsection{Dimensional crossovers}
Looking at this evolution from a broader perspective, 
\marginpar{\em Dimensional crossovers are a necessary consequence of stripe 
physics.}
there are many consequences that can be
understood based entirely on the notion that the effective dimensionality 
of the coherent electronic motion is temperature dependent. At high 
temperatures, before local stripe order occurs, the electronic motion 
is largely incoherent---{\it i.e} the physics is entirely local. 
Below $T^*_{stripe}$, the motion crosses over from quasi 0D to 
quasi 1D behavior.\footnote{It is intuitively clear that kinetic 
energy driven stripe formation 
should lead to increased hole mobility, as is observed, but how the famous 
$T$-linear resistivity can emerges from local quasi-0D physics is not yet 
clear.  See, however, Refs.~\citen{EK1,orbitalK,kelosalamos,newdhl}.}
Here, significant
$\vec k$ space structure of various response  functions is expected, and 
there may well emerge a degree of coherence and  possibly pseudogaps, 
but the electron is not an elementary excitation, 
so broad spectral functions and non-Fermi liquid behavior should be the rule. 
Then, at a still lower temperature, a 1D to 3D crossover occurs as coherent 
electronic motion between stripes becomes possible. At this point coherent 
quasiparticles come to dominate the single particle spectrum, and more 
familiar metallic and/or superconducting physics will emerge. If
the spin gap is larger than this crossover temperature 
(as it presumably is in underdoped materials), then this 
crossover occurs in the neighborhood of $T_c$.  However, if the
spin gap is small, then the dimensional crossover will likely occur 
at temperatures well above $T_c$, and $T_c$ itself will have a more 
nearly BCS character, as discussed in Section \ref{dimensional}---this 
seems to be crudely what happens in the overdoped materials
\cite{johnsonoverdoped}. Since once there are well developed quasiparticles, 
there is every reason to expect them to be able to move coherently between 
planes, there is actually no substantial region of quasi 2D behavior 
expected.  Although it may be hard, without a macroscopically oriented 
stripe array, to study the dimensional crossover by measuring in-plane 
response functions, the dimensional crossover can  be studied by 
comparing in-plane to out-of plane behavior.\footnote{Much of
the successful phenomenology of dimensional crossover developed 
in conjunction with the interlayer pairing mechanism of superconductivity 
\cite{interlayer} is explained in this way in the context of a stripe 
theory.} 

\subsubsection{The cuprates as quasi-1D superconductors}
When $T^{*}_{stripe} >> T^{*}_{pair} >> T_{c}$, 
the model of a  quasi-one dimensional superconductor introduced in 
Section~\ref{dimensional} is applicable in the entire temperature 
range below $T^{*}_{stripe}$.
The application of these results to the overdoped side is suspect,
since that is where all of these energy scales appear to crash into 
each other. 

The temperature dependence of the spectral response of 
\marginpar{\em ARPES and stripes}
a quasi-one dimensional superconductor may be described as follows: 
At temperatures high compared to both the Josephson coupling and the spin gap,
the system behaves as a collection of independent (gapless) Luttinger liquids.
Spin-charge separation holds, so that an added hole dissolves into a spin part
and a charge part. Consequently the spectral response exhibits broad EDC's 
and sharp MDC's.\footnote{See Section~\ref{1D} for a description of EDC's 
and MDC's.} In the intermediate temperature regime (below the spin gap), 
spin-charge separation still holds, and the ARPES response still exhibits 
fractionalized spectra, but with a pseudogap.
In the low temperature phase, Josephson coupling between stripes confines 
spin and charge excitations, restoring the electron as an elementary
excitation, and a sharp coherent peak emerges from the incoherent background, 
with weight proportional to the coupling between stripes.

There is a wealth of ARPES data on BSCCO, a material which lends itself 
more to surface probes than to diffraction. However, as mentioned previously, the
presently available evidence of stripes in this material is compelling, but not
definitive, so it requires a leap of faith to interpret the ARPES data in terms of
stripes.  The best evidence of stripes comes from STM data which is suggestive of 
local stripe correlations
\cite{hoffman,aharon}. Since STM observes a static  modulation, any stripes observed
in STM can certainly be considered static as far as ARPES is concerned.
\footnote {Unfortunately,there is currently little
direct experimental information  concerning the temperature dependence of the stripe
order in BSCCO,  although what neutron scattering evidence does exist\cite{mookbscco}, 
suggests that substantial 
stripe correlations survive to temperatures well above the superconducting 
$T_c$.} As long as the stripes have integrity over a length scale
at least as large as $\xi_s = v_s/\Delta_s$, it is possible for the 
stripes to support superconducting pairing through the spin gap proximity 
effect.

\marginpar{\em ARPES spectra from the antinodal region resemble a quasi-1D
superconductor.} At any rate, many features of the ARPES spectra, especially those
for $\vec k$ in the antinodal region of the Brillouin zone (near $(\pi,0)$) in BSCCO
are unlike anything seen in a conventional metal, and highly reminiscent of a quasi-1D
superconductor. Above
$T_c$, ARPES spectra reveal sharp MDC's and broad EDC's. We take this\cite{frac} as
evidence of electron  fractionalization.   Below $T_c$, 
a well defined quasiparticle peak emerges\cite{fedorov}, 
whose features are strikingly similar to those derived in this model.
The quasiparticle peak is nearly dispersionless along the $(0,0)$ to $(\pi,0)$
direction,  and within experimental bounds its energy and lifetime are temperature 
independent. The only strongly temperature dependent part of the spectrum 
is the intensity associated with the superconducting peak. The temperature
dependence of the intensity is consistent with its being proportional
to a fractional power of the local condensate fraction or the superfluid density.
Similar behavior has been measured now in an untwinned 
single crystal of YBCO\cite{ybcoanis} as well.

The most dramatic
\marginpar{\em Stripes and superconductivity
involve the same regions of $k$-space}
signatures of superconducting phenomena in ARPES experiments, 
both the development of the gap {\it and} the striking onset of the 
superconducting peak with phase coherence, occur in the {\it same} 
regions of $k$-space most associated with stripes:  Specifically,
an array  of ``horizontal'' charge stripes embedded in a locally
antiferromagnetic environment\cite{mats,marku,zacherhanke,hankearpes} 
has most of its low energy spectral weight concentrated near
the $(\pi,0)$ regions of $k$-space. Similarly,
the strongest gap develops in the $(\pi,0)$ regions,
and in both BSCCO and YBCO, 
the only dramatic change in the ARPES response upon entering the
superconducting state is the coherent peak seen in these same regions.  

The ARPES spectrum from the nodal region ($\vec k$ near $(\pi/2,\pi/2)$) 
is less obviously one dimensional in character, although  
the nodal spectrum is certainly {\em consistent} with the existence of 
stripes, as has been demonstrated in several model
calculations\cite{marku,sachdev33,sachdevnodal,mats,zacherhanke}.  
However, to a large extent, the spectrum in the nodal region is 
insensitive to stripe correlations.\cite{sachdevnodal}
Nodal quasiparticles are certainly important for the low temperature 
properties of the superconducting state.  Moreover, there is indirect evidence 
that they dominate the in-plane transport above $T_c$.   
But the fact that the ARPES spectrum in the nodal direction does not change\cite{Valla2} in any dramatic 
fashion from above to below $T_c$, as one would have deduced even from the
simplest BCS considerations, suggests that they do not play a direct role in the mechanism of
superconductivity.  This observation, however, must not be accepted unconditionally.
 There is an apparent  contradiction between the smooth evolution of the spectral
function observed in ARPES and  the evolution inferred from  macroscopic transport
experiments\cite{onglifetime,bonnlifetime};  the latter suggest that a
catastrophic change in the nodal quasiparticle lifetime occurs in the immediate
neighborhood of $T_c$.

\subsubsection{Inherent competition}
\label{sec:inherent}
Finally, it should be made clear that a stripes based mechanism of high 
temperature superconductivity {\em predicts} competition between stripes and 
superconductivity:
\marginpar{\em  We, too, think stripes compete with superconductivity.}
static stripes may be good for pairing, but are certainly bad for the 
Josephson coupling (superfluid stiffness) between stripes.
On the other hand, fluctuating stripes produce
better Josephson coupling, but weaker pairing.  
The dependence of the gap on stripe fluctuations finds its
origin in the spin gap proximity effect, where
the development of the spin gap
hinges on the one dimensionality of the
electronic degrees of freedom\cite{spin gap},
whereas stripe fluctuations cause the system to be
more two dimensional. 
In addition, as described in Section~\ref{smectic}, 
stripe fluctuations work against $2k_F$ CDW order along
a stripe, but strengthen the Josephson coupling.

This is consistent with the
empirical phase diagram: on the underdoped side there is a large gap, 
small superfluid stiffness, small transition temperature, 
and static stripes have been observed. With increasing
doping, stripes fluctuate more, reducing the pairing gap, but increasing the
Josephson coupling between stripes. This is a specific example
of the doping dependent crossover scenario proposed in
Refs.~\citen{nature,crossovers}, in which underdoped cuprates have a strong 
pairing scale but weak phase stiffness and $T_c$ is determined
more or less by $T_{\theta}$,  whereas the overdoped cuprates
have a strong phase stiffness but weak pairing scale 
and $T_c$ is more closely associated with $T^{*}_{pair}$.  
Optimal doping is a crossover between a dominantly phase ordering transition
and a dominantly pairing transition.

\subsection{Some open questions}

As has been stressed by many authors, the cuprate superconductors are 
exceedingly complex
\marginpar{\em Concerning negative results:
``Accentuate the positive.'' }
systems. Crisp theoretical statements can be made concerning the 
behavior of simplified models of these systems, but it is probably
ultimately impossible to make clean predictions about whether the 
results will actually be found in any given material. 
We are therefore reliant on experiment to establish 
certain basic empirical {\em facts}. In this subsection, we
will discuss some of the fundamental issues of fact that are 
pertinent to the stripe scenario presented above, and make a few 
comments about the present state of knowledge concerning them.  
A word of caution is in order before we begin: 
positive results have clearer implications than negative results.  
Especially in these complicated materials, there can be many reasons 
to fail to see an effect.

\subsubsection{Are stripes universal in the cuprate superconductors?}

If stripes are not, in some sense, universal in the high temperature 
superconductors, then they cannot be, in any sense, essential to the 
mechanism of high temperature superconductivity. So an important 
experimental issue is whether stripes are universal in the cuprate 
superconductors.

The evidence from neutron scattering is discussed in Section \ref{1241}:
Incommensurate (IC) spin peaks (whether elastic or inelastic) have been detected
throughout the doping range of superconductivity in the lanthanum compounds.
In YBCO, IC spin peaks are seen with inelastic scattering, but
it is presently unclear how much of that scattering
intensity should be associated with stripe fluctuations,
and how much should be associated with the 
``resonance peak''.  
Neutron scattering has produced some evidence\cite{mookbscco} of IC spin
peaks in BSCCO, but this result is controversial\cite{bscco}. 
No such peaks have been reported in TlBaCaCuO or HgBaCaCuO, 
although little or no neutron scattering has yet been done on
crystals of these materials.

CDW order turns out to be much harder to observe, even when 
we know it is there. Charge stripe order has only been observed directly 
in {\LNSCO}\cite{ichikawa} and very underdoped {\YBCO}\cite{mookchg}, 
although the general argument presented in Section \ref{coulfrust} implies that it must
occur wherever  spin stripe order exists. 
Given 
the difficulty in observing the charge order where we know it exists, 
we consider an important open question to be: Where
does charge stripe order exist in the general 
phase diagram of the cuprate superconductors?  

As mentioned before,  STM experiments point to local charge stripes in
BSCCO, both with\cite{hoffman} and without\cite{aharon} a magnetic field.
But there is nowhere near enough systematic data to 
know whether charge stripes are ubiquitous as a function of 
doping and in all the superconducting cuprates, how pronounced it is, 
and over what range of temperatures significant stripe correlations exist, 
even where we know they exist at low temperatures. Perhaps, in the future, 
this issue can be addressed further with STM,  or even with
ARPES or new and improved X-ray scattering experiments.

\subsubsection{Are stripes an
unimportant low temperature complication?}

There is a general tendency for increasingly subtle forms of order to 
appear as systems are cooled---involving residual low energy degrees of 
freedom that remain after the correlations that are the central features of 
the physics have developed.  (A classic example of this is 
transitions involving ordering of the nuclear moments at ultra-low 
temperatures in a metal.)  While such forms of order are fascinating 
in their own right, one would not, typically, view them as important 
aspects of the basic materials physics of the studied system. There 
is a body of thought that holds that the various forms of stripe order 
that have been observed are in this class of phenomena---interesting 
side shows, but not the main event. It is also true that 
actual, static stripe order has only been observed under rather 
restrictive conditions---mostly in highly underdoped materials or 
materials with significantly depressed superconducting $T_{c}$'s, and 
at temperatures less than or of order the optimal superconducting $T_{c}$.

To be central to the physics of high temperature superconductivity, 
charge stripes must occur at high enough energies and temperatures 
that they are relevant to zeroth order.  Specifically, we want to 
look for evidence that local stripes persist up to temperatures which 
are greater than or equal to $T_{c}$.  If stripes are universal, 
then there must be a characteristic crossover scale below which 
significant stripe correlations emerge---clearly, at high enough 
temperature, no significant self-organization is possible.  
Undoubtedly, there is a high energy scale associated with one or more 
pseudogap crossovers in many underdoped materials---can we associate 
some of this crossover with the scale at which local stripe 
correlations become significant?  If so, then manifestly stripes are a 
central player in the drama.  If not, and if no still higher energy 
scale can be identified  at which stripe physics begins, it would 
become increasingly difficult to envisage a starring role for stripes 
in the physics of the cuprates.

This issue has not been unambiguously resolved. There is 
substantial (yet not definitive) evidence that local stripe order 
persists to rather high temperatures. Evidence of local stripe order 
from observed\cite{basovando} infrared active phonon modes has been 
seen to persist to at least 300K in highly underdoped {\LSCO}.  
Phonon anomalies, which have been tentatively associated with stripes, 
have been observed in neutron scattering experiments in slightly underdoped 
{\YBCO} up to comparable temperatures \cite{mook}. Still more indirect 
evidence also abounds. This is a key question, and much more work is 
necessary to resolve it.

\subsubsection{Are the length and time scales reasonable?}
As emphasized above, to understand the mechanism of high temperature 
superconductivity, we are primarily concerned with mesoscopic physics, 
on length scales a few times the superconducting coherence length and 
time scales a few times $\hbar/\Delta_{0}$. So the real question we 
want to address is: Does stripe order exist on these length and time 
scales? Given that it is so difficult to determine where long range 
charge stripe order occurs, it is clearly still more complicated to 
determine where substantial stripey short range order occurs, or even 
precisely how much short range order is sufficient.

\subsubsection{Are stripes conducting or insulating?}

The earliest theoretical studies which predicted stripes 
as a general feature of doped antiferromagnets envisaged insulating 
stripes\cite{zaanengunnarsson,schulzstripes,machida89}. These stripes are
conceptually close  relatives of conventional CDW's in that they are obtained as a
Fermi surface  instability due to near perfect nesting of the Fermi surface.  Such 
stripes have no low lying fermionic excitations.  
This perspective has led to an interesting theory of superconductivity
which relies on stripe defects for charge transport \cite{zaanenSC}.

The strongest evidence that charge stripes are incompressible,
and therefore insulating, 
comes from plotting the magnetic 
incommensurability against the doping concentration. If this 
relationship is strictly linear, 
it implies that the concentration of holes on a 
stripe does not change, but rather the only effect of further doping 
is to change the concentration of stripes in a plane, bringing the 
stripes closer together. 
The data for LSCO are close to linear in the range $.024 \leq x \leq .12$,
despite the change in orientation from diagonal to vertical at 
$x=.05$,\cite{birgeneausmallx,birgeneautinyx} but the small deviation from
linearity below $x=.06$ does exceed the error bars.
At present, the data leave open 
the possibility that the relationship is not strictly linear, and is also consistent
with compressible (metallic)  stripes throughout the doping range where they are
observed.   (See, {\it e.g.}, Fig. 7 of Ref.~\citen{yamada}.)

Most of the other experiments we have mentioned support the notion that 
the stripes are intrinsically metallic. Of course, the observed coexistence 
of static stripe order and superconductivity is a strong indicator of this, 
as presumably 
it would be hard to attribute long distance charge transport
to stripe motion.\footnote{One could envisage stripe defect motion which
transports charge perpendicular to the stripes, \cite{zaanenSC} but certainly 
the effective number of carriers due to this effect must be small.}
The situation is most dramatic in nonsuperconducting LSCO with 
$0.02 < x < 0.05$, where the stripes are 
ordered\cite{birgeneautinyx,birgeneausmallx}, 
and far enough separated that the intrinsic properties of an 
individual stripe must surely determine the electronic structure---the 
mean stripe spacing\cite{birgeneautinyx} grows to be as large as 350$\AA$ 
or so for $x=0.02$.\footnote{This is equivalent to 64 (orthorhombic)
lattice constants, $b^{*}_{ortho}=5.41$\AA.\cite{lsco1d}}
These materials exhibit\cite{ando1,ando1d} a
metallic (linear in
$T$) temperature dependence of the resistivity down to moderate temperatures. 
More remarkably, as shown\cite{ando1} by Ando {\it et al.}, although the 
magnitude of the resistance is large compared to the quantum of resistance 
at all temperatures, when interpreted in terms of a model in which the 
conduction occurs along dilute, metallic stripes, the inferred electron
mobility within a stripe is nearly the same as that observed in optimally 
doped LSCO!  

\subsubsection{Are stripes good or bad for superconductivity?}

Striking empirical evidence which suggests that stripes and superconductivity
\marginpar{\em The Uemura plot and the Yamada plot may be about the same physics.}
are related comes from the Yamada plot \cite{yamada}, 
which reports $T_c$ {\it vs.} the incommensurability seen in neutron 
scattering, {\it i.e.} the inverse spacing between stripes.  
First noted in LSCO, the relationship is remarkably linear for
the underdoped region of the lanthanum compounds\cite{yamada}.
For far separated stripes, the transition
temperature is depressed. As the stripes move closer together, and the
Josephson coupling between them increases, $T_c$ increases. In addition, the 
similarities between the Yamada plot and the Uemura plot\cite{uemura}, which 
shows a linear relationship between $T_c$ and the superfluid density, 
indirectly imply that the Josephson coupling between stripes 
plays an important role in determining the macroscopic superfluid density.

It has been argued that since stripes compete with superconductivity, 
they cannot be involved in the mechanism of superconductivity\cite{baskaran}.  
(We would point out that, at the very least, such competition
must imply that stripes and superconductivity are strongly connected.)
The empirics are presently unclear on the issue. There is 
some  evidence 
that static stripes compete with 
superconductivity, whereas fluctuating stripes enhance it.  
In instances where stripes are pinned, $T_c$ is generally suppressed, 
such as with Nd doping, Zn doping, or at the $1/8$ anomaly. 
An exception to this trend occurs in the LCO family, which exhibits 
its highest $T_c$ for a static stripe ordered material. Recently, Ichikawa 
{\it et al.}\cite{ichikawa} have argued that it is spin stripe order, 
rather than charge stripe order, which competes with superconductivity.  
Whatever the details, the gross trend in materials other than LCO seems 
to be that the highest transition temperatures are achieved for dopings 
that presumably do not support actual (static) stripe order.  
It is also worth noting that in LSCO\cite{aeppli} and 
YBCO\cite{daimook}, neutron scattering shows a gap developing 
in the incommensurate magnetic fluctuations at $T_c$, perhaps 
indicating that superconductivity  favors fluctuating stripes.  

On the other hand, $T_c$ is a nonmonotonic function of $x$, and pretty 
clearly determined by the lesser of two distinct energy scales.  
But the superconducting gap, as deduced from low temperature tunnelling 
or ARPES experiments deep in the superconducting state, is a monotonically
decreasing function of $x$. It is generally believed that stripe correlations 
are similarly strongest when $x$ is small and vanish with sufficient 
overdoping, although in truth the direct experimental evidence for this 
intuitively obvious statement is not strong. 
Thus, there is at least a generally positive correlation between the 
degree of local stripe order and the most obvious scale characterizing 
pairing. This leads us to our next question:

\subsubsection{Do stripes produce pairing?}
It is well known that the physics of an antiferromagnet is 
kinetic energy driven, and phase coherence must be kinetic energy driven 
when $T_{pair} >> T_c$, since spatial fluctuations of the phase drive pair 
currents. But can pair formation be kinetic energy driven?
In particular, do stripes produce pairing?  
As reviewed in Section~\ref{numerical}, 
numerical studies do find pairing in ``fat'' 1D systems.

However,
there is no experiment we can point to that proves the pairing is either
\marginpar{\em No smoking gun}
kinetic energy driven\footnote{The brilliantly conceived high precision
measurements of the optical conductivity of van der Marel and 
collaborators\cite{marel}, and more recently by Bontemps and 
collaborators\cite{nicolkinetic}, are highly suggestive in this regard.  
In optimally doped BSCCO, they observe a strongly temperature dependent 
change in the optical spectral weight integrated  up to frequencies two
orders of magnitude greater than $T_c$---if interpreted in terms of the 
single band sum rule, this observation implies a decrease
in the kinetic energy upon entering the superconducting state 
of a magnitude comparable to reasonable estimates of the condensation 
energy. This is very striking, since in a BCS superconductor, the kinetic 
energy would {\em increase} by precisely this amount. However, neither the 
single band sum rule, nor the notion of a condensation energy are 
unambiguously applicable in the present problem. This is {\em the best} 
existing evidence that the mechanism of superconductivity is kinetic energy 
driven, but it is not yet evidence that would stand up in court.}  
or due to stripes. Nor is it clear what such an experiment would be. 
There are ways to falsify the conjecture that stripes produce pairing, 
such as a demonstration that stripes are not in some sense ubiquitous in 
the cuprates, or a demonstration that pairing generally precedes local charge
stripe formation as the temperature is lowered. 
We have discussed many 
predictions which find some support in experiments, such as the fact that 
static stripes are good for pairing but bad for phase coherence, and vice
versa, and the systematics of the superconducting  coherence peak.  
But these interpretations are not necessarily unique. Much of the 
phenomenology is {\em consistent} with a spin gap proximity effect mechanism of
pairing,  but we see no smoking gun.  

\subsubsection{Do stripes really make the electronic structure quasi-1D?}

Does the existence of stripes provide a sufficient excuse to treat the 
cuprates as self-organized quasi-1D conductors?  If so, then we can 
apply many of the insights we have obtained directly, and without apology 
to the interpretation of experiment.  As has been
discussed in previous sections, and in considerably more detail in other
places\cite{dimxover,frac,spin gap,pnas,marku,mats,hankearpes}, there are
many
striking  experiments in the cuprates that can be simply and naturally
understood in 
this way. 
But do they actually affect the electronic structure so 
profoundly as to render it quasi-1D?

The most direct evidence comes from the ARPES results of Shen and 
collaborators\cite{lnsco} on the stripe ordered material, \LNSCO. 
These experiments reveal a remarkable confinement of the majority of the 
electronic spectral weight inside a dramatically
1D Fermi surface. This experiment probes fairly high energy excitations, 
and so demonstrates a profound effect of an ordered stripe array on all 
aspects of the electronic structure. More generally, studies have shown
\cite{dimxover,frac,mats,hankearpes} that many of the most striking
features of
the ARPES spectra of the cuprates are readily rationalized on the basis 
of an assumed, locally quasi-1D electronic structure.

Transport measurements are macroscopic, so even if locally the electronic
structure is strongly quasi-1D, the effects of stripe meandering, 
domain formation, and disorder will always produce a substantially reduced
effective anisotropy at long distances.  From this perspective, the
order 1, strongly temperature dependent transport anisotropies observed by
Ando and collaborators\cite{ando1d} in 
\marginpar{\em Macroscopic anisotropy}
LSCO and YBCO provide tangible evidence of a strong susceptibility of 
the electron liquid in the copper oxide planes to develop anisotropies in 
tensor response functions.  
Less direct, but even more dramatic
evidence that stripes make the electron dynamics quasi-1D has been
adduced from Hall effect measurements on the stripe ordered
material,
\LNSCO, by Noda {\it et al.}\cite{noda11} 
They have observed that the
Hall coefficient, $R_H$, which is relatively weakly temperature dependent
above the stripe ordering transition temperature, $T_{co}$, drops
dramatically for $T < T_{co}$, such that $R_H\to 0$ as $T\to 0$ for doped
hole concentration, $x
\le 1/8$, and $R_H$ tends to a reduced but finite value for $x > 1/8$.   This
observation was initially interpreted\cite{noda11} as evidence that
ordered stripes prevent coherent transverse motion of electrons within the
copper oxide plane;  this interpretation  was later shown to be
not entirely correct\cite{prl}, although the basic conclusion 
that the stripes render the electron dynamics quasi-one dimensional
is probably sound.  Further evidence that stripe formation inhibits
transverse electronic motion is strongly suggested by the observed
suppression of c-axis coherent charge motion in the stripe ordered state
of the same materials\cite{tajimaprl}.

However, it would be very desirable to develop new strategies to directly 
address this issue.  For instance, a defect, such as a twin bounary, could
purposely be introduced to locally aline the stripe orientation, and the
induced electronic anisotropy then be detected with STM.

\subsubsection{What about overdoping?}
On the underdoped side of the phase diagram of the cuprates, the energy 
scales of $T^{*}_{stripe}$, $T^{*}_{pair}$, and $T_{c}$ are generally 
sufficiently separated to make the application of many of these ideas 
plausible. Yet on the overdoped side, the energy scales seem to come 
crashing into each other, depressing $T_c$.  Furthermore, on the overdoped 
side, we have $T_{\theta} > T_{pair}$, in violation of a common assumption 
we have made throughout this article. The very existence of stripes on the 
overdoped side is questionable.
The Uemura and Yamada plot are not 
satisfied there. If there are no stripes, and yet there is superconductivity, 
this does not bode well for a stripes based mechanism. Indeed, it is easier 
to believe that a mean field like solution is crudely applicable on the 
overdoped side, where $T_c$ is closer to $T_{pair}$ than it is to $T_{\theta}$.

One possibility is that the superconducting state far on the overdoped side 
(especially, where $T_c$ is low and the normal state ARPES spectrum begins 
to look more Fermi liquid-like) is best approached in terms of a Fermi 
surface instability and a BCS-Eliashberg mechanism, while on the underdoped 
side it is best viewed from a stripes perspective. In keeping with the 
multiscale approach advocated above, it may be no simple matter to unify these
approaches in a smooth way. 

However, there is an attractive possibility that is worth
mentioning here. As we have mentioned, in a stripe liquid, 
so long as the characteristic stripe fluctuations frequency, 
$\hbar\bar\omega$, is small compared to the superconducting gap
scale, the stripes can be treated as quasi-static for the purposes 
of understanding the mechanism of pairing. Conversely, 
when $\hbar\bar\omega \gg \Delta_0$, the stripe fluctuations
can be integrated out to yield an effectively homogeneous system 
with an induced interaction between electrons. Indeed, it has previously 
been proposed\cite{castro4} that stripe fluctuations themselves are a
candidate for the ``glue'' that mediates an effective
attraction between electrons. It is easy to imagine that 
$\hbar\bar\omega/\Delta_0$ is a strongly increasing function of $x$.  
A sort of unification of the two limits could be achieved
if stripe fluctuations play the role of the intermediate boson 
which mediates the pairing in highly overdoped materials, 
while in underdoped and optimally doped materials 
the system can be broken up into quasi-1D ladders, 
which exhibit the spin gap proximity effect.

\subsubsection{How large is the regime of substantial fluctuation 
superconductivity?}

This important question is fundamentally ill-defined.  
It is important, because its answer determines the point of view we take 
with regard to a number of key experiments.  
But it is ill-defined in the following sense: in the neighborhood of 
any phase transition, there is a region above $T_c$ where substantial 
local order exists, but how broad the fluctuation region is
said to be depends on exactly how ``substantial'' is defined, or measured. 
There has been an enormous amount written on this subject already, so we 
will just make a few brief observations.

Because in one dimension, phase fluctuations always reduce the 
superconducting $T_c$ to zero, in a quasi-one dimensional 
superconductor ({\it i.e.} in the limit of large anisotropy), there
is necessarily a parametrically large fluctuation regime between the mean field 
transition temperature and the actual ordering temperature.
   
The finite frequency superfluid density measured in BSCCO\cite{corson}
with $T_{c}=74K$ shows a local superfluid density 
persists up to at least
$90K$, indicative of fluctuation superconductivity in that regime. 
Both microwave absorption\cite{bonnandhardy} and thermal
expansivity measurements\cite{meingast} on optimally doped YBCO detect 
significant critical superconducting fluctuations within $\pm 10$\% of $T_c$. 
All of these experiments are well accounted for in terms of the critical 
properties of a phase-only (XY) model, and are not
well described as Gaussian fluctuations of a Landau-Ginzberg theory. 
Thus, there is no question that there is a well defined magnitude of
the order parameter, and substantial local superconducting order for
at least 10K to 20K above $T_c$, and a correspondingly broad range of
substantial phase fluctuations below $T_c$.

There are, however, some experiments that suggest that substantial 
local pairing persists in a much broader range of temperatures.\cite{diapseudo} 
Nernst measurements\cite{ong,ongxu} have detected vortex-like
signals up to $100K$ above $T_c$ in LSCO and YBCO, {\it i.e.} 
to temperatures up to 5 times $T_c$! In both cases, however, the final 
word has yet to be spoken concerning the proper
interpretation of these intriguing experiments.\cite{shivajihuse}
ARPES\cite{shenscience,ding} and tunnelling\cite{renner} studies 
find that the gap in BSCCO persists up to $100K$ above $T_c$, {\it i.e.} 
to temperatures of order two or more times $T_c$.  

Finally, there are preliminary indications that there may be substantial
local superconducting order in severely underdoped materials in which no
macroscopic indications of superconductivity appear at any temperature.
Presumably, if this is the case, long range phase coherence has been
suppressed in these materials  by quantum phase
fluctuations\cite{doniach} which proliferate due to the small bare
superfluid stiffness and the poor screening of the Coulomb potential.
In particular, experiments on films of severely underdoped
nonsuperconducting YBCO have revealed that a metastable superconducting
state can be induced by photodoping.  This has permitted the patterning
of small scale superconducting structures, in which it has been
shown\cite{drew} that substantial Josephson coupling between two
superconducting regions can persist even when they are separated by as
much as $1000\AA$.
This ``anomalous proximity
effect'' implies that there is a substantial pair field susceptibility
in this nonsuperconducting material.

\subsubsection{What about phonons?}

Phonons are clearly strongly coupled to the electron 
gas 
in the cuprates.
\marginpar{\em This is a good question.}
Certainly, when there is charge order of any sort, it is unavoidable that it 
induces (or is induced by) lattice distortions. Manifestly, phonons will
enhance any electronic tendency to phase separation or stripe 
formation\cite{castro3}. They will also tend to make any stripes ``heavy,'' 
and so suppress quantum fluctuations---likely, this leads to a depression of 
superconductivity. There is a dramatic isotope effect anomaly 
seen\cite{crawford} in some materials when the doped hole density, 
$x=1/8$;  presumably, this is related to just such a phonon-induced pinning 
of the stripe order\cite{tranqnature}. Recently, there has been considerable
controversy generated by the suggestion\cite{lanzara} that certain features 
of the ARPES spectrum of a wide class of cuprates reflect the effects of 
strong electron-phonon coupling. This is clearly an area in which much work 
remains to be done. In our opinion, other than in 1D, the effects
of electron-phonon coupling in a strongly correlated electron gas 
is an entirely unsolved problem. 

\subsubsection{What are the effects of quenched disorder?}

We have said essentially nothing about the effects of quenched 
disorder on the materials of interest, although the materials 
are complicated, and disorder is always present. There are 
even some theories which consider the disorder to be essential to the 
mechanism of high temperature superconductivity \cite{phillips}. A 
strong case against this proposition is made by the observation that 
as increasingly well ordered materials are produced, including some 
which are stoichiometric and so do not have any of the intrinsic 
disorder associated with a random alloy, the superconducting 
properties are not fundamentally altered, and that if anything 
$T_{c}$ and the superfluid density both seem to rise very slightly
as disorder is decreased.

However, other properties of the system are manifestly sensitive to 
disorder.  Since disorder couples to spatial symmetry breaking order 
parameters in the same way that a random field couples to a magnetic 
order parameter, it is generally a relevant perturbation. Among 
other things,  this 
means that none of the stripe orders discussed above will ever 
occur as true long range order, and the putative transitions are 
rounded and rendered glassy\cite{rome,zachar5,schmalianglass,hasselmann}. 
So even the supposedly sharp statements 
discussed above are only sharp, in practice, if we can study such 
highly perfect crystals that they approximate the disorder${\to}0$ limit.  
This is a general problem.  Progress has been made in recent years in 
growing more and more perfect single crystals of particular 
stoichiometric superconductors.  Clearly, advances in this area are 
an essential component of the ongoing effort to unravel the physics 
of these materials.

\section*{Acknowledgements}
We would especially like to acknowledge the profound
influence on our understanding of every aspect of the physics discussed in
this paper of discussions with our colleagues and collaborators, John
Tranquada, Vadim Oganesyan, and Eduardo Fradkin. We also want to explicitly
aknowledge the extremely helpful suggestions and critiques we obtained from  
J.~W.~Allen, Y.~Ando, N.~P.~Armitage, A.~Auerbach, D.~A.~Bonn, R.~J.~Birgeneau, 
E.~Dagotto,  A.~H.~Castro Neto, E.~Fradkin, S.~Sachdev, 
D.~J.~Scalapino, S.~L.~Sondhi, J.~M.~Tranquada, and O.~Zachar. 
Finally, we have benefited greatly from discussion of the ideas 
presented herein with more colleagues than we can hope to acknowledge, 
but we feel we must at least acknowledge our intellectual debts to G.~Aeppli, 
J.~W.~Allen, P.~B.~Allen, Y.~Ando, D.~N.~Basov, M.~R.~Beasley, A.~H.~Castro~Neto, 
S.~Chakravarty, E.~Daggatto, P.~C.~Dai, J.~C.~S.~Davis, C.~Di Castro, 
R.~C.~Dynes, H.~Eisaki, A.~Finkelstein, 
M.~P.~A.~Fisher, T.~H.~Geballe, M.~Granath, P.~D.~Johnson, C.~Kallin, 
A.~Kapitulnik, H.-Y.~Kee, Y.-B.~Kim, 
R.~B.~Laughlin, D-H.~Lee, Y.~S.~Lee, K.~A.~Moller, H.~A.~Mook, 
C.~Nayak, Z.~Nussinov, N.~P.~Ong, J.~Orenstein, 
S.~Sachdev, D.~J.~Scalapino, J.~R.~Schrieffer, Z-X.~Shen, S.~L.~Sondhi, 
B.~I.~Spivak, T.~Timusk, T.~Valla, J.~Zaanen, O.~Zachar, S-C.~Zhang, and 
X.~J.~Zhou. 
This work was supported, in part, by NSF grant DMR-0110329 at UCLA,
DOE grant DE-FG03-00ER45798 at UCLA and BNL, and NSF grant DMR-97-12765
and the Office of the Provost at Boston University.

\listofsymbols

\bibliographystyle{unsrt}
\bibliography{htscchapter}

\begin{thebibliography}{100}
\bibitem{ando7}
Preliminary evidence of the existence of nematic order in
  La$_{2-x}$Sr$_x$CuO$_4$ and YBa$_2$Cu$_3$O$_{7-\delta}$ can be found in
  \cite{ando1d}.
\end{thebibliography}

\end{document}